\documentclass{jpp}
\usepackage{amsmath, amssymb, mathabx, mathrsfs, braket, comment, xspace, upgreek, esint, utilde, color, longtable}

\newcommand{\eq}[1]{(\ref{#1})}
\newcommand{\Sec}[1]{section~\ref{#1}}
\newcommand{\Secs}[1]{sections~\ref{#1}}
\newcommand{\App}[1]{appendix~\ref{#1}}
\newcommand{\Tab}[1]{table~\ref{#1}}

\newcommand{\mc}[1]{\mathcal{#1}}
\newcommand{\mcc}[1]{\mathfrak{#1}}
\newcommand{\msf}[1]{\mathsf{#1}}
\newcommand{\mcu}[1]{\mathscr{#1}}
\newcommand{\oper}[1]{\widehat{#1}}

\newcommand{\latin}[1]{\textit{#1}\xspace}
\newcommand{\ie}{i.e.\xspace}
\newcommand{\perse}{\latin{per~se}}
\newcommand{\adhoc}{\latin{ad~hoc}}

\newcommand{\m}[1]{\(\smash{#1}\)}
\newcommand{\mt}[1]{\texorpdfstring{\({#1}\)}{}}

\DeclareMathOperator{\tr}{tr}
\DeclareMathOperator{\re}{re}
\DeclareMathOperator{\im}{im}
\DeclareMathOperator{\pv}{pv}
\DeclareMathOperator{\sgn}{sgn}

\newcommand{\ee}{\mathrm{e}}
\newcommand{\ii}{\mathrm{i}}
\newcommand{\dd}{\mathrm{d}}
\newcommand{\pd}{\partial}
\newcommand{\del}{\nabla}
\newcommand{\herm}{\text{H}} 
\newcommand{\aherm}{\text{A}} 
\newcommand{\const}{\text{const}}
\newcommand{\placeholder}{\sqbullet}

\renewcommand{\vec}[1]{{\boldsymbol{#1}}}
\newcommand{\matr}[1]{{\boldsymbol{#1}}}
\newcommand{\boper}[1]{\oper{\boldsymbol{#1}}}

\newcommand{\favr}[1]{\langle #1 \rangle}
\newcommand{\cb}[1]{(#1)}
\newcommand{\cbb}[1]{\lbrace #1 \rbrace}
\newcommand{\poissonx}[1]{\cbb{#1}_\indexst}

\newcommand{\poissonz}[1]{\cbb{#1}}

\newcommand{\avr}[1]{\overline{#1}}
\newcommand{\osc}[1]{\widetilde{#1}}
\newcommand{\micro}[1]{\utilde{#1}}
\newcommand{\macro}[1]{\underline{#1}}
\newcommand{\macrob}[1]{\braket{#1}}
\newcommand{\env}[1]{{\breve{#1}}}

\newcommand{\wsymb}[1]{\text{symb}\,#1}
\newcommand{\wsymbx}[1]{\text{symb}_\indexst #1}

\newcommand{\woperx}[1]{\text{oper}_\indexst #1}
\newcommand{\wsymbX}[1]{\text{symb}_X #1}
\newcommand{\woperX}[1]{\text{oper}_X #1}
\newcommand{\fourier}[1]{\mathring{#1}}

\newcommand{\vg}{v_{\text{g}}}
\newcommand{\vvg}{\vec{v}_{\text{g}}}
\newcommand{\avg}{\avr{v}_{\text{g}}}
\newcommand{\avvg}{\avr{\vec{v}}_{\text{g}}}
\newcommand{\hilspace}{\mcu{H}}
\newcommand{\hilspacex}{\hilspace_\indexst}
\newcommand{\hilspaceX}{\hilspace_X}

\newcommand{\collision}{\mc{C}}
\newcommand{\entropy}{\sigma}
\newcommand{\invXi}{\smash{\boper{\Xi}}^{-1}}
\newcommand{\Psiosc}{\smash{\osc{\vec{\Psi}}}}
\newcommand{\Psiq}{\micro{\osc{\vec{\Psi}}}}
\newcommand{\Psim}{\macro{\osc{\vec{\Psi}}}}
\newcommand{\matrU}{\matrst{U}}
\newcommand{\matru}{\matr{\mcc{W}}}
\newcommand{\sigmax}{\sigma_{\st{x}}}
\newcommand{\sigmak}{\sigma_{\st{k}}}

\newcommand{\st}[1]{\msf{#1}}
\newcommand{\stn}{{\st{n}}}
\newcommand{\matrst}[1]{{\boldsymbol{\st{#1}}}}
\newcommand{\vecst}[1]{\vec{\st{#1}}}
\newcommand{\operst}[1]{\oper{\st{#1}}}
\newcommand{\boperst}[1]{\boper{\vecst{#1}}}
\newcommand{\indexst}{{\st{x}}}
\newcommand{\metric}{\st{g}}
\newcommand{\mmetric}{\mcc{g}}

\newcommand{\ldx}{\overset{{\scriptscriptstyle \leftarrow}}{\pd}_\vecst{x}}
\newcommand{\ldp}{\overset{{\scriptscriptstyle \leftarrow}}{\pd}_\vecst{k}}
\newcommand{\rdx}{\overset{{\scriptscriptstyle \rightarrow}}{\pd}_\vecst{x}}
\newcommand{\rdp}{\overset{{\scriptscriptstyle \rightarrow}}{\pd}_\vecst{k}}

\newcommand{\ldX}{\overset{{\scriptscriptstyle \leftarrow}}{\pd}_\vec{X}}
\newcommand{\ldP}{\overset{{\scriptscriptstyle \leftarrow}}{\pd}_\vec{K}}
\newcommand{\rdX}{\overset{{\scriptscriptstyle \rightarrow}}{\pd}_\vec{X}}
\newcommand{\rdP}{\overset{{\scriptscriptstyle \rightarrow}}{\pd}_\vec{K}}

\newcommand{\citesup}{\citep[supplementary material]{my:quasiop1}\xspace}

\allowdisplaybreaks
\definecolor{darkred}{rgb}{0.75, 0, 0}

\usepackage[colorlinks=true, linkcolor=darkred, citecolor=darkred, hypertexnames=false]{hyperref}

\shorttitle{Quasilinear theory}
\shortauthor{I. Y. Dodin}
\title{Quasilinear theory for inhomogeneous plasma}
\author{I. Y. Dodin\corresp{\email{idodin@princeton.edu}}}
\affiliation{
\aff{}Princeton Plasma Physics Laboratory, Princeton, NJ 08543, USA
\aff{}Department of Astrophysical Sciences, Princeton University, Princeton, NJ 08544, USA
}

\begin{document}
\bibliographystyle{jpp}
\maketitle

\begin{abstract}
This paper presents quasilinear theory (QLT) for classical plasma interacting with inhomogeneous turbulence. The particle Hamiltonian is kept general; for example, relativistic, electromagnetic, and gravitational effects are subsumed. A Fokker--Planck equation for the dressed `oscillation-center' distribution is derived from the Klimontovich equation and captures quasilinear diffusion, interaction with the background fields, and ponderomotive effects simultaneously. The local diffusion coefficient is manifestly positive-semidefinite. Waves are allowed to be off-shell (\ie not constrained by a dispersion relation), and a collision integral of the Balescu--Lenard type emerges in a form that is not restricted to any particular Hamiltonian. This operator conserves particles, momentum, and energy, and it also satisfies the \m{H}-theorem, as usual. As a spin-off, a general expression for the spectrum of microscopic fluctuations is derived. For on-shell waves, which satisfy a quasilinear wave-kinetic equation, the theory conserves the momentum and energy of the wave--plasma system. The action of nonresonant waves is also conserved, unlike in the standard version of QLT. Dewar's oscillation-center QLT of electrostatic turbulence (1973, Phys. Fluids \textbf{16}, 1102) is proven formally as a particular case and given a concise formulation. Also discussed as examples are relativistic electromagnetic and gravitational interactions, and QLT for gravitational waves is proposed.
\end{abstract}

\setcounter{tocdepth}{2}
\tableofcontents

\section{Introduction}
\label{sec:intro}

\subsection{Background}

Electromagnetic waves are present in plasmas naturally, and they are also launched into plasmas using external antennas, for example, for plasma heating and current drive \citep{book:stix, ref:fisch87, ref:pinsker01}. Nonlinear effects produced by these waves are often modeled within the quasilinear (QL) approximation, meaning that the nonlinearities are retained in the low-frequency (`average') dynamics but neglected in the high-frequency dynamics. Two separate paradigms exist within this approach. 

In the \textit{first paradigm}, commonly known as `the' QL theory (QLT), the focus is made on resonant interactions. Nonresonant particles are considered as a background that is homogeneous in spatial \citep{ref:vedenov61b, ref:drummond62, ref:kennel66, ref:rogister68, ref:rogister69} or generalized coordinates \citep{ref:kaufman72, ref:eriksson94, ref:catto17}; then the oscillating fields can be described in terms of global modes. This approach has the advantage of simplicity, but its applications are limited in that real plasmas are never actually homogeneous in any predefined variables (and, furthermore, tend to exhibit nonlinear instabilities in the presence of intense waves). The `ponderomotive' dynamics determined by the gradients of the wave and plasma parameters is lost in this approach; then, spurious effects can emerge and have to be dealt with \citep{ref:lee18}. 

The \textit{second paradigm} successfully captures the ponderomotive dynamics by introducing effective Hamiltonians for the particle average motion \citep{ref:gaponov58, ref:motz67, ref:cary81, ref:kaufman87, my:itervar}. But as usual in perturbation theory \citep{book:lichtenberg}, those Hamiltonians are by default singular for resonant interactions. Thus, such models have limited reach as well, and remarkable subtleties are still found even in basic QL problems. For example, it is still debated \citep{ref:ochs21, phd:ochs21} to which extent the QL effects that remove resonant particles while capturing their energy \citep{ref:fisch92} also remove charge along with the resonant particles thereby driving plasma rotation \citep{ref:fetterman08}. This state of affairs means, arguably, that a clear comprehensive theory of QL wave--plasma interactions remains to be developed --- a challenge that must be faced.

The first framework that subsumed both resonant and nonresonant interactions in inhomogeneous plasmas was proposed by \citet{ref:dewar73} for electrostatic turbulence in nonmagnetized plasma and is known as `oscillation-center' (OC) QLT. It was later extended by \citet{ref:mcdonald85b} to nonrelativistic magnetized plasma. However, both of these models are partly heuristic and limited in several respects. For example, they are bounded by the limitations of the variational approach used therein, and they separate resonant particles from nonresonant particles somewhat arbitrarily (see also \citep{ref:ye92b}). Both models also assume specific particle Hamiltonians and require that waves be governed by a QL wave-kinetic equation (WKE), \ie be only \textit{weakly} dissipative, or `on-shell'. (Somewhat similar formulations were also proposed, independently and without references to the OC formalism, in \citep{ref:weibel81, ref:yasseen83, ref:yasseen86}.) This means that collisions and microscopic fluctuations are automatically excluded. Attempts to merge QLT and the WKE with theory of plasma collisions were made \citep{ref:rogister68, ref:schlickeiser14, ref:yoon16} but have not yielded a local theory applicable to inhomogeneous plasma. In particular, the existing models rely on global-mode decompositions and treat complex frequencies heuristically. Thus, the challenge stands. 

Related problems are also of interest in the context of gravitostatic interactions \citep{ref:chavanis12, ref:hamilton20, ref:magorrian21}, where inhomogeneity of the background fields cannot be neglected in principle \citep{book:binney}. (To our knowledge, OC QLT analogs have not been considered in this field.) Similar challenges also arise in QLT of dispersive gravitational waves \citep{foot:mygwquasi, my:gwponder}. Hence, one cannot help but wonder whether a specific form of the particle Hamiltonian really matters for developing QLT or it is irrelevant and therefore should not be assumed. Since basic theory of linear waves is independent of Maxwell's equations \citep{book:tracy, my:amc, my:nonloc}, a general QLT might be possible too, and it might be easier to develop than a zoo of problem-specific models.

\subsection{Outline}

Here, we propose a general QLT that allows for plasma inhomogeneity and is not restricted to any particular Hamiltonian or interaction field. By starting with the Klimontovich equation, we derive a model that captures QL diffusion, interaction with background fields, and ponderomotive effects simultaneously. The local diffusion coefficient in this model is manifestly positive-semidefinite. Waves are allowed to be off-shell, and a collision integral of the Balescu--Lenard type emerges for general Hamiltonian interactions. This operator conserves particles, momentum, and energy, and it also satisfies the \m{H}-theorem, as usual. As a spin-off, a general expression for the spectrum of microscopic fluctuations of the interaction field is derived. For on-shell waves governed by the WKE, the theory conserves the momentum and energy of the wave--plasma system. The action of nonresonant waves is also conserved, unlike in the standard version of QLT.\footnote{The standard QLT (as in, for example, \citep{ref:drummond62}) does not properly conserve energy--momentum either, even though it is formally conservative (see \Sec{sec:ihomg}).} Dewar's OC QLT of electrostatic turbulence \citep{ref:dewar73} is proven formally as a particular case and given a concise formulation. Also discussed as examples are relativistic electromagnetic and gravitational interactions, and QLT for gravitational waves is proposed. Overall, our formulation interconnects many known results that in the past were derived independently and reproduces them within a unifying framework.

This progress is made by giving up the traditional Fourier--Laplace approach. The author takes the stance that the global-mode language is unnatural for inhomogeneous-plasma problems (\ie all real-plasma problems). A fundamental theory must be local. Likewise, the variational approach that is used sometimes in QL calculations is not universally advantageous, especially for describing dissipation. Instead of those methods, we use operator analysis and the Weyl symbol calculus, as has also been proven fruitful in other recent studies of ponderomotive effects and turbulence \citep{phd:ruiz17, my:qponder, my:wkeadv} and linear-wave theory \citep{my:quasiop1}. No logical leaps are made in this paper other than assuming the QL approximation \perse and a certain ordering.\footnote{We treat the traditional QL approximation as a given mathematical model. We seek to push this model to its limits rather than to examine its validity, which is a separate issue. For discussions on the validity of the QL approximation, see \citep{ref:besse11, ref:escande18, ref:crews22}.} In a nutshell, we treat the commonly known QL-diffusion coefficient as a nonlocal operator, and we systematically approximate it using the Weyl symbol calculus. It is the nonlocality of this operator that gives rise to ponderomotive effects and ensures the proper conservation laws. The existing concept of `adiabatic diffusion' \citep{book:galeev85, book:stix} captures some of that, but systematic application of operator analysis yields a more general, more accurate, and more rigorous theory.

The author hopes not that this paper is an entertaining read. However, the paper was intended as self-contained, maximally structured, and easily searchable, so readers interested in specific questions could find and understand answers without having to read the whole paper. The text is organized as follows. In \Sec{sec:primer}, we present a primer on the Weyl symbol calculus and the associated notation. In \Sec{sec:model}, we formulate our general model. In \Sec{sec:prelim0}, we introduce the necessary auxiliary theorems. In \Sec{sec:dop}, we derive a QL model for plasma interacting with prescribed waves. The waves may or may not be on-shell or self-consistent. (Their origin and dynamics are not addressed in \Sec{sec:dop}.) In \Sec{sec:th}, we consider interactions with self-consistent waves. In particular, we separate out microscopic fluctuations, calculate their average distribution, and derive the corresponding collision operator. In \Sec{sec:onshell}, we assume that the remaining macroscopic waves are on-shell, rederive the WKE, and show that our QL model combined with the WKE is conservative. In \Sec{sec:Boltz}, we discuss the general properties that our model predicts for plasmas in thermal equilibrium. In \Sec{sec:examples}, we show how to apply our theory to nonrelativistic electrostatic interactions, relativistic electromagnetic interactions, Newtonian gravity, and relativistic gravity. In \Sec{sec:conclusions}, we summarize our results. Auxiliary calculations are presented in appendices \ref{sec:avrwig}--\ref{app:emcons}, and \App{app:notation} summarizes our notation. This notation is extensive and may not be particularly intuitive. Thus, readers are encouraged to occasionally scout \Sec{sec:examples} for examples even before fully absorbing the preceding sections.

An impatient reader can also skip calculations entirely and consult only the summaries of the individual sections (\ref{eq:summath}, \ref{sec:summodel}, \ref{sec:sumprelim}, \ref{sec:Deq}, \ref{sec:sumth}, \ref{sec:sumshell}, \ref{sec:sumbol}; they are mostly self-contained) and then proceed to the examples in \Sec{sec:examples}. However, the main point of this work is not just the final results \perse (surely, some readers will find them obvious) but also that they are derived with minimal assumptions and rigorously, which makes them reliable. A reader may also notice that we rederive some known results along the way, for example, basic linear-wave theory and the WKE. This is done for completeness and, more importantly, with the goal to present all pieces of the story within a unified notation. 

\section{A math primer}
\label{sec:primer}

Here, we summarize the machinery to be used in the next sections. This machinery is not new, but a brief overview is in order at least to introduce the necessary notation. A more comprehensive summary, with proofs, can be found in \citesup. For extended discussions, see \citep{book:tracy, phd:ruiz17, ref:mcdonald88, ref:littlejohn86}.

\subsection{Weyl symbol calculus on spacetime}
\label{sec:weyl1}

\subsubsection{Basic notation}
\label{sec:bc}

We denote the time variable as \(t\), space coordinates as \(\vec{x} \equiv \cb{x^1, x^2, \ldots, x^n}\), spacetime coordinates as \(\vecst{x} \equiv \cb{\st{x}^0, \st{x}^1, \ldots, \st{x}^n}\), where \(\st{x}^0 \doteq t\) and \(\st{x}^i \doteq x^i\). The symbol \(\doteq\) denotes definitions, and Latin indices from the middle of the alphabet (\(i, j, \ldots\)) range from 1 to \(n\) unless specified otherwise. We assume the spacetime-coordinate domain to be \m{\mathbb{R}^\stn}.\footnote{This excludes periodic boundary conditions, albeit not entirely (\Sec{sec:basm}). Other than that, the spacetime metric can still be non-Euclidean, as illustrated by an application to relativistic gravity in \Sec{sec:gw}. See also the footnotes on pages \pageref{ft:rigas} and \pageref{ft:action}.} Functions on \(\vecst{x}\) form a Hilbert space \m{\hilspacex} with an inner product that we define as
\begin{equation}\label{eq:inner1}
\braket{\xi|\psi} \doteq \int \dd\vecst{x}\, \xi^*(\vecst{x}) \psi(\vecst{x}).
\end{equation}
The symbol \(^*\) denotes complex conjugate,
\begin{equation}
\dd\vecst{x} 
\doteq \dd\st{x}^0\, \dd\st{x}^1 \ldots \dd \st{x}^n
= \dd t\, \dd x^1 \ldots \dd x^n
\end{equation}
(a similar convention is assumed also for other multi-dimensional variables used below), and integrals in this paper are taken over \((-\infty, \infty)\) unless specified otherwise. Operators on \(\hilspacex\) will be denoted with carets, and we will use indexes \(_\herm\) and \(_\aherm\) to denote their Hermitian and anti-Hermitian parts. For a given operator \m{\operst{A}}, one has \m{\operst{A}=\operst{A}_\herm + \ii \operst{A}_\aherm},
\begin{equation}\label{eq:ha}
\operst{A}_\herm = \operst{A}_\herm^\dag \doteq \frac{1}{2}\,(\operst{A} + \operst{A}^\dag),
\qquad
\operst{A}_\aherm = \operst{A}_\aherm^\dag \doteq \frac{1}{2\ii}\,(\operst{A} - \operst{A}^\dag),
\end{equation}
where \(^\dag\) denotes the Hermitian adjoint with respect to the inner product \eq{eq:inner1}. The case of a more general inner product is detailed in \citesup.

\subsubsection{Vector fields}
\label{sec:matr}

For multi-component fields \m{\vec{\psi} \equiv \cb{\psi^1, \psi^2, \ldots, \psi^M}^\intercal} (our \m{^\intercal} denotes the matrix transpose), or `row vectors' (actually, tuples), we \textit{define} the dual `column vectors' \m{\vec{\psi}^\dag \equiv \cb{\psi_1^*, \psi_2^*, \ldots, \psi_M^*}} via \m{\vec{\psi}^\dag \doteq \matr{\metric}\vec{\psi}^*}. The matrix \m{\matr{\metric}} is assumed to be real, diagonal, invertible, and constant; other than that, it can be chosen as suits a specific problem. (For example, a unit matrix may suffice.) This induces the standard rule of index manipulation
\begin{equation}
\psi_i = \metric_{ij}\psi^j, 
\qquad
\psi^i = \metric^{ij}\psi_j,
\qquad
i, j = 1, 2, \ldots, M,
\end{equation}
where \m{\metric_{ij}} are elements of \m{\matr{\metric}} and \m{\metric^{ij}} are elements of \m{\matr{\metric}^{-1}}. Summation over repeating indices is assumed. The rules of matrix multiplication apply to row and column vectors as usual. Then, for \m{\vec{\psi} \equiv \cb{\psi^1, \psi^2, \ldots, \psi^M}^\intercal} and \m{\vec{\xi} \equiv \cb{\xi^1, \xi^2, \ldots, \xi^M}^\intercal}, the quantity \m{\vec{\psi}\vec{\xi}} is a matrix with elements \m{\psi^i\xi^j}, \m{\vec{\psi}\vec{\xi}^\dag} is a matrix with elements \m{\psi^i\xi_j^*} and \m{\vec{\xi}^\dag\vec{\psi}} is its scalar trace:\footnote{A common notation is \m{\vec{\xi}^\dag\vec{\psi} = \vec{\xi} \cdot \vec{\psi}}, but we reserve the dot-product notation for a scalar product of different quantities (\Sec{sec:braket}).}
\begin{equation}\label{eq:cdot}
\vec{\xi}^\dag\vec{\psi}
= \tr(\vec{\psi}\vec{\xi}^\dag) 
= \xi_i^* \psi^i 
= \metric_{ij} \xi^{i*} \psi^j,
\qquad
i, j = 1, 2, \ldots, M.
\end{equation}
(Similarly, for \m{\vec{\chi} \equiv \cb{\chi_1, \chi_2, \ldots, \chi_M}} and \m{\vec{\eta} \equiv \cb{\eta_1, \eta_2, \ldots, \eta_M}}, \m{\vec{\eta}\vec{\chi}} is a matrix with elements \m{\eta_i \chi_j}.) We use the \eq{eq:cdot} to define a Hilbert space \m{\hilspacex^M} of \m{M}-dimensional vector fields on \(\vecst{x}\), specifically, by adopting the inner product
\begin{equation}\label{eq:innerv}
\braket{\vec{\xi}|\vec{\psi}} \doteq \int \dd\vecst{x}\, \vec{\xi}^\dag(\vecst{x}) \vec{\psi}(\vecst{x}).
\end{equation}
Below, the distinction between \(\hilspacex^M\) and \(\hilspacex\) will be assumed but not emphasized. Also note that \eq{eq:cdot} yields
\begin{equation}
\vec{\xi}^\dag(\vec{\psi}\vec{\psi}^\dag)\vec{\xi} 
= (\vec{\xi}^\dag \vec{\psi})(\vec{\psi}^\dag \vec{\xi})
= |\vec{\xi}^\dag \vec{\psi}|^2 \ge 0
\end{equation}
for any \m{\vec{\xi}} and \m{\vec{\psi}}. Thus, dyadic matrices of the form \m{\vec{\psi}\vec{\psi}^\dag} are positive-semidefinite, even though \m{\vec{\psi}^\dag\vec{\psi}} may be negative (when \m{\matr{\metric}} is not positive-definite).

For general matrices, the indices can be raised and lowered using \m{\matr{\metric}} and \m{\matr{\metric}^{-1}} as usual. The Hermitian adjoint \m{\matrst{A}^\dag} for a given matrix \m{\vecst{A}} is defined such that \m{(\matrst{A}^\dag \vec{\xi})^\dag\vec{\psi} = \vec{\xi}^\dag(\matrst{A}\vec{\psi})} for any \m{\vec{\psi}} and \m{\vec{\xi}}, which means
\begin{equation}
(\matrst{A}^\dag)_j{}^i = (\matrst{A})^{i*}{}_j \equiv \st{A}^{i*}{}_j,
\qquad
i, j = 1, 2, \ldots, M.
\end{equation}
The Hermitian and anti-Hermitian parts are defined as
\begin{equation}
\matrst{A}_\herm = \matrst{A}_\herm^\dag \doteq \frac{1}{2}\,(\matrst{A} + \matrst{A}^\dag),
\qquad
\matrst{A}_\aherm = \matrst{A}_\aherm^\dag \doteq \frac{1}{2\ii}\,(\matrst{A} - \matrst{A}^\dag),
\end{equation}
so \m{\matrst{A} = \matrst{A}_\herm + \ii \matrst{A}_\aherm}. For one-dimensional matrices (scalars), one has \(\matrst{A} = \st{A}\),
\begin{equation}\label{eq:reim}
\st{A}_\herm = \st{A}_\herm^* = \re \st{A}, 
\qquad
\st{A}_\aherm = \st{A}_\aherm^* = \im \st{A},
\end{equation}
where \m{\re} and \m{\im} denote the real part and the imaginary part, respectively. We also define matrix operators \m{\boperst{A}} as matrices of the corresponding operators \m{\operst{A}^i{}_j}. Because \m{\matr{\metric}} is constant, index manipulation applies as usual. Also as usual, one has
\begin{equation}
\boperst{A}_\herm = \boperst{A}_\herm^\dag \doteq \frac{1}{2}\,(\boperst{A} + \boperst{A}^\dag),
\qquad
\boperst{A}_\aherm = \boperst{A}_\aherm^\dag \doteq \frac{1}{2\ii}\,(\boperst{A} - \boperst{A}^\dag),
\end{equation}
and \m{\boperst{A} = \boperst{A}_\herm + \ii \boperst{A}_\aherm}, where \(^\dag\) is the Hermitian adjoint with respect to the inner product \eq{eq:innerv}.

\subsubsection{Bra--ket notation}
\label{sec:braket}

Let us define the following operators that are Hermitian under the inner product \eq{eq:inner1}:
\begin{equation}\label{eq:xwk}
\operst{x}^0 \equiv \oper{t} \doteq t, 
\quad 
\operst{x}^i \equiv \oper{x}^i \doteq x^i, 
\quad
\operst{k}_0 \equiv -\oper{\omega} \doteq -\ii\pd_t,
\quad
\oper{k}_i \doteq -\ii\pd_i,
\end{equation}
where \(\pd_0 \equiv \pd_t \doteq \pd/\pd x^0\) and \(\pd_i \doteq \pd/\pd x^i\). Accordingly, 
\begin{equation}
\boperst{x} \equiv \cb{\operst{x}^0, \operst{x}^1, \ldots, \operst{x}^{n}} 
= \cb{\oper{t}, \boper{x}}, 
\qquad
\boperst{k} \equiv \cb{\operst{k}_0, \operst{k}_1, \ldots, \operst{k}_{n}} 
= \cb{-\oper{\omega}, \boper{k}}
\end{equation}
are understood as the spacetime-position operator and the corresponding wavevector operator, which will also be expressed as follows:
\begin{equation}\label{eq:XK}
\boperst{x} = \vecst{x}, 
\qquad
\boperst{k} = - \ii\pd_\vecst{x}.
\end{equation}
Also note the commutation property, where \m{\delta_i^j} is the Kronecker symbol:\footnote{\label{ft:rigas}Spaces with periodic boundary conditions require a different approach \citep{ref:rigas11}, so they are not considered here (yet see \Sec{sec:basm}). That said, for a system that is large enough, the boundary conditions are unimportant; then the toolbox presented here is applicable as is.}
\begin{equation}
[\operst{x}^i, \operst{k}_j] = \ii \delta^i_j,
\qquad
i, j = 0, 1, \ldots, n.
\end{equation}

The eigenvectors of the operators \eq{eq:XK} will be denoted as `kets' \m{\ket{\vecst{x}}} and \m{\ket{\vecst{k}}}:\footnote{More precisely, \m{\ket{\vecst{x}}} is the ket \m{\ket{\mcc{e}(\boperst{x}; \vecst{x})}} that is an eigenvector of each \m{\operst{x}^i} with the corresponding eigenvalues being \m{\st{x}^i}. A similar comment applies to \m{\ket{\vecst{k}}}.}
\begin{equation}\label{eq:ketxk}
\boperst{x} \ket{\vecst{x}} = \vecst{x} \ket{\vecst{x}},
\qquad
\boperst{k} \ket{\vecst{k}} = \vecst{k} \ket{\vecst{k}},
\end{equation}
and we assume the usual normalization:
\begin{equation}\label{eq:ketxknorm}
\braket{\vecst{x}_1|\vecst{x}_2} = \delta(\vecst{x}_1 - \vecst{x}_2),
\qquad
\braket{\vecst{k}_1|\vecst{k}_2} = \delta(\vecst{k}_1 - \vecst{k}_2),
\end{equation}
where \(\delta\) is the Dirac delta function. Both sets \(\cbb{\ket{\vecst{x}}, \vecst{x} \in \mathbb{R}^\stn}\) and \(\cbb{\ket{\vecst{k}}, \vecst{k} \in \mathbb{R}^\stn}\), where \(\stn \doteq n + 1\), form a complete basis on \(\hilspacex\), and the eigenvalues of these operators form an extended \textit{real} phase space \(\cb{\vecst{x}, \vecst{k}}\), where
\begin{equation}
\vecst{x} \equiv \cb{t, \vec{x}},
\qquad
\vecst{k} \equiv \cb{-\omega, \vec{k}}.
\end{equation}
The notation \m{\vecst{k} \cdot \vecst{s} \doteq -\omega \tau + \vec{k} \cdot \vec{s}} will be assumed for any \(\vecst{s} \equiv \cb{\tau, \vec{s}}\), and \m{\vec{k} \cdot \vec{s} \doteq k_i s^i}. In particular, for any \(\psi\) and constant \(\vecst{s}\), one has
\begin{equation}\label{eq:eshift}
\exp(\ii \boperst{k} \cdot \vecst{s}) \psi(\vecst{x}) = \exp(\vecst{s} \cdot \pd_\vecst{x}) \psi(\vecst{x}) = \psi(\vecst{x} + \vecst{s}),
\end{equation}
as seen from comparing the Taylor expansions of the latter two expressions. (A generalization of this formula is discussed in \Sec{sec:shift}.) Also,
\begin{equation}
\braket{\vecst{x}|\vecst{k}} = \braket{\vecst{k}|\vecst{x}}^* = (2\upi)^{-\stn/2} \exp(\ii \vecst{k} \cdot \vecst{x}),
\end{equation}
and
\begin{equation}
\int \dd\vecst{x}\,\ket{\vecst{x}}\bra{\vecst{x}} = \operst{1},
\qquad
\int \dd\vecst{k}\,\ket{\vecst{k}}\bra{\vecst{k}} = \operst{1}.
\end{equation}
Here, `bra' \(\bra{\vecst{x}}\) is the one-form dual to \(\ket{\vecst{x}}\), \(\bra{\vecst{k}}\) is the one-form dual to \(\ket{\vecst{k}}\), and \(\operst{1}\) is the unit operator. Any field \(\psi\) on \(\vecst{x}\) can be viewed as the \(\vecst{x}\)~representation (`coordinate representation') of \(\ket{\psi}\), \ie the projection of an abstract ket vector \(\ket{\psi} \in \hilspacex\) on \m{\ket{\vecst{x}}}: 
\begin{equation}
\psi(\vecst{x}) = \braket{\vecst{x}|\psi}.
\end{equation}
Similarly, \(\braket{\vecst{k}|\psi}\) is the \(\vecst{k}\)~representation (`spectral representation') of \(\ket{\psi}\), or the Fourier image of~\(\psi\):
\begin{equation}\label{eq:psik}
\fourier{\psi}(\vecst{k}) \doteq  \braket{\vecst{k}|\psi} = \frac{1}{(2\upi)^{\stn/2}}\int \dd\vecst{x}\,
\ee^{-\ii \vecst{k} \cdot \vecst{x}}\psi(\vecst{x}).
\end{equation}

\subsubsection{Wigner--Weyl transform}
\label{sec:wwt}

For a given operator \(\operst{A}\) and a given field \(\psi\), \(\operst{A}\psi\) can be expressed in the integral form
\begin{equation}
\operst{A}\psi(\vecst{x}) = \int \dd \vecst{x}' \braket{\vecst{x}|\operst{A}|\vecst{x}'}\psi(\vecst{x}'),
\end{equation}
where \m{\braket{\vecst{x}|\operst{A}|\vecst{x}'}} is a function of \(\cb{\vecst{x}, \vecst{x}'}\). This is called the \(\vecst{x}\)~representation (`coordinate representation') of \m{\operst{A}}. Equivalently, \(\operst{A}\) can be given a phase-space, or Weyl, representation, \ie expressed through a function of the phase-space coordinates, \(\st{A}(\vecst{x}, \vecst{k})\):\footnote{Analytic continuation to complex arguments is possible, but by default, \m{\vecst{x}} and \m{\vecst{k}} are real.}
\begin{equation}\label{eq:woperx}
\operst{A} = \frac{1}{(2\upi)^{\stn}} \int \dd \vecst{x} \,\dd\vecst{k}\,\dd \vecst{s}\,
\ket{\vecst{x} + \vecst{s}/2} 
\st{A}(\vecst{x}, \vecst{k})\,\ee^{\ii \vecst{k} \cdot \vecst{s}}
\bra{\vecst{x}-\vecst{s}/2}
\equiv \woperx{\st{A}}.
\end{equation}
The function \(\st{A}(\vecst{x}, \vecst{k})\), called the Weyl symbol (or just `symbol') of \(\operst{A}\), is given~by
\begin{equation}\label{eq:wsymbx}
\st{A}(\vecst{x}, \vecst{k}) \doteq \int \dd\vecst{s}
\braket{\vecst{x}+\vecst{s}/2 | \operst{A} | \vecst{x}-\vecst{s}/2}
\ee^{-\ii \vecst{k} \cdot \vecst{s}} \equiv \wsymbx{\operst{A}}.
\end{equation}
The \(\vecst{x}\) and phase-space representations are connected by the Fourier transform:
\begin{eqnarray}\label{eq:Mxx}
\braket{\vecst{x}| \operst{A} | \vecst{x}'} = \frac{1}{(2\upi)^\stn}\int\dd \vecst{k}\,
\ee^{\ii\vecst{k} \cdot(\vecst{x} - \vecst{x}')} \,
\st{A}\left(\frac{\vecst{x} + \vecst{x}'}{2}, \vecst{k}\right).
\end{eqnarray}
This also leads to the following notable properties of Weyl symbols:
\begin{eqnarray}\label{eq:Mxx2}
\braket{\vecst{x}| \operst{A} | \vecst{x}} = 
\frac{1}{(2\upi)^{\stn}}\int \dd \vecst{k}\,\st{A}(\vecst{x}, \vecst{k}),
\qquad
\braket{\vecst{k}| \operst{A} | \vecst{k}} = 
\frac{1}{(2\upi)^{\stn}}\int \dd \vecst{x}\,\st{A}(\vecst{x},\vecst{k}).
\end{eqnarray}

An operator unambiguously determines its symbol, and \latin{vice versa}. We denote this isomorphism as \m{\operst{A} \leftrightarrow \st{A}}. The mapping \m{\operst{A} \mapsto \st{A}} is called the Wigner transform, and \m{\st{A} \mapsto \operst{A}} is called the Weyl transform. For uniformity, we call them the direct and inverse Wigner--Weyl transform. The isomorphism \m{\leftrightarrow} is natural in that it has the following properties:
\begin{eqnarray}\label{eq:natural}
 \operst{1} \leftrightarrow 1, 
 \quad 
 \boperst{x} \leftrightarrow \vecst{x}, 
 \quad 
 \boperst{k} \leftrightarrow \vecst{k},
 \quad
h(\boperst{x}) \leftrightarrow h(\vecst{x}),
\quad  
h(\boperst{k}) \leftrightarrow h(\vecst{k}),
\quad
\operst{A}^\dag \leftrightarrow \st{A}^*,
\end{eqnarray}
where \(h\) is any function and \m{\operst{A}} is any operator. The product of two operators maps to the so-called Moyal product, or star product, of their symbols \citep{ref:moyal49}:
\begin{eqnarray}\label{eq:moyalx}
\operst{A}\operst{B}\,
\leftrightarrow 
\,
\st{A}(\vecst{x}, \vecst{k}) \star \st{B}(\vecst{x}, \vecst{k})
\doteq
\st{A}(\vecst{x}, \vecst{k})\ee^{\ii\oper{\mc{L}}_\indexst/2}\st{B}(\vecst{x}, \vecst{k}),
\end{eqnarray}
which is associative:
\begin{eqnarray}
\label{eq:moyal3}
\operst{A}\operst{B}\operst{C}\,
\leftrightarrow 
\,
(\st{A} \star \st{B}) \star \st{C} = \st{A} \star (\st{B} \star \st{C}) \equiv \st{A} \star \st{B} \star \st{C}.
\end{eqnarray}
Here, \m{\oper{\mc{L}}_\indexst \doteq \ldx \cdot \rdp - \ldp \cdot \rdx}, and the arrows indicate the directions in which the derivatives act. For example, \m{\st{A}\oper{\mc{L}}_\indexst\st{B}} is just the canonical Poisson bracket on \m{\cb{\vecst{x}, \vecst{k}}}:
\begin{equation}\label{eq:poissonx}
\st{A} \oper{\mc{L}}_\indexst \st{B} 
= \poissonx{\st{A}, \st{B}} \doteq
- \frac{\pd \st{A}}{\pd t}\frac{\pd \st{B}}{\pd \omega}
+ \frac{\pd \st{A}}{\pd \omega}\frac{\pd \st{B}}{\pd t}
+ \frac{\pd \st{A}}{\pd x^i}\frac{\pd \st{B}}{\pd k_i}
- \frac{\pd \st{A}}{\pd k_i}\frac{\pd \st{B}}{\pd x^i}.
\end{equation}
These formulas readily yield
\begin{eqnarray}\label{eq:hk}
h(\boperst{x})\operst{k}_\alpha
\leftrightarrow 
\st{k}_\alpha h(\vecst{x}) + \frac{\ii}{2}\,\pd_\alpha h(\vecst{x}),
\qquad
\operst{k}_\alpha h(\boperst{x})
\leftrightarrow 
\st{k}_\alpha h(\vecst{x}) - \frac{\ii}{2}\,\pd_\alpha h(\vecst{x}),
\end{eqnarray}
also \(h(\boperst{k}) \ee^{\ii\vecst{K} \cdot\boperst{x}} \leftrightarrow h(\vecst{k})\star \ee^{\ii\vecst{K} \cdot\vecst{x}} = h(\vecst{k} + \vecst{K}/2) \ee^{\ii\vecst{K} \cdot\vecst{x}}\), etc. Another notable formula to be used below, which flows from \eq{eq:Mxx2} and \eq{eq:moyal3}, is
\begin{eqnarray}\label{eq:Mxx3}
\braket{\vecst{x}| \operst{A}\operst{B}\operst{C} | \vecst{x}} = 
\frac{1}{(2\upi)^{\stn}}\int \dd \vecst{k}\,(\st{A} \star \st{B} \star \st{C})(\vecst{x}, \vecst{k}).
\end{eqnarray}

The Moyal product is particularly handy when \(\pd_{\vecst{x}} \pd_{\vecst{k}} \sim \epsilon \ll 1\). Such \(\epsilon\) is often called the geometrical-optics parameter. Since \(\oper{\mc{L}}_\indexst = \mc{O}(\epsilon)\), one can express the Moyal product as an asymptotic series in powers of \(\epsilon\):
\begin{equation}
 \star = \operst{1} + \ii \oper{\mc{L}}_\indexst/2 - \oper{\mc{L}}_\indexst^2/8 + \ldots
\end{equation}

\subsubsection{Weyl expansion of operators}
\label{sec:wexp}

Operators can be approximated by approximating their symbols \citep{my:quasiop1, ref:mcdonald88}. If \m{\operst{A}} is approximately local in \(\vecst{x}\) (\ie if \m{\operst{A}\psi(\vecst{x})} is determined by values \(\psi(\vecst{x} + \vecst{s})\) only with small enough \(\vecst{s}\)), its symbol can be adequately represented by the first few terms of the Taylor expansion in \(\vecst{k}\):
\begin{equation}
\st{A}(\vecst{x}, \vecst{k}) = \st{A}(\vecst{x}, \vecst{0}) + \vec{\Uptheta}_0(\vecst{x}) \cdot \vecst{k} + \ldots,
\qquad
\vec{\Uptheta}_0(\vecst{x}) \doteq (\pd_\vecst{k} \st{A}(\vecst{x}, \vecst{k}))_{\vecst{k}=\vecst{0}}.
\end{equation}
Application of \m{\woperx{}} to this formula leads to
\begin{equation}\label{eq:A2}
\operst{A} \approx \st{A}(\boperst{x}, \vecst{0}) + \frac{1}{2}\,(\boper{\Uptheta}_0 \cdot \boperst{k} + \boperst{k} \cdot \boper{\Uptheta}_0) + \ldots,
\end{equation}
where \(\boper{\Uptheta}_0 \doteq \vec{\Uptheta}_0(\boperst{x})\). One can also rewrite \eq{eq:A2} using the commutation property
\begin{equation}
[\boperst{k}, \boper{\Uptheta}_0] = -\ii(\pd_{\vecst{x}} \cdot \vec{\Uptheta}_0)(\boperst{x}).
\end{equation}
In the \(\vecst{x}\)~representation, this leads to
\begin{equation}
\operst{A} = \st{A}(\vecst{x}, \vecst{0}) - \ii \vec{\Uptheta}_0(\vecst{x}) \cdot \pd_{\vecst{x}} 
- \frac{\ii}{2}\,(\pd_{\vecst{x}} \cdot \vec{\Uptheta}_0(\vecst{x})) + \ldots
\end{equation}

The effect of a nonlocal operator on eikonal (monochromatic or quasimonochromatic) fields can be approximated similarly. Suppose \(\psi = \ee^{\ii\theta}\env{\psi}\), where the dependence of \(\avr{\vecst{k}} \doteq \pd_{\vecst{x}}\theta\) and \(\env{\psi}\) on \(\vecst{x}\) is slower than that of \(\theta\) by factor \(\epsilon \ll 1\). Then, \(\operst{A}\psi = \ee^{\ii\theta}\operst{A}' \env{\psi}\), where \(\operst{A}' \doteq \ee^{-\ii\theta(\operst{x})}\operst{A}\ee^{\ii\theta(\operst{x})}\), and the symbol of \(\operst{A}'\) can be approximated as follows:
\begin{equation}\label{eq:symbexp}
\st{A}'(\vecst{x}, \vecst{k}) = \st{A}(\vecst{x}, \avr{\vecst{k}}(\vecst{x}) + \vecst{k}) + \mc{O}(\epsilon^2).
\end{equation}
By expanding this in \(\vecst{k}\) and applying \m{\woperx{}}, one obtains
\begin{eqnarray}\label{eq:Mexp}
\operst{A}' = \st{A}(\vecst{x}, \avr{\vecst{k}}(\vecst{x})) - \ii \vec{\Uptheta}(\vecst{x}) \cdot \pd_{\vecst{x}}
- \frac{\ii}{2}\,(\pd_{\vecst{x}} \cdot \vec{\Uptheta}(\vecst{x})) + \mc{O}(\epsilon^2),
\end{eqnarray}
where \(\vec{\Uptheta}(\vecst{x}) \doteq (\pd_\vecst{k} \st{A}(\vecst{x}, \vecst{k}))_{\vecst{k}=\avr{\vecst{k}}(\vecst{x})}\). Neglecting the \(\mc{O}(\epsilon^2)\) corrections in this formula leads to what is commonly known as the geometrical-optics approximation \citep{my:quasiop1}.

\subsubsection{Wigner functions}
\label{sec:wigfun}

Any ket \(\ket{\psi}\) generates a dyadic \(\ket{\psi}\bra{\psi}\). In quantum mechanics, such dyadics are known as density operators (of pure states). For our purposes, though, it is more convenient to define the density operator in a slightly different form, namely, as
\begin{equation}\label{eq:stWdop}
\operst{W}_\psi \doteq (2\upi)^{-\stn} \ket{\psi}\bra{\psi}.
\end{equation}
The symbol of this operator, \m{\st{W}_{\psi} = \wsymbx{\operst{W}_\psi}}, is a real function called the Wigner function. It is given by
\begin{align}
\st{W}_\psi(\vecst{x}, \vecst{k}) 
& = \frac{1}{(2\upi)^\stn} \int \dd\vecst{s} 
\braket{\vecst{x} + \vecst{s}/2 | \psi} \braket{\psi | \vecst{x} - \vecst{s}/2} \ee^{-\ii \vecst{k} \cdot \vecst{s}}
\notag\\
& = \frac{1}{(2\upi)^\stn} \int \dd\vecst{s}\, 
\psi(\vecst{x} + \vecst{s}/2) \psi^*(\vecst{x} - \vecst{s}/2)\,\ee^{-\ii \vecst{k} \cdot \vecst{s}},
\label{eq:aux10}
\end{align}
which is manifestly real and can be understood as the (inverse) Fourier image of
\begin{eqnarray}\label{eq:C1}
\st{C}_\psi(\vecst{x}, \vecst{s}) 
\doteq \psi(\vecst{x} + \vecst{s}/2) \psi^*(\vecst{x} - \vecst{s}/2) 
= \int \dd\vecst{k}\,\st{W}_\psi(\vecst{x}, \vecst{k})\,\ee^{\ii \vecst{k} \cdot \vecst{s}}.
\end{eqnarray}

Any function bilinear in~\m{\psi} and~\m{\psi^*} can be expressed through \m{\st{W}_\psi}. Specifically, for any operators~\m{\operst{L}} and~\m{\operst{R}}, one has
\begin{align}
(\operst{L}\psi(\vecst{x}))(\operst{R}\psi(\vecst{x}))^*
& = 
\braket{\vecst{x}|\operst{L}|\psi} 
\braket{\psi|\operst{R}^\dag |\vecst{x}}
\notag\\
& = (2\upi)^\stn 
\braket{\vecst{x}|\operst{L}\operst{W}_\psi\operst{R}^\dag|\vecst{x}}
\notag\\
& = \textstyle \int \dd\vecst{k}\,\st{L}(\vecst{x}, \vecst{k}) \star \st{W}_\psi(\vecst{x}, \vecst{k}) \star \st{R}^*(\vecst{x}, \vecst{k}),
\label{eq:LR1}
\end{align}
where \(\st{L}\) and \(\st{R}\) are the corresponding symbols and \eq{eq:Mxx2} was used along with \eq{eq:moyal3}. As a corollary, and as also seen from \eq{eq:Mxx2}, one has
\begin{equation}\label{eq:wigfprop1}
|\psi(\vecst{x})|^2 = \int \dd \vecst{k}\,\st{W}_\psi(\vecst{x}, \vecst{k}),
\qquad
|\fourier{\psi}(\vecst{k})|^2 = \int \dd \vecst{x}\,\st{W}_\psi(\vecst{x}, \vecst{k}).
\end{equation}
As a reminder, \(\psi(\vecst{x}) = \braket{\vecst{x}|\psi}\) and \(\fourier{\psi}(\vecst{k}) \doteq \braket{\vecst{k}|\psi}\) is the Fourier image of \(\psi\) \eq{eq:psik}, so \m{|\psi(\vecst{x})|^2} and \m{|\fourier{\psi}(\vecst{k})|^2} can be loosely understood as the densities of quanta (associated with the field \m{\psi}) in the \m{\vecst{x}}-space and the \m{\vecst{k}}-space, respectively. Because of \eq{eq:wigfprop1}, \m{\st{W}_\psi} is commonly attributed as a quasiprobability distribution of wave quanta in phase space. (The prefix `quasi' is added because \m{\st{W}_\psi} can be negative.) In case of real fields, which satisfy \(\braket{\vecst{x}|\psi} = \braket{\psi|\vecst{x}}\), one also has
\begin{equation}
\st{W}_\psi(\vecst{x}, \vecst{k}) = \st{W}_\psi(\vecst{x}, -\vecst{k}).
\end{equation}
Of particular importance are Wigner functions averaged over a sufficiently large phase-space volume \(\Delta\vecst{x}\,\Delta\vecst{k} \gtrsim 1\). The average Wigner function \m{\avr{\st{W}}_\psi} is a \textit{local} property of the field (as opposed to, say, the field's global Fourier spectrum) and satisfies (\App{sec:avrwig})
\begin{equation}
\avr{\st{W}}_\psi \ge 0.
\end{equation}

\subsubsection{Generalization to vector fields}
\label{sec:vecf}

In case of vector (tuple) fields \(\vec{\psi} = \cb{\psi^1, \psi^2, \ldots, \psi^M}^\intercal\), kets are column vectors, \(\ket{\vec{\psi}} = \cb{\ket{\psi^1}, \ket{\psi^2}, \ldots, \ket{\psi^M}}\), and bras are row vectors, \(\bra{\vec{\psi}} = \cb{\bra{\psi^1}, \bra{\psi^2}, \ldots, \bra{\psi^M}}\). The operators acting on such kets and bras are matrices of operators. The Weyl symbol of a matrix operator is defined as the matrix of the corresponding symbols. As a result, the symbol of a Hermitian adjoint of a given operator is the Hermitian adjoint of the symbol of that operator:
\begin{equation}
\smash{\boperst{A}}^\dag \leftrightarrow \smash{\matrst{A}}^\dag,
\end{equation}
and as a corollary, the symbol of a Hermitian matrix operator is a Hermitian matrix.

In particular, the density operator of a given vector field \(\vec{\psi}\) is a matrix operator
\begin{equation}\label{eq:boperstW}
\boperst{W}_{\vec{\psi}} \doteq (2\upi)^{-\stn} \ket{\vec{\psi}}\bra{\vec{\psi}}.
\end{equation}
The symbol of this operator, \m{\matrst{W}_{\vec{\psi}} = \wsymbx{\boperst{W}_{\vec{\psi}}}}, is a Hermitian matrix function\footnote{By construction, \m{\boperst{W}_{\vec{\psi}}} is a matrix with mixed indices, \m{(\boperst{W}_{\vec{\psi}})^i{}_j}. In \Secs{sec:D} and \ref{sec:wuw}, we also operate with a Wigner matrix that has two upper indices. Because the field of interest is real there, the dagger \m{^\dag} in \eq{eq:Wtenx} is assumed omitted in that case.}
\begin{equation}\label{eq:Wtenx}
\matrst{W}_{\vec{\psi}}(\vecst{x}, \vecst{k}) 
= \frac{1}{(2\upi)^\stn} \int \dd\vecst{s}\, 
\vec{\psi}(\vecst{x} + \vecst{s}/2) \vec{\psi}^\dag(\vecst{x} - \vecst{s}/2)\,\ee^{-\ii \vecst{k} \cdot \vecst{s}},
\end{equation}
called the Wigner matrix. (It is also called the `Wigner tensor' when \m{\vec{\psi}} is a true vector rather than a tuple.) It can be understood as the (inverse) Fourier image of
\begin{eqnarray}\label{eq:C2}
\matrst{C}_{\vec{\psi}}(\vecst{x}, \vecst{s}) 
\doteq \vec{\psi}(\vecst{x} + \vecst{s}/2) \vec{\psi}^\dag(\vecst{x} - \vecst{s}/2) 
= \int \dd\vecst{k}\,\matrst{W}_{\vec{\psi}}(\vecst{x}, \vecst{k})\,\ee^{\ii \vecst{k} \cdot \vecst{s}}.
\end{eqnarray}
The analog of \eq{eq:LR1} is (\App{app:LRs})
\begin{subequations}\label{eq:LRs}
\begin{eqnarray}
(\boperst{L}\vec{\psi}(\vecst{x}))(\boperst{R}\vec{\psi}(\vecst{x}))^\dag
= \textstyle \int \dd\vecst{k}\,\matrst{L}(\vecst{x}, \vecst{k}) \star 
\matrst{W}_{\vec{\psi}}(\vecst{x}, \vecst{k}) \star \matrst{R}^\dag(\vecst{x}, \vecst{k}).
\label{eq:LR2}
\end{eqnarray}
The Wigner matrix averaged over a sufficiently large phase-space volume \(\Delta\vecst{x}\,\Delta\vecst{k} \gtrsim 1\) is a local property of the field, and it is positive-semidefinite (\App{sec:avrwig}).

For real fields, one also has
\begin{equation}
\matrst{W}_{\vec{\psi}}(\vecst{x}, \vecst{k}) 
= \matrst{W}_{\vec{\psi}}^\intercal(\vecst{x}, -\vecst{k}) 
= \matrst{W}_{\vec{\psi}}^*(\vecst{x}, -\vecst{k}),
\end{equation}
and \eq{eq:LRs} yields the following corollary at \m{\epsilon \to 0}, when \m{\star} becomes the usual product (\App{app:LRs}):
\begin{eqnarray}
(\boperst{L}\vec{\psi})^\dag \boperst{R}\vec{\psi}
= \int \dd\vecst{k}\,\tr\big(\matrst{W}_{\vec{\psi}}(\matrst{L}^\dag\matrst{R})_\herm\big).
\label{eq:LRtr}
\end{eqnarray}
\end{subequations}
The generalizations of the other formulas from the previous sections are obvious.

\subsection{Weyl symbol calculus on phase space}
\label{sec:weyl2}

\subsubsection{Notation}
\label{sec:notz}

Consider a Hamiltonian system with coordinates \(\vec{x} \equiv \cb{x^1, x^2, \ldots, x^n}\) and canonical momenta \(\vec{p} \equiv \cb{p_1, p_2, \ldots, p_n}\). Together, these variables comprise the phase-space coordinates \(\vec{z} \equiv \cb{\vec{x}, \vec{p}}\), \ie
\begin{equation}
\vec{z} \equiv \cb{z^1, \ldots, z^{2n}} 
= \cb{x^1, \ldots, x^n, p_1, \ldots, p_n}.
\end{equation}
Components of \(\vec{z}\) will be denoted with Greek indices ranging from 1 to \(2n\).\footnote{However, the index \m{\sigma} is reserved as a tag for individual particles and waves.}

Hamilton's equations for \(z^\alpha\) can be written as \(\dot{z}^{\alpha} = \poissonz{z^{\alpha}, H}\), or equivalently, as
\begin{equation}\label{eq:hamz}
\dot{z}^{\alpha} = J^{\alpha\beta}\,\pd_\beta H.
\end{equation}
Here, \(H = H(t, \vec{z})\) is a Hamiltonian, \(\pd_\beta \doteq \pd/\pd z^\beta\), 
\begin{equation}\label{eq:poissonz}
\poissonz{A, B} \doteq J^{\alpha\beta}\, (\pd_\alpha A) (\pd_\beta B)
\end{equation}
is the Poisson bracket on \m{\vec{z}}, \(J^{\alpha\beta}\) is the canonical Poisson structure:
\begin{equation}\label{eq:Jdef}
\matr{J} = -\matr{J}^\intercal 
= \left(
\begin{array}{cc}
 \matr{0}_n & \matr{1}_n \\
 -\matr{1}_n & \matr{0}_n \\
\end{array}
\right),
\end{equation}
\(\matr{0}_n\) is an \(n\)-dimensional zero matrix, and \(\matr{1}_n\) is an \(n\)-dimensional unit matrix. The corresponding equation for the probability distribution \(f(t, \vec{z})\) is
\begin{equation}\label{eq:LE}
\pd_t f = \poissonz{H, f}.
\end{equation}
Solutions of \eq{eq:LE} and other functions of the extended-phase-space coordinates \(\vec{X} \equiv \cb{t, \vec{z}}\) can be considered as vectors in the Hilbert space \(\hilspaceX\) with the usual inner product\footnote{Note that the inner product \eq{eq:inner2} is different from \eq{eq:inner1}. Still, we use the same notation assuming it will be clear from the context which inner product is used in each given case.}
\begin{equation}\label{eq:inner2}
\braket{\xi|\psi} \doteq \int \dd\vec{X}\, \xi^*(\vec{X}) \psi(\vec{X}).
\end{equation}
Assuming the notation \(N \doteq \dim \vec{X} = 2n + 1\), one has
\begin{equation}
\dd\vec{X} 
\doteq \dd X^1\, \dd X^2 \ldots \dd X^N
= \dd t\, \dd x^1 \ldots \dd x^n\,\dd p_1, \ldots , \dd p_n.
\end{equation}

Let us introduce the position operator on \(\vec{z}\),
\begin{equation}
\boper{z} \doteq
\cb{
\underbrace{x^1, \ldots, x^n}_{\boper{x}}, 
\underbrace{p_1, \ldots, p_n}_{\boper{p}}
},
\end{equation}
and the momentum operator on \(\vec{z}\),
\begin{equation}
\boper{q} \equiv \cb{
\underbrace{-\ii \pd_1, \ldots, -\ii\pd_n}_{\boper{k}}, 
\underbrace{-\ii\pd^1, \ldots, -\ii\pd^n}_{\boper{r}}
},
\end{equation}
where \(\pd_i \doteq \pd/\pd x^i\) but \(\pd^i \doteq \pd/\pd p_i\); that is, \(\boper{z} = \cb{\boper{x}, \boper{p}}\), \(\boper{q} = \cb{\boper{k}, \boper{r}}\), and
\begin{equation}
\oper{z}^{\alpha}\doteq z^{\alpha},
\qquad
\oper{q}_{\alpha}\doteq -\ii\pd_{\alpha}.
\end{equation}
Then, much like in \Sec{sec:weyl1}, one can also introduce the position and momentum operators on the extended phase space \(\vec{X}\):
\begin{equation}
\boper{X} = \cb{\oper{t}, \boper{z}} = \cb{\oper{t}, \boper{x}, \boper{p}},
\qquad
\boper{K} = \cb{-\oper{\omega}, \boper{q}} = \cb{-\oper{\omega}, \boper{k}, \boper{r}}.
\end{equation}
Assuming the convention that Latin indices from the beginning of the alphabet \(\cb{a, b, c, \ldots}\) range from 0 to \(2n\), and \(\pd_a \doteq \pd/\pd X^a\), one can compactly express this as
\begin{equation}
\oper{X}^a = X^a,
\qquad
\oper{K}_a = - \ii\pd_a.
\end{equation}

The eigenvectors of these operators will be denoted \(\ket{\vec{X}}\) and \(\ket{\vec{K}}\):
\begin{equation}
\boper{X} \ket{\vec{X}} = \vec{X} \ket{\vec{X}},
\qquad
\boper{K} \ket{\vec{K}} = \vec{K} \ket{\vec{K}},
\end{equation}
and we assume the usual normalization:
\begin{equation}\label{eq:norm}
\braket{\vec{X}_1|\vec{X}_2} = \delta(\vec{X}_1 - \vec{X}_2),
\qquad
\braket{\vec{K}_1|\vec{K}_2} = \delta(\vec{K}_1 - \vec{K}_2).
\end{equation}
Both sets \(\cbb{\ket{\vec{X}}, \vec{X} \in \mathbb{R}^N}\) and \(\cbb{\ket{\vec{K}}, \vec{K} \in \mathbb{R}^N}\) form a complete basis on \(\hilspaceX\), and the eigenvalues of these operators form a real extended phase space \(\cb{\vec{X}, \vec{K}}\), where
\begin{equation}
\vec{X} \equiv \cb{t, \vec{z}},
\qquad
\vec{K} \equiv \cb{-\omega, \vec{q}}.
\end{equation}
Particularly note the following formula, which will be used below:
\begin{equation}\label{eq:Jqq}
J^{\alpha\beta}\oper{q}_\alpha q_\beta 
= 
\left(
\begin{array}{cc}
 \boper{k} & \boper{r}
\end{array}
\right)
\left(
\begin{array}{cc}
 \matr{0}_n & \matr{1}_n \\
 -\matr{1}_n & \matr{0}_n \\
\end{array}
\right)
\left(
\begin{array}{c}
 \vec{k}\\
 \vec{r}\\
\end{array}
\right)
= \boper{k} \cdot \vec{r} - \boper{r} \cdot \vec{k}.
\end{equation}

\subsubsection{Wigner--Weyl transform}
\label{sec:wwtX}

One can construct the Weyl symbol calculus on the extended phase space \(\vec{X}\) just like it is done on spacetime \(\vecst{x}\) in \Sec{sec:weyl1}, with an obvious modification of the notation. The Wigner--Weyl transform is defined as
\begin{eqnarray}
& \displaystyle A(\vec{X}, \vec{K}) = \int \dd\vec{S} 
\braket{\vec{X} + \vec{S}/2 | \oper{A} | \vec{X} - \vec{S}/2} 
\ee^{-\ii \vec{K} \cdot \vec{S}}
\equiv \wsymbX{\oper{A}},
\label{eq:WWT}
\\
& \displaystyle \oper{A} = \frac{1}{(2\upi)^N} \int \dd\vec{X}\,\dd\vec{K}\,\dd\vec{S}
\ket{\vec{X}+\vec{S}/2} A(\vec{X}, \vec{K}) \bra{\vec{X} - \vec{S}/2}
\ee^{\ii \vec{K} \cdot \vec{S}}
\equiv \woperX{A}.
\label{eq:invWWT}
\end{eqnarray}
(Notice the change in the font and in the index compared to \eq{eq:wsymbx} and \eq{eq:woperx}.) The corresponding Moyal product is denoted \(\bigstar\) (as opposed to \(\star\) introduced earlier):
\begin{equation}\label{eq:bigstar}
A \bigstar B = A(\vec{X}, \vec{K})\,\ee^{\ii\oper{\mc{L}}_X/2}B(\vec{X}, \vec{K}),
\end{equation}
where \(\oper{\mc{L}}_X \doteq \ldX \cdot \rdP - \ldP \cdot \rdX\) can be expressed as follows:
\begin{equation}\label{eq:poissonX}
A \oper{\mc{L}}_X B \doteq
- \frac{\pd A}{\pd t}\frac{\pd B}{\pd \omega}
+ \frac{\pd A}{\pd \omega}\frac{\pd B}{\pd t}
+ \frac{\pd A}{\pd x^i}\frac{\pd B}{\pd k_i}
- \frac{\pd A}{\pd k_i}\frac{\pd B}{\pd x^i}
+ \frac{\pd A}{\pd p^i}\frac{\pd B}{\pd r_i}
- \frac{\pd A}{\pd r_i}\frac{\pd B}{\pd p^i}.
\end{equation}

If an operator \m{\oper{A}} is local in \m{\vec{p}}, its \m{\vec{X}}~representation and \m{\vecst{x}}~representation satisfy
\begin{eqnarray}
\braket{t, \vec{x}, \vec{p}|\boper{A}|t', \vec{x}', \vec{p}'}
= \braket{t, \vec{x}|\boper{A}|t', \vec{x}'}\delta(\vec{p} - \vec{p}'),
\end{eqnarray}
and therefore the Weyl symbol of \m{\oper{A}} is the same irrespective of whether the operator is considered on \m{\hilspaceX} or on \m{\hilspacex}. In this case, we will use a unifying notation \m{\wsymb{\oper{A}}} instead of \m{\wsymbX{\oper{A}}} and \m{\wsymbx{\oper{A}}}. 

\subsubsection{Wigner functions and Wigner matrices}
\label{sec:WX}

The density operator of a given scalar field \(\psi\) is given by
\begin{equation}\label{eq:densopX}
\oper{W}_{\psi} \doteq (2\upi)^{-N} \ket{\psi}\bra{\psi}.
\end{equation}
The symbol of this operator, \m{W_{\psi} = \wsymbX{\oper{W}_\psi}}, is a real function called the Wigner function. It is given by
\begin{equation}\label{eq:Wigf}
W_{\psi}(\vec{X}, \vec{K}) 
= \frac{1}{(2\upi)^N} \int \dd\vec{S}\, 
\psi(\vec{X} + \vec{S}/2) \psi^{*}(\vec{X} - \vec{S}/2)\,\ee^{-\ii \vec{K} \cdot \vec{S}},
\end{equation}
which can be understood as the (inverse) Fourier image of
\begin{eqnarray}\label{eq:Ccor}
C_\psi(\vec{X}, \vec{S}) 
\doteq \psi(\vec{X} + \vec{S}/2) \psi^*(\vec{X} - \vec{S}/2) 
= \int \dd\vec{K}\,W_\psi(\vec{X}, \vec{K})\,\ee^{\ii \vec{K} \cdot \vec{S}}.
\end{eqnarray}
In particular, one has
\begin{eqnarray}\label{eq:redW}
\int \dd\vec{r}\,\wsymbX{\oper{W}_\psi}
= \wsymbx{\operst{W}_\psi}(\vec{p}),
\end{eqnarray}
where the right-hand side is \m{\st{W}_\psi} given by \eq{eq:aux10}, with \(\vec{p}\) treated as a parameter. Also, for real fields,
\begin{equation}
W_\psi(\vec{X}, \vec{K}) = W_\psi(\vec{X}, -\vec{K}).
\end{equation}

The density operator of a given vector field \(\vec{\psi} = \cb{\psi^1, \psi^2, \ldots, \psi^M}\) is a matrix operator
\begin{equation}\label{eq:boperW}
\boper{W}_{\vec{\psi}} \doteq (2\upi)^{-N} \ket{\vec{\psi}}\bra{\vec{\psi}}.
\end{equation}
The symbol of this operator, or the Wigner matrix, is a Hermitian matrix function
\begin{equation}\label{eq:wigtX}
\matr{W}_{\vec{\psi}}(\vec{X}, \vec{K}) 
= \frac{1}{(2\upi)^N} \int \dd\vec{S}\, 
\vec{\psi}(\vec{X} + \vec{S}/2) \vec{\psi}^\dag(\vec{X} - \vec{S}/2)\,\ee^{-\ii \vec{K} \cdot \vec{S}},
\end{equation}
which can be understood as the (inverse) Fourier image of
\begin{eqnarray}\label{eq:Ccor2}
\matr{C}_{\vec{\psi}}(\vec{X}, \vec{S}) 
\doteq \vec{\psi}(\vec{X} + \vec{S}/2) \vec{\psi}^\dag(\vec{X} - \vec{S}/2) 
= \int \dd\vec{K}\,\matr{W}_{\vec{\psi}}(\vec{X}, \vec{K})\,\ee^{\ii \vec{K} \cdot \vec{S}}.
\end{eqnarray}
In particular, one has
\begin{eqnarray}\label{eq:WWxx}
\int \dd\vec{r}\,\wsymbX{\boper{W}_{\vec{\psi}}}
= \wsymbx{\boperst{W}_{\vec{\psi}}}(\vec{p}),
\end{eqnarray}
where the right-hand side is \m{\matrst{W}_{\vec{\psi}}} given by \eq{eq:Wtenx}, with \(\vec{p}\) treated as a parameter. Also, for real fields,
\begin{equation}\label{eq:Wtstar}
\matr{W}_{\vec{\psi}}(\vec{X}, \vec{K}) 
= \matr{W}_{\vec{\psi}}^\intercal(\vec{X}, -\vec{K}) 
= \matr{W}_{\vec{\psi}}^*(\vec{X}, -\vec{K}).
\end{equation}

Like those on \m{\cb{\vecst{x}, \vecst{k}}}, the Wigner matrices (Wigner functions) on \m{\cb{\vec{X}, \vec{K}}} become positive-semidefinite (non-negative), and characterize local properties of the corresponding fields, when averaged over a sufficiently large phase-space volume \(\Delta\vec{X}\,\Delta\vec{K} \gtrsim 1\).

\subsection{Summary of \Sec{sec:primer}}
\label{eq:summath}

In summary, we have introduced a generic \m{n}-dimensional physical space \m{\vec{x}}, the dual \m{n}-dimensional wavevector space \m{\vec{k}}, the corresponding \m{\stn}-dimensional (\m{\stn = n + 1}) spacetime \m{\vecst{x} \equiv \cb{t, \vec{x}}}, and the dual \m{\stn}-dimensional wavevector space \m{\vecst{k} \equiv \cb{-\omega, \vec{k}}}. We have also introduced an \m{n}-dimensional momentum space \m{\vec{p}}, the corresponding \m{2n}-dimensional phase space \m{\vec{z} \equiv \cb{\vec{x}, \vec{p}}}, the \m{N}-dimensional (\m{N = 2n + 1}) extended space \m{\vec{X} \equiv \cb{t, \vec{z}} \equiv \cb{t, \vec{x}, \vec{p}}}, and the dual \m{N}-dimensional wavevector space \m{\vec{K} \equiv \cb{-\omega, \vec{q}} \equiv \cb{-\omega, \vec{k}, \vec{r}}}, where \m{\vec{r}} is the \m{n}-dimensional wavevector space dual to \m{\vec{p}}. We have also introduced the \m{2N}-dimensional phase space \m{\cb{\vec{X}, \vec{K}}}. Each of the said variables has a corresponding operator associated with it, which is denoted with a caret. For example, \m{\boper{x}} is the operator of position in the \m{\vec{x}} space, and \m{\boper{k} = -\ii\pd_{\vec{x}}} is the corresponding wavevector operator.

Functions on \m{\vecst{x}} form a Hilbert space \m{\hilspacex}, and the corresponding bra-ket notation is introduced as usual. Any operator \m{\operst{A}} on \m{\hilspacex} can be represented by its Weyl symbol \m{\st{A}(\vecst{x}, \vecst{k})}. The correspondence between operators and their symbols, \m{\operst{A} \leftrightarrow \st{A}}, is determined by the Wigner--Weyl transform and is natural in the sense that \eq{eq:natural} is satisfied. In particular, \m{\operst{A}\operst{B} \leftrightarrow \st{A} \star \st{B}}, where \m{\star} is the Moyal product on \m{\cb{\vecst{x}, \vecst{k}}}. When the geometrical-optics parameter is negligible (\m{\epsilon \to 0}), one has \m{\operst{A} = \st{A}(\boperst{x}, \boperst{k})} and the Moyal product becomes the usual product. Similarly, functions on \m{\vec{X}} form a Hilbert space \m{\hilspaceX}, the corresponding bra-ket notation is also introduced as usual, any operator \m{\oper{A}} on \m{\hilspaceX} can be represented by its Weyl symbol \m{A(\vec{X}, \vec{K})}, and \m{\oper{A}\oper{B} \leftrightarrow A \bigstar B}. An operator that is local in \m{\vec{p}} has the same symbol irrespective of whether it is considered on \m{\hilspacex} or on \m{\hilspaceX}.

Any given field \m{\psi} generates the corresponding density operator \m{(2\upi)^{-\stn} \ket{\psi}\bra{\psi}} and its symbol called the Wigner function (Wigner matrix, if the field is a vector). If the density operator is considered on \m{\hilspacex}, it is denoted \m{\operst{W}_\psi} and the corresponding Wigner function is denoted \m{\st{W}_\psi(\vecst{x}, \vecst{k})}. If the density operator is considered on \m{\hilspaceX}, it is denoted \m{\oper{W}_\psi} and the corresponding Wigner function is denoted \m{W_\psi(\vec{X}, \vec{K})}. The two Wigner functions are related via \m{\int \dd\vec{r}\,W_\psi(t, \vec{x}, \vec{p}, \omega, \vec{k}, \vec{r}) = \st{W}_\psi(t, \vec{x}, \omega, \vec{k}; \vec{p})}, where \m{\vec{p}} enters \m{\st{W}_\psi} as a parameter, if at all. If averaged over a sufficiently large phase-space volume, the Wigner functions (matrices) are non-negative (positive-semidefinite) and characterize local properties of the corresponding fields.

\section{Model}
\label{sec:model}

Here, we introduce the general assumptions and the key ingredients of our theory.

\subsection{Basic assumptions}
\label{sec:basm}

\subsubsection{Ordering}
\label{sec:ordering}

Let us consider particles governed by a Hamiltonian \(H = \avr{H} + \osc{H}\) such that
\begin{eqnarray}\label{eq:Hasm}
\osc{H} = \mc{O}(\varepsilon) \ll \avr{H} = \mc{O}(1).
\end{eqnarray}
In other words, \(\osc{H}\) serves as a small perturbation to the leading-order Hamiltonian \(\avr{H}\). The system will be described in canonical variables \m{\vec{z} \equiv \cb{\vec{x}, \vec{p}} \in \mathbb{R}^{2n}}. Let us also assume that the system is close to being homogeneous in \(\vec{x}\). This includes two conditions. First, we require that the external fields are weak (yet see \Sec{sec:qlapp}), meaning
\begin{eqnarray}\label{eq:order}
\pd_{\vec{x}}\avr{H} \sim \kappa_x\avr{H} = \mc{O}(\epsilon),
\qquad
\pd_{\vec{p}}\avr{H} \sim \kappa_p\avr{H} = \mc{O}(1), 
\end{eqnarray}
where \(\epsilon \ll 1\) is a small parameter, \(\kappa_x\) and \(\kappa_p\) are the characteristic inverse scales in the \(\vec{x}\) and \(\vec{p}\) spaces, respectively, and the bar denotes local averaging.\footnote{An exception will be made for eikonal waves, specifically, for quantities evaluated on the local wavevector \(\avr{\vecst{k}} \equiv \cb{-\avr{\omega}, \avr{\vec{k}}}\).} Hence, the particle momenta \(\vec{p}\) are close to being local invariants. Second, the \textit{statistical} properties of \m{\osc{H}} are also assumed to vary in \(\vec{x}\) slowly. These properties can be characterized using the density operator of the perturbation Hamiltonian,
\begin{eqnarray}\label{eq:WHH}
\oper{W} \doteq (2\upi)^{-N} \ket{\osc{H}}\bra{\osc{H}},
\end{eqnarray}
and its symbol, the (real) Wigner function, as in \eq{eq:aux10}:
\begin{eqnarray}\label{eq:WH}
W(\vec{X}, \vec{K}) 
= \frac{1}{(2\upi)^N}
\int \dd\vec{S}\,\osc{H}(\vec{X} + \vec{S}/2) \osc{H}(\vec{X} - \vec{S}/2)\,\ee^{-\ii \vec{K} \cdot \vec{S}}.
\end{eqnarray}
Specifically, we will use the \textit{average} Wigner function, \(\avr{W}\), which represents the Fourier spectrum of the symmetrized autocorrelation function of~\(\osc{H}\):
\begin{eqnarray}\label{eq:avCH}
\avr{C}(\vec{X}, \vec{S})
\doteq \avr{\osc{H}(\vec{X} + \vec{S}/2) \osc{H}(\vec{X} - \vec{S}/2)}
= \int \dd\vec{K}\,\avr{W}(\vec{X}, \vec{K})\,\ee^{\ii \vec{K} \cdot \vec{S}}.
\end{eqnarray}
The averaging is performed over sufficiently large volume of \(\vec{x}\) to eliminate rapid oscillations and also over phase-space volumes \(\Delta\vec{X}\,\Delta\vec{K} \gtrsim 1\), which guarantees \(\avr{W}\) to be non-negative and local (\Sec{sec:WX}). The function \(\avr{W}\) can be understood as a measure of the phase-space density of wave quanta when the latter is well defined (\Sec{sec:onshell}). 

We will assume\footnote{As a reminder, the notation \m{A = \mc{O}(\epsilon)} does not rule out the possibility that \m{A/\epsilon} is small. Also note that the terms `\(\sim\)' and `of order' in this paper mean the same as `\(\mc{O}\)'.}
\begin{eqnarray}\label{eq:wigorder}
\pd_t \avr{W} = \mc{O}(\epsilon),
\qquad
\pd_\vec{x} \avr{W} = \mc{O}(\epsilon),
\qquad
\pd_\vec{p} \avr{W} = \mc{O}(1).
\end{eqnarray}
That said, we will also allow (albeit not require) for oscillations to be constrained by a dispersion relation. In this case, \(\avr{W}\, \propto\, \delta (\omega -\avr{\omega}(t, \vec{x}))\), so \eq{eq:wigorder} \perse is not satisfied; then we assume a similar ordering for \(\int \dd\omega\,\avr{W}\) instead. Also note that in application to the standard QLT of homogeneous turbulence \citep[chapter~16]{book:stix}, \(\epsilon\) is understood as the geometrical-optics parameter characterizing the smallness of the linear-instability growth rates. (We discuss the ordering further in the end of \Sec{sec:avrf}.)

\subsubsection{Quasilinear approximation}
\label{sec:qlapp}

The particle-motion equations can be written as
\begin{eqnarray}
\dot{z}^{\alpha} = \poissonz{z^{\alpha}, \avr{H} + \osc{H}} 
= v^{\alpha} + u^{\alpha},
\end{eqnarray}
where \(v^{\alpha}\) and \(u^{\alpha}\) are understood as the unperturbed phase-space velocity and the perturbation to the phase-space
velocity, respectively:
\begin{eqnarray}\label{eq:uv}
v^{\alpha} \doteq J^{\alpha\beta} \pd_{\beta}\avr{H},
\qquad
u^{\alpha} \doteq J^{\alpha\beta}\pd_{\beta}\osc{H}.
\end{eqnarray}
The notation \(v^i\) (with \(i = 1, 2, \ldots n\)) will also be used for the \textit{spatial part} of the phase-space velocity \(v^{\alpha}\), \ie for the true velocity \perse. Likewise, \(\vec{v}\) will be used to denote either the phase-space velocity vector or the spatial velocity vector depending on the context. Also note that a slightly different definition of \(\vec{v}\) will be used starting from \Sec{sec:Deq}.

The corresponding Klimontovich equation for the particle distribution \(f(t, \vec{z})\) is
\begin{eqnarray}\label{eq:liouv}
\pd_t f = \poissonz{\avr{H} + \osc{H}, f}.
\end{eqnarray}
(If collisions are not of interest, \eq{eq:liouv} can as well be understood as the Vlasov equation. Also, a small collision term can be included \adhoc; see the comment in the end of \Sec{sec:avrf}.) Let us search for \(f\) in the form
\begin{eqnarray}
f = \avr{f} + \osc{f},
\qquad
\avr{\osc{f}} = 0.
\end{eqnarray}
The equations for \m{\avr{f}} and \m{\osc{f}} are obtained as the average and oscillating parts of \eq{eq:liouv}, and we neglect the nonlinearity in the equation for \m{\osc{f}}, following the standard QL approximation \citep[chapter~16]{book:stix}. Then, one obtains
\begin{eqnarray}
\pd_t\avr{f} = \poissonz{\avr{H}, \avr{f}} + \avr{\poissonz{\osc{H}, \osc{f}}},\label{eq:fbareq}
\\
\pd_t\osc{f} = \poissonz{\avr{H}, \osc{f}} + \poissonz{\osc{H}, \avr{f}}.\label{eq:fteq}
\end{eqnarray}

A comment is due here regarding plasmas in strong fields and magnetized plasmas in particular. Our formulation can be applied to such plasmas in canonical angle--action variables \m{\cb{\vec{\phi}, \vec{\msf{J}}}}. For fast angle variables, the ordering \eq{eq:order} is not satisfied and the Weyl symbol calculus is inapplicable as~is (see the footnote on p.~\pageref{ft:rigas}). Such systems can be accommodated by representing the distribution function as a Fourier series in \m{\vec{\phi}} and treating the individual-harmonic amplitudes separately as slow functions of the remaining coordinates. Then, our averaging procedure subsumes averaging over \m{\vec{\phi}}, so the averaged quantities are \m{\vec{\phi}}-independent and \eq{eq:order} is reinstated. In particular, magnetized plasmas can be described using guiding-center variables. Although not canonical by default \citep{ref:littlejohn83}, they can always be cast in a canonical form, at least in principle \citep{ref:littlejohn79}. Examples of canonical guiding-center variables are reviewed in \citep{ref:cary09}. To make the connection with the homogeneous-plasma theory, one can also order the canonical pairs of guiding-center variables such that they would describe the gyromotion, the parallel motion, and the drifts separately \citep{ref:wong00}. This readily leads to results similar to those in \citep{ref:catto17}. Further discussions on this topic are left to future papers.

\subsection{Equation for \mt{\osc{f}}}
\label{sec:ft}

Let us consider solutions of \eq{eq:fteq} as a subclass of solutions of the more general equation
\begin{eqnarray}\label{eq:aux71}
\pd_\tau\osc{f} = \oper{L}\osc{f} + \mcu{F},
\qquad
\mcu{F}(\vec{X}) \doteq \poissonz{\osc{H}, \avr{f}}.
\end{eqnarray}
Here, we have introduced an auxiliary second `time' \m{\tau}, the operator
\begin{eqnarray}\label{eq:Loper}
\oper{L} \doteq -\pd_t + \poissonz{\avr{H}, \placeholder} 
= -\pd_t + J^{\alpha\beta}(\pd_{\alpha}\avr{H})\pd_{\beta}
= -\pd_t - v^{\lambda}\pd_{\lambda}
= -V^a \pd_a
\end{eqnarray}
(here and further, \(\placeholder\) denotes a placeholder), and \(\vec{V}(\vec{X}) \equiv \cb{1, \vec{v}(t, \vec{z})}\) is the unperturbed velocity in the \(\vec{X}\) space. Note that
\begin{eqnarray}\label{eq:VV}
\pd_a V^a = \pd_{\lambda}v^{\lambda} = 0
\end{eqnarray}
due to the incompressibility of the phase flow. Hence, \m{[\pd_a, V^a] = 0}, so \m{\oper{L}} is anti-Hermitian.

Let us search for a solution of \eq{eq:aux71} in the form\footnote{Using the auxiliary variable \m{\tau} allows us to express the propagator as a regular exponential, rather than ordered exponential, even for \(t\)-dependent \m{\avr{H}}, because \m{\oper{L}} is independent of \m{\tau}.}
\begin{eqnarray}\label{eq:fL}
\osc{f}(\tau, \vec{X}) = \ee^{\oper{L} \tau} \xi (\tau, \vec{X}).
\end{eqnarray}
Then, \(\pd_\tau\osc{f} = \oper{L}\osc{f} + \ee^{\oper{L} \tau}\pd_\tau\xi\), so \(\pd_\tau\xi = \ee^{-\oper{L} \tau}\mcu{F}(\vec{X})\) and therefore
\begin{eqnarray}
\xi(\tau, \vec{X}) = \ee^{-\oper{L} \tau_0} \xi_0(\vec{X}) + \int_{\tau_0}^\tau \dd \tau'\,\ee^{-\oper{L} \tau'} \mcu{F}(\vec{X}),
\end{eqnarray}
where \m{\xi_0(\vec{X}) \doteq \osc{f}(\tau_0, \vec{X})}. Hence, one obtains
\begin{eqnarray}
\osc{f}(\tau, \vec{X}) 
= \ee^{\oper{L}(\tau - \tau_0)} \xi_0(\vec{X})
+ \int_{\tau_0}^\tau \dd \tau'\,\ee^{-\oper{L} (\tau' - \tau)} \mcu{F}(\vec{X}),
\end{eqnarray}
or equivalently, using \(\tau'' \doteq \tau - \tau'\),
\begin{eqnarray}\label{eq:ft2}
\osc{f}(\tau, \vec{X}) = g_0(\tau, \vec{X}) 
+ \int_0^{\tau - \tau_0} \dd\tau''\,\oper{T}_{\tau''} \mcu{F}(\vec{X}).
\end{eqnarray}
Here, \m{g_0} is a solution of \(\pd_\tau g_0 = \oper{L} g_0\), specifically,
\begin{eqnarray}\label{eq:gt0}
g_0(\tau, \vec{X}) \doteq \oper{T}_{\tau - \tau_0} \xi_0(\vec{X}),
\qquad
g_0(\tau_0, \vec{X}) = \osc{f}(\tau_0, \vec{X}),
\end{eqnarray} 
and we have also introduced
\begin{eqnarray}\label{eq:Tr2}
\oper{T}_\tau \doteq \ee^{\oper{L} \tau} = \ee^{-\tau V^a\pd_a}.
\end{eqnarray}
Because \m{\oper{L}} is anti-Hermitian, the operator \m{\oper{T}_\tau} is unitary, and comparison with \eq{eq:eshift} shows that it can be recognized as a shift operator. For further details, see \Sec{sec:shift}. 

Using \m{\oper{T}_\tau}, one can express \eq{eq:ft2} as
\begin{eqnarray}\label{eq:ft3}
\osc{f} = g_0 + \oper{\mcu{G}}\mcu{F}, 
\qquad 
\textstyle
\oper{\mcu{G}} \doteq \int_0^{\tau-\tau_0} \dd\tau'\,\oper{T}_{\tau'},
\end{eqnarray}
where \(\oper{\mcu{G}}\) is the Green's operator understood as the right inverse of the operator \m{\pd_\tau - \oper{L}}, or on the space of \m{\tau}-independent functions, \m{\pd_t - \poissonz{\avr{H}, \placeholder}}. Let us rewrite this operator as \m{\oper{\mcu{G}} = \oper{\mcu{G}}_< + \oper{\mcu{G}}_>}, where
\begin{eqnarray}
\oper{\mcu{G}}_< = \int_0^{\tau-\tau_0} \dd\tau'\,\ee^{-\nu \tau'}\oper{T}_{\tau'},
\qquad
\oper{\mcu{G}}_> = \int_0^{\tau-\tau_0} \dd\tau'\,(1 - \ee^{-\nu \tau'})\oper{T}_{\tau'},
\end{eqnarray}
and \(\nu\) is a positive constant. Note that \(\oper{\mcu{G}}_<\) is well defined at \(\tau_0 \to -\infty\), meaning that \(\oper{\mcu{G}}_<\mcu{F}\) is well defined for any physical (bounded) field \(\mcu{F}\).\footnote{Unlike classic plasma-wave theory, this approach does not involve spectral decomposition, so there is no need to consider fields that are exponential in time on the \textit{whole} interval \m{(-\infty, \infty)}.} Thus, so is \m{g_0 + \oper{\mcu{G}}_>\mcu{F}}. Let us take \(\tau_0 \to -\infty\) and then take \(\nu \to 0+\). (Here, \(0+\) denotes that \(\nu\) must remain positive, \ie the \textit{upper} limit is taken.) Then, \eq{eq:ft3} can be expressed as
\begin{eqnarray}\label{eq:ft4}
\osc{f} = g + \oper{G}\mcu{F}, 
\qquad 
g \doteq \lim_{\nu \to 0+} \lim_{\tau_0 \to -\infty} (g_0 + \oper{\mcu{G}}_>\mcu{F}).
\end{eqnarray}
Here, we introduced an `effective' Green's operator \m{\oper{G} \doteq \lim_{\nu \to 0+} \lim_{\tau_0 \to -\infty} \oper{\mcu{G}}_<}, \ie
\begin{eqnarray}\label{eq:greeneff}
\oper{G} \doteq \lim_{\nu \to 0+} \int_0^\infty \dd\tau\,\ee^{-\nu \tau}\oper{T}_{\tau}.
\end{eqnarray}
This operator will be discussed in \Sec{sec:G}, and \(g\) will be discussed in \Sec{sec:g}. Meanwhile, note that because \m{\tau} is just an auxiliary variable, we will be interested in solutions independent of \(\tau\). In particular, this means that \m{\osc{f}(\tau_0, \vec{X}) = \osc{f}(\vec{X})}, so \m{\xi_0(\vec{X}) = \osc{f}(\vec{X})}, so \eq{eq:gt0} leads to
\begin{eqnarray}\label{eq:g0def}
g_0(\tau, \vec{X}) 
= \oper{T}_{\tau - \tau_0} \osc{f}(\vec{X}).
\end{eqnarray}

\subsection{Equation for \mt{\avr{f}}}
\label{sec:avrf}

Using \eq{eq:ft3}, one can rewrite \eq{eq:fteq} for \(\avr{f}\) as follows:
\begin{eqnarray}
\pd_t\avr{f} = \poissonz{\avr{H}, \avr{f}}
+ \avr{\poissonz{\osc{H}, g}} 
+ \avr{\poissonz{\osc{H}, \oper{G}\poissonz{\osc{H}, \avr{f}}}}
.\end{eqnarray}
Notice that
\begin{equation}
\poissonz{\osc{H}, g}
= - \poissonz{g, \osc{H}}
= -\pd_{\alpha} (J^{\alpha\beta}\, g \pd_{\beta} \osc{H})
= -\pd_{\alpha} (u^\alpha g)
\end{equation}
and also
\begin{align}
\poissonz{\osc{H}, \oper{G}\poissonz{\osc{H}, \avr{f}}}
& = \pd_{\beta}(
J^{\alpha\beta} (\pd_{\alpha}\osc{H})
\oper{G}\poissonz{\osc{H}, \avr{f}}
)
\notag \\
& = \pd_{\beta}(J^{\alpha\beta}
(\pd_{\alpha}\osc{H})
\oper{G}
(J^{\mu \nu} (\pd_{\mu}\osc{H})(\pd_{\nu}\avr{f})))
\notag \\
& = \pd_{\beta}(u^{\beta}\oper{G}(u^{\nu}\pd_{\nu}\avr{f})).
\label{eq:aux3}
\end{align}
The field \(u^\alpha\) enters here as a multiplication factor and can be considered as an operator:
\begin{eqnarray}\label{eq:u}
\oper{u}^\alpha\psi(\vec{X}) \doteq u^\alpha(\vec{X})\psi(\vec{X}).
\end{eqnarray}
Then, \eq{eq:aux3} can be compactly represented as
\begin{eqnarray}
\poissonz{\osc{H}, \oper{G}\poissonz{\osc{H}, \avr{f}}} 
= \pd_{\alpha} (\oper{u}^{\alpha} \oper{G} \oper{u}^{\beta} \pd_{\beta} \avr{f}).
\end{eqnarray}
We will also use the notation
\begin{eqnarray}\label{eq:ddt}
\dd_t \doteq \pd_t + v^{\gamma}\pd_{\gamma} = \pd_t - \poissonz{\avr{H}, \placeholder}.
\end{eqnarray}
This leads to the following equation for \(\avr{f}\):
\begin{eqnarray}\label{eq:ql1}
\dd_t \avr{f}
= \pd_{\alpha}(\oper{D}^{\alpha\beta} \pd_{\beta}\avr{f}) + \Gamma, 
\end{eqnarray}
where we introduced the following average quantities:
\begin{eqnarray}\label{eq:GD}
\oper{D}^{\alpha\beta} \doteq \avr{\oper{u}^{\alpha}\oper{G}\oper{u}^{\beta}},
\qquad
\Gamma \doteq - \pd_\alpha(\avr{u^{\alpha} g}),
\end{eqnarray}

Our goal is to derive explicit approximate expressions for the quantities \eq{eq:GD} and to rewrite \eq{eq:ql1} in a more tractable form using the assumptions introduced in \Sec{sec:basm}. We will use\footnote{Starting with \Sec{sec:Deq}, we will assume \m{\dd_t\avr{f} \sim \epsilon\varepsilon^2 \avr{f}} instead.}
\begin{eqnarray}
\pd_t\avr{f}  \sim \poissonz{\avr{H}, \avr{f}} = \mc{O}(\epsilon),
\qquad
\dd_t\avr{f} = \mc{O}(\varepsilon^2), 
\label{eq:asmff}
\end{eqnarray}
and we will keep terms of order \(\epsilon\), \(\varepsilon^2\), and \m{\epsilon \varepsilon^2} in the equation for \m{\avr{f}}, while terms of order \m{\varepsilon^4}, \m{\epsilon^2 \varepsilon^2}, and higher will be neglected. This implies the ordering
\begin{eqnarray}\label{eq:ordering}
\varepsilon^2 \ll \epsilon \ll \varepsilon \ll 1.
\end{eqnarray}
As a reminder, \m{\varepsilon} is a linear measure of the characteristic amplitude of oscillations, and \m{\epsilon} is the geometrical-optics parameter, which is proportional to the inverse scale of the plasma inhomogeneity in spacetime. As usual then, linear dissipation is assumed to be of order \m{\epsilon}. This model implies the assumption that collisionless dissipation is much stronger than collisional dissipation, which is to emerge as an effect quadratic in \m{\osc{f}} (\Sec{sec:th}). Furthermore, the inverse plasma parameter\footnote{By the plasma parameter we mean the number of particles within the Debye sphere.} will be assumed to be of order \m{\epsilon}, so the collision operator for \m{\avr{f}} (\Sec{sec:cop}) will be of order \m{\epsilon\varepsilon^2}. Within the assumed accuracy, this operator must be retained, while the dynamics of \m{\osc{f}} is considered linear and therefore collisionless. Alternatively, one can switch from the Klimontovich description to the Vlasov--Boltzmann description and introduce an \adhoc order-\m{\epsilon} collision operator directly in \eq{eq:liouv}. This will alter the Green's operator, but the conceptual formulation would remain the same, so it will not be considered separately in detail.

\subsection{Summary of \Sec{sec:model}}
\label{sec:summodel}

Our QL model is defined as usual, except: (i)~we allow for a general particle Hamiltonian~\m{H}; (ii)~we use the Klimontovich equation rather than the Vlasov equation to retain collisions; (iii)~we use \textit{local} averaging (denoted with overbar) and allow for weak inhomogeneity of all averaged quantities; (iv)~we retain the initial conditions~\m{g} for the oscillating part of the distribution function (defined as in \eq{eq:ft4} but yet to be calculated explicitly). Then, the average part of the distribution function satisfies
\begin{eqnarray}\label{eq:vfs}
\pd_t\avr{f} - \poissonz{\avr{H}, \avr{f}} = \pd_{\alpha}(\oper{D}^{\alpha\beta} \pd_{\beta}\avr{f}) + \Gamma, 
\end{eqnarray}
where \m{\oper{D}^{\alpha\beta} \doteq \avr{\oper{u}^{\alpha}\oper{G}\oper{u}^{\beta}}}, \m{\Gamma \doteq - \pd_\alpha(\avr{u^{\alpha} g})}, \m{u^{\alpha}} is the wave-driven perturbation of the phase-space velocity (see \eq{eq:uv}), \m{\oper{u}^{\alpha}} is the same quantity considered as an operator on \m{\hilspaceX} (see \eq{eq:u}), and \m{\oper{G} } is the `effective' Green's operator given by
\begin{eqnarray}
\oper{G} \doteq \lim_{\nu \to 0+} \int_0^\infty \dd\tau\,\ee^{-\nu \tau-\tau V^a\pd_a}.
\end{eqnarray}
Also, \m{\pd_{\alpha} \equiv \pd/\pd z^\alpha}, and \m{\poissonz{\placeholder, \placeholder}} is the Poisson bracket on the particle phase space \m{\vec{z}}. The equation for \m{\avr{f}} used in the standard QLT is recovered from \eq{eq:vfs} by neglecting \m{\Gamma} and the spatial gradients (in particular, the whole Poisson bracket) and also by approximating the operator \m{\oper{D}^{\alpha\beta}} with a local function of \m{\vec{z}}.

\section{Preliminaries}
\label{sec:prelim0}

Before we start calculating the functions in \eq{eq:vfs} explicitly, let us get some preliminaries out of the way. In this section, we discuss the shift operators~\m{\oper{T}_\tau} (\Sec{sec:shift}), approximate the operator~\m{\oper{G}} (\Sec{sec:G}), and develop a model for the function \(g\) that encodes the initial conditions for~\m{\osc{f}} (\Sec{sec:g}). 

\subsection{Shift operator}
\label{sec:shift}

Here, we derive some properties of the shift operator \m{\oper{T}_\tau} introduced in \Sec{sec:ft}.

\subsubsection{\mt{\oper{T}_\tau} as a shift}
\label{eq:shift1}

Here, we formally prove (an admittedly obvious fact) that
\begin{eqnarray}\label{eq:Tr12}
\oper{T}_\tau \psi(\vec{X}) = \psi(\vec{X} - \vec{\ell}_\tau(\vec{X})),
\qquad
\textstyle
\ell_{\tau}^{a}(\vec{X}) \doteq \int_0^{\tau} \dd t\, V^{a}(\vec{Y}(t, \vec{X})),
\end{eqnarray}
where the `characteristics' \m{Y^a} solve\footnote{In terms of \m{t'\doteq t - \tau}, \eq{eq:yc} has a more recognizable form \m{\dd Y^{a}/\dd t' = V^{a}(\vec{Y})}, with \m{Y^a(t' = t) = X^a}.}
\begin{eqnarray}\label{eq:yc}
\frac{\dd Y^{a}}{\dd\tau} = -V^{a}(\vec{Y}),
\qquad
Y^a(\tau = 0) = X^a,
\end{eqnarray}
and thus \(\ell_{\tau}^{a}\) can be Taylor-expanded in \(\tau\) as
\begin{eqnarray}\label{eq:ad}
\ell_{\tau}^{a}(\vec{X}) = \tau V^{a} - \frac{1}{2}\,\tau^2 V^{b} \pd_{b} V^{a} + \ldots,
\qquad
V^{a} \equiv V^{a}(\vec{X}).
\end{eqnarray}

As the first step to proving \eq{eq:Tr12}, let us Taylor-expand \(V^{a}\) around a fixed point \(\vec{X}_1\):
\begin{eqnarray}
V^{a} = V^{a}_1 + (\pd_{b} V^{a}_1)\,\delta X^{b} + \ldots,
\qquad
\delta X^{a}\doteq X^{a} - X_1^{a},
\end{eqnarray}
where \(V^a_1 \equiv V^a(\vec{X}_1)\). If one neglects the \textit{first} and higher derivatives of \(V^{a}\), one obtains
\begin{eqnarray}
\oper{T}_{\tau}\psi(\vec{X})
\approx \ee^{-\tau V^{a}_1\pd_{a}}\psi(\vec{X})
= \psi(\vec{X} - \tau \vec{V}_1).
\end{eqnarray}
By taking the limit \(\vec{X}_1 \to \vec{X}\), which corresponds to \(\vec{V}_1 \to \vec{V}\), one obtains
\begin{eqnarray}
\oper{T}_{\tau}\psi(\vec{X}) = \psi(\vec{X} - \tau \vec{V}) + \mc{O}(\tau^2).
\label{eq:aux17}
\end{eqnarray}
Similarly, if one neglects the \textit{second} and higher derivatives of \(V^{a}\), one obtains\footnote{We use the Zassenhaus formula
\m{\ee^{\oper{A} + \oper{B}} = 
\ee^{\oper{A}}\,
\ee^{\oper{B}}\,
\ee^{-[\oper{A},\oper{B}]/2}
\ee^{[\oper{B}, [\oper{A}, \oper{B}]]/3+[\oper{A}, [\oper{A}, \oper{B}]]/6}
\cdots}\vphantom{\Big|}}
\begin{align}
\oper{T}_{\tau}\psi(\vec{X})
& = \ee^{
- \tau (V^{a}_1 + (\pd_{b}V^{a}_1) \delta X^{b}+\ldots) \pd_{a}
} \psi(\vec{X})
\notag \\
& \approx \ee^{- \tau (\pd_{b} V^{a}_1) \delta X^{b} \pd_{a}}
\,\ee^{-\tau V^{a}_1 \pd_{a}}
\,\ee^{-\frac{1}{2}[
-\tau(\pd_{b}V^{a}_1) \delta X^{b}\pd_{a},
-\tau V^{c}_1\pd_{c}
]} \psi(\vec{X})
\notag \\
& \approx \ee^{
-\tau (\pd_{b} V^{a}_1) \delta X^{b} \pd_{a}}
\,\ee^{-\tau V^{a}_1 \pd_{a}}
\,\ee^{
\frac{1}{2}\tau^2 V^{c}_1(\pd_{b}V^{a}_1)
[\pd_{c}, \delta X^{b}\pd_{a}]
}\psi(\vec{X})
\notag \\
& \approx \ee^{-\tau (\pd_{b}V^{a}_1)\delta X^{b}\pd_{a}}
\,\ee^{-\tau V^{a}_1 \pd_{a}}
\,\ee^{\frac{1}{2}\tau^2 V^{b}_1 (\pd_{b}V^{a}_1) \pd_{a}}
\psi(\vec{X})
\notag \\
& \approx \ee^{-\tau (\pd_{b}V^{a}_1) \delta X^{b} \pd_{a}}
\,\ee^{-\tau V^{a}_1\pd_{a}
+ \frac{1}{2}\tau^2 V^{b}_1(\pd_{b}V^{a}_1)\pd_{a}}\psi(\vec{X})
\notag \\
& \approx \ee^{-\tau (\pd_{b}V^{a}_1)\delta X^{b}\pd_{a}}
\psi(\vec{X} - \tau \vec{V}_1 + \textstyle\frac{1}{2}\,\tau^2 V^{b}_1 \pd_{b}\vec{V}_1).
\end{align}
In the limit \(\vec{X}_1 \to \vec{X}\), when \m{\ee^{-\tau(\pd_{b}V^{a}_1)\delta X^{b} \pd_{a}} \to 1} and \(\vec{V}_1 \to \vec{V}\), one obtains
\begin{eqnarray}
\oper{T}_{\tau}\psi(\vec{X})
= \psi\left(\vec{X} - \tau \vec{V}(\vec{X}) + \frac{1}{2}\,\tau^2 (\vec{V} \cdot \pd_{\vec{X}}) \vec{V}\right) + \mc{O}(\tau^3).
\label{eq:aux16}
\end{eqnarray}
In conjunction with \eq{eq:ad}, equations \eq{eq:aux17} and \eq{eq:aux16} show agreement with the sought result \eq{eq:ad} within the assumed accuracy. One can also retain \(\msf{m}\) derivatives of \(\vec{V}\) and derive the corresponding approximations similarly. Then the error will be \m{\mc{O}(\tau^{\msf{m} + 2})}. 

For an order-one time interval \(\tau\), one can split this interval on \(N_\tau \gg 1\) subintervals of small duration \(\tau/N_\tau\) and apply finite-\(\msf{m}\) formulas (for example, \eq{eq:aux17} or \eq{eq:aux16}) to those. Then the total error scales as \m{\mc{O}(N_\tau^{-\msf{m} - 1})} and the exact formula \eq{eq:Tr12} is obtained at \(N_\tau \to \infty\).

\subsubsection{Symbol of \mt{\oper{T}_\tau}}
\label{sec:Tsymb}

Using the bra-ket notation, \eq{eq:Tr12} can be written as
\begin{eqnarray}
\braket{\vec{X} |\oper{T}_{\tau} | \psi} = \braket{\vec{X} - \vec{\ell}_{\tau}(\vec{X})|\psi}.
\end{eqnarray}
Thus, \m{\bra{\vec{X}} \oper{T}_{\tau} = \bra{\vec{X} - \vec{\ell}_{\tau}(\vec{X})}},~so
\begin{eqnarray}
\braket{\vec{X}_1|\oper{T}_{\tau}|\vec{X}_2}
= \braket{\vec{X}_1 - \vec{\ell}_{\tau}(\vec{X}_1)|\vec{X}_2} 
= \delta (\vec{X}_1 - \vec{X}_2 - \vec{\ell}_{\tau}(\vec{X}_1)). 
\end{eqnarray}
Using \eq{eq:WWT}, one obtains the Weyl symbol of \m{\oper{T}_{\tau}} in the form
\begin{eqnarray}\label{eq:aux15}
T_{\tau}(\vec{X}, \vec{K}) 
= \int \dd\vec{S}\, \ee^{-\ii \vec{K} \cdot \vec{S}}\,
\delta (\vec{S} - \vec{\ell}_{\tau}(\vec{X} + \vec{S}/2)).
\end{eqnarray}
From \eq{eq:ad}, one has
\begin{align}
\ell^a_{\tau}(\vec{X} + \vec{S}/2) 
& = \tau V^{a}(\vec{X} + \vec{S}/2) - (\tau^2/2)\,V^{b} \pd_{b} V^{a} + \mc{O}(\epsilon^2) 
\notag\\
& = \tau V^{a} + (\tau/2) (\pd_b V^{a}) S^b - (\tau^2/2)\,V^{b} \pd_{b} V^{a} + \mc{O}(\epsilon^2) 
\notag\\
& = {M^a}_b V^{b}\tau  + {m^a}_b S^b + \mc{O}(\epsilon^2),
\label{eq:mmm}
\end{align}
where we introduced a matrix \m{\matr{M} \doteq \matr{1} - \matr{m}}, or explicitly,
\begin{eqnarray}
{M^a}_b \doteq \delta^a_b - {m^a}_b, 
\qquad
{m^a}_b \doteq (\tau/2)(\pd_b V^{a}).
\end{eqnarray}
Let us express the term \m{\mc{O}(\epsilon^2)} in \eq{eq:mmm} as \m{-{M^a}_b\mu^b}. Then,
\begin{align}
\delta (\vec{S} - \vec{\ell}_{\tau}(\vec{X} + \vec{S}/2))
& = \delta (\vec{S} - \matr{M}\vec{V}\tau - \vec{m}\vec{S} + \matr{M}\vec{\mu})
\notag\\
& = \delta (\matr{M}(\vec{S} - \vec{V}\tau + \vec{\mu}))
\notag\\
& = \delta (\vec{S} - \vec{V}\tau + \vec{\mu})/|\det \matr{M}|.
\end{align}
Because \m{\matr{m} = \mc{O}(\epsilon)}, the well-known formula yields \m{\det \matr{M} = 1 + \tr\matr{m} + \mc{O}(\epsilon^2)}. But \(\tr\matr{m} = 0\) by \eq{eq:VV}, so
\begin{eqnarray}
\delta (\vec{S} - \vec{\ell}_{\tau}(\vec{X} + \vec{S}/2)) = \delta (\vec{S} - \vec{V}\tau + \vec{\mu})  + \mc{O}(\epsilon^2).
\end{eqnarray}
The last term \m{\mc{O}(\epsilon^2)} is insignificant and can be neglected right away, so \eq{eq:aux15} leads to
\begin{eqnarray}\label{eq:TsymbM0}
T_{\tau}(\vec{X}, \vec{K}) \approx \exp(\ii \tau \Omega(\vec{X}, \vec{K}) + \ii \vec{K} \cdot \vec{\mu}),
\end{eqnarray}
where we have introduced the following notation:
\begin{eqnarray}
\label{eq:Oms}
\Omega(\vec{X}, \vec{K}) \doteq - \vec{K} \cdot \vec{V}(\vec{X})
= \omega - q_{\alpha} v^{\alpha} 
= \omega - \vec{k} \cdot \vec{v} + \mc{O}(\epsilon).
\end{eqnarray}
By definition, \m{\vec{\mu}} is a polynomial of \m{\tau} with coefficients that are of order \m{\epsilon^2} and therefore small. But because \m{\tau} can be large, and because \m{\vec{\mu}} is under the exponent, this makes \m{T_{\tau}} potentially sensitive to this term, so we retain it (for now).

\subsection{Effective Green's operator}
\label{sec:G}

The effective Green's operator \eq{eq:greeneff} can be understood as the right inverse of the operator (cf. \Sec{sec:ft})
\begin{eqnarray}\label{eq:aux4}
\oper{L}_\text{eff} \doteq \lim_{\nu \to 0+} (\pd_t - \poissonz{\avr{H}, \placeholder} + \nu),
\end{eqnarray}
so we denote it also as \(\oper{G} = \oper{L}^{-1}_\text{eff}\) (which is admittedly abuse of notation). Because \m{\oper{L}_\text{eff}} has real \(\vec{X}\)~representation by definition, the \(\vec{X}\)~representation of \m{\oper{G}} is real too. In particular, \m{\braket{\vec{X} + \vec{S}/2 | \oper{G} | \vec{X} - \vec{S}/2}} is real, hence
\begin{eqnarray}\label{eq:Gstar}
G(\vec{X}, -\vec{K}) = G^*(\vec{X}, \vec{K})
\end{eqnarray}
by definition of the Weyl symbol \eq{eq:WWT}. As a corollary, the derivative of \(G(\vec{X}, \vec{K})\) with respect to the \(a\)th component of the whole second argument, denoted \(G^{|a}\), satisfies
\begin{eqnarray}\label{eq:Gstar2}
(G^{|a}(\vec{X}, \vec{K}))^* = -G^{|a}(\vec{X}, -\vec{K}).
\end{eqnarray}
Also note that \(\oper{G}\) can be expressed as
\begin{eqnarray}
\oper{G} = 
\lim_{\nu \to 0+} \ii (\oper{\omega} - \dot{x}^i \oper{k}_i - \dot{p}_i \oper{r}^i + \ii \nu)^{-1}
\end{eqnarray}
(the notation `\m{\lim_{\nu \to 0+} A(\omega + \ii \nu)}' will also be shortened as `\m{A(\omega + \ii 0)}'), whence
\begin{eqnarray}\label{eq:Gr}
\frac{\pd G}{\pd r^i} = \mc{O}(\dot{p}_i) = \mc{O}(\epsilon). 
\end{eqnarray}

Due to \eq{eq:TsymbM0}, the leading-order approximation of the symbol of the operator \eq{eq:greeneff} is \(G(\vec{X}, \vec{K}) = G_0(\Omega(\vec{X}, \vec{K}))\), where
\begin{eqnarray}\label{eq:G00}
G_0(\Omega) \doteq 
\lim_{\nu \to 0+}\int_0^{\infty}\dd\tau\, \ee^{-\nu \tau + \ii\Omega \tau}
= \upi\,\delta(\Omega) + \ii\,\pv\frac{1}{\Omega}
\end{eqnarray}
and the (standard) notation \(\pv (1/\Omega)\) is defined as follows:
\begin{eqnarray}\label{eq:pv}
\pv\frac{1}{\Omega} \doteq \lim_{\nu \to 0+} \frac{\Omega}{\nu^2 + \Omega^2}.
\end{eqnarray}
This means, in particular, that for any \(A\), one has
\begin{align}
\mc{J}[A, G_0]
& \doteq \int \dd\vec{K}\,A(\vec{X}, \vec{K}) G_0(\Omega(\vec{X}, \vec{K}))
\notag\\
& =\upi \int \dd\vec{K}\,A(\vec{X}, \vec{K}) \delta(\Omega (\vec{X}, \vec{K}))
+ \ii \fint \dd\vec{K}\,\frac{A(\vec{X}, \vec{K})}{\Omega(\vec{X}, \vec{K})}, 
\end{align}
where \(\fint\) is a principal-value integral. Also usefully, \m{\avr{G}_0 = G_0} and 
\begin{align}
\pd_a \mc{J}[A, G_0]
& = \int \dd\vec{K}\,A(\vec{X}, \vec{K}) G_0'(\Omega(\vec{X}, \vec{K}))\,\pd_a\Omega(\vec{X}, \vec{K})
\notag\\
& = -(\pd_a V^b(\vec{X})) \int \dd\vec{K}\,K_b A(\vec{X}, \vec{K}) G_0'(\Omega(\vec{X}, \vec{K}))
\notag\\
& = -(\pd_a V^b(\vec{X}))\,\frac{\eth}{\pd \Omega} \int \dd\vec{K}\,K_b A(\vec{X}, \vec{K}) G_0(\Omega (\vec{X}, \vec{K})), 
\end{align}
where the notation \(\eth /\pd \lambda \equiv \eth_\lambda\) is defined, for any \m{\lambda} and \(Q\), as follows:
\begin{eqnarray}\label{eq:eth}
\frac{\eth}{\pd \lambda} \int Q(\lambda) \doteq \left(\frac{\pd}{\pd \vartheta}\,\int Q(\lambda + \vartheta)\right)_{\vartheta = 0}.
\end{eqnarray}

Now let us reinstate the term \(\vec{\mu}\) in \eq{eq:TsymbM0}. It is readily seen (\App{app:dJ}) that although \(\vec{\mu}\) may significantly affect \m{T_{\tau}} \perse, its effect on \(\mc{J}[A, G]\) is small, namely,
\begin{eqnarray}\label{eq:dJ}
\mc{J}[A, G] - \mc{J}[A, G_0] = \mc{O}(\epsilon^2).
\end{eqnarray}
Below, we apply this formulation to \m{A = \mc{O}(\varepsilon^2)}, in which case \eq{eq:dJ} becomes \m{\mc{J}[A, G] - \mc{J}[A, G_0] = \mc{O}(\epsilon^2\varepsilon^2)}. Such corrections are negligible within our model, so from now on we adopt
\begin{eqnarray}\label{eq:G0}
G(\vec{X}, \vec{K}) \approx \avr{G}(\vec{X}, \vec{K}) \approx G_0(\Omega(\vec{X}, \vec{K})).
\end{eqnarray}

\subsection{Initial conditions}
\label{sec:g}

Consider the function \(g\) from \eq{eq:ft4}. Using \eq{eq:g0def}, the latter can be written as follows:
\begin{eqnarray}\label{eq:gmm1}
g = 
\lim_{\nu \to 0+} \lim_{\tau_0 \to -\infty} 
\left(\oper{T}_{\tau - \tau_0} \osc{f}(\vec{X}) + \int_0^{\tau-\tau_0} \dd\tau'\,(1 - \ee^{-\nu \tau'})\oper{T}_{\tau'}\mcu{F}(\vec{X})\right).
\end{eqnarray}
Because \((1 - \ee^{-\nu \tau})\) is smooth and \m{\oper{T}_{\tau}\mcu{F}} is rapidly oscillating, the second term in the external parenthesis is an oscillatory function of \(\tau_0\) with the average negligible at \m{\nu \to 0}. But the whole expression in these parenthesis is independent of \(\tau_0\) at large \(\tau_0\) (\Sec{sec:ft}). Thus, it can be replaced with its own average over \(\tau_0\), denoted \m{\favr{\ldots}_{\tau_0}}. Because there is no \m{\nu}-dependence left in this case, one can also omit \m{\lim_{\nu \to 0+}}. That gives
\begin{eqnarray}
g = \lim_{\tau_0 \to -\infty} \favr{\oper{T}_{\tau - \tau_0} \osc{f}(\vec{X})}_{\tau_0}.
\end{eqnarray}
Using
\begin{eqnarray}
f(\vec{X}) = \sum_\sigma \delta(\vec{z} - \vec{z}_\sigma(t)),
\qquad
\mcc{f}(\vec{X}) \doteq \sum_\sigma \delta(\vec{z} - \avr{\vec{z}}_\sigma(t)),
\end{eqnarray}
where the sum is taken over individual particles, one can write\footnote{Taylor-expanding delta functions is admittedly a questionable procedure, but here it is understood as a shorthand for Taylor-expanding integrals of \m{f}.}
\begin{eqnarray}
\textstyle
f(\vec{X}) \approx \mcc{f}(\vec{X})
- \sum_\sigma \osc{\vec{z}}_\sigma(\vec{X}) \pd_{\vec{z}}\delta(\vec{z} - \avr{\vec{z}}_\sigma(\vec{X})),
\end{eqnarray}
where \m{\vec{z}_\sigma \doteq \vec{z}_\sigma - \avr{\vec{z}}_\sigma} are the \m{\osc{H}}-driven small deviations from the particle unperturbed trajectories \m{\avr{\vec{z}}_\sigma}. Then, \m{\avr{f} = \avr{\mcc{f}}(\vec{X})}, and the linearized perturbation \m{\osc{f} \doteq f - \avr{f}} is given by
\begin{eqnarray}
\textstyle
\osc{f}(\vec{X}) = \underbrace{\,
\mcc{f}(\vec{X})
-\avr{\mcc{f}}(\vec{X})
}_{\micro{\osc{f}}}
\underbrace{\textstyle
- \sum_\sigma \osc{\vec{z}}_\sigma(\vec{X}) \pd_{\vec{z}}\delta(\vec{z} - \avr{\vec{z}}_\sigma(\vec{X}))
}_{\macro{\osc{f}}}.
\end{eqnarray}

By definition, the unperturbed trajectories \m{\avr{\vec{z}}_\sigma} satisfy \m{\oper{L} \delta(\vec{z} - \avr{\vec{z}}_\sigma(\vec{X})) = 0}, where \m{\oper{L}} as in \eq{eq:Loper}; thus,
\begin{eqnarray}
\oper{T}_{\tau - \tau_0} \micro{\osc{f}} = \ee^{\oper{L} (\tau - \tau_0)} \micro{\osc{f}} = \micro{\osc{f}}.
\end{eqnarray}
Also, \m{\favr{\oper{T}_{\tau - \tau_0} \macro{\osc{f}}}_{\tau_0} = 0}, because \m{\osc{\vec{z}}_\sigma} are oscillatory functions of \(\vec{X}\) that is slowly evolved by \m{\oper{T}_{\tau - \tau_0}}. Hence, \m{g} is the microscopic part of the unperturbed distribution function:
\begin{eqnarray}
g = \micro{\osc{f}} = \mcc{f}(\vec{X}) - \avr{\mcc{f}}(\vec{X}).
\end{eqnarray}
This indicates that the term \m{\Gamma} defined in \eq{eq:GD} is due to collisional effects. We postpone discussing these effects until \Sec{sec:th}, so \m{\Gamma} will be ignored for now.

\subsection{Summary of \Sec{sec:prelim0}}
\label{sec:sumprelim}

The main result of this section is that the Weyl symbol of the effective Green's operator \m{\oper{G}} can be approximated within the assumed accuracy as follows:
\begin{eqnarray}
G(\vec{X}, \vec{K}) \approx G_0(\Omega(\vec{X}, \vec{K})), 
\qquad
\Omega(\vec{X}, \vec{K}) \doteq - \vec{K} \cdot \vec{V}(\vec{X}).
\end{eqnarray}
Here, \m{\vec{V}} is the unperturbed velocity in the \m{\vec{X}} space, so \m{\Omega(\vec{X}, \vec{K}) = \omega - \vec{k} \cdot \vec{v} + \mc{O}(\epsilon)}, where \m{\vec{v}} is the unperturbed velocity in the \m{\vec{x}} space, and
\begin{eqnarray}
G_0(\Omega) = \upi\,\delta(\Omega) + \ii\,\pv\frac{1}{\Omega},
\qquad 
\pv\frac{1}{\Omega} \doteq \lim_{\nu \to 0+} \frac{\Omega}{\nu^2 + \Omega^2}.
\end{eqnarray}
We also show that the term \m{\Gamma} defined in \eq{eq:GD} is due to collisional effects. We postpone discussing these effects until \Sec{sec:th}, so \m{\Gamma} will be ignored for now.

\section{Interaction with prescribed fields}
\label{sec:dop}

In this section, we explore the effect of the diffusion operator \m{\oper{D}^{\alpha\beta}}. The oscillations will be described by \m{\avr{W}} as a prescribed function, so they are allowed (yet not required) to be `off-shell', \ie do not have to be constrained by a dispersion relation. Examples of off-shell fluctuations include driven near-field oscillations, evanescent waves, and microscopic fluctuations (see also \Sec{sec:th}). We will first derive the symbol of \m{\oper{D}^{\alpha\beta}} and, using this symbol, approximate the diffusion operator with a differential operator (\Sec{sec:D}). Then, we will calculate the coefficients in the approximate expression for \m{\oper{D}^{\alpha\beta}} (\Secs{sec:wuw} and \ref{sec:phi}). Finally, we will introduce the concept of the OC distribution (\Sec{sec:OC}) and summarize and simplify the resulting equations (\Sec{sec:Deq}).

\subsection{Expansion of the dispersion operator}
\label{sec:D}

The (effective) Green's operator can be represented through its symbol \(G\) using \eq{eq:invWWT}:
\begin{eqnarray}
\oper{G} = \frac{1}{(2\upi)^N} \int \dd\vec{X}\,\dd\vec{K}\,\dd\vec{S}
\ket{\vec{X} + \vec{S}/2} G(\vec{X}, \vec{K}) \bra{\vec{X} - \vec{S}/2}
\ee^{\ii \vec{K} \cdot \vec{S}}.
\end{eqnarray}
The corresponding representation of \(\oper{u}^\alpha\) is even simpler, because the symbol of \(\oper{u}^\alpha\) is independent of \(\vec{K}\):\footnote{One can also derive \eq{eq:uop2} formally from \eq{eq:invWWT}.}
\begin{eqnarray}\label{eq:uop2}
\oper{u}^\alpha = \int \dd\vec{X} \ket{\vec{X}} u^\alpha(\vec{X}) \bra{\vec{X}}.
\end{eqnarray}
Let us also introduce the Wigner matrix of \(u^\alpha\), denoted \m{W_{\vec{u}}^{\alpha\beta}}, and its inverse Fourier transform \m{C_{\vec{u}}^{\alpha\beta}} as in \Sec{sec:WX}. Using these together with \eq{eq:norm}, one obtains
\begin{align}
(2\upi)^N \oper{u}^{\alpha}\oper{G}\oper{u}^{\beta}
& = \int \dd\vec{X}'\,\dd\vec{X}''\,\dd\vec{X}\,\dd\vec{K}\,\dd\vec{S}\,
u^{\alpha}(\vec{X}') u^{\beta}(\vec{X}'')
G(\vec{X}, \vec{K})\,
\ee^{\ii \vec{K} \cdot \vec{S}}
\notag\\& \qquad\times
\ket{\vec{X}'}\braket{\vec{X}' | \vec{X} + \vec{S}/2}
\braket{\vec{X} - \vec{S}/2 | \vec{X}''}
\bra{\vec{X}''}
\notag\\
& = \int \dd\vec{X}\,\dd\vec{K}\,\dd\vec{S}\,
{C}_{\vec{u}}^{\alpha\beta}(\vec{X}, \vec{S})
G(\vec{X}, \vec{K})\,\ee^{\ii \vec{K} \cdot \vec{S}}
\ket{\vec{X} + \vec{S}/2}\bra{\vec{X} - \vec{S}/2}
\notag\\
& = \int \dd\vec{X}'\,\dd\vec{K}'\,\dd\vec{K}''\,\dd\vec{S}'\,
{W}_{\vec{u}}^{\alpha\beta}(\vec{X}', \vec{K}')
G(\vec{X}', \vec{K}'')\,\ee^{\ii (\vec{K}' + \vec{K}'') \cdot \vec{S}'}
\notag\\ & \qquad\times
\ket{\vec{X}' + \vec{S}'/2}\bra{\vec{X}'-\vec{S}'/2}.
\label{eq:uGu}
\end{align}
Then, by taking \m{\wsymbX{}} of \eq{eq:uGu}, one finds that the symbol of \m{\oper{D}^{\alpha\beta}} is a convolution of \m{\avr{W}_{\vec{u}}^{\alpha\beta}} and \m{G} (\App{app:Dconv}):
\begin{eqnarray}
D^{\alpha\beta}(\vec{X}, \vec{K}) = \int \dd\vec{K}'\,
\avr{W}_{\vec{u}}^{\alpha\beta}(\vec{X}, \vec{K}')
G(\vec{X}, \vec{K} - \vec{K}').
\label{eq:Dconv}
\end{eqnarray}

Let us Taylor-expand the symbol \eq{eq:Dconv} in \(\vec{K}\):
\begin{align}
D^{\alpha\beta}(\vec{X}, \vec{K}) \approx & \int \dd\vec{K}'\,
\avr{W}_{\vec{u}}^{\alpha\beta}(\vec{X}, \vec{K}') G(\vec{X}, -\vec{K}')
\notag\\
& + K_c \int \dd\vec{K}'\,\avr{W}_{\vec{u}}^{\alpha\beta}(\vec{X}, \vec{K}') G^{|c}(\vec{X}, -\vec{K}')
+\mc{O}(K_a K_b G^{|ab}). \label{eq:Dexp}
\end{align}
As a reminder, \(G^{|a}(\vec{X}, -\vec{K}) = - \pd^a G(\vec{X}, -\vec{K})\) denotes the derivative of \(G\) with respect to (the \(a\)th component of) the whole second argument, \(-K_a\), and
\begin{eqnarray}\label{eq:KG}
K_a\,\frac{\pd G}{\pd K_a}
= \omega\,\frac{\pd G}{\pd \omega} 
+ k_i\,\frac{\pd G}{\pd k_i}
+ r^i\,\frac{\pd G}{\pd r^i}.
\end{eqnarray}
Upon application of \m{\woperX{}}, \(\omega\) gets replaced (roughly) with \(\ii \pd_t = \mc{O}(\epsilon)\) and \(k_i\) gets replaced (also roughly) with \(-\ii\pd_i = \mc{O}(\epsilon)\). By \eq{eq:Gr}, the last term in \eq{eq:KG} is of order \(\epsilon\) too. This means that the contribution of the whole \m{K_a \pd^a G} term to the equation for \m{\avr{f}} is of order \(\epsilon\). The standard QLT neglects this contribution entirely, \ie adopts \m{D^{\alpha\beta}(\vec{X}, \vec{K}) \approx D^{\alpha\beta}(\vec{X}, \vec{0})}, in which case the diffusion operator becomes just a local function of phase-space variables, \m{\oper{D}^{\alpha\beta} \approx D^{\alpha\beta}(\vec{X}, \vec{0})}. In this work, we retain corrections to the first order in \(\vec{X}\), \ie keep the second term in \eq{eq:Dexp} as well, while neglecting the higher-order terms as usual. 

Within this model, one can rewrite \eq{eq:Dexp} as follows:
\begin{eqnarray}\label{eq:dd}
D^{\alpha\beta}(\vec{X}, \vec{K}) \approx D_0^{\alpha\beta}(\vec{X}) + K_c \Theta^{\alpha \beta c}(\vec{X}).
\end{eqnarray}
Here, we used \eq{eq:Gstar} and introduced
\begin{eqnarray}
& \displaystyle
D_0^{\alpha\beta}(\vec{X}) 
\doteq \int \dd\vec{K}\,\avr{W}_{\vec{u}}^{\alpha\beta}(\vec{X}, \vec{K})G^*(\vec{X}, \vec{K}),
\label{eq:D0}
\\
& \displaystyle
\Theta^{\alpha \beta c}(\vec{X}) 
\doteq -\int \dd\vec{K}\,
\avr{W}_{\vec{u}}^{\alpha\beta}(\vec{X}, \vec{K})
(G^{|c}(\vec{X}, \vec{K}))^*,
\label{eq:T0}
\end{eqnarray}
which satisfy (\App{app:DTheta})
\begin{eqnarray}\label{eq:DTheta}
D_0^{\alpha\beta}(\vec{X}) = (D_0^{\alpha\beta}(\vec{X}))^*,
\qquad
\Theta^{\alpha\beta c}(\vec{X}) = - (\Theta^{\alpha\beta c}(\vec{X}))^*.
\end{eqnarray}
The first-order Weyl expansion of \m{\oper{D}^{\alpha\beta}} is obtained by applying \m{\woperX{}} to \eq{eq:dd}. Namely, for any \(\psi\), one has (cf.\ \Sec{sec:wexp})
\begin{eqnarray}\label{eq:Dopa}
\oper{D}^{\alpha\beta}\psi \approx D_0^{\alpha\beta} \psi 
- \ii \Theta^{\alpha \beta c} \pd_c\psi 
- \frac{\ii}{2}\,(\pd_c\Theta^{\alpha \beta c})\psi.
\end{eqnarray}
What remains now is to calculate the functions \m{D_0^{\alpha\beta}} and \m{\Theta^{\alpha \beta c}} explicitly.

\subsection{Wigner matrix of the velocity oscillations}
\label{sec:wuw}

To express \m{\oper{D}^{\alpha\beta}} through the Wigner function \(W\) of the perturbation Hamiltonian (\Sec{sec:ordering}), we need to express \m{W_{\vec{u}}^{\alpha\beta}} through~\(W\). Recall that \m{W_{\vec{u}}^{\alpha\beta}} is the symbol of the density operator \m{\oper{W}_{\vec{u}}^{\alpha\beta} \doteq (2\upi)^{-N} \ket{u^\alpha}\bra{u^\beta}} (\Sec{sec:WX}). By definition \eq{eq:uv}, one has \m{u^{\alpha} = \ii J^{\alpha \mu} \oper{q}_{\mu} \osc{H}}, where \m{\oper{q}_\alpha \doteq - \ii\pd_\alpha} (\Sec{sec:notz}). Then,
\begin{eqnarray}
\oper{W}_{\vec{u}}^{\alpha\beta}
= (2\upi)^{-N} J^{\alpha \mu}\oper{q}_{\mu}\ket{\osc{H}}\bra{\osc{H}}\oper{q}_{\nu} J^{\beta \nu}
= J^{\alpha\mu} J^{\beta \nu}\oper{q}_{\mu} \oper{W} \oper{q}_{\nu},
\end{eqnarray}
where \(\oper{W}\) is the density operator whose symbol is \(W\). By applying \m{\wsymbX{}}, one obtains
\begin{eqnarray}
W_{\vec{u}}^{\alpha\beta} = J^{\alpha \mu}J^{\beta \nu} 
(q_{\mu} \bigstar W \bigstar q_{\nu}), 
\end{eqnarray}
where \(\bigstar\) is the Moyal product \eq{eq:bigstar}. Using formulas analogous to \eq{eq:hk} in the \m{\cb{\vec{X}, \vec{K}}} space, one obtains
\begin{align}
q_{\mu} \bigstar W \bigstar q_{\nu}
& = \left(q_{\mu} W -\frac{\ii}{2} \frac{\pd W}{\pd z^{\mu}} \right) \bigstar q_{\nu}
\notag\\
& = q_{\mu}q_{\nu} W
- \frac{\ii}{2}\,q_{\nu}\,\frac{\pd W}{\pd z^{\mu}}
+ \frac{\ii}{2} \frac{\pd}{\pd z^{\nu}}
\left(q_{\mu} W_h - \frac{\ii}{2}\frac{\pd W}{\pd z^{\mu}}\right)
\notag\\
& = q_{\mu}q_{\nu} W_h 
+ \frac{\ii}{2} \left(q_{\mu}\frac{\pd W}{\pd z^{\nu}} - q_{\nu}\frac{\pd W}{\pd z^{\mu}}\right)
+ \frac{1}{4}\frac{\pd^2 W}{\pd z^{\mu}\pd z^{\nu}}.
\end{align}
Hence, \(W_{\vec{u}}^{\alpha\beta}\) and \(W\) are connected via the following exact formula:
\begin{eqnarray}
W_{\vec{u}}^{\alpha\beta}
= J^{\alpha \mu}J^{\beta \nu}
\left(
q_{\mu} q_{\nu} W 
- \frac{\ii}{2} 
\left(
 q_{\nu}\,\frac{\pd W}{\pd z^{\mu}}
-q_{\mu}\,\frac{\pd W}{\pd z^{\nu}}
\right) 
+ \frac{1}{4} \frac{\pd^2 W}{\pd z^{\mu} \pd z^{\nu}}
\right).
\label{eq:WuW}
\end{eqnarray}

\subsection{Nonlinear potentials}
\label{sec:phi}

Due to \eq{eq:DTheta}, one has \(D_0^{\alpha\beta} = \re D_0^{\alpha\beta}\). Using this together with \eq{eq:D0}, \eq{eq:WuW}, \eq{eq:G0}, and \eq{eq:G00}, one obtains
\begin{align}
D_0^{\alpha\beta} 
& = J^{\alpha \mu}J^{\beta \nu} \re \int \dd\vec{K}
\left(\upi\,\delta(\Omega) - \ii\, \pv\frac{1}{\Omega}\right).
\notag\\
& \hspace{2.5cm}\times 
\left(
q_{\mu} q_{\nu} \avr{W}
- \frac{\ii}{2}\left(
  q_{\nu}\,\frac{\pd\avr{W}}{\pd z^{\mu}}
- q_{\mu}\,\frac{\pd\avr{W}}{\pd z^{\nu}}
\right) 
+ \frac{1}{4} \frac{\pd^2\avr{W}}{\pd z^{\mu}\pd z^{\nu}}
\right),
\end{align}
with notation as in \eq{eq:reim}. This can be written as \m{D_0^{\alpha\beta} = \st{D}^{\alpha\beta} + \varrho^{\alpha\beta} + \varsigma^{\alpha\beta}}, where
\begin{eqnarray}\label{eq:stD}
\st{D}^{\alpha\beta} \doteq J^{\alpha\mu}J^{\beta\nu}
\int \dd\vec{K}\, \upi\,\delta(\Omega)\, q_{\mu}q_{\nu}\avr{W},
\end{eqnarray}
and we also introduced
\begin{eqnarray}\label{eq:varrho}
& \displaystyle
\varrho^{\alpha\beta} \doteq
- \frac{1}{2}\,J^{\alpha\mu}J^{\beta\nu}
\fint \dd\vec{K} \left(
  q_{\nu}\,\frac{\pd \avr{W}}{\pd z^{\mu}}
- q_{\mu}\,\frac{\pd \avr{W}}{\pd z^{\nu}}
\right)
\frac{1}{\Omega},
\\
& \displaystyle
\varsigma^{\alpha\beta} \doteq
\frac{1}{4}\,
J^{\alpha\mu} J^{\beta\nu}
\int \dd\vec{K}\,\upi\,\delta(\Omega)\,\frac{\pd^2 \avr{W}}{\pd z^{\mu} \pd z^{\nu}}.
\end{eqnarray}
As shown in \App{app:contr}, the contributions of these two functions to \eq{eq:ql1} are
\begin{eqnarray}\label{eq:rhoest}
\frac{\pd}{\pd z^{\alpha}}
\left(\varrho^{\alpha\beta}\,
\frac{\pd\avr{f}}{\pd z^{\beta}}
\right) = \mc{O}(\epsilon\varepsilon^2),
\qquad
\frac{\pd}{\pd z^{\alpha}} \left(
\varsigma^{\alpha\beta}\,\frac{\pd\avr{f}}{\pd z^{\beta}}
\right) = \mc{O}(\epsilon^2\varepsilon^2).
\end{eqnarray}
Thus, \m{\varrho^{\alpha\beta}} must be retained and \m{\varsigma^{\alpha\beta}} must be neglected, which leads to
\begin{eqnarray}
D_0^{\alpha\beta} \approx \st{D}^{\alpha\beta} + \varrho^{\alpha\beta}.
\end{eqnarray}

The function \m{\Theta^{\alpha \beta c} = \ii\, \im \Theta^{\alpha \beta c}} can be written as follows:
\begin{eqnarray}\notag
\Theta^{\alpha \beta c}(\vec{X})
= \ii J^{\alpha \mu}J^{\beta \nu}
\int \dd\vec{K}\,q_{\mu}q_{\nu}\avr{W}(\vec{X}, \vec{K})\,\frac{\pd}{\pd K_c}\,\pv \frac{1}{\Omega (\vec{X}, \vec{K})}
= -\ii V^c(\vec{X})\, \Uptheta^{\alpha\beta}(\vec{X}),
\end{eqnarray}
where we introduced
\begin{eqnarray}\label{eq:Upth}
\Uptheta^{\alpha\beta} \doteq J^{\alpha \mu}J^{\beta \nu} \frac{\eth}{\pd \Omega}
\fint \dd\vec{K}\,\frac{q_{\mu}q_{\nu}\avr{W}}{\Omega}
\end{eqnarray}
and \m{\eth} is defined as in \eq{eq:eth}. Then finally, one can rewrite \eq{eq:Dopa} as follows:
\begin{align}
\oper{D}^{\alpha\beta}\psi 
& \approx (\st{D}^{\alpha\beta} + \varrho^{\alpha\beta})\psi 
-\Uptheta^{\alpha\beta} V^c \pd_c\psi - \frac{1}{2}\,V^c(\pd_c\Uptheta^{\alpha\beta})\psi 
\notag\\
& = (\st{D}^{\alpha\beta} + \varrho^{\alpha\beta})\psi 
- \Uptheta^{\alpha\beta}
(\pd_t + v^{\lambda}\pd_{\lambda})\psi 
-\frac{1}{2}\,((\pd_t+v^{\lambda}\pd_{\lambda})\Uptheta^{\alpha\beta})\psi,
\label{eq:aux11}
\end{align}
where we used \eq{eq:VV}. With some algebra (\App{app:DDD}), and assuming the notation
\begin{eqnarray}
\label{eq:phi}
\Phi = - J^{\mu \nu}\,\frac{\pd}{\pd z^{\mu}}
\fint \dd\vec{K}\,\frac{q_{\nu}\avr{W}}{2\Omega},
\end{eqnarray}
one finds that \eq{eq:aux11} leads to
\begin{eqnarray}\label{eq:DDD}
\pd_{\alpha}(\oper{D}^{\alpha\beta}\pd_{\beta}\avr{f})
= \pd_{\alpha}(\st{D}^{\alpha\beta}\pd_{\beta}\avr{f})
- \frac{1}{2}\,\dd_t\pd_{\alpha}(\Uptheta^{\alpha\beta} \pd_{\beta}\avr{f})
+ \poissonz{\Phi, \avr{f}}.
\end{eqnarray}
Hence, \eq{eq:ql1} becomes (to the extent that \(\Gamma\) is negligible; see \Sec{sec:pd})
\begin{eqnarray}\label{eq:ql1F}
\dd_t \avr{f} + \frac{1}{2}\,\dd_t\pd_{\alpha}(\Uptheta^{\alpha\beta} \pd_{\beta}\avr{f}) - \poissonz{\Phi, \avr{f}}
= \pd_{\alpha}(\st{D}^{\alpha \beta} \pd_{\beta} \avr{f}).
\end{eqnarray}
The functions \m{\Uptheta^{\alpha\beta}}, \m{\Phi}, and \m{\st{D}^{\alpha \beta}} that determine the coefficients in this equation are fundamental and, for the lack of a better term, will be called nonlinear potentials.

\subsection{Oscillation-center distribution}
\label{sec:OC}

Let us introduce
\begin{eqnarray}
F \doteq \avr{f} + \frac{1}{2}\,\pd_{\alpha}(\Uptheta^{\alpha\beta} \pd_{\beta}\avr{f}).
\label{eq:Fdef}
\end{eqnarray}
Then, using \eq{eq:DDD}, one can rewrite \eq{eq:ql1F} as\footnote{The difference between \m{F} and \m{\avr{f}} is related to the concept of so-called adiabatic diffusion \citep{book:galeev85, book:stix}, which captures some but not all adiabatic effects.}
\begin{eqnarray}
\pd_t F - \poissonz{\mc{H}, F} =
\pd_{\alpha}(\st{D}^{\alpha \beta} \pd_{\beta} F),
\label{eq:FC}
\end{eqnarray}
where corrections \(\mc{O}(\varepsilon^4)\) have been neglected and we introduced \m{\mc{H} \doteq \avr{H} + \Phi}. As a reminder, the nonlinear potentials in \eq{eq:FC} are as follows:
\begin{align}
\label{eq:stDF}
\st{D}^{\alpha\beta} & = J^{\alpha\mu}J^{\beta\nu}
\int \dd\vec{K}\, \upi\,\delta(\Omega)\, q_{\mu}q_{\nu}\avr{W},
\\
\label{eq:UpthF}
\Uptheta^{\alpha\beta} & = J^{\alpha \mu}J^{\beta \nu} \frac{\eth}{\pd \Omega}
\fint \dd\vec{K}\,\frac{q_{\mu}q_{\nu}\avr{W}}{\Omega},
\\
\label{eq:phiF}
\Phi & = - J^{\mu \nu}\,\frac{\pd}{\pd z^{\mu}}
\fint \dd\vec{K}\,\frac{q_{\nu}\avr{W}}{2\Omega}.
\end{align}

Equations \eq{eq:Fdef}--\eq{eq:phiF} form a closed model that describes the evolution of the average distribution \m{\avr{f}} in turbulence with prescribed~\(\avr{W}\). In particular, \eq{eq:FC} can be interpreted as a Liouville-type equation for~\(F\) as an effective, or `dressed', distribution. The latter can be understood as the distribution of `dressed' particles called OCs. Then, \(\mc{H}\) serves as the OC Hamiltonian, \m{\st{D}^{\alpha\beta}} is the phase-space diffusion coefficient, \m{\Phi} is the ponderomotive energy, \m{\Omega = \omega - q_\alpha v^\alpha}, and \m{v^\alpha \doteq J^{\alpha\beta}\pd_\beta\avr{H}}. Within the assumed accuracy, one can redefine \m{v^\alpha} to be the OC velocity rather than the particle velocity; specifically,\footnote{The advantage of the amended definition \eq{eq:vOC} is that it will lead to \textit{exact} conservation laws of our theory, as to be discussed in \Sec{sec:cons}.}
\begin{eqnarray}
v^\alpha \doteq J^{\alpha\beta}\pd_\beta\mc{H}
= J^{\alpha\beta}\pd_\beta\avr{H} + \mc{O}(\epsilon^2).
\label{eq:vOC}
\end{eqnarray}
Then, the presence of \m{\delta(\Omega)} in \eq{eq:stDF} signifies that OCs diffuse in phase space in response to waves they are resonant with. Below, we use the terms `OCs' and `particles' interchangeably except where specified otherwise. 

That said, the interpretation of OCs as particle-like objects is limited. Single-OC motion equations are not introduced in our approach. (They would have been singular for resonant interactions.) Accordingly, the transformation \eq{eq:Fdef} of the distribution function \m{\avr{f} \mapsto F} is not \textit{derived} from a coordinate transformation but rather is \textit{fundamental}. As a result, particles and OCs live in the same phase space, but the `dynamics of OCs' can be irreversible (\Sec{sec:hthOC}). This qualitatively distinguishes our approach from the traditional OC theory \citep{ref:dewar73} and from the conceptually similar gyrokinetic theory \citep{ref:littlejohn81, ref:cary09}, where coordinate transformations are central.

\subsection{\mt{H}-theorem}
\label{sec:hthOC}

Because \m{\avr{W}} is non-negative (\Sec{sec:wigfun}), \m{\st{D}^{\alpha \beta}} is positive-semidefinite; that is,
\begin{eqnarray}\label{eq:Dspsp}
\st{D}^{\alpha\beta} \psi_\alpha \psi_\beta  
= \int \dd\vec{K}\, \upi\,\delta(\Omega)\, a^2 \avr{W} \ge 0,
\qquad
a \doteq J^{\alpha\mu} \psi_\alpha q_{\mu}
\end{eqnarray}
for any real \m{\psi}. This leads to the following theorem. Consider the OC entropy defined as
\begin{eqnarray}\label{eq:Hth2tot}
\mcu{S} \doteq - \int \dd\vec{z}\,F(t, \vec{z})\ln F(t, \vec{z}).
\end{eqnarray}
According to \eq{eq:FC}, \m{\mcu{S}} satisfies
\begin{align}
\frac{\dd\mcu{S}}{\dd t}
& = - \int \dd\vec{z}\,\frac{\dd(F \ln F)}{\dd F}\left(
\poissonz{\mc{H}, F} + \pd_{\alpha}(\st{D}^{\alpha \beta} \pd_{\beta} F)
\right)
\notag\\
& = 
- \int \dd\vec{z}\,\frac{\dd(F \ln F)}{\dd F}\,J^{\alpha\beta}(\pd_\alpha \mc{H})(\pd_\beta F)
- \int \dd\vec{z}\,(1 + \ln F)\,\pd_{\alpha}(\st{D}^{\alpha \beta} \pd_{\beta} F)
\notag\\
& = 
- \int \dd\vec{z}\,J^{\alpha\beta}(\pd_\alpha \mc{H}) \pd_\beta(F \ln F)
- \int \dd\vec{z}\,\ln F\,\pd_{\alpha}(\st{D}^{\alpha \beta} \pd_{\beta} F)
\notag\\
& 
= \int \dd\vec{z}\,(J^{\alpha\beta} \pd^2_{\alpha\beta} \mc{H})\,F \ln F
+ \int \dd\vec{z}\,\st{D}^{\alpha \beta} (\pd_\alpha\ln F) (\pd_{\beta} \ln F) F.
\end{align}
The first integral vanishes due to \m{J^{\alpha\beta} \pd^2_{\alpha\beta} = 0}. The second integral is non-negative due to \eq{eq:Dspsp}. Thus, 
\begin{eqnarray}\label{eq:HSth}
\frac{\dd\mcu{S}}{\dd t} \ge 0,
\end{eqnarray}
which is recognized as the \m{H}-theorem \citep[section~4]{book:landau10} for QL OC dynamics.

\subsection{Summary of \Sec{sec:dop}}
\label{sec:Deq}

From now on, we assume that the right-hand side of \eq{eq:FC} scales not as \m{\mc{O}(\varepsilon^2)} but as \m{\mc{O}(\epsilon\varepsilon^2)}, either due to the scarcity of resonant particles or, for QL diffusion driven by microscopic fluctuations (\Sec{sec:th}), due to the plasma parameter's being large. Also, the spatial derivatives can be neglected within the assumed accuracy in the definition of \(F\) \eq{eq:Fdef} and on the right-hand side of \eq{eq:FC}. Using this together with \eq{eq:Jqq}, and with \eq{eq:poissonz} for the Poisson bracket, our results can be summarized as follows. 

QL evolution of a particle distribution in a prescribed wave field is governed by\footnote{Remember that here we neglect \m{\Gamma} \eq{eq:GD}, which is a part of the collision operator to be reinstated in \Sec{sec:th}.}
\begin{eqnarray}\label{eq:OCFeq}
\frac{\pd F}{\pd t}
- \frac{\pd \mc{H}}{\pd \vec{x}} \cdot \frac{\pd F}{\pd \vec{p}}
+ \frac{\pd \mc{H}}{\pd \vec{p}} \cdot \frac{\pd F}{\pd \vec{x}}
= \frac{\pd}{\pd \vec{p}} \cdot \left(\vecst{D}\,\frac{\pd F}{\pd \vec{p}}\right).
\end{eqnarray}
The OC distribution \m{F} is defined as
\begin{eqnarray}
F = \avr{f} + \frac{1}{2}\,\frac{\pd}{\pd \vec{p}} \cdot 
\left(\vec{\Uptheta}\, \frac{\pd \avr{f}}{\pd \vec{p}}\right),
\label{eq:Fdef2}
\end{eqnarray}
so the density of OCs is the same as the locally averages density of the true particles:
\begin{eqnarray}\label{eq:density}
\mc{N} \doteq \int \dd\vec{p}\,F = \int \dd\vec{p}\,\avr{f}.
\end{eqnarray}
The function \m{\mc{H}} is understood as the OC Hamiltonian. It is given by
\begin{eqnarray}
\mc{H} \doteq \avr{H} + \Phi,
\label{eq:mcH}
\end{eqnarray}
where \m{\avr{H}} is the average Hamiltonian, which may include interaction with background fields, and \m{\Phi} is the ponderomotive potential. The nonlinear potentials that enter \eq{eq:OCFeq} can be calculated to the zeroth order in \(\epsilon\) and are given by\footnote{See \Sec{sec:examples} for examples and \Sec{sec:Phi} for the explanation on how \m{\Phi} is related to \m{\Delta}, which is yet to be introduced. Also note that in combination with \eq{eq:mcH}, equation \eq{eq:Phisc} generalizes the related results from \citep{ref:kentwell87b, ref:fraiman95, my:lens}.}
\begin{align}
\matrst{D}
& = \int \dd\omega\,\dd\vec{k}\, \upi\,\vec{k}\vec{k} \avr{\st{W}}(t, \vec{k} \cdot \vec{v}, \vec{k}; \vec{p}) ,
\label{eq:Dsc}
\\
\matr{\Uptheta}
& =\frac{\pd}{\pd \vartheta} \fint \dd\omega\,\dd\vec{k}\,
 \left.
\frac{\vec{k} \vec{k} \avr{\st{W}}}{\omega - \vec{k} \cdot \vec{v} + \vartheta}
\right|_{\vartheta=0},
\label{eq:uptheta2}
\\
\Phi & 
= \frac{1}{2}\frac{\pd}{\pd \vec{p}} \cdot
\fint \dd\omega\,\dd\vec{k}\,\frac{\vec{k} \avr{\st{W}}}{\omega - \vec{k} \cdot \vec{v}},
\label{eq:Phisc}
\end{align}
where \m{\vec{k} \vec{k}} is a dyadic matrix with two lower indices, and the same conventions apply as in \Sec{sec:matr}. Also, \m{\vec{v}} is hereby redefined as the OC spatial velocity, namely,
\begin{eqnarray}
\vec{v} \doteq \pd_\vec{p}\mc{H}
= \pd_\vec{p}\avr{H} + \mc{O}(\epsilon^2).
\label{eq:vOC2}
\end{eqnarray}
The function \m{\avr{\st{W}}} is defined as
\begin{eqnarray}\label{eq:Wst}
\avr{\st{W}}(t, \vec{x}, \omega, \vec{k}; \vec{p}) \doteq \int \dd\vec{r}\,\avr{W}(t, \vec{x}, \vec{p}, \omega, \vec{k}, \vec{r}),
\end{eqnarray}
where \(\avr{W}\) is the average Wigner function \eq{eq:WH} of the perturbation Hamiltonian, \ie the spectrum of its symmetrized autocorrelation function \eq{eq:avCH}. Due to \eq{eq:redW}, it can be understood as the average of \m{\st{W} \doteq \wsymbx{\operst{W}}} (where \m{\operst{W}} is defined in \eq{eq:WHH}), \ie as the Wigner function of the perturbation Hamiltonian with \m{\vec{p}} treated as a parameter. As such, \m{\avr{\st{W}}} is non-negative, so \m{\matrst{D}} is positive-semidefinite. This leads to an \m{H}-theorem (proven similarly to \eq{eq:HSth}) for the entropy density \m{\entropy \doteq - \int \dd\vec{p}\,F\ln F}:
\begin{eqnarray}\label{eq:HthD}
\left(\frac{\dd\entropy}{\dd t}\right)_{\st{D}} \ge 0,
\qquad
\left(\frac{\pd \psi}{\pd t}\right)_{\st{D}} \doteq \frac{\pd}{\pd \vec{p}} \cdot \left(\vecst{D}\,\frac{\pd \psi}{\pd \vec{p}}\right).
\end{eqnarray}

Also note that for homogeneous turbulence in particular, where \m{\avr{\st{W}}} is independent of~\m{\vec{x}}, \eq{eq:wigfprop1} yields that
\begin{eqnarray}
\int \dd\omega \,\avr{\st{W}}(t, \vec{x}, \omega, \vec{k}; \vec{p}) 
= \frac{1}{\mcu{V}_n} \int \dd\omega \,\dd\vec{x}\,\avr{\st{W}}(t, \vec{x}, \omega, \vec{k}; \vec{p}) 
= \frac{1}{\mcu{V}_n}\,\avr{|\fourier{\osc{H}}(t, \vec{k}, \vec{p})|^2},
\label{eq:Wfourr}
\end{eqnarray}
where \m{\mcu{V}_n} is the plasma volume (the index \m{n} denotes the number of spatial dimensions) and \m{\fourier{\osc{H}}} is the spatial spectrum of \m{\osc{H}} as defined in \eq{eq:psik}. 

Equation \eq{eq:OCFeq} can be used to calculate the ponderomotive force \m{\pd_t\int\dd\vec{p}\,\avr{f}} that a given wave field imparts on a plasma. This potentially resolves the controversies mentioned in \citep{ref:kentwell87}. We will revisit this subject for on-shell waves in \Sec{sec:cons}.

\section{Interaction with self-consistent fields}
\label{sec:th}

Here, we explain how to calculate the function \(\avr{\st{W}}\) in the presence of microscopic fluctuations (nonzero \(g\)). In particular, we reinstate the term \m{\Gamma} that was omitted in \Sec{sec:dop}. We also show that a collision operator of the Balescu--Lenard type emerges from our theory within a general interaction model. This calculation can be considered as a generalization of that in \citep{ref:rogister68} for homogeneous plasmas. Another related calculation was proposed in \citep{ref:chavanis12} in application to potential interactions in inhomogeneous systems using action--angle variables, with global averaging over the angles. (See also \citep{ref:mynick88} for a related calculation in action--angle variables based on the Fokker--Planck approach.) In contrast, our model holds for any Hamiltonian interactions via any vector fields and allows for weak inhomogeneities in canonical coordinates.

\subsection{Interaction model}
\label{sec:plasmaL}

Let us assume that particles interact via an \(M\)-component real field \(\vec{\Psi} \equiv \cb{\Psi^1, \Psi^2, \ldots, \Psi^M}^\intercal\). It is treated below as a column vector; hence the index \m{^\intercal}. (A complex field can be accommodated by considering its real and imaginary parts as separate components.) We split this field into the average part \m{\avr{\vec{\Psi}}} and the oscillating part \m{\osc{\vec{\Psi}}}. The former is considered given. For the latter, we assume the action integral of this field without plasma in the form
\begin{eqnarray}\label{eq:Lagr0}
S_0 = \int \dd\vecst{x}\,\mcc{L}_0,
\qquad
\mcc{L}_0 = \frac{1}{2}\,\smash{\osc{\vec{\Psi}}}^\dag\boper{\Xi}_0\osc{\vec{\Psi}}
\end{eqnarray}
(see \Sec{sec:examples} for examples), where \m{\boper{\Xi}_0} is a Hermitian operator\footnote{\label{ft:action}The field action often has the form \m{S_0 = \frac{1}{2}\int \dd\vecst{x}\,\sqrt{\mmetric}\,(\matr{\mmetric}\smash{\osc{\vec{\Psi}}}^*)\boper{\Xi}_0 \osc{\vec{\Psi}}}, where \m{\matr{\mmetric}(\vecst{x})} is a spacetime metric, \m{\mmetric \doteq |\det\matr{\mmetric}\,|}, and \m{\boper{\Xi}_0} is Hermitian with respect to the inner product \m{\favr{\vec{\xi}|\vec{\psi}}_\mmetric \doteq \int \dd\vecst{x}\,\sqrt{\mmetric}\,(\matr{\mmetric}\vec{\xi}^*)\vec{\psi}}. Using \m{\smash{\osc{\vec{\Psi}}}' \doteq \mmetric^{1/4}\osc{\vec{\Psi}}} and \m{\smash{\boper{\Xi}}_0' \doteq \mmetric^{1/4}\matr{\mmetric}\boper{\Xi}_0 \mmetric^{-1/4}}, one can cast this action in the form \eq{eq:Lagr0}, with \m{\smash{\boper{\Xi}}_0'} that is Hermitian with respect to the inner product \eq{eq:innerv}.} and \m{\smash{\osc{\vec{\Psi}}}^\dag = \smash{\osc{\vec{\Psi}}}^\intercal} is a row vector dual to \m{\osc{\vec{\Psi}}}. Plasma is allowed to consist of multiple species, henceforth denoted with index \m{s}. Because \m{\osc{\vec{\Psi}}} is assumed small, the generic Hamiltonian for each species \m{s} can be Taylor-expanded in~\m{\osc{\vec{\Psi}}} and represented in a generic form
\begin{eqnarray}\label{eq:Hs}
H_s(t, \vec{x}, \vec{p}) \approx H_{0s} + \boper{\alpha}_s^\dag \osc{\vec{\Psi}} 
+ \frac{1}{2}\,(\boper{L}_s\osc{\vec{\Psi}})^\dag (\boper{R}_s\osc{\vec{\Psi}})
\end{eqnarray}
(see \Sec{sec:examples} for examples), which can be considered as a second-order Taylor expansion of the full Hamiltonian in~\m{\osc{\vec{\Psi}}}. Here, \m{H_{0s} \equiv H_{0s}(t, \vec{x}, \vec{p})} is independent of \m{\osc{\vec{\Psi}} \equiv \osc{\vec{\Psi}}(t, \vec{x})}, \m{\boper{\alpha}_s \equiv \cb{\oper{\alpha}_{s,1}, \oper{\alpha}_{s,2}, \ldots, \oper{\alpha}_{s,M}}^\intercal} is a column vector whose elements \m{\oper{\alpha}_{s,i}} are linear operators on \m{\hilspacex}, and the dagger is added so that \m{\boper{\alpha}_s^\dag} could be understood as a row vector whose elements \m{\oper{\alpha}_{s,i}^\dag} act on the individual components of the field; \ie \m{\boper{\alpha}_s^\dag \osc{\vec{\Psi}} \equiv \oper{\alpha}^\dag_{s,i}\osc{\Psi}^i}. We let \m{\boper{\alpha}_s} be nonlocal in \(t\) and \(\vec{x}\) (for example, \m{\boper{\alpha}_s} can be a spacetime derivative or a spacetime integral), and we also let \m{\boper{\alpha}_s} depend on \m{\vec{p}} parametrically,~so
\begin{eqnarray}
\wsymb \boper{\alpha}_s = \vec{\alpha}_s(t, \vec{x}, \omega, \vec{k}; \vec{p}).
\end{eqnarray}
The matrix operators \m{\boper{L}_s} and \m{\boper{R}_s} and their symbols \m{\matr{L}_s} and \m{\matr{R}_s} are understood similarly.

The Lagrangian density of the oscillating-field--plasma system is
\begin{eqnarray}\label{eq:Lp}
\mcc{L}_p = \frac{1}{2}\,\smash{\osc{\vec{\Psi}}}^\dag\boper{\Xi}_0 \osc{\vec{\Psi}}
+ \sum_s \sum_{\sigma_s} (\vec{p}_{\sigma_s} \cdot \dot{\vec{x}}_{\sigma_s} - H_s(t, \vec{x}_{\sigma_s}, \vec{p}_{\sigma_s}))
\delta(\vec{x} - \vec{x}_{\sigma_s}(t)),
\end{eqnarray}
where the sum is taken over individual particles. Note that
\begin{align}
\sum_{\sigma_s} H_s(t, \vec{x}_{\sigma_s}, \vec{p}_{\sigma_s})
\delta(\vec{x} - \vec{x}_{\sigma_s}(t))
& = \int \dd\vec{p}\, \sum_{\sigma_s} \delta(\vec{z} - \vec{z}_{\sigma_s}(t)) H_s(t, \vec{x}, \vec{p})
\notag\\
& = \int \dd\vec{p}\, f_s(t, \vec{x}, \vec{p}) H_s(t, \vec{x}, \vec{p}),
\end{align}
so the \m{\osc{\vec{\Psi}}}-dependent part of the system action can be written as \m{S = \int \dd\vecst{x}\,\mcc{L}} with
\begin{eqnarray}\label{eq:abr}
& \displaystyle
\mcc{L} = \frac{1}{2}\,\smash{\osc{\vec{\Psi}}}^\dag \boper{\Xi}_p \osc{\vec{\Psi}}
- \sum_s \int \dd\vec{p}\,\osc{f}_s\boper{\alpha}_s^{\dag}\osc{\vec{\Psi}},
\\
& \displaystyle
\boper{\Xi}_p \doteq \boper{\Xi}_0 - \sum_s \int \dd\vec{p}\,\smash{\boper{L}}_s^\dag f_s \boper{R}_s.
\end{eqnarray}
(The contribution of \m{\osc{f}_s} to the second term in \eq{eq:abr} has been omitted because it averages to zero at integration over spacetime and thus does not contribute to \m{S}.) This `abridged' action is not sufficient to describe the particle motion, but it is sufficient to describe the dynamics of \m{\osc{\vec{\Psi}}} at given \m{f_s}, as discussed below. The operator \m{\boper{\Xi}_p} can be considered Hermitian without loss of generality, because its anti-Hermitian part does not contribute to \m{S}. Also, we assume that unless either of \m{\boper{L}} and \m{\boper{R}} is zero, the high-frequency field has no three-wave resonances, so terms cubic in \m{\osc{\vec{\Psi}}} can be neglected in \m{S};\footnote{This is tacitly assumed already in \eq{eq:Hs}, where cubic terms are neglected. Also note that three-wave interactions that involve resonances between low-frequency oscillations of \m{F_s} and two high-frequency waves, like Raman scattering \citep{my:shpulse}, are still allowed.} then,
\begin{eqnarray}
\boper{\Xi}_p \approx \boper{\Xi}_0 - \sum_s \int \dd\vec{p}\,(\smash{\boper{L}}_s^\dag F_s \boper{R}_s)_\herm.
\end{eqnarray}
Using the same assumption, one can also adopt
\begin{eqnarray}\label{eq:Hs1}
\avr{H}_s = H_{0s} + \frac{1}{2}\,\avr{(\boper{L}_s\osc{\vec{\Psi}})^\dag (\boper{R}_s\osc{\vec{\Psi}})},
\qquad
\osc{H}_s \approx \boper{\alpha}_s^\dag \osc{\vec{\Psi}},
\end{eqnarray}
because in the absence of three-wave resonances, the \textit{oscillating} part of \m{(\boper{L}_s\osc{\vec{\Psi}})^\dag (\boper{R}_s\osc{\vec{\Psi}})} contributes only \m{\mc{O}(\varepsilon^4)} terms to the equation for \m{F_s}.

\subsection{Field equations}
\label{sec:feq}

The Euler--Lagrange equation for \m{\osc{\vec{\Psi}}} derived from \eq{eq:abr} is
\begin{eqnarray}
\boper{\Xi}_p \osc{\vec{\Psi}} = \sum_s \int \dd\vec{p}\,\boper{\alpha}_s \osc{f}_s.
\end{eqnarray}
Then, to the extent that the linear approximation for \m{\tilde{f}_s} is sufficient (see below), one finds that the oscillating part of the field satisfies
\begin{eqnarray}
\boper{\Xi}_p \osc{\vec{\Psi}}
- \sum_s \int \dd\vec{p}\,\boper{\alpha}_s \oper{G}_s
\poissonz{\boper{\alpha}_s^\dag \osc{\vec{\Psi}}, \avr{f}_s}
= \sum_s \int \dd\vec{p}\,\boper{\alpha}_s g_s,
\label{eq:aux31}
\end{eqnarray}
where we used \eq{eq:ft4}. Note that the right-hand side of \eq{eq:aux31} is determined by microscopic fluctuations \(g_s(t, \vec{x}, \vec{p})\) (\Sec{sec:g}). Equation \eq{eq:aux31} can also be expressed as
\begin{eqnarray}\label{eq:Ug}
\boper{\Xi}\osc{\vec{\Psi}} = \sum_s \int \dd\vec{p}\,\boper{\alpha}_s g_s,
\end{eqnarray}
where \m{\boper{\Xi}} is understood as the plasma dispersion operator and is given by
\begin{eqnarray}\label{eq:Xidef}
\boper{\Xi} \doteq 
\boper{\Xi}_p
- \sum_s \int \dd\vec{p}\,\boper{\alpha}_s \oper{G}_s 
\poissonz{\boper{\alpha}_s^\dag \,\placeholder\,, \avr{f}_s},
\end{eqnarray}
where \m{\placeholder} is a placeholder. The general solution of \eq{eq:Ug} can be written as
\begin{eqnarray}\label{eq:aux32}
\osc{\vec{\Psi}} = \Psim + \Psiq,
\qquad 
\Psiq = \sum_s \int \dd\vec{p}\,\invXi\boper{\alpha}_s g_s.
\end{eqnarray}
Here, \m{\invXi} is the right inverse of \m{\boper{\Xi}} (meaning \m{\boper{\Xi} \invXi = \boper{1}} yet \m{\invXi\boper{\Xi} \ne \boper{1}}) such that \m{\Psiq} vanishes at zero \(g\).\footnote{Most generally, the problem of finding \m{\invXi} is the standard problem of calculating the field produced by a given radiation~source.} The rest of the solution, \m{\Psim}, is the macroscopic field that satisfies
\begin{eqnarray}\label{eq:Psime}
\boper{\Xi}\Psim = \vec{0}.
\end{eqnarray}
In the special case when the dispersion operator is Hermitian (\m{\boper{\Xi} = \boper{\Xi}_\herm}), \eq{eq:Psime} also flows from the `adiabatic' macroscopic part of the action \m{S}, namely,
\begin{eqnarray}\label{eq:sad}
S_{\text{ad}} \doteq \frac{1}{2}\int \dd\vecst{x}\,\smash{\Psim}^\dag\boper{\Xi}_\herm\Psim.
\end{eqnarray}

Because we have assumed a linear model for \m{f_s} in \eq{eq:aux31}, \m{\Psim} is decoupled from \m{\Psiq}, and hence the dynamics of \m{\Psim} turns out to be collisionless. This is justified, because collisional dissipation is assumed to be much slower that collisionless dissipation (\Sec{sec:avrf}). One can reinstate collisions in \eq{eq:Psime} by modifying \m{\oper{G}_s} \adhoc, if necessary. Alternatively, one can avoid separating \m{\Psim} and \m{\Psiq} and, instead, derive an equation for the average Wigner matrix of the whole \m{\osc{\vec{\Psi}}} \citep{ref:mcdonald91}. However, this approach is beyond QLT, so it is not considered in this paper.

\subsection{Dispersion matrix}
\label{sec:Xi}

As readily seen from the definition \eq{eq:Xidef}, the operator \m{\boper{\Xi}} can be expressed as
\begin{eqnarray}\label{eq:Xiop2}
\boper{\Xi} =
\boper{\Xi}_p
- \ii\oper{k}_j \sum_s \int \dd\vec{p}\,\boper{\alpha}_s \oper{G}_s 
\boper{\alpha}_s^\dag \,\frac{\pd F_s}{\pd p_j} + \mc{O}(\epsilon, \varepsilon^2).
\end{eqnarray}
The corrections caused by nonzero \(\epsilon\) and \(\varepsilon\) in this formula will be insignificant for our purposes, so they will be neglected. In particular, this means that \m{G_s \doteq \wsymbX \oper{G}_s} can be adopted in the form independent of \m{\vec{r}} (\Sec{sec:G}):
\begin{eqnarray}
G_s \approx 
\frac{\ii}{\omega - \vec{k} \cdot \vec{v}_s + \ii 0}
= \upi\,\delta(\omega - \vec{k} \cdot \vec{v}_s) + \ii\,\pv\frac{1}{\omega - \vec{k} \cdot \vec{v}_s}.
\end{eqnarray}
Then, \m{\oper{G}_s} can be considered as an operator on \m{\hilspacex} with \(\vec{p}\) as a parameter, and \m{\wsymb \oper{G}_s = G_s(t, \vec{x}, \omega, \vec{k}; \vec{p})}. Also, 
\begin{eqnarray}
\wsymb{(\boper{\alpha}_s \oper{G}_s \boper{\alpha}_s^\dag)}
= \vec{\alpha}_s \star G_s \star \vec{\alpha}_s^\dag
\approx \vec{\alpha}_s G_s \vec{\alpha}_s^\dag.
\end{eqnarray}
This readily yields the `dispersion matrix' \m{\matr{\Xi} \doteq \wsymbx{\boper{\Xi}}}:
\begin{eqnarray}\label{eq:Xis}
& \displaystyle
\matr{\Xi}(\omega, \vec{k}) \approx
\matr{\Xi}_p(\omega, \vec{k})
+ \sum_s \int \dd\vec{p}\,
\frac{\vec{\alpha}_s(\omega, \vec{k}; \vec{p})\vec{\alpha}_s^\dag(\omega, \vec{k}; \vec{p})}{\omega - \vec{k} \cdot \vec{v}_s + \ii 0}\,
\vec{k} \cdot \frac{\pd F_s(\vec{p})}{\pd \vec{p}},
\\
& \displaystyle
\matr{\Xi}_p(\omega, \vec{k}) \approx \matr{\Xi}_0(\omega, \vec{k}) 
- \sum_s \int \dd\vec{p}\,\matr{\wp}_s(\omega, \vec{k}; \vec{p})F_s(\vec{p})
\label{eq:Xiwps}
\end{eqnarray}
(see \Sec{sec:examples} for examples). Here, \m{\vec{\alpha}_s \vec{\alpha}_s^\dag} is a dyadic matrix, and the arguments \(t\) and \(\vec{x}\) are henceforth omitted for brevity. Also, we introduced the operators \m{\boper{\wp}_s = \boper{\wp}_s^\dag} and their symbols \m{\matr{\wp}_s = \matr{\wp}_s^\dag} as
\begin{eqnarray}
\boper{\wp}_s \doteq (\smash{\boper{L}}_s^\dag\boper{R})_\herm,
\qquad
\matr{\wp}_s \doteq \wsymb \boper{\wp}_s \approx (\matr{L}_s^\dag \matr{R}_s)_\herm.
\label{eq:defwp}
\end{eqnarray}

The appearance of \(+ \ii 0\) in the denominator in \eq{eq:Xis} is related to the Landau rule. (Remember that as arguments of Weyl symbols, \(\omega\) and \(\vec{k}\) are real by definition.) The Hermitian and anti-Hermitian parts of the dispersion matrix are
\begin{align}
\matr{\Xi}_\herm(\omega, \vec{k}) 
& \approx
\matr{\Xi}_p(\omega, \vec{k})
+ \sum_s \fint \dd\vec{p}\,
\frac{\vec{\alpha}_s(\omega, \vec{k}; \vec{p})\vec{\alpha}_s^\dag(\omega, \vec{k}; \vec{p})}{\omega - \vec{k} \cdot \vec{v}_s}\,
\vec{k} \cdot \frac{\pd F_s(\vec{p})}{\pd \vec{p}},
\label{eq:Xih}
\\
\matr{\Xi}_\aherm(\omega, \vec{k}) 
& \approx
-\upi \sum_s \int \dd\vec{p}\,
\vec{\alpha}_s(\omega, \vec{k}; \vec{p})\vec{\alpha}_s^\dag(\omega, \vec{k}; \vec{p})
\delta(\omega - \vec{k} \cdot \vec{v}_s)\,
\vec{k} \cdot \frac{\pd F_s(\vec{p})}{\pd \vec{p}}.
\label{eq:Xiah}
\end{align}
Assuming the notation \m{\matr{\Xi}^{-\dag} \doteq (\matr{\Xi}^\dag)^{-1}}, the inverse dispersion matrix can be expressed as
\begin{eqnarray}
\matr{\Xi}^{-1} 
= \matr{\Xi}^{-1} \matr{\Xi}^{\dag} \matr{\Xi}^{-\dag}
= \matr{\Xi}^{-1} \matr{\Xi}_\herm \matr{\Xi}^{-\dag} 
- \ii \matr{\Xi}^{-1} \matr{\Xi}_\aherm \matr{\Xi}^{-\dag}.
\end{eqnarray}
Because \m{\matr{\Xi}^{-\dag} = (\matr{\Xi}^{-1})^\dag}, this leads to the following formulas, which we will need later:
\begin{eqnarray}
(\matr{\Xi}^{-1})_\herm = \matr{\Xi}^{-1} \matr{\Xi}_\herm \matr{\Xi}^{-\dag},
\qquad
(\matr{\Xi}^{-1})_\aherm = -\matr{\Xi}^{-1} \matr{\Xi}_\aherm \matr{\Xi}^{-\dag}.
\label{eq:invXi}
\end{eqnarray}

\subsection{Spectrum of microscopic fluctuations}
\label{sec:WPsi}

Other objects to be used below are the density operators of the oscillating fields:
\begin{eqnarray}
\boperst{W}_{\Psim} \doteq (2\upi)^{-\stn}\,\ket{\Psim}\bra{\Psim},
\qquad
\boperst{W}_{\Psiq} \doteq (2\upi)^{-\stn}\,\ket{\Psiq}\bra{\Psiq},
\end{eqnarray}
and the corresponding average Wigner matrices on \m{\cb{\vecst{x}, \vecst{k}}}. The former, \m{\matrU \doteq \avr{\matrst{W}}_{\Psim}}, is readily found by definition \eq{eq:Wtenx}, and the latter, \m{\matru \doteq \avr{\matrst{W}}_{\Psiq}}, is calculated as follows. Let us consider \m{g_s(t, \vec{x}, \vec{p})} as a ket in \m{\hilspacex}, with \m{\vec{p}} as a parameter. Then, \eq{eq:aux32} readily yields
\begin{eqnarray}
\boperst{W}_{\Psiq} = \frac{1}{(2\upi)^\stn}\sum_{s,s'} \int \dd\vec{p}\,\dd\vec{p}'\,
\invXi\boper{\alpha}_s(\vec{p}) \ket{g_s(\vec{p})}
\bra{g_{s'}(\vec{p}')}\boper{\alpha}^\dag_{s'}(\vec{p}')\smash{\boper{\Xi}}^{-\dag}.
\end{eqnarray}
By applying \m{\wsymbx{}} to this, one obtains
\begin{eqnarray}
\matru(\omega, \vec{k}) = \sum_{s,s'} \int \dd\vec{p}\,\dd\vec{p}'\,
\matr{\Xi}^{-1} \star \matr{\alpha}_s(\vec{p}) 
\star \matr{\mcc{G}}_{ss'}(\vec{p}, \vec{p}') \star
\matr{\alpha}^\dag_{s'}(\vec{p}') \star \matr{\Xi}^{-\dag},
\label{eq:u1}
\end{eqnarray}
where most arguments are omitted for brevity and (\App{app:Wmcc})
\begin{align}
\mcc{G}_{ss'}(\vec{p}, \vec{p}') 
& \doteq  \frac{1}{(2\upi)^\stn} \int \dd\vecst{s}\,\ee^{- \ii\vecst{k}\cdot\vecst{s}}\,
\avr{g_{s}(\vecst{x} + \vecst{s}/2, \vec{p})\,
g_{s'}(\vecst{x} - \vecst{s}/2, \vec{p}')}
\notag\\
& \approx 
\frac{1}{(2\upi)^n}\,\delta_{ss'}\delta(\vec{p} - \vec{p}') 
\delta(\omega - \vec{k} \cdot \vec{v}_s)F_s(\vec{p}),
\label{eq:mccG}
\end{align}
assuming corrections due to inter-particle correlations are negligible. Then, \eq{eq:u1} gives
\begin{align}
\matru(\omega, \vec{k}) = \frac{1}{(2\upi)^{n}} \sum_{s'}\int \dd\vec{p}'\,&\delta(\omega - \vec{k} \cdot \vec{v}'_{s'})F_{s'}(\vec{p}')
\notag\\
&
\times
\matr{\Xi}^{-1}(\omega, \vec{k})
(\vec{\alpha}_{s'}\vec{\alpha}_{s'}^\dag)(\omega, \vec{k}; \vec{p}')
\matr{\Xi}^{-\dag}(\omega, \vec{k}),
\label{eq:ufina}
\end{align}
where \m{\vec{v}'_{s'} \doteq \vec{v}_{s'}(t, \vec{x}, \vec{p}')}. It is readily seen from \eq{eq:ufina} that \m{\matru} is positive-semidefinite. One can also recognize \eq{eq:ufina} as a manifestation of the dressed-particle superposition principle \citep{ref:rostoker64}. Specifically, \eq{eq:ufina} shows that the contributions of individual particles to \m{\matru} are additive and affected by the plasma collective response, \ie by the difference between \m{\matr{\Xi}} and the vacuum dispersion matrix \m{\matr{\Xi}_0}.

Using \eq{eq:ufina}, one can also find other averages quadratic in the field via (cf.\ \eq{eq:LR2})
\begin{eqnarray}
\avr{(\boperst{L}\Psiq)(\boperst{R}\Psiq)^\dag}
\approx \int \dd\omega\,\dd\vec{k}\,
(\matrst{L} \matru \matrst{R}^\dag)(\omega, \vec{k}),
\label{eq:LRu}
\end{eqnarray}
where \m{\boperst{L}} and \m{\boperst{R}} are any linear operators and \m{\matrst{L}} and \m{\matrst{R}} are their symbols; for example,
\begin{eqnarray}
\avr{\Psiq(t, \vec{x})\Psiq^\dag(t, \vec{x})} \approx \int \dd\omega\,\dd\vec{k}\,\matru(\omega, \vec{k}).
\label{eq:psipsi}
\end{eqnarray}
Because of this, we loosely attribute \m{\matru} as the spectrum of microscopic oscillations, but see also \Sec{sec:fdt}, where an alternative notation is introduced and a fluctuation--dissipation theorem is derived from \eq{eq:ufina} for plasma in thermal equilibrium. See also \Sec{sec:examples} for specific examples.

\subsection{Nonlinear potentials}
\label{sec:trc}

From \eq{eq:aux32}, the oscillating part of the Hamiltonian \eq{eq:Hs1} can be split into the macroscopic part and the microscopic part as \m{\osc{H}_s = \macro{\osc{H}}_s + \micro{\osc{H}}_s}, \m{\macro{\osc{H}}_s = \boper{\alpha}_s^\dag \Psim}, and
\begin{eqnarray}\label{eq:Hmm}
\micro{\osc{H}}_s(\vec{p}) = \sum_{s'} \int \dd\vec{p}'\,\oper{\mc{X}}_{ss'}(\vec{p}, \vec{p}')g_{s'}(\vec{p}').
\end{eqnarray}
Here, \m{\oper{\mc{X}}_{ss'}} is an operator on \m{\hilspacex} given by
\begin{eqnarray}\label{eq:opUps}
\oper{\mc{X}}_{ss'}(\vec{p}, \vec{p}') \doteq \boper{\alpha}_s^\dag(\vec{p})\invXi \boper{\alpha}_{s'}(\vec{p}'),
\end{eqnarray}
with the symbol
\begin{eqnarray}
\mc{X}_{ss'}(\omega, \vec{k}; \vec{p}, \vec{p}') \approx
\vec{\alpha}_s^\dag(\omega, \vec{k}; \vec{p})
\matr{\Xi}^{-1}(\omega, \vec{k})
\vec{\alpha}_{s'}(\omega, \vec{k}; \vec{p}')
\label{eq:mcYsymb}
\end{eqnarray}
 (see \Sec{sec:examples} for examples). The corresponding average Wigner functions on \m{\cb{\vecst{x}, \vecst{k}}} are \m{\avr{\st{W}}_s = \avr{\st{W}}_s^{\text{(m)}} + \avr{\st{W}}_s^{(\mu)}}, where the index `m' stands for `macroscopic' and the index `\m{\mu}' stands for `microscopic'. Because the dependence on \m{t} and \m{\vec{x}} is slow, one can approximate them as follows:
\begin{subequations}
\begin{eqnarray}
& \avr{\st{W}}_s^{\text{(m)}} \approx 
\matr{\alpha}_s^\dag(\omega, \vec{k}; \vec{p})
\matrU(\omega, \vec{k})
\matr{\alpha}_s(\omega, \vec{k}; \vec{p}),
\label{eq:WsW}\\
& \avr{\st{W}}_s^{(\mu)} \approx
\matr{\alpha}_s^\dag(\omega, \vec{k}; \vec{p})
\matru(\omega, \vec{k})
\matr{\alpha}_s(\omega, \vec{k}; \vec{p}).
\end{eqnarray}
\end{subequations}
The matrix \m{\matrU} is positive-semidefinite as an average Wigner tensor (\Sec{sec:vecf}), and so is \m{\matru} (\Sec{sec:WPsi}). Hence, both \m{\avr{\st{W}}_s^{\text{(m)}}} and \m{\avr{\st{W}}_s^{(\mu)}} are non-negative. Using \eq{eq:ufina}, one can also rewrite the Wigner function of \m{\micro{\osc{H}}_s} more compactly as
\begin{eqnarray}\label{eq:WZF}
\avr{\st{W}}_s^{(\mu)} (\omega, \vec{k}; \vec{p}) = (2\upi)^{-n} \sum_{s'}\int \dd\vec{p}'\,
\delta(\omega - \vec{k} \cdot \vec{v}'_{s'})
|\mc{X}_{ss'}(\omega, \vec{k}; \vec{p}, \vec{p}')|^2\,F_{s'}(\vec{p}').
\end{eqnarray}

Now we can represent the nonlinear potentials \eq{eq:Dsc}--\eq{eq:Phisc} as
\begin{eqnarray}
\matrst{D}_s = \matrst{D}_s^{\text{(m)}} + \matrst{D}_s^{(\mu)},
\qquad
\matr{\Uptheta}_s = \matr{\Uptheta}_s^{\text{(m)}} + \matr{\Uptheta}_s^{(\mu)},
\qquad
\Phi_s & = \Phi_s^{\text{(m)}} + \Phi_s^{(\mu)}.
\end{eqnarray}
Here, the index \m{^{\text{(m)}}} denotes contributions from \m{\avr{\st{W}}_s^{\text{(m)}}} and the index \m{^{(\mu)}} denotes contributions from \m{\avr{\st{W}}_s^{(\mu)}}. Specifically,
\begin{align}
\matrst{D}_s^{\text{(m)}}
& = \int \dd\vec{k}\, \upi\, \vec{k} \vec{k} \avr{\st{W}}_s^{\text{(m)}}(\vec{k} \cdot \vec{v}_s, \vec{k}; \vec{p}),
\\
\matr{\Uptheta}_s^{\text{(m)}}
& =\frac{\pd}{\pd \vartheta} \fint \dd\omega\,\dd\vec{k}\,
 \left.
\frac{\vec{k} \vec{k} \avr{\st{W}}_s^{\text{(m)}}(\omega, \vec{k}; \vec{p})}{\omega - \vec{k} \cdot \vec{v}_s + \vartheta}
\right|_{\vartheta=0},
\\
\Phi_s^{\text{(m)}} 
& = \frac{1}{2}\frac{\pd}{\pd \vec{p}} \cdot
\fint \dd\omega\,\dd\vec{k}\,\frac{\vec{k} \avr{\st{W}}_s^{\text{(m)}}(\omega, \vec{k}; \vec{p})}{\omega - \vec{k} \cdot \vec{v}_s}.
\label{eq:Phim}
\end{align}
Here, \m{\avr{\st{W}}_s^{\text{(m)}}} is a non-negative function \eq{eq:WsW}, so \m{\matrst{D}_s^{\text{(m)}}} is positive-semidefinite and leads to an \m{H}-theorem similar to \eq{eq:HthD}. One also has
\begin{align}
\matrst{D}_s^{(\mu)}
& = \sum_{s'}\int \frac{\dd\vec{k}}{(2\upi)^n}\,\dd\vec{p}'\,\upi\, \vec{k} \vec{k} \,
\delta(\vec{k} \cdot \vec{v}_s - \vec{k} \cdot \vec{v}'_{s'})\,
|\mc{X}_{ss'}(\vec{k} \cdot \vec{v}_{s}, \vec{k}; \vec{p}, \vec{p}')|^2\,F_{s'}(\vec{p}'),
\label{eq:Dmu}
\\
\matr{\Uptheta}_s^{(\mu)}
& = \sum_{s'} \frac{\pd}{\pd \vartheta} \fint \frac{\dd\vec{k}}{(2\upi)^n}\,\dd\vec{p}'\,
 \left.
\frac{\vec{k} \vec{k} F_{s'}(\vec{p}')}{\vec{k} \cdot (\vec{v}'_{s'} - \vec{v}_s) + \vartheta}\,
|\mc{X}_{ss'}(\vec{k} \cdot \vec{v}'_{s'}, \vec{k}; \vec{p}, \vec{p}') |^2
\right|_{\vartheta=0},
\\
\Phi_s^{(\mu)}
& = \sum_{s'}\frac{\pd}{\pd \vec{p}} \cdot
\fint \frac{\dd\vec{k}}{(2\upi)^n}\,\dd\vec{p}'\,
\frac{\vec{k} F_{s'}(\vec{p}')}{2\vec{k} \cdot (\vec{v}'_{s'} - \vec{v}_s)}
\,
|\mc{X}_{ss'}(\vec{k} \cdot \vec{v}'_{s'}, \vec{k}; \vec{p}, \vec{p}') |^2.
\end{align}
The functions \m{\matr{\Uptheta}_s^{(\mu)}} and \m{\Phi_s^{(\mu)}} scale as \m{\avr{\st{W}}_s^{(\mu)}}, \ie as \m{\epsilon\varepsilon^2} (\Sec{sec:avrf}). Their contribution to \eq{eq:OCFeq} is of order \m{\epsilon\matr{\Uptheta}_s^{(\mu)}} and \m{\epsilon\Phi_s^{(\mu)}}, respectively, so it scales as \m{\epsilon^2\varepsilon^2} and therefore is negligible within our model. In contrast, \m{\matrst{D}_s^{(\mu)}} must be retained alongside with \m{\matrst{D}_s^{\text{(m)}}}. This is because although weak, macroscopic fluctuations can resonate with particles from the bulk distribution, while the stronger macroscopic fluctuations are assumed to resonate only with particles from the tail distribution, which are few.

\subsection{Oscillation-center Hamiltonian}
\label{sec:Phi}

Within the assumed accuracy, the OC Hamiltonian is \m{\mc{H}_s = \avr{H}_s + \Phi_s^{\text{(m)}}}, and \m{\avr{H}_s} is given by \eq{eq:Hs1}. Combined with the general theorem \eq{eq:LRtr}, the latter readily yields \m{\avr{H}_s = H_{0s} + \phi_s}, where
\begin{eqnarray}
\phi_s \approx \frac{1}{2}\int \dd\omega\,\dd\vec{k}\,
\tr\big(\matrU\matr{\wp}_s\big)
\label{eq:auxphi}
\end{eqnarray}
and the contribution of \m{\matru} has been neglected. Because both \m{\Phi_s^{\text{(m)}}} and \m{\phi_s} are quadratic in \m{\osc{\vec{\Psi}}} and enter \m{\mc{H}_s} only in the combination \m{\Delta_s \doteq \Phi_s^{\text{(m)}} + \phi_s}, it is convenient to attribute the latter as the `total' ponderomotive energy. Using \eq{eq:Phim} in combination with \eq{eq:WsW}, one can express it as follows:
\begin{eqnarray}
\Delta_s = \frac{1}{2}\frac{\pd}{\pd \vec{p}} \cdot
\fint \dd\omega\,\dd\vec{k}\,\vec{k}\, 
\frac{\vec{\alpha}_s^\dag \matrU \vec{\alpha}_s}{\omega - \vec{k} \cdot \vec{v}_s}
+ \frac{1}{2}\int \dd\omega\,\dd\vec{k}\,\tr\big(\matrU\matr{\wp}_s\big)
\label{eq:Phifin}
\end{eqnarray}
(see \Sec{sec:examples} for examples). Notably,
\begin{eqnarray}\label{eq:kchi}
\Delta_s = -\frac{1}{2} \frac{\delta}{\delta F_s}\fint \dd\omega\,\dd\vec{k}\, \tr(\matr{\Xi}_\herm \matrU) = -\frac{\delta S_{\text{ad}}}{\delta F_s},
\end{eqnarray}
where \m{\delta/\delta F_s} denotes a functional derivative and \m{S_{\text{ad}}} is the adiabatic action defined in \eq{eq:sad}. Equation \eq{eq:kchi} is a generalization of the well-known `\m{K}--\m{\chi} theorem' \citep{ref:kaufman84, ref:kaufman87}. Loosely speaking, it says that the coefficient connecting \m{\Delta_s} with \m{\matrU} is proportional to the linear polarizability of an individual particle of type~\m{s} \citep{my:nonloc, my:kchi}. (`\m{K}' in the name of this theorem is the same as our \m{\Delta_s}, and `\m{\chi}' is the linear susceptibility.) Also, the OC Hamiltonian and the OC velocity can be expressed as 
\begin{eqnarray}
\mc{H}_s = H_{0s} + \Delta_s,
\qquad
\vec{v} = \pd_\vec{p} H_{0s} + \pd_\vec{p}\Delta_s.
\end{eqnarray}

\subsection{Polarization drag}
\label{sec:pd}

Within the assumed accuracy, the OC distribution can be expressed as
\begin{eqnarray}
F_s = \avr{f}_s + \frac{1}{2}\,\frac{\pd}{\pd \vec{p}} \cdot 
\left(\vec{\Uptheta}_s^{\text{(m)}}\, \frac{\pd \avr{f}_s}{\pd \vec{p}}\right),
\end{eqnarray}
and \eq{eq:OCFeq} becomes
\begin{eqnarray}\label{eq:OCFeqC}
& \displaystyle
\frac{\pd F_s}{\pd t}
- \frac{\pd \mc{H}_s}{\pd \vec{x}} \cdot \frac{\pd F_s}{\pd \vec{p}}
+ \frac{\pd \mc{H}_s}{\pd \vec{p}} \cdot \frac{\pd F_s}{\pd \vec{x}}
= \frac{\pd}{\pd \vec{p}} \cdot \left(\vecst{D}_s^{\text{(m)}}\,\frac{\pd F_s}{\pd \vec{p}}\right)
+ \collision_s,
\\
& \displaystyle
\label{eq:cop1}
\collision_s
\doteq \frac{\pd}{\pd \vec{p}} \cdot \left(\vecst{D}_s^{(\mu)}\,\frac{\pd F_s}{\pd \vec{p}}\right) + \Gamma_s,
\end{eqnarray}
where we have reinstated the term \m{\Gamma_s} introduced in \Sec{sec:avrf}. As a collisional term, \m{\Gamma_s} is needed only to the zeroth order in \m{\epsilon}, so
\begin{eqnarray}
& \displaystyle
\Gamma_s = \avr{\poissonz{\osc{H}_s, g_s}} 
\approx \pd_{\vec{p}} \cdot (\avr{g_s \pd_{\vec{x}}\osc{H}_s})
\equiv \pd_{\vec{p}} \cdot  \vec{\zeta}_s,
\qquad
\vec{\zeta}_s = \ii \avr{\braket{\vecst{x}|\boper{k} \osc{H}_s} \braket{\vecst{x}|g_s}}.
\end{eqnarray}
Correlating with \m{g_s} is only the microscopic part of \m{\osc{H}_s}, so using \eq{eq:Hmm} one obtains
\begin{eqnarray}
\vec{\zeta}_s 
= \ii \sum_{s'} \int \dd\vec{p}'\,\langle\vecst{x}|\boper{k}\oper{\mc{X}}_{ss'}(\vec{p}, \vec{p}')
\avr{\ket{g_{s'}(\vec{p}')}\bra{g_s(\vec{p})}}  \vecst{x}\rangle.
\end{eqnarray}
Next, let us use \eq{eq:Mxx2} and \m{\vec{\zeta} = \re \vec{\zeta}} to express this result as follows:
\begin{align}
\vec{\zeta}_s
& = \ii \sum_{s'} \int \dd\omega\,\dd\vec{k}\,\dd\vec{p}'\,\vec{k} \star \mc{X}_{ss'}(\omega, \vec{k}, \vec{p}, \vec{p}')
\star \matr{\mcc{G}}_{s's}(\omega, \vec{k}, \vec{p}', \vec{p})
\notag\\
& \approx \ii \int \dd\omega\,\dd\vec{k}\,\dd\vec{p}'\,\vec{k} \mc{X}_{ss'}(\omega, \vec{k}, \vec{p}, \vec{p}')
 \,(2\upi)^{-n}\delta(\vec{p} - \vec{p}')\,\delta(\omega - \vec{k} \cdot \vec{v}_s)\,F_s(\vec{p})
\notag\\
& = - \int \frac{\dd\vec{k}}{(2\upi)^n}\,\vec{k} \im \mc{X}_{ss'}(\vec{k} \cdot \vec{v}_s, \vec{k}, \vec{p}, \vec{p})\,F_s(\vec{p}),
\end{align}
where we have approximated \m{\star} with the usual product and substituted \eq{eq:mccG}.  Hence,
\begin{eqnarray}\label{eq:gammaf}
\Gamma_s \approx -\pd_\vec{p} \cdot (\vec{\mcc{F}}_s F_s),
\end{eqnarray}
where \m{\vec{\mcc{F}}_s} can be interpreted as the polarization drag (\ie the average force that is imposed on an OC by its dress) and is given by
\begin{eqnarray}\label{eq:drag0}
\vec{\mcc{F}}_s = \int \frac{\dd\vec{k}}{(2\upi)^n}\,\vec{k} \im \mc{X}_{ss'}(\vec{k} \cdot \vec{v}_s, \vec{k}, \vec{p}, \vec{p}).
\end{eqnarray}
Using \eq{eq:mcYsymb}, one also rewrite this as follows:
\begin{subequations}
\begin{align}
\vec{\mcc{F}}_s
& \approx \int \frac{\dd\vec{k}}{(2\upi)^n}\,\vec{k}\,
(\vec{\alpha}_s^\dag (\matr{\Xi}^{-1})_\aherm \vec{\alpha}_s)(\vec{k} \cdot \vec{v}_s, \vec{k}; \vec{p})
\label{eq:auxFd}\\
& \approx -\int \frac{\dd\vec{k}}{(2\upi)^n}\,\vec{k}\,
(\vec{\alpha}_s^\dag \matr{\Xi}^{-1} \matr{\Xi}_\aherm \matr{\Xi}^{-\dag}\vec{\alpha}_s)(\vec{k} \cdot \vec{v}_s, \vec{k}; \vec{p}),
\end{align}
\end{subequations}
where we have substituted \eq{eq:invXi} for \m{(\matr{\Xi}^{-1})_\aherm}. With \eq{eq:Xiah} for \m{\matr{\Xi}_\aherm}, this yields
\begin{align}
\vec{\mcc{F}}_s \approx \sum_{s'}\int \frac{\dd\vec{k}}{(2\upi)^n}\,\dd\vec{p}'\,&
\upi\,\delta(\vec{k} \cdot \vec{v}_s - \vec{k} \cdot \vec{v}'_{s'})\,\vec{k}\vec{k}\cdot\frac{\pd F_{s'}(\vec{p}')}{\pd \vec{p}'}
\notag\\
& \times \vec{\alpha}_s^\dag(\vec{k} \cdot \vec{v}_s, \vec{k}; \vec{p})
\matr{\Xi}^{-1}(\vec{k} \cdot \vec{v}_s, \vec{k})
\vec{\alpha}_{s'}(\vec{k} \cdot \vec{v}_s, \vec{k}; \vec{p}')
\notag\\
& 
\times \vec{\alpha}_{s'}^\dag(\vec{k} \cdot \vec{v}_s, \vec{k}; \vec{p}')
\matr{\Xi}^{-\dag}(\vec{k} \cdot \vec{v}_s, \vec{k})
\vec{\alpha}_s(\vec{k} \cdot \vec{v}_s, \vec{k}; \vec{p}).
\end{align}
The product of the last two lines equals \m{|\mc{X}_{ss'}(\vec{k} \cdot \vec{v}_{s}, \vec{k}; \vec{p}, \vec{p}')|^2}. Hence,
\begin{eqnarray}\label{eq:drag}
\vec{\mcc{F}}_s =  
\sum_{s'}\int \frac{\dd\vec{k}}{(2\upi)^n}\,\dd\vec{p}'\,
\upi\,\delta(\vec{k} \cdot \vec{v}_s - \vec{k} \cdot \vec{v}'_{s'})\,
|\mc{X}_{ss'}(\vec{k} \cdot \vec{v}_s, \vec{k}; \vec{p}, \vec{p}')|^2\,\vec{k}\vec{k}\cdot
\frac{\pd F_{s'}(\vec{p}')}{\pd \vec{p}'}.
\end{eqnarray}

\subsection{Collision operator}
\label{sec:cop}

By combining \eq{eq:gammaf} for \m{\Gamma_s} with \eq{eq:Dmu} for \m{\matrst{D}_s^{(\mu)}}, one can express \m{\collision_s}~as
\begin{align}\label{eq:cop2}
\collision_s
= \frac{\pd}{\pd \vec{p}}\cdot
\sum_{s'}\int \frac{\dd\vec{k}}{(2\upi)^n}\,\dd\vec{p}'\,&
\upi \,\delta(\vec{k} \cdot \vec{v}_s - \vec{k} \cdot \vec{v}'_{s'})\,
|\mc{X}_{ss'}(\vec{k} \cdot \vec{v}_{s}, \vec{k}; \vec{p}, \vec{p}')|^2
\notag\\
& \times \vec{k}\vec{k} \cdot
\left(
\frac{\pd F_s(\vec{p})}{\pd \vec{p}}\,F_{s'}(\vec{p}')
- F_s(\vec{p})\,\frac{\pd F_{s'}(\vec{p}')}{\pd \vec{p}'}
\right),
\end{align}
where \(\mc{X}_{ss'}\) is given by \eq{eq:mcYsymb}. One can recognize this as a generalization of the Balescu--Lenard collision operator \citep[section~11.11]{book:krall} to interactions via a general multi-component field \(\vec{\Psi}\). Specific examples can be found in \Sec{sec:examples}.

It is readily seen that \m{\collision_s} conserves particles, \ie
\begin{eqnarray}\label{eq:copcl0}
\int \dd\vec{p}\,\collision_s = 0,
\end{eqnarray}
and vanishes in thermal equilibrium (\Sec{sec:centr}). Other properties of \m{\collision_s} are determined by the properties of the coupling coefficient \m{\mc{X}_{ss'}}, which are as follows. Note that
\begin{eqnarray}
|\mc{X}_{ss'}(\omega, \vec{k}; \vec{p}, \vec{p}')|^2 =
\mc{Q}_{ss'}(\omega, \vec{k}; \vec{p}, \vec{p}') 
+ \mc{R}_{ss'}(\omega, \vec{k}; \vec{p}, \vec{p}')/2,
\end{eqnarray}
where we introduced
\begin{align}
\mc{Q}_{ss'}(\omega, \vec{k}; \vec{p}, \vec{p}') & \doteq 
(|\mc{X}_{ss'}(\omega, \vec{k}; \vec{p}, \vec{p}')|^2
+ 
|\mc{X}_{s's}(\omega, \vec{k}; \vec{p}', \vec{p})|^2)/2,
\\
\mc{R}_{ss'}(\omega, \vec{k}; \vec{p}, \vec{p}') 
& \doteq 
|\mc{X}_{ss'}(\omega, \vec{k}; \vec{p}, \vec{p}')|^2
- |\mc{X}_{s's}(\omega, \vec{k}; \vec{p}', \vec{p})|^2.
\end{align}
To calculate \m{\mc{R}_{ss'}}, note that \eq{eq:mcYsymb} yields
\begin{align}
|\mc{X}_{ss'}(\omega, \vec{k}; \vec{p}, \vec{p}')|^2 
& \approx
|\vec{\alpha}_s^\dag(\omega, \vec{k}; \vec{p})
\matr{\Xi}^{-1}(\omega, \vec{k})
\vec{\alpha}_{s'}(\omega, \vec{k}; \vec{p}')|^2,
\notag\\
|\mc{X}_{s's}(\omega, \vec{k}; \vec{p}', \vec{p})|^2 
& \approx
|\vec{\alpha}_s^\dag(\omega, \vec{k}; \vec{p})
\matr{\Xi}^{-\dag}(\omega, \vec{k})
\vec{\alpha}_{s'}(\omega, \vec{k}; \vec{p}')|^2,
\end{align}
whence one obtains
\begin{align}
\mc{R}_{ss'}(\omega, \vec{k}; \vec{p}, \vec{p}') 
\approx 4\im\big(&
\vec{\alpha}_s^\dag(\omega, \vec{k}; \vec{p})
(\matr{\Xi}^{-1})_\herm(\omega, \vec{k})
\vec{\alpha}_{s'}(\omega, \vec{k}; \vec{p}')
\notag\\
& \vec{\alpha}_{s'}^\dag(\omega, \vec{k}; \vec{p}')
(\matr{\Xi}^{-1})_\aherm(\omega, \vec{k})
\vec{\alpha}_{s'}(\omega, \vec{k}; \vec{p}')
\big).
\end{align}
The operators \m{(\matr{\Xi}^{-1})_\herm}, \m{(\matr{\Xi}^{-1})_\aherm}, and \m{\boper{\alpha}_s} (for all \m{s}) have been introduced for real fields, so their matrix elements in the coordinate representation are real. Then, the corresponding symbols satisfy \m{\matr{A}(-\omega, -\vec{k}) = \matr{A}^*(\omega, \vec{k})}, where \m{\matr{A}} is any of the three symbols. This gives
\begin{eqnarray}\label{eq:Delta}
\mc{R}_{ss'}(\vec{k}; \vec{p}, \vec{p}')
\doteq
\mc{R}_{ss'}(\vec{k} \cdot \vec{v}_s, \vec{k}; \vec{p}, \vec{p}')
= -\mc{R}_{ss'}(-\vec{k}; \vec{p}, \vec{p}').
\end{eqnarray}
Because the rest of the integrand in \eq{eq:cop2} is even in \m{\vec{k}}, \eq{eq:Delta} signifies that \m{\mc{R}_{ss'}} does not contribute to \m{\collision_s}. Thus, \m{\mc{X}_{ss'}} in \eq{eq:cop2} can as well be replaced with \m{\mc{Q}_{ss'}}:
\begin{align}\label{eq:cop21}
\collision_s
= \frac{\pd}{\pd \vec{p}} \cdot
\sum_{s'}\int \frac{\dd\vec{k}}{(2\upi)^n}\,\dd\vec{p}'\,&
\upi \,\delta(\vec{k} \cdot \vec{v}_s - \vec{k} \cdot \vec{v}'_{s'})\,
\mc{Q}_{ss'}(\vec{k} \cdot \vec{v}_{s}, \vec{k}; \vec{p}, \vec{p}')
\notag\\
& \times \vec{k}\vec{k} \cdot
\left(
\frac{\pd F_s(\vec{p})}{\pd \vec{p}}\,F_{s'}(\vec{p}')
- F_s(\vec{p})\,\frac{\pd F_{s'}(\vec{p}')}{\pd \vec{p}'}
\right).
\end{align}
In this representation, the coupling coefficient in \m{\collision_s} is manifestly symmetric,
\begin{eqnarray}
\mc{Q}_{ss'}(\omega, \vec{k}; \vec{p}, \vec{p}') 
= \mc{Q}_{s's}(\omega, \vec{k}; \vec{p}', \vec{p}),
\end{eqnarray}
which readily leads to momentum and energy conservation (\App{app:ccons}):\footnote{Remember that \m{\vec{v}_s} is defined as the OC velocity in the above formulas (\Sec{sec:Deq}). If \m{\vec{v}_s} is treated as the particle velocity instead, then \m{\mc{H}_s} in \eq{eq:copcl} should be replaced with \m{\avr{H}_s}. Both options are admissible within the assumed accuracy, but the former option is preferable because it leads to other conservation laws that are exact within our model (\Sec{sec:cons}).}
\begin{eqnarray}\label{eq:copcl}
\sum_s \int \dd\vec{p}\,\vec{p}\collision_s = 0,
\qquad
\sum_s \int \dd\vec{p}\,\mc{H}_s\collision_s = 0.
\end{eqnarray}
The collision operator \m{\collision_s} also satisfies the \m{H}-theorem (\App{app:Hth}):
\begin{eqnarray}\label{eq:Hth1}
\left(\frac{\dd\entropy}{\dd t}\right)_{\text{coll}} \ge 0,
\end{eqnarray}
where the entropy density \m{\entropy} is defined as
\begin{eqnarray}\label{eq:Hth2}
\entropy \doteq -\sum_s \int \dd\vec{p}\,F_s(\vec{p})\ln F_s(\vec{p})
\end{eqnarray}
and \m{(\pd_t F_s)_{\text{coll}} \doteq \collision_s}. Note that these properties are not restricted to any particular~\m{\mc{H}_s}. Also note that if applied in proper variables (\Sec{sec:qlapp}), our formula \eq{eq:cop21} can describe collisions in strong background fields. This topic, including comparison with the relevant literature, is left to future work.

\subsection{Summary of \Sec{sec:th}}
\label{sec:sumth}

Let us summarize the above general results (for examples, see \Sec{sec:examples}). We consider species \m{s} governed by a Hamiltonian of the form
\begin{eqnarray}
H_s = H_{0s} + \boper{\alpha}_s^\dag(\vec{p}) \osc{\vec{\Psi}} 
+ \frac{1}{2}\,(\boper{L}_s\osc{\vec{\Psi}})^\dag (\boper{R}_s\osc{\vec{\Psi}}),
\end{eqnarray}
where \m{\osc{\vec{\Psi}}} is a real oscillating field (of any dimension \m{M}), which generally consists of a macroscopic part \m{\Psim} and a microscopic part \m{\Psiq}. The term \m{H_{0s}} is independent of \m{\osc{\vec{\Psi}}}, and the operators \m{\smash{\boper{\alpha}}_s^\dag}, \m{\boper{L}_s}, and \m{\boper{R}_s} may be nonlocal in \(t\) and \(\vec{x}\) and may depend on the momentum \(\vec{p}\) parametrically. The dynamics of this system averaged over the fast oscillations can be described in terms of the OC distribution function
\begin{eqnarray}
F_s = \avr{f}_s + \frac{1}{2}\,\frac{\pd}{\pd \vec{p}} \cdot 
\left(\vec{\Uptheta}_s\, \frac{\pd \avr{f}_s}{\pd \vec{p}}\right)
\end{eqnarray}
(the index \m{^{\text{(m)}}} is henceforth omitted for brevity), which is governed by the following equation of the Fokker--Planck type:
\begin{eqnarray}\label{eq:FS}
\frac{\pd F_s}{\pd t}
- \frac{\pd \mc{H}_s}{\pd \vec{x}} \cdot \frac{\pd F_s}{\pd \vec{p}}
+ \frac{\pd \mc{H}_s}{\pd \vec{p}} \cdot \frac{\pd F_s}{\pd \vec{x}}
= \frac{\pd}{\pd \vec{p}} \cdot \left(\vecst{D}_s\,\frac{\pd F_s}{\pd \vec{p}}\right)
+ \collision_s.
\end{eqnarray}
Here, \m{\mc{H}_s = H_{0s} + \Delta_s} is the OC Hamiltonian, \m{\vec{\Uptheta}_s} is the dressing function, and \m{\Delta_s} is the total ponderomotive energy (\ie the part of the OC Hamiltonian that is quadratic in \m{\osc{\vec{\Psi}}}), so \m{\vec{v}_s(t, \vec{x}, \vec{p}) \doteq \pd_\vec{p}\mc{H}_s} is the OC velocity. Specifically,
\begin{subequations}\label{eq:trfW}
\begin{align}
\matrst{D}_s
& = \int \dd\vec{k}\,\upi \,\vec{k}\vec{k} \avr{\st{W}}_s(t, \vec{x}, \vec{k} \cdot \vec{v}_s, \vec{k}; \vec{p}),
\label{eq:Dpdiff}
\\
\matr{\Uptheta}_s
& =\frac{\pd}{\pd \vartheta} \fint \dd\omega\,\dd\vec{k}\,\vec{k}\vec{k}
 \left.
\frac{\avr{\st{W}}_s(t, \vec{x}, \omega, \vec{k}; \vec{p})}{\omega - \vec{k} \cdot \vec{v}_s + \vartheta}
\right|_{\vartheta=0},
\\
\Delta_s
& = \frac{1}{2}\frac{\pd}{\pd \vec{p}} \cdot
\fint \dd\omega\,\dd\vec{k}\,\vec{k}\,\frac{\avr{\st{W}}_s(t, \vec{x}, \omega, \vec{k}; \vec{p})}{\omega - \vec{k} \cdot \vec{v}_s}
\notag\\
& \hphantom{\,=\,}
+\frac{1}{2}\int \dd\omega\,\dd\vec{k}\,\tr\big(\matrU\matr{\wp}_s\big)(t, \vec{x}, \omega, \vec{k}; \vec{p}).
\end{align}
\end{subequations}
Here, \m{\avr{\st{W}}_s = \matr{\alpha}_s^\dag \matrU \matr{\alpha}_s} is a scalar function, the average Wigner matrix \m{\matrU} is understood as the Fourier spectrum of the symmetrized autocorrelation matrix of the macroscopic oscillations:
\begin{equation}\label{eq:Usum}
\matrst{U}(t, \vec{x}, \omega, \vec{k}) 
= \int \frac{\dd\tau}{2\upi}\,\frac{\dd\vec{s}}{(2\upi)^{n}}\,\, 
\avr{
{\macro{\osc{\vec{\Psi}}}}(t + \tau/2, \vec{x} + \vec{s}/2)\, 
\smash{\macro{\osc{\vec{\Psi}}}}^\dag(t - \tau/2, \vec{x} - \vec{s}/2)
}
\,\ee^{\ii\omega \tau - \ii \vec{k} \cdot \vec{s}},
\end{equation}
with \m{n \doteq \dim \vec{x}}. Also, the vector \m{\vec{\alpha}_s(t, \vec{x}, \omega, \vec{k}; \vec{p})} is the Weyl symbol of \m{\boper{\alpha}_s} as defined in \eq{eq:wsymbx}, \m{\matr{\wp}_s(t, \vec{x}, \omega, \vec{k}; \vec{p}) \approx (\matr{L}_s^\dag \matr{R}_s)_\herm}, \m{\matr{L}_s} and \m{\matr{R}_s} are the Weyl symbols of \m{\boper{L}_s} and \m{\boper{R}_s}, respectively, and \m{_\herm} denotes the Hermitian part. The matrix \m{\matrst{D}_s} is positive-semidefinite and satisfies an \m{H}-theorem of the form \eq{eq:HthD}. Also, \m{\Delta_s} satisfies the `\m{K}--\m{\chi} theorem':
\begin{eqnarray}
\Delta_s = -\frac{1}{2} \frac{\delta}{\delta F_s}\fint \dd\omega\,\dd\vec{k}\, \tr(\matr{\Xi}_\herm \matrU).
\end{eqnarray}
The matrix \(\matr{\Xi}\) characterizes the collective plasma response to the field \(\osc{\vec{\Psi}}\) and is given by 
\begin{align}
\matr{\Xi}\approx
\matr{\Xi}_0
+ \sum_s \int \dd\vec{p}\,
\Bigg(
\frac{\vec{\alpha}_s(\vec{p})\,\vec{\alpha}_s^\dag(\vec{p})}{\omega - \vec{k} \cdot \vec{v}_s(\vec{p}) + \ii 0}\,
\vec{k}\cdot \frac{\pd F_s(\vec{p})}{\pd \vec{p}}
- \matr{\wp}_s(\vec{p})F_s(\vec{p}) 
\Bigg).
\label{eq:Xifinn}
\end{align}
Here, the arguments \m{(t, \vec{x}, \omega, \vec{k})} are omitted for brevity, \m{\vec{\alpha}_s\vec{\alpha}_s^\dag} is a dyadic matrix, and \m{\matr{\Xi}_0} is the symbol of the Hermitian dispersion operator \m{\boper{\Xi}_0} that governs the field \m{\osc{\vec{\Psi}}} in the absence of plasma. Specifically, \m{\boper{\Xi}_0} is defined such that the field Lagrangian density without plasma is \m{\mcc{L}_0 = \smash{\osc{\vec{\Psi}}}^\dag \boper{\Xi}_0 \osc{\vec{\Psi}}/2}.

The spectrum of microscopic fluctuations (specifically, the spectrum of the symmetrized autocorrelation function of the microscopic field \m{\Psiq}) is a positive-semidefinite matrix function and given by
\begin{align}
\matru(\omega, \vec{k}) = \frac{1}{(2\upi)^{n}} \sum_{s'}\int \dd\vec{p}'\,&\delta(\omega - \vec{k} \cdot \vec{v}'_{s'})F_{s'}(\vec{p}')
\notag\\
&
\times
\matr{\Xi}^{-1}(\omega, \vec{k})
(\vec{\alpha}_{s'}\vec{\alpha}_{s'}^\dag)(\omega, \vec{k}; \vec{p}')
\matr{\Xi}^{-\dag}(\omega, \vec{k}),
\end{align}
where \m{\vec{v}'_{s'} \doteq \vec{v}_{s'}(\vec{p}')}. (The dependence on \m{t} and \m{\vec{x}} is assumed too but not emphasized.) The microscopic fluctuations give rise to a collision operator of the Balescu--Lenard type:
\begin{align}
\collision_s
= \frac{\pd}{\pd \vec{p}}\cdot
\sum_{s'}\int &\frac{\dd\vec{k}}{(2\upi)^n}\,\dd\vec{p}'\,
\upi \,\delta(\vec{k} \cdot \vec{v}_s - \vec{k} \cdot \vec{v}'_{s'})\,
\mc{Q}_{ss'}(\vec{k} \cdot \vec{v}_{s}, \vec{k}; \vec{p}, \vec{p}')
\notag\\
& \times \vec{k}\vec{k} \cdot
\left(
\frac{\pd F_s(\vec{p})}{\pd \vec{p}}\,F_{s'}(\vec{p}')
- F_s(\vec{p})\,\frac{\pd F_{s'}(\vec{p}')}{\pd \vec{p}'}
\right),
\end{align}
where the coupling coefficient \m{\mc{Q}_{ss'}(\omega, \vec{k}; \vec{p}, \vec{p}') 
= \mc{Q}_{s's}(\omega, \vec{k}; \vec{p}', \vec{p})} is given by
\begin{eqnarray}
& \displaystyle
\mc{Q}_{ss'}(\omega, \vec{k}; \vec{p}, \vec{p}') = 
(|\mc{X}_{ss'}(\omega, \vec{k}; \vec{p}, \vec{p}')|^2
+ 
|\mc{X}_{s's}(\omega, \vec{k}; \vec{p}', \vec{p})|^2)/2,
\\
& 
\mc{X}_{ss'}(\omega, \vec{k}; \vec{p}, \vec{p}') \approx
\vec{\alpha}_s^\dag(\omega, \vec{k}; \vec{p})
\matr{\Xi}^{-1}(\omega, \vec{k})
\vec{\alpha}_{s'}(\omega, \vec{k}; \vec{p}').
\end{eqnarray}
The operator \m{\collision_s} satisfies the \m{H}-theorem and conserves particles, momentum, and energy:
\begin{eqnarray}\notag
\int \dd\vec{p}\,\collision_s = 0,
\qquad 
\sum_s \int \dd\vec{p}\,\vec{p}\collision_s = 0,
\qquad
\sum_s \int \dd\vec{p}\,\mc{H}_s\collision_s = 0.
\end{eqnarray} 

\section{Interaction with on-shell waves}
\label{sec:onshell}

Here, we discuss QL interaction of plasma with `on-shell' waves, \ie waves constrained by dispersion relations. To motivate the assumptions that will be adopted, and also to systematically introduce our notation, we start with briefly overviewing theory of linear waves in dispersive media \citep{book:tracy, book:whitham}, including monochromatic waves (\Sec{sec:mono}), conservative eikonal waves (\Sec{sec:nondis}), general eikonal waves (\Sec{sec:eikonal}), and general broadband waves described by the WKE (\Sec{sec:broad}). After that, we derive conservation laws for the total momentum and energy, which are exact within our model (\Sec{sec:cons}). All waves in this section are considered macroscopic, so we adopt a simplified notation \m{\Psim \equiv \osc{\vec{\Psi}}} and the index \m{^{(\text{m})}} will be omitted.

\subsection{Monochromatic waves}
\label{sec:mono}

Conservative (nondissipative) waves can be described using the least-action principle \(\delta S = 0\). Assuming the notation as in \Sec{sec:feq}, the action integral can be expressed as \(S = \int \dd\vecst{x}\,\mcc{L}\) with the Lagrangian density given~by
\begin{eqnarray}
\mcc{L} = \frac{1}{2}\,\smash{\osc{\vec{\Psi}}}^\dag\boper{\Xi}_\herm\osc{\vec{\Psi}}.
\end{eqnarray}
First, let us assume a homogeneous stationary medium, so \m{\matr{\Xi}_\herm(t, \vec{x}, \omega, \vec{k}) = \matr{\Xi}_\herm(\omega, \vec{k})}. Because we assume real fields,\footnote{A complex field can be accommodated by considering its real and imaginary parts as separate components.} \m{\braket{t_1, \vec{x}_1 |\boper{\Xi}|t_2, \vec{x}_2}} is real for all \m{\cb{t_1, \vec{x}_1, t_2, \vec{x}_2}}, one also has
\begin{eqnarray}\label{eq:Xiprop}
\matr{\Xi}_\herm(-\omega, -\vec{k})
= \matr{\Xi}_\herm^*(\omega, \vec{k})
= \matr{\Xi}_\herm^\intercal(\omega, \vec{k}),
\end{eqnarray}
where the latter equality is due to \m{\matr{\Xi}_\herm^\dag(\omega, \vec{k}) = \matr{\Xi}_\herm(\omega, \vec{k})}.

Because \(\matr{\Xi}_\herm(\omega, \vec{k})\) is Hermitian, it has \(M \doteq \dim\matr{\Xi}_\herm\) orthonormal eigenvectors~\(\vec{\eta}_b\):
\begin{eqnarray}
\label{eq:Lamdef1}
\matr{\Xi}_\herm(\omega, \vec{k})\vec{\eta}_b(\omega, \vec{k}) = \Lambda_b(\omega, \vec{k})\vec{\eta}_b(\omega, \vec{k}),
\qquad
\vec{\eta}_b^\dag(\omega, \vec{k})\vec{\eta}_{b'}(\omega, \vec{k}) = \delta_{b,b'}.
\end{eqnarray}
Here \m{\Lambda_b} are the corresponding eigenvalues, which are real and satisfy
\begin{eqnarray}\label{eq:labetet}
\Lambda_b(\omega, \vec{k}) = \vec{\eta}_b^\dag(\omega, \vec{k})\matr{\Xi}_\herm(\omega, \vec{k})\vec{\eta}_b(\omega, \vec{k}).
\end{eqnarray}
Due to \eq{eq:Xiprop}, one has
\begin{eqnarray}
\Lambda_b(-\omega, -\vec{k})
= \text{eigv}_b\, (\matr{\Xi}_\herm(\omega, \vec{k}))^\intercal
= \text{eigv}_b\, \matr{\Xi}_\herm(\omega, \vec{k})
= \Lambda_b(\omega, \vec{k}),
\end{eqnarray}
where \(\text{eigv}_b\) stands for the \(b\)th eigenvalue. Using this together with \eq{eq:Xiprop}, one obtains from \eq{eq:Lamdef1} that
\begin{eqnarray}
\matr{\Xi}_\herm^*(\omega, \vec{k})\vec{\eta}_b(-\omega, -\vec{k})
= \Lambda_b(\omega, \vec{k})\vec{\eta}_b(-\omega, -\vec{k}),
\end{eqnarray}
whence
\begin{eqnarray}\label{eq:etastar}
\vec{\eta}_b(-\omega, -\vec{k}) = \vec{\eta}_b^*(\omega, \vec{k}).
\end{eqnarray}

Let us consider a monochromatic wave of the form
\begin{eqnarray}\label{eq:mono}
\osc{\vec{\Psi}}(t, \vec{x}) = \re(\ee^{-\ii \avr{\omega}t + \ii \avr{\vec{k}} \cdot \vec{x}} \,\env{\vec{\Psi}}),
\end{eqnarray}
with real frequency \(\avr{\omega}\), real wavevector \(\avr{\vec{k}}\), and complex amplitude \(\env{\vec{\Psi}}\). For such a wave, the action integral can be expressed as \m{S = \int \dd\vecst{x}\,\avr{\mcc{L}}}, where the average Lagrangian density \(\avr{\mcc{L}}\) is given by\footnote{Here we use that for any oscillating \m{a = \re (\ee^{\ii \theta}\env{a})} and \m{b = \re(\ee^{\ii \theta}\env{b})}, one has \m{\avr{a b} = \re (\env{a}^* \env{b})/2} and that \m{\smash{\env{\vec{\Psi}}}^\dag \matr{\Xi}_\herm(\avr{\omega}, \avr{\vec{k}}) \env{\vec{\Psi}}} is real because \m{\matr{\Xi}_\herm(\avr{\omega}, \avr{\vec{k}})} is Hermitian.}
\begin{eqnarray}
\avr{\mcc{L}} = \frac{1}{2}\,\avr{\smash{\osc{\vec{\Psi}}}^\dag \boper{\Xi}_\herm \osc{\vec{\Psi}}}
= \frac{1}{4}\,\re(\smash{\env{\vec{\Psi}}}^\dag \matr{\Xi}_\herm(\avr{\omega}, \avr{\vec{k}}) \env{\vec{\Psi}})
= \frac{1}{4}\,\smash{\env{\vec{\Psi}}}^\dag \matr{\Xi}_\herm(\avr{\omega}, \avr{\vec{k}}) \env{\vec{\Psi}}.
\label{eq:Laux}
\end{eqnarray}
Let us decompose \(\env{\vec{\Psi}}\) in the basis formed by the eigenvectors \m{\vec{\eta}_b}, that is, as
\begin{eqnarray}\label{eq:Psideco}
\env{\vec{\Psi}} = \sum_b \vec{\eta}_b \env{a}^b.
\end{eqnarray}
Then, \eq{eq:Laux} becomes
\begin{eqnarray}
\avr{\mcc{L}} = \frac{1}{4}\sum_b \Lambda_b(\avr{\omega}, \avr{\vec{k}})\,|\env{a}^b|^2.
\end{eqnarray}
The real and imaginary parts of the amplitudes \(\env{a}^b\) can be treated as independent variables. This is equivalent to treating \(\env{a}^{b*}\) and \(\env{a}^b\) as independent variables, so one arrives at the following Euler--Lagrange equations:
\begin{eqnarray}
0 = \frac{\delta S[\env{\vec{a}}^*, \env{\vec{a}}]}{\delta \env{a}^{b*}}
= \frac{1}{4}\,\Lambda_b(\avr{\omega}, \avr{\vec{k}})\env{a}^b,
\qquad
0 = \frac{\delta S[\env{\vec{a}}^*, \env{\vec{a}}]}{\delta \env{a}^b}
= \frac{1}{4}\,\env{a}^{b*}\Lambda_b(\avr{\omega}, \avr{\vec{k}}).
\end{eqnarray}
Hence the \(b\)th mode with a nonzero amplitude \(\env{a}^b\) satisfies the dispersion relation
\begin{eqnarray}
0 = \Lambda_b(\avr{\omega}, \avr{\vec{k}})
= \Lambda_b(-\avr{\omega}, -\avr{\vec{k}}).
\label{eq:Ldr}
\end{eqnarray}
Equation \eq{eq:Ldr} determines a dispersion surface in the \(\vecst{k}\) space where the waves can have nonzero amplitude. This surface is sometimes called a shell, so waves constrained by a dispersion relation are called on-shell. Also note that combining \eq{eq:Ldr} with \eq{eq:Lamdef1} yields that on-shell waves satisfy
\begin{eqnarray}\label{eq:zeroeta}
\matr{\Xi}_\herm(\avr{\omega}, \avr{\vec{k}})\vec{\eta}_b(\avr{\omega}, \avr{\vec{k}}) = \vec{0},
\qquad
\vec{\eta}_b^\dag(\avr{\omega}, \avr{\vec{k}})\matr{\Xi}_\herm(\avr{\omega}, \avr{\vec{k}}) = \vec{0},
\end{eqnarray}
which are two mutually adjoint representations of the same equation.

Below, we consider the case when \eq{eq:Ldr} is satisfied only for one mode at a time, so summation over \(b\) and the index \(b\) itself can be omitted. (A more general case is discussed, for example, in \citep{my:quasiop1}.) Then, \m{\env{\vec{\Psi}} = \vec{\eta}(\avr{\omega}, \avr{\vec{k}})\env{a}},
\begin{eqnarray}
\avr{\mcc{L}} = \frac{1}{4}\,\Lambda(\avr{\omega}, \avr{\vec{k}})|\env{a}|^2, 
\end{eqnarray}
and \(\avr{\omega}\) is connected with \(\avr{\vec{k}}\) via \m{\avr{\omega} = w(\avr{\vec{k}})}, where \(w(\vec{k}) = -w(-\vec{k})\) is the function that solves \(\Lambda (w(\vec{k}), \vec{k}) = 0\). Also importantly, \eq{eq:zeroeta} ensures that
\begin{align}
\pd_\placeholder \Lambda(\avr{\omega}, \avr{\vec{k}})
& = ((\pd_\placeholder \vec{\eta}^\dag) \matr{\Xi}_\herm\vec{\eta} + 
 \vec{\eta}^\dag(\pd_\placeholder \matr{\Xi}_\herm)\vec{\eta} + 
 \vec{\eta}^\dag\matr{\Xi}_\herm(\pd_\placeholder \vec{\eta}))\big|_{(\omega, \vec{k}) = (\avr{\omega}, \avr{\vec{k}})}
\notag\\
& = (\vec{\eta}^\dag(\pd_\placeholder \matr{\Xi}_\herm)\vec{\eta})\big|_{(\omega, \vec{k}) = (\avr{\omega}, \avr{\vec{k}})},
\label{eq:derzer}
\end{align}
where \m{\placeholder} can be replaced with any variable.

\subsection{Conservative eikonal waves}
\label{sec:nondis}

\subsubsection{Basic properties}
\label{sec:bpe}

In case of a quasimonochromatic eikonal wave and, possibly, inhomogeneous non-stationary plasma, one can apply the same arguments as in \Sec{sec:mono} except the above equalities are now satisfied up to \m{\mc{O}(\epsilon)}. For a single-mode wave, one has
\begin{eqnarray}\label{eq:Psic}
\osc{\vec{\Psi}}(t, \vec{x}) = \re(\osc{\vec{\Psi}}_{\text{c}}(t, \vec{x})) + \mc{O}(\epsilon),
\qquad
\osc{\vec{\Psi}}_{\text{c}} = \ee^{\ii \theta(t, \vec{x})} \vec{\eta}(t, \vec{x}) \env{a}(t, \vec{x}), 
\end{eqnarray}
where the local frequency and the wavevector,
\begin{eqnarray}\label{eq:wkr}
\avr{\omega} \doteq - \pd_t\theta,
\qquad
\avr{\vec{k}} \doteq \pd_{\vec{x}} \theta
\end{eqnarray}
are slow functions of \m{\cb{t, \vec{x}}}, and so is \m{\vec{\eta}(t, \vec{x}) \doteq \vec{\eta}(t, \vec{x}, \avr{\omega}(t, \vec{x}), \avr{\vec{k}}(t, \vec{x}))}, which satisfies \eq{eq:Lamdef1}.~Then,
\begin{eqnarray}\label{eq:La}
\avr{\mcc{L}} = \frac{1}{4}\,\Lambda(t, \vec{x}, \avr{\omega}, \avr{\vec{k}})\,|\env{a}(t, \vec{x})|^2 + \mc{O}(\epsilon).
\end{eqnarray}
Within the leading-order theory, the term \m{\mc{O}(\epsilon)} is neglected.\footnote{Corrections to the lowest-order dispersion relation produce the so-called spin Hall effect; see \citep{my:quasiop1, my:covar} for an overview and \citep{ref:bliokh15, my:qdiel, my:spinhall, ref:andersson21} for examples. These corrections are beyond the accuracy of the model considered, so they will be ignored.} Then, the least action principle
\begin{eqnarray}
0 = \frac{\delta S[\theta, \env{\vec{a}}^*, \env{\vec{a}}]}{\delta \env{a}^{b*}}
\approx \frac{1}{4}\,\Lambda_b(\avr{\omega}, \avr{\vec{k}})\env{a}^b,
\qquad
0 = \frac{\delta S[\theta, \env{\vec{a}}^*, \env{\vec{a}}]}{\delta \env{a}^b}
\approx \frac{1}{4}\,\env{a}^{b*}\Lambda_b(\avr{\omega}, \avr{\vec{k}})
\end{eqnarray}
leads to the same (but now local) dispersion relation as for monochromatic waves, \m{\Lambda(t, \vec{x}, \avr{\omega}, \avr{\vec{k}}) = 0}. This shows that quasimonochromatic waves are also on-shell, and thus they satisfy \eq{eq:derzer} as well. Also notice that the dispersion relation can now be understood as a Hamilton--Jacobi equation for the eikonal phase \m{\theta}:
\begin{eqnarray}\label{eq:HJ}
\Lambda(t, \vec{x}, -\pd_t \theta, \pd_{\vec{x}}\theta) = 0.
\end{eqnarray}

Like in the previous section, let us introduce the function \m{w} that solves 
\begin{eqnarray}\label{eq:wdef}
\Lambda(t, \vec{x}, w(t, \vec{x}, \vec{k}), \vec{k}) = 0
\end{eqnarray}
and therefore satisfies
\begin{eqnarray}\label{eq:wsym}
w(t, \vec{x}, \vec{k}) = -w(t, \vec{x}, -\vec{k}).
\end{eqnarray}
Differentiating \eq{eq:wdef} with respect to \(t\), \(\vec{x}\), and \(\vec{k}\) leads to
\begin{subequations}\label{eq:Lambdaw}
\begin{align}
& \pd_{t} \Lambda + (\pd_\omega \Lambda)\pd_{t}w = 0,
\\
& \pd_{\vec{x}} \Lambda + (\pd_\omega \Lambda)\pd_{\vec{x}}w = 0,
\\
& \pd_{\vec{k}} \Lambda + (\pd_\omega \Lambda)\pd_{\vec{k}}w = 0,
\label{eq:vvg1}
\end{align}
\end{subequations}
where the derivatives of \(\Lambda\) are evaluated at \m{(t, \vec{x}, w(t, \vec{x}, \vec{k}), \vec{k})}. In particular, \eq{eq:vvg1} gives
\begin{eqnarray}\label{eq:vvg2}
\vvg \doteq \frac{\pd w}{\pd \vec{k}} = - \frac{\pd_{\vec{k}} \Lambda}{\pd_\omega \Lambda},
\end{eqnarray}
for the group velocity \m{\vvg}, whose physical meaning is to be specified shortly.

Because \(\theta\) is now an additional dynamical variable, one also obtains an additional Euler--Lagrange equation:
\begin{eqnarray}\label{eq:act}
0 = \delta_{\theta} S[\theta, \env{a}^*, \env{a}] = \pd_t\mc{I} + \pd_{\vec{x}} \cdot \vec{\mc{J}},
\end{eqnarray}
where \(\mc{I}\) is called the action density and \(\vec{\mc{J}}\) is the action flux density:
\begin{align}
& \mc{I} \doteq \frac{\pd \mcc{L}}{\pd \omega}
= \frac{|\env{a}|^2}{4}\frac{\pd \Lambda}{\pd \omega} 
= \frac{|\env{a}|^2}{4}\,\vec{\eta}^\dag\,\frac{\pd \matr{\Xi}_\herm}{\pd \omega}\, \vec{\eta},
\label{eq:mcIdef}\\
& \mc{J}^i \doteq -\frac{\pd \mcc{L}}{\pd k_i}
= - \frac{|\env{a}|^2}{4}\frac{\pd \Lambda}{\pd k_i}
= - \frac{|\env{a}|^2}{4}\,\vec{\eta}^\dag\,\frac{\pd \matr{\Xi}_\herm}{\pd k_i}\,\vec{\eta},
\label{eq:mcJdef}
\end{align}
where we used \eq{eq:derzer} and the derivatives are evaluated on \m{(t, \vec{x}, w(t, \vec{x}, \avr{\vec{k}}(t, \vec{x})), \avr{\vec{k}}(t, \vec{x}))}. Using \eq{eq:vvg2}, one can also rewrite \eq{eq:mcJdef} as
\begin{eqnarray}\label{eq:avvg}
\vec{\mc{J}} = \avvg \mc{I},
\qquad
\avvg(t, \vec{x}) \doteq \vvg(t, \vec{x}, \avr{\vec{k}}(t, \vec{x})).
\end{eqnarray}
(The arguments \(\cb{t, \vec{x}}\) will be omitted from now on for brevity. We will also use \((\vec{k})\) as a shorthand for \((w(\vec{k}), \vec{k})\) where applicable.) Then, \eq{eq:act} becomes
\begin{eqnarray}
\pd_t\mc{I} + \pd_{\vec{x}} \cdot (\avvg\mc{I}) = 0,\label{eq:act2}
\end{eqnarray}
which can be a understood as a continuity equation for quasiparticles (`photons' or, more generally, `wave quanta') with density \m{\mc{I}} and fluid velocity \m{\avvg} (see also \Sec{sec:rays}). Thus, if an eikonal wave satisfies the least-action principle, its total action \m{\int \dd\vec{x}\,\mc{I}} (`number of quanta') is an invariant. This conservation law can be attributed to the fact that the wave Lagrangian density \m{\avr{\mcc{L}}} depends on derivatives of~\m{\theta} but not on~\m{\theta} \perse.

Also notice the following. By expanding \eq{eq:La} in \m{\pd_t \theta} around \m{\pd_t \theta = -w(t, \vec{x}, \pd_{\vec{x}}\theta)}, which is satisfied on any solution, one obtains
\begin{eqnarray}
\avr{\mcc{L}} \approx 
-\frac{1}{4}\,(\pd_t \theta + w(t, \vec{x}, \pd_{\vec{x}}\theta))\pd_\omega \Lambda\,|\env{a}|^2
= -(\pd_t \theta + w(t, \vec{x}, \pd_{\vec{x}}\theta))\mc{I},
\end{eqnarray}
where we used that \m{\avr{\mcc{L}}(t, \vec{x}, -\pd_t\theta, \pd_{\vec{x}}\theta) = 0} due to \eq{eq:HJ}. Then, one arrives at the canonical form of the action integral \citep{ref:hayes73}
\begin{eqnarray}\label{eq:hayes}
S[\mc{I}, \theta] = - \int \dd t\,\dd\vec{x}\,(\pd_t \theta + w(t, \vec{x}, \vec{k}))\mc{I}.
\end{eqnarray}
From here, \m{\delta_{\mc{I}} S = 0} yields the dispersion relation in the Hamilton--Jacobi form \m{\pd_t \theta + w(t, \vec{x}, \vec{k}) = 0}, and \m{\delta_{\theta} S = 0} yields the action conservation \eq{eq:act2}.

\subsubsection{Ray equations}
\label{sec:rays}

By \eq{eq:wkr}, one has the so-called consistency relations:
\begin{eqnarray}
\pd_t \avr{k}_i + \pd_i \avr{\omega} = 0,
\qquad
\pd_i \avr{k}_j = \pd_j \avr{k}_i.
\end{eqnarray}
These lead to
\begin{align}
\bigg(\frac{\pd}{\pd t} & + \avvg \cdot \frac{\pd}{\pd \vec{x}}\bigg) \avr{k}_i(t, \vec{x})
= - \frac{\pd w(t, \vec{x}, \avr{\vec{k}}(t, \vec{x}))}{\pd x^i} + \avvg \cdot \frac{\pd \avr{k}_i(t, \vec{x})}{\pd \vec{x}}
\notag\\
& = - \left(\frac{\pd w(t, \vec{x}, \vec{k})}{\pd x^i}
\right)_{\vec{k} = \avr{\vec{k}}(t, \vec{x})}
- \avg^j\,\frac{\pd \avr{k}_j(t, \vec{x})}{\pd x^i} + \avg^j\, \frac{\pd \avr{k}_i(t, \vec{x})}{\pd x^j}
\notag\\
& = - \left(\frac{\pd w(t, \vec{x}, \vec{k})}{\pd x^i}\right)_{\vec{k} = \avr{\vec{k}}(t, \vec{x})},
\label{eq:kdw}
\end{align}
and similarly,
\begin{align}
\bigg(\frac{\pd}{\pd t} & + \avvg \cdot \frac{\pd}{\pd \vec{x}}\bigg) \avr{\omega}(t, \vec{x})
= \bigg(\frac{\pd}{\pd t} + \avvg \cdot \frac{\pd}{\pd \vec{x}}\bigg) w(t, \vec{x}, \avr{\vec{k}}(t, \vec{x}))
\notag\\
& = \left(\frac{\pd w(t, \vec{x}, \vec{k})}{\pd t} + \avg^i\, \frac{\pd w(t, \vec{x}, \vec{k})}{\pd x^i}\right)_{\vec{k} = \avr{\vec{k}}(t, \vec{x})}
+  \avg^i \left(\frac{\pd}{\pd t} + \avvg \cdot \frac{\pd}{\pd \vec{x}}\right) \avr{k}_i(t, \vec{x})
\notag\\
& = \left(\frac{\pd w(t, \vec{x}, \vec{k})}{\pd t}\right)_{\vec{k} = \avr{\vec{k}}(t, \vec{x})},
\label{eq:wdw}
\end{align}
where we used \eq{eq:kdw}. Using the convective derivative associated with the group velocity,
\begin{eqnarray}\label{eq:conv2}
\dd/\dd t \equiv \dd_t \doteq \pd_t + (\avvg \cdot \pd_\vec{x}),
\end{eqnarray}
one can rewrite these compactly as
\begin{eqnarray}\label{eq:kwdw}
\frac{\dd\avr{k}_i(t, \vec{x})}{\dd t} = - \left(\frac{\pd w(t, \vec{x}, \vec{k})}{\pd x^i}\right)_{\vec{k} = \avr{\vec{k}}(t, \vec{x})},
\quad
\frac{\dd\avr{\omega}(t, \vec{x})}{\dd t} = \left(\frac{\pd w(t, \vec{x}, \vec{k})}{\pd t}\right)_{\vec{k} = \avr{\vec{k}}(t, \vec{x})}.
\end{eqnarray}
One can also represent \eq{eq:kwdw} as \textit{ordinary} differential equations for \m{\avr{\vec{k}}(t) \doteq \avr{\vec{k}}(t, \avr{\vec{x}}(t))} and \m{\avr{\omega}(t) \doteq \avr{\omega}(t, \avr{\vec{x}}(t))}, where \m{\avr{\vec{x}}(t)} are the `ray trajectories' governed by
\begin{eqnarray}\label{eq:xi}
\frac{\dd\avr{x}^i(t)}{\dd t} = \vg^i(t, \avr{\vec{x}}(t), \avr{\vec{k}}(t)).
\end{eqnarray}
Specifically, together with \eq{eq:xi}, equations \eq{eq:kwdw} become Hamilton's equations also known as the ray equations:
\begin{eqnarray}\label{eq:rays}
\frac{\dd\avr{x}^i}{\dd t} = \frac{\pd w(t, \avr{\vec{x}}, \avr{\vec{k}})}{\pd \avr{k}_i},
\qquad
\frac{\dd \avr{k}_i}{\dd t} =- \frac{\pd w(t, \avr{\vec{x}}, \avr{\vec{k}})}{\pd \avr{x}^i},
\qquad
\frac{\dd \avr{\omega}}{\dd t} = \frac{\pd w(t, \avr{\vec{x}}, \avr{\vec{k}})}{\pd t},
\end{eqnarray}
where \m{\avr{\vec{x}}} is the coordinate, \m{\hbar\avr{\vec{k}}} is the momentum, \m{\hbar\avr{\omega}} is the energy, \m{\hbar w} is the Hamiltonian, and the constant factor \m{\hbar} can be anything. If \m{\hbar} is chosen to be the Planck constant, then \eq{eq:rays} can be interpreted as the motion equations of individual wave quanta, for example, photons. Hamilton's equations for `true' particles, such as electrons and ions, are also subsumed under \eq{eq:rays} in that they can be understood as the ray equations of the particles considered as quantum-matter waves in the semiclassical limit. 

Also notably, \eq{eq:rays} can be obtained by considering the point-particle limit of \eq{eq:hayes} \citep{my:qlagr}. Specifically, adopting \m{\mc{I}(t, \vec{x}) \,\propto\, \delta(\vec{x} - \avr{\vec{x}}(t))} and taking the integral in \eq{eq:hayes} by parts leads to a canonical action \m{S\, \propto\, \int \dd t\,(\avr{\vec{k}} \cdot \dot{\avr{\vec{x}}} - w(t, \avr{\vec{x}}, \avr{\vec{k}}))}, whence Hamilton's equations follow as usual.

\subsubsection{Wave momentum and energy}
\label{sec:wenm}

Using \eq{eq:act2} and \eq{eq:conv2}, one arrives at the following equality for any given field \m{\msf{X}}:
\begin{align}
\pd_t(\msf{X} \mc{I})+\pd_\vec{x} \cdot (\msf{X} \mc{I} \avvg)
& =(\pd_t\msf{X})\mc{I}+\msf{X}(\pd_t\mc{I})+[\pd_\vec{x} \cdot (\mc{I} \avvg)]\msf{X}+\mc{I}(\avvg\cdot \pd_\vec{x} )\msf{X} 
\notag\\
&=\mc{I}[\pd_t+(\avvg\cdot \pd_\vec{x})]\msf{X}+\msf{X}[\pd_t\mc{I}+\pd_\vec{x} \cdot (\mc{I}\avvg)]
\notag\\
&=\mc{I}\,\dd_t\msf{X}.
\label{eq:mcX}
\end{align}
For \m{\msf{X} = \avr{k}_i} and \m{\msf{X} = \avr{\omega}}, \eq{eq:mcX} yields, respectively,
\begin{subequations}\label{eq:meeq}
\begin{align}
\pd_t P_{\text{w}, i} + \pd_{\vec{x}} \cdot (\avvg P_{\text{w}, i}) & = - \mc{I}\pd_i w,
\label{eq:meq}
\\
\pd_t\mc{E}_{\text{w}} + \pd_{\vec{x}} \cdot (\avvg\mc{E}_{\text{w}}) & = \mc{I}\pd_t w,
\label{eq:eeq}
\end{align}
\end{subequations}
where we used \eq{eq:kwdw} and introduced the following notation:
\begin{eqnarray}\label{eq:eikenmom}
\vec{P}_{\text{w}} \doteq \avr{\vec{k}}\mc{I},
\qquad
\mc{E}_{\text{w}} \doteq \avr{\omega}\mc{I}.
\end{eqnarray}
When a medium is homogeneous along \m{x^i}, \eq{eq:meq} yields \m{\int \dd\vec{x}\,P_{\text{w},i} = \const}. Likewise, when a medium is stationary, \eq{eq:eeq} yields \m{\int \dd\vec{x}\,\mc{E}_{\text{w}} = \const}. Hence, by definition, \m{\vec{P}_{\text{w}}} and \m{\mc{E}_{\text{w}}} are the densities of the wave canonical momentum and energy, at least up to a constant factor~\m{\kappa}.\footnote{Therefore, in a zero-dimensional wave, where \m{\int \dd\vec{x}} can be omitted, conservation of the total action \m{\mc{I}} implies conservation of \m{\mc{E}_{\text{w}}/\omega}, which is a well-known adiabatic invariant of a discrete harmonic oscillator with a slowly varying frequency \cite[section~49]{book:landau1}.}  A proof that \m{\kappa = 1} can be found, for example, in \citep{my:amc}. In \Sec{sec:cons}, we will show this using different arguments. 

\subsection{Non-conservative eikonal waves}
\label{sec:eikonal}

In a medium with nonzero \m{\matr{\Xi}_\aherm}, where waves are non-conservative, the wave properties are defined as in the previous section but the wave action evolves differently. The variational principle is not easy to apply in this case (however, see \citep{my:nonloc}), so a different approach will be used to derive the action equation. A more straightforward but less intuitive approach can be found in \citep{my:quasiop1, ref:mcdonald88}.

\subsubsection{Monochromatic waves}
\label{sec:homg}

First, consider a homogeneous stationary medium and a `monochromatic' (exponentially growing at a constant rate) wave field in the form
\begin{eqnarray}\label{eq:mono2}
\osc{\vec{\Psi}}(t, \vec{x}) = \re(\ee^{-\ii \avr{\omega}t + \ii \avr{\vec{k}} \cdot \vec{x}}\,\env{\vec{\Psi}}_{\text{c}}),
\qquad
\env{\vec{\Psi}}_{\text{c}} = \ee^{\avr{\gamma}t} \times \const,
\end{eqnarray}
where the constants \m{\avr{\omega}} and \m{\avr{\vec{k}}} are, as usual, the real frequency and wavenumber, and \m{\avr{\gamma}} is the linear growth rate, which can have either sign. Then, \eq{eq:Psime} becomes
\begin{eqnarray}\label{eq:PsiXig}
\vec{0} = \matr{\Xi}(\avr{\omega} + \ii\avr{\gamma}, \avr{\vec{k}})\env{\vec{\Psi}}_{\text{c}}
= \matr{\Xi}_\herm(\avr{\omega}, \avr{\vec{k}})\env{\vec{\Psi}}_{\text{c}} + \ii(\avr{\gamma}
\pd_\omega \matr{\Xi}_\herm(\avr{\omega}, \avr{\vec{k}}) + \matr{\Xi}_\aherm(\avr{\omega}, \avr{\vec{k}})
)\env{\vec{\Psi}}_{\text{c}} + \mc{O}(\epsilon^2),
\end{eqnarray}
where we assume that \m{\matr{\Xi}} is a smooth function of \m{\omega} and also that both \m{\matr{\Xi}_\aherm} and \m{\avr{\gamma}} are \m{\mc{O}(\epsilon)}. Like in \Sec{sec:bpe}, we adopt \m{\env{\vec{\Psi}}_{\text{c}} = \vec{\eta}\env{a} + \mc{O}(\epsilon)}, where the polarization vector \m{\vec{\eta}} is the relevant eigenvector of \m{\matr{\Xi}_\herm}. Then, by projecting \eq{eq:PsiXig} on \m{\vec{\eta}}, one obtains
\begin{eqnarray}\label{eq:PsiXig2}
0 = \Lambda(\avr{\omega}, \avr{\vec{k}})\env{a} + \ii(\avr{\gamma}
\pd_\omega \Lambda + \vec{\eta}^\dag\matr{\Xi}_\aherm \vec{\eta}
)\big|_{(\omega, \vec{k}) = (\avr{\omega}, \avr{\vec{k}})}\env{a} + \mc{O}(\epsilon^2),
\end{eqnarray}
where \m{\Lambda = \vec{\eta}^\dag\matr{\Xi}_\herm\vec{\eta}} is the corresponding eigenvalue of \m{\matr{\Xi}_\herm} and we used \eq{eq:derzer}. Let us neglect \m{\mc{O}(\epsilon^2)}, divide \eq{eq:PsiXig2} by \m{\env{a}}, and consider the real and imaginary parts of the resulting equation separately:
\begin{eqnarray}
\Lambda(\avr{\omega}, \avr{\vec{k}}) = 0,
\qquad
(\avr{\gamma} \pd_\omega \Lambda + \vec{\eta}^\dag\matr{\Xi}_\aherm \vec{\eta}
)\big|_{(\omega, \vec{k})} = 0.
\end{eqnarray}
The former is the same dispersion relation for \m{\avr{\omega}} as for conservative waves, and the latter yields \(\avr{\gamma} = \gamma(\vec{\avr{k}})\), where
\begin{eqnarray}\label{eq:gamma1}
\gamma(\vec{k}) 
\doteq - \frac{\vec{\eta}^\dag \matr{\Xi}_\aherm \vec{\eta}}{\pd_{\omega}\Lambda}.
\end{eqnarray}
Because \m{|\env{a}|\, \propto\, \ee^{\avr{\gamma}t}}, one can write the amplitude equation as
\begin{eqnarray}\label{eq:act31}
\pd_t |\env{a}|^2 = 2 \avr{\gamma} |\env{a}|^2.
\end{eqnarray}
One can also define the action density \m{\mc{I}} as in \Sec{sec:bpe} and rewrite \eq{eq:act31} in terms of that. Because \m{\mc{I} = |\env{a}|^2 \times \const}, one obtains
\begin{eqnarray}\label{eq:act3}
\pd_t \mc{I} = 2 \avr{\gamma} \mc{I}.
\end{eqnarray}

\subsubsection{Non-monochromatic waves}
\label{sec:ihomg}

When weak inhomogeneity and weak dissipation coexist, their effect on the action density is additive, so \eq{eq:act2} and \eq{eq:act3} merge into a general equation
\begin{eqnarray}\label{eq:act4} 
\pd_t\mc{I} + \pd_\vec{x} (\avvg \mc{I}) = 2 \avr{\gamma} \mc{I}.
\end{eqnarray}
(A formal derivation of \eq{eq:act4}, which uses the Weyl expansion \eq{eq:Mexp} and projection of the field equation on the polarization vector, can be found in \citep{my:quasiop1}.) Then, \eq{eq:mcX} is modified as follows:
\begin{eqnarray}
\pd_t(\msf{X} \mc{I})+\pd_\vec{x} \cdot (\msf{X} \mc{I} \avvg)
= \mc{I}\,\dd_t\msf{X} + 2 \avr{\gamma} \msf{X} \mc{I},
\label{eq:mcX2}
\end{eqnarray}
and the equations \eq{eq:meeq} for the wave momentum and energy \eq{eq:eikenmom} become
\begin{subequations}\label{eq:peme}
\begin{align}
\pd_t P_{\text{w}, i} + \pd_{\vec{x}} \cdot (\avvg P_{\text{w}, i}) & = 2 \avr{\gamma}P_{\text{w}, i} - \mc{I}\pd_i w,
\\
\pd_t\mc{E}_{\text{w}} + \pd_{\vec{x}} \cdot (\avvg\mc{E}_{\text{w}}) & = 2 \avr{\gamma}\mc{E}_{\text{w}} + \mc{I}\pd_t w.
\end{align}
\end{subequations}

A comment is due here regarding the relation between \eq{eq:act4} and the amplitude equation \eq{eq:act31} that is commonly used in the standard QLT for homogeneous plasma (for example, see~(2.21) in \citep{ref:drummond62}). In a nutshell, the latter is incorrect, even when \m{\pd_\vec{x} = 0}. Because \m{\avr{f}} is time-dependent, waves do not grow or decay exponentially. Rather, they can be considered as geometrical-optics (WKB) waves, and unlike in \Sec{sec:homg}, the ratio \m{|\env{a}|^2/\mc{I}} generally evolves at a rate comparable to \m{\avr{\gamma}}. The standard QLT remains conservative only because it also incorrectly replaces \eq{eq:ft2} with its stationary-plasma limit (\m{\epsilon = 0}) and the two errors cancel each other. These issues are less of a problem for waves in not-too-hot plasmas (\eg Langmuir waves), because in such plasmas, changing the distribution functions does not significantly affect the dispersion relations and thus \m{|\env{a}|^2/\mc{I}} does in fact approximately remain constant. See also the discussion in \Sec{sec:eikw}.

\subsection{General waves}
\label{sec:broad}

Let us now discuss a more general case that includes broadband waves. The evolution of such waves can be described statistically in terms of their average Wigner matrix \m{\matrU}. This matrix also determines the function \m{\avr{\st{W}}_s} that is given by \eq{eq:WsW} and enters the nonlinear potentials \eq{eq:trfW}. Below, we derive the general form of \m{\matrU} in terms of the phase-space action density \m{J} and the governing equation for \m{J} (\Secs{sec:wigvec}--\ref{sec:wke}). Then, we also express the function \m{\avr{\st{W}}_s} through \m{J} (\Sec{sec:wf}). Related calculations can also be found in \citep{ref:mcdonald85, phd:ruiz17}.

\subsubsection{Average Wigner matrix of an eikonal wave field}
\label{sec:wigvec}

Let us start with calculating the average Wigner matrix of an eikonal field~\m{\osc{\vec{\Psi}}} of the form \eq{eq:Psic} (see also \App{sec:wfune}). Using \m{\osc{\vec{\Psi}} = (\Psiosc_{\text{c}} + \Psiosc_{\text{c}}^*)/2}, it can be readily expressed through the average Wigner functions of the complexified field\footnote{Field complexification is discussed, for example, in \citep{ref:brizard93}.} \m{\Psiosc_{\text{c}}} and of its complex conjugate:
\begin{eqnarray}
\matrU
\approx (\avr{\matrst{W}}_{\Psiosc_{\text{c}}} + \avr{\matrst{W}}_{\Psiosc_{\text{c}}^*})/4
\equiv (\matrU_{\text{c}+} + \matrU_{\text{c}-})/4.
\end{eqnarray}
For \m{\Psiosc_{\text{c}} = \env{a}\vec{\eta}(\avr{\omega}, \avr{\vec{k}})\ee^{\ii \theta}}, where the arguments \(\cb{t, \vec{x}}\) are omitted for brevity, one has
\begin{eqnarray}
\matrU_{\text{c}} \equiv \matrU_{\text{c}+}
\approx (\vec{\eta}\vec{\eta}^\dag)(\avr{\omega}, \avr{\vec{k}}) |\env{a}|^2\int \frac{\dd\tau\,\dd\vec{s}}{(2\upi)^\stn}\,
\ee^{\ii \theta (t + \tau/2, \vec{x} + \vec{s}/2)}
\ee^{- \ii\theta(t - \tau/2, \vec{x} - \vec{s}/2)}
\ee^{\ii \omega \tau - \ii \vec{k} \cdot \vec{s}},
\end{eqnarray}
where we neglected the dependence of \m{\env{a}} and \m{\vec{\eta}} on \m{\cb{t, \vec{x}}} because it is weak compared to that of \m{\ee^{\pm \ii\theta}}. By Taylor-expanding \m{\theta}, one obtains
\begin{eqnarray}\notag
\matrU_{\text{c}}
\approx (\vec{\eta}\vec{\eta}^\dag)(\avr{\omega}, \avr{\vec{k}}) |\env{a}|^2
\int \frac{\dd\tau\,\dd\vec{s}}{(2\upi)^\stn}\,
\ee^{\ii (\omega - \avr{\omega}) \tau - \ii (\vec{k} - \avr{\vec{k}}) \cdot \vec{s}}
=
(\vec{\eta}\vec{\eta}^\dag)(\avr{\omega}, \avr{\vec{k}}) |\env{a}|^2 
\delta(\omega - \avr{\omega})
\delta(\vec{k} - \avr{\vec{k}}).
\end{eqnarray}
For \m{\Psiosc_{\text{c}}^* = \env{a}^*\vec{\eta}^*(\avr{\omega}, \avr{\vec{k}})\ee^{-\ii \theta}}, which can also be written as \m{\Psiosc_{\text{c}}^* = \env{a}^*\vec{\eta}(-\avr{\omega}, -\avr{\vec{k}})\ee^{-\ii \theta}} due to \eq{eq:etastar}, the result is the same up to replacing \m{\avr{\omega} \to - \avr{\omega}} and \m{\avr{\vec{k}} \to -\avr{\vec{k}}}. Also notice that
\begin{align}
\delta(\omega \mp \avr{\omega}) \delta(\vec{k} \mp \avr{\vec{k}})
& = \delta(\omega \mp w(\avr{\vec{k}})) \delta(\vec{k} \mp \avr{\vec{k}})
\notag\\
& = \delta(\omega \mp w(\pm \vec{k})) \delta(\vec{k} \mp \avr{\vec{k}})
\notag\\
& = \delta(\omega - w(\vec{k})) \delta(\vec{k} \mp \avr{\vec{k}}),
\end{align}
so one can rewrite \m{\matrU_{\text{c}\pm}} as follows:
\begin{eqnarray}
\matrU_{\text{c}\pm} =
\vec{\eta}(\vec{k})
\vec{\eta}^\dag(\vec{k}) |\env{a}|^2
\delta(\omega - w(\vec{k}))
\delta(\vec{k} \mp \avr{\vec{k}}),
\end{eqnarray}
where \m{(\vec{k}) \equiv (w(\vec{k}), \vec{k})}. Thus finally,
\begin{eqnarray}
\matrU(\omega, \vec{k})
\approx 
\vec{\eta}(\vec{k})\vec{\eta}^\dag(\vec{k})\,
|\env{a}|^2
(\delta(\vec{k} - \avr{\vec{k}}) + \delta(\vec{k} + \avr{\vec{k}}))
\delta(\omega - w(\vec{k}))/4.
\end{eqnarray}

\subsubsection{Average Wigner matrix of a general wave}
\label{sec:wifinc}

Assuming the background medium is sufficiently smooth, a general wave field can be represented as a superposition of eikonal fields:
\begin{eqnarray}
\textstyle
\osc{\vec{\Psi}} = \re \Psiosc_{\text{c}},
\qquad
\Psiosc_{\text{c}} = \sum_\sigma \Psiosc_{\sigma,\text{c}},
\qquad
\Psiosc_{\sigma,\text{c}} = \env{a}_\sigma \ee^{\ii\theta_\sigma}.
\end{eqnarray}
As a quadratic functional, its average Wigner matrix \m{\matrU} equals the sum of the average Wigner matrices \m{\matrU_\sigma} of the individual eikonal waves:
\begin{eqnarray}\label{eq:iaxU}
\textstyle
\matrU 
= \sum_\sigma \matrU_\sigma
= \sum_\sigma(\matrU_{\sigma,\text{c}+} + \matrU_{\sigma,\text{c}-})/4,
\end{eqnarray}
where \m{\matrU_{\sigma,\text{c}+} \equiv \matrU_{\sigma,\text{c}}} and \m{\matrU_{\sigma,\text{c}-}} are the average Wigner matrices of \m{\Psiosc_{\sigma,\text{c}}} and \m{\Psiosc_{\sigma,\text{c}}^*}, respectively:
\begin{eqnarray}
\matrU_{\sigma,\text{c}\pm} = \vec{\eta}(\vec{k})\vec{\eta}^\dag(\vec{k})|\env{a}_\sigma|^2
\delta(\vec{k} \mp \avr{\vec{k}}_\sigma)
\delta(\omega - w(\vec{k})).
\end{eqnarray}
Equation \eq{eq:iaxU} can also be expressed as
\begin{eqnarray}
\textstyle
\matrU 
= (\matrU_{\text{c}+} + \matrU_{\text{c}-})/4,
\qquad
\matrU_{\text{c}\pm} = \sum_\sigma \matrU_{\sigma,\text{c}\pm},
\end{eqnarray}
where \m{\matrU_{\text{c}\pm}} are the average Wigner matrices of \m{\Psiosc_{\text{c}}} and \m{\Psiosc_{\text{c}}^*}, respectively:
\begin{eqnarray}
\matrU_{\text{c}\pm} = \vec{\eta}(\vec{k})\vec{\eta}^\dag(\vec{k}) h_{\text{c}\pm}(\vec{k}) \delta(\omega - w(\vec{k})),
\qquad
h_{\text{c}\pm}(\vec{k}) \doteq
\textstyle \sum_{\sigma} |\env{a}_\sigma|^2\delta(\vec{k} \mp \avr{\vec{k}}_\sigma).
\label{eq:Winc1}
\end{eqnarray}
Because \m{h_{\text{c}-}(\vec{k}) = h_{\text{c}+}(-\vec{k}) \equiv h_{\text{c}}(-\vec{k})}, the matrix \m{\matrU} can also be written as follows:
\begin{eqnarray}\label{eq:UW}
\textstyle
\matrU(\omega, \vec{k}) = 
(\vec{\eta} \vec{\eta}^\dag)(\vec{k})\,(h(\vec{k}) + h(-\vec{k}))\delta(\omega - w(\vec{k})),
\end{eqnarray}
where \m{h(\vec{k}) \doteq h_{\text{c}}(\vec{k})/4} is given by
\begin{eqnarray}
\textstyle
h(\vec{k}) = \frac{1}{4} \sum_{\sigma} |\env{a}_\sigma|^2 \delta(\vec{k} - \avr{\vec{k}}_\sigma) \ge 0.
\label{eq:hdef}
\end{eqnarray}
This shows that for broadband waves comprised of eikonal waves, \m{\matrU} has the same form as for an eikonal wave except \m{h(\vec{k})} is not necessarily delta-shaped.

\subsubsection{Phase-space action density and the wave-kinetic equation}
\label{sec:wke}

The wave equation for the complexified field \m{\Psiosc_{\text{c}}} can be written in the invariant form as \m{\boper{\Xi}\ket{\Psiosc_{\text{c}}} = \ket{\vec{0}}}. Multiplying it by \m{\bra{\Psiosc_{\text{c}}}} from the right leads to
\begin{equation}
\boper{\Xi}\boper{\matrU}_{\text{c}} = \boper{0}, 
\qquad 
\boper{\matrU}_{\text{c}} \doteq (2\upi)^{-\stn}\,\avr{
\ket{\Psiosc_{\text{c}}}\bra{\Psiosc_{\text{c}}}
\vphantom{\osc{\vec{\Psi}}}
}.
\end{equation}
This readily yields an equation for the Wigner matrix: \m{\matr{\Xi} \star \matrU_{\text{c}} = \matr{0}}. Let us integrate this equation over \m{\omega} to make the left-hand side a smooth function of \m{\cb{t, \vec{x}, \vec{k}}}. Let us also take the trace of the resulting equation to put it in a scalar form:
\begin{eqnarray}
\textstyle
\tr \int \dd\omega\,\matr{\Xi} \star \matrU_{\text{c}} = 0.
\label{eq:trwke}
\end{eqnarray}
As usual, we assume \(\matr{\Xi} = \matr{\Xi}_\herm + \ii \matr{\Xi}_\aherm\) with \(\matr{\Xi}_\aherm = \mc{O}(\epsilon) \ll \matr{\Xi}_\herm = \mc{O}(1)\) for generic \(\cb{\vecst{x}, \vecst{k}}\). The integrand in \eq{eq:trwke} can be written as \m{\matr{\Xi} \star \matrU_{\text{c}} = \matr{\Xi}\ee^{\ii\oper{\mc{L}}_\indexst/2}\matrU_{\text{c}}}, and its expansion in the differential operator \m{\oper{\mc{L}}_\indexst} \eq{eq:poissonx} contains derivatives of all orders. High-order derivatives on \m{\matrU_{\text{c}}} are not negligible \perse, because for on-shell waves this function is delta-shaped. However, using integration by parts, one can reapply all derivatives with respect to \m{\omega} to \m{\matr{\Xi}} and take the remaining derivatives (with respect to \m{t}, \m{\vec{x}}, and \m{\vec{k}}) outside the integral. Then it is seen that each power \m{m} of \m{\oper{\mc{L}}_\indexst} in the expansion of \m{\matr{\Xi}\ee^{\ii\oper{\mc{L}}_\indexst/2}\matrU_{\text{c}}} contributes \m{\mc{O}(\epsilon^m)} to the integral. Let us neglect terms with \m{m \ge 2} and use \eq{eq:Winc1}. Hence, one obtains\footnote{\citet{ref:mcdonald85} first Taylor-expand \m{\matr{\Xi} \star \matrU_{\text{c}}} and then integrate over~\m{\omega}. Strictly speaking, that is incorrect (because \m{\matr{\Xi} \star \matrU_{\text{c}}} is not smooth), but the final result is the same.}
\begin{align}
0 & \approx
\tr \int \dd\omega\,\left(\matr{\Xi}_\herm \matrU_{\text{c}} + \ii \matr{\Xi}_\aherm \matrU_{\text{c}} + \frac{\ii}{2}\,\poissonx{\matr{\Xi}_\herm, \matrU_{\text{c}}}\right)
\notag\\
& \approx
(\vec{\eta}^\dag \matr{\Xi}_\herm \vec{\eta} + \ii \vec{\eta}^\dag \matr{\Xi}_\aherm \vec{\eta})h_{\text{c}}
+ \frac{\ii}{2}\,\tr \int \dd\omega\left(
\frac{\pd \matr{\Xi}_\herm}{\pd \st{x}^i}\frac{\pd \matrU_{\text{c}}}{\pd \st{k}_i}
- \frac{\pd \matr{\Xi}_\herm}{\pd \st{k}_i}\frac{\pd \matrU_{\text{c}}}{\pd \st{x}^i}
\right).
\end{align}
Let us also re-express this as follows, using \eq{eq:labetet} and \eq{eq:gamma1}:
\begin{align}
0 & \approx
\left(\Lambda - \ii\gamma\, \frac{\pd\Lambda}{\pd\omega}\right)h_{\text{c}}
- \frac{\ii}{2} \tr \int \dd\omega\,\frac{\pd}{\pd \omega}\left(\frac{\pd \matr{\Xi}_\herm}{\pd t}\,\matrU_{\text{c}}\right)
+ \frac{\ii}{2}\frac{\pd}{\pd t}\tr\int \dd\omega\,\frac{\pd \matr{\Xi}_\herm}{\pd \omega}\,\matrU_{\text{c}}
\notag\\
& \hspace{1.15in} + \frac{\ii}{2}\frac{\pd}{\pd k_i} \tr \int \dd\omega\,\frac{\pd \matr{\Xi}_\herm}{\pd x^i}\,\matrU_{\text{c}}
- \frac{\ii}{2}\frac{\pd}{\pd x^i}\tr\int \dd\omega\,\frac{\pd \matr{\Xi}_\herm}{\pd k_i}\,\matrU_{\text{c}}.
\label{eq:wke0}
\end{align}
Clearly,
\begin{eqnarray}
\int \dd\omega\,\frac{\pd}{\pd \omega}\left(\frac{\pd \matr{\Xi}_\herm}{\pd t}\,\matrU_{\text{c}}\right) = 0.
\end{eqnarray}
To simplify the remaining terms, we proceed as follows. As a Hermitian matrix, \m{\matr{\Xi}_\herm} can be represented in terms of its eigenvalues \m{\Lambda_b} and eigenvectors \m{\vec{\eta}_b } as \m{\matr{\Xi}_\herm = \Lambda_b \vec{\eta}_b \vec{\eta}_b^\dag}. For \m{\matrU_{\text{c}}}, let us use \eq{eq:Winc1} again, where \m{\vec{\eta}} is one of the vectors \m{\vec{\eta}_b}, say, \m{\vec{\eta} \equiv \vec{\eta}_0}. (Accordingly, \m{\Lambda \equiv \Lambda_0}.) Then, for any \m{\placeholder \in \lbrace \omega, x^i, k_i \rbrace}, one has
\begin{align}
\tr\int \dd\omega\,\frac{\pd \matr{\Xi}_\herm}{\pd \placeholder}\,\matrU_{\text{c}}
& = \frac{\pd \Lambda_b}{\pd \placeholder}\, |\vec{\eta}_b^\dag \vec{\eta}|^2 h_{\text{c}} 
+ \Lambda_b\,\Big(\vec{\eta}^\dag\frac{\pd \vec{\eta}_b}{\pd\placeholder}\Big)(\vec{\eta}_b^\dag\vec{\eta})h_{\text{c}}
+ \Lambda_b\,(\vec{\eta}^\dag \vec{\eta}_b)\Big(\frac{\pd \vec{\eta}_b^\dag}{\pd\placeholder}\,\vec{\eta}\Big)h_{\text{c}}
\notag\\
& = \frac{\pd \Lambda_b}{\pd \placeholder}\,(\delta_{b,0})^2h_{\text{c}}
+ \Lambda_b\,\Big(\vec{\eta}^\dag\frac{\pd \vec{\eta}_b}{\pd\placeholder}\Big)\delta_{b,0} h_{\text{c}}
+ \Lambda_b\,\delta_{b,0}\Big(\frac{\pd \vec{\eta}_b^\dag}{\pd\placeholder}\,\vec{\eta}\Big)h_{\text{c}}
\notag\\
& = \frac{\pd \Lambda}{\pd \placeholder}\,h_{\text{c}} 
+ \Big(\vec{\eta}^\dag\frac{\pd\vec{\eta}}{\pd\placeholder} + \frac{\pd \vec{\eta}^\dag}{\pd\placeholder}\,\vec{\eta}\Big)\,\Lambda h_{\text{c}}
\notag\\
& = \frac{\pd \Lambda}{\pd \placeholder}\,h_{\text{c}} 
+ \frac{\pd(\vec{\eta}^\dag\vec{\eta})}{\pd\placeholder}\,\Lambda h_{\text{c}}
\notag\\
& = \frac{\pd \Lambda}{\pd \placeholder}\,h_{\text{c}},
\end{align}
where we used \m{\vec{\eta}_b^\dag\vec{\eta} = \delta_{b,0}} and, in particular, \m{\vec{\eta}^\dag\vec{\eta} = 1}. Then, \eq{eq:wke0} can be written as
\begin{eqnarray}\label{eq:LUE}
\Lambda h_{\text{c}} - 2\ii \mcu{E} = 0,
\end{eqnarray}
where
\begin{eqnarray}
\mcu{E} = 2\gamma\left(\frac{\pd\Lambda}{\pd\omega}\,h\right)
-\frac{\pd}{\pd t}\left(\frac{\pd\Lambda}{\pd\omega}\,h\right)
-\frac{\pd}{\pd k_i}\left(\frac{\pd\Lambda}{\pd x^i}\,h\right)
+\frac{\pd}{\pd x^i}\left(\frac{\pd\Lambda}{\pd k_i}\,h\right).
\end{eqnarray}

The real part of \eq{eq:LUE} gives \m{\Lambda = 0}, which is the dispersion relation. The imaginary part of \eq{eq:LUE} gives \m{\mcu{E} = 0}. To understand this equation, let us rewrite \m{\mcu{E}} as
\begin{eqnarray}
\mcu{E} = 2\gamma J
-\frac{\pd J}{\pd t}
+\frac{\pd}{\pd k_i}\left(\frac{\pd w}{\pd x^i}\,J\right)
-\frac{\pd}{\pd x^i}\left(\frac{\pd w}{\pd k_i}\,J\right).
\end{eqnarray}
Here, we introduced
\begin{eqnarray}
J(\vec{k}) \doteq h(\vec{k})\,\pd_{\omega}\Lambda(\vec{k}),
\qquad
\Lambda(\vec{k}) \doteq \Lambda(w(\vec{k}), \vec{k}),
\label{eq:Jinc}
\end{eqnarray}
which, according to \eq{eq:Lambdaw}, satisfy
\begin{subequations}\label{eq:Jhwk0}
\begin{align}
J \pd_t w(\vec{k}) & = -h (\pd_t \Lambda)(\vec{k}),
\label{eq:Jhwt}
\\
J \pd_{\vec{x}} w(\vec{k}) & = -h (\pd_{\vec{x}} \Lambda)(\vec{k}),
\label{eq:Jhwx}
\\
J \pd_{\vec{k}} w(\vec{k}) & = -h (\pd_{\vec{k}} \Lambda)(\vec{k}).
\label{eq:Jhwk}
\end{align}
\end{subequations}
Note that using \eq{eq:hdef}, one can also express \m{J} as
\begin{eqnarray}
\textstyle
J = \sum_{\sigma} \big(\frac{1}{4}\,|\env{a}_\sigma|^2 \pd_{\omega}\Lambda(\vec{k}_\sigma)\big)\delta(\vec{k} - \avr{\vec{k}}_\sigma)
= \sum_{\sigma} \mc{I}_\sigma \delta(\vec{k} - \avr{\vec{k}}_\sigma),
\end{eqnarray}
where \m{\mc{I}_\sigma} are the action densities \eq{eq:mcIdef} of the individual eikonal waves that comprise the total wave field (\Sec{sec:wifinc}). In particular, \m{\int\dd\vec{k}\,J = \sum_\sigma \mc{I}_\sigma}, which is the total action density. Therefore, the function \m{J} can be interpreted as the \textit{phase-space} action density. In terms of \m{J}, the equation  \m{\mcu{E} = 0} can be written as 
\begin{eqnarray}\label{eq:wkeJ}
\frac{\pd J}{\pd t} + \frac{\pd w}{\pd k_i}\,\frac{\pd J}{\pd x^i} 
- \frac{\pd w}{\pd x^i}\,\frac{\pd J}{\pd k_i}
= 2 \gamma J.
\end{eqnarray}
This equation, called the WKE, serves the same role in QL wave-kinetic theory as the Vlasov equation serves in plasma kinetic theory.\footnote{The term `WKE' is also used for the equation that describes nonlinear interactions of waves in statistically homogeneous media, or `wave--wave collisions' \citep{book:zakharov-b}. That is not what we consider here. Inhomogeneities are essential in our formulation, and the QL WKE is linear (in \m{J}) by definition of the QL approximation. That said, the Weyl symbol calculus that we use can facilitate derivations of wave--wave collision operators as well \citep{my:wcol}.} Unlike the field equation used in the standard QLT \citep{ref:drummond62}, \eq{eq:wkeJ} exactly conserves the action of nonresonant waves, \ie those with \m{\gamma = 0}. Also note that \eq{eq:act4} for eikonal waves can be deduced from \eq{eq:wkeJ} as a particular case by assuming the ansatz
\begin{eqnarray}\label{eq:Jdelta}
J(t, \vec{x}, \vec{k}) = \mc{I}(t, \vec{x})\delta(\vec{k} - \avr{\vec{k}}(t, \vec{x}))
\end{eqnarray}
and integrating over \m{\vec{k}}. In other words, eikonal-wave theory can be understood as the `cold-fluid' limit of wave-kinetic theory.

\subsubsection{Function \mt{\avr{\st{W}}_s} in terms of \mt{J}}
\label{sec:wf}

Here we explicitly calculate the function \eq{eq:WsW} that determines the nonlinear potentials~\eq{eq:trfW}. Using \eq{eq:UW}, one obtains
\begin{eqnarray}
\avr{\st{W}}_s(\omega, \vec{k}; \vec{p}) = |\vec{\alpha}_s^\dag\vec{\eta}|^2(\vec{k}; \vec{p})\,(h(\vec{k}) + h(-\vec{k}))\,\delta(\omega - w(\vec{k}))
\ge 0,
\end{eqnarray}
where \m{(\vec{k}; \vec{p}) \equiv (w(\vec{k}), \vec{k}; \vec{p})}. By definition of \m{\boper{\alpha}_s}, the function \m{\braket{t_1, \vec{x}_1 | \boper{\alpha}_s |t_2, \vec{x}_2}} is real for all \m{\cb{t_1, \vec{x}_1}} and \m{\cb{t_2, \vec{x}_2}}, so \m{\vec{\alpha}_s(-\omega, -\vec{k}) = \vec{\alpha}_s^*(\omega, \vec{k})} by definition of the Weyl symbol \eq{eq:wsymbx}. Together with \eq{eq:etastar}, this gives \(\smash{|\vec{\alpha}_s^\dag\vec{\eta}|^2(\omega, \vec{k}; \vec{p})} = \smash{|\vec{\alpha}_s^\dag \vec{\eta}|^2(-\omega, -\vec{k}; \vec{p})}\), so
\begin{subequations}\label{eq:aeta}
\begin{eqnarray}
|\vec{\alpha}_s^\dag\vec{\eta}|^2 \equiv 
|\vec{\alpha}_s^\dag\vec{\eta}|^2(\vec{k}; \vec{p}) = |\vec{\alpha}_s^\dag\vec{\eta}|^2(-\vec{k}; \vec{p}),
\end{eqnarray}
and similarly,
\begin{eqnarray}
|\vec{\eta}^\dag \vec{\wp}_s \vec{\eta}|^2 \equiv  
|\vec{\eta}^\dag \vec{\wp}_s \vec{\eta}|^2(\vec{k}; \vec{p}) 
= |\vec{\eta}^\dag \vec{\wp}_s \vec{\eta}|^2(-\vec{k}; \vec{p}).
\end{eqnarray}
\end{subequations}
This also means that \(\smash{\avr{\st{W}}_s(\omega, \vec{k}; \vec{p})} = \smash{\avr{\st{W}}_s(-\omega, -\vec{k}; \vec{p})}\). Then finally, using \eq{eq:Jinc}, one can express this function through the phase-space action density:
\begin{eqnarray}\label{eq:Winc22}
&\avr{\st{W}}_s(\omega, \vec{k}; \vec{p}) 
= |\vec{\alpha}_s^\dag\vec{\eta}|^2
(\varsigma_{\vec{k}} J(\vec{k}) + \varsigma_{-\vec{k}}J(-\vec{k}))
\,\delta(\Lambda(\omega, \vec{k})),
\\
&\varsigma_{\vec{k}} \doteq \sgn \pd_{\omega}\Lambda(\vec{k}) = \sgn(J(\vec{k})/h(\vec{k})) = \sgn J(\vec{k}).
\label{eq:varsig}
\end{eqnarray}

\subsection{Conservation laws}
\label{sec:cons}

Let us rewrite \eq{eq:wkeJ} together with \eq{eq:FS} in the `divergence' form:
\begin{align}
\frac{\pd J}{\pd t} + \frac{\pd (\vg^i J)}{\pd x^i}
- \frac{\pd}{\pd k_i}\left(\frac{\pd w}{\pd x^i}\,J\right)
& = 2\gamma J,
\label{eq:divJ}
\\
\frac{\pd F_s}{\pd t}
+ \frac{\pd (v_s^i\,F_s)}{\pd x^i}
- \frac{\pd}{\pd p_i} \left(\frac{\pd \mc{H}_s}{\pd x^i} \,F_s\right)
& = \frac{\pd}{\pd p_i} \left(\st{D}_{s,ij}\,\frac{\pd F_s}{\pd p_j}\right)
+ \collision_s.
\label{eq:divF}
\end{align}
Using \eq{eq:Winc1}, the diffusion matrix \m{\st{D}_{s,ij}} can be represented as follows:
\begin{eqnarray}
\st{D}_{s,ij} = 2 \upi \int \dd\vec{k}\,k_i k_j\,
|\vec{\alpha}_s^\dag \vec{\eta}|^2\,
\frac{J(\vec{k})}{\pd_\omega \Lambda(\vec{k})}\,\delta(\vec{k} \cdot \vec{v}_s - w(\vec{k})).
\end{eqnarray}
Also, by substituting \eq{eq:Xiah} into \eq{eq:gamma1}, one finds
\begin{eqnarray}\label{eq:gamma2}
\gamma 
= \upi \sum_s 
\int\dd\vec{p}\,\frac{|\vec{\alpha}_s^\dag \vec{\eta}|^2}{\pd_{\omega}\Lambda(\vec{k})}\,
\delta(w(\vec{k}) - \vec{k} \cdot \vec{v}_s(\vec{p}))\,\vec{k} \cdot \frac{\pd F_s(\vec{p})}{\pd \vec{p}}.
\end{eqnarray}
Together with \eq{eq:Jhwk0}, these yield the following notable corollaries. First of all, if \m{\matr{\Xi}_0}, \m{|\vec{\alpha}_s^\dag \vec{\eta}|^2}, and \m{\vec{\eta}^\dag\matr{\wp}_s\vec{\eta}} are independent of \m{\vec{x}},\footnote{Having \m{\vec{x}}-dependence in \m{\matr{\Xi}_0}, \m{|\vec{\alpha}_s^\dag \vec{\eta}|^2}, or \m{\vec{\eta}^\dag\matr{\wp}_s\vec{\eta}} would signify interaction with external fields not treated self-consistently. Such fields could exchange momentum with the wave--plasma system, so the momentum of the latter would not be conserved. A similar argument applies to the temporal dependence of these coefficients vs.\ energy conservation considered below.} one has for each \m{l} that (\App{app:emcons1})
\begin{align}
\frac{\pd}{\pd t} \left(
\sum_s \int\dd\vec{p}\,p_l F_s +
\int \dd\vec{k}\,k_l J
\right)
& + \frac{\pd}{\pd x^i}
\left(
\sum_s \int\dd\vec{p}\, p_l v_s^i F_s
+ \int \dd\vec{k}\,k_l \vg^i J
\right)
\notag\\
& 
+ \frac{\pd}{\pd x^l}\sum_s \int\dd\vec{p}\, \Delta_s F_s
= -\sum_s\int\dd\vec{p}\,\frac{\pd H_{0s}}{\pd x^l}\,F_s.
\label{eq:mcons}
\end{align}
This can be viewed as a momentum-conservation theorem, because at \m{\pd_l H_{0s} = 0}, one has
\begin{eqnarray}\label{eq:mcons1}
\sum_s \int\dd\vec{x}\,\dd\vec{p}\,p_l F_s + \int \dd\vec{x}\,\dd\vec{k}\,k_l J = \const.
\end{eqnarray}
Also, the ponderomotive force on a plasma is readily found from \eq{eq:mcons} as the sum of the terms quadratic in the wave amplitude (after \m{F_s} has been expressed through \m{\avr{f}_s}). Similarly, if \m{\matr{\Xi}_0}, \m{|\vec{\alpha}_s^\dag \vec{\eta}|^2}, and \m{\vec{\eta}^\dag\matr{\wp}_s\vec{\eta}} are independent of \m{t}, one has (\App{app:emcons2})
\begin{align}
\frac{\pd}{\pd t} \left(
\sum_s \int\dd\vec{p}\,H_{0s} F_s +
\int \dd\vec{k}\,w J
\right)
& + \frac{\pd}{\pd x^i}
\left(
\sum_s \int\dd\vec{p}\, H_{0s} v_s^i F_s
+ \int \dd\vec{k}\,w \vg^i J
\right)
\notag\\
& 
+ \frac{\pd}{\pd x^i}\sum_s \int\dd\vec{p}\, \Delta_s v_s^i F_s
= \sum_s\int\dd\vec{p}\,\frac{\pd H_{0s}}{\pd t}\,F_s.
\label{eq:econs}
\end{align}
This can be viewed as an energy-conservation theorem, because at \m{\pd_t H_{0s} = 0}, one has
\begin{eqnarray}\label{eq:encons1}
\sum_s \int\dd\vec{x}\,\dd\vec{p}\,H_{0s} F_s + \int \dd\vec{x}\,\dd\vec{k}\,w J = \const.
\end{eqnarray}
Related equations are also discussed in \citep{my:amc, ref:dewar77}.

The individual terms in \eq{eq:mcons} and \eq{eq:econs} can be interpreted as described in \Tab{tab:inter}. The results of \Sec{sec:wenm} are reproduced as a particular case for the eikonal-wave ansatz \eq{eq:Jdelta}.\footnote{There is no ambiguity in the definition of the wave momentum and energy in this case (\ie \m{\kappa = 1}), because \eq{eq:mcons1} and \eq{eq:encons1} connect those with the momentum and energy of particles (OCs), which are defined unambiguously.} In particular, note that electrostatic \textit{waves} carry nonzero momentum density \m{\int \dd\vec{k}\,\vec{k} J} just like any other waves, even though the electrostatic \textit{field} of these waves carries no momentum. The momentum is stored in the particle motion in this case (\Sec{sec:onshelles}), and it is pumped there via either temporal dependence \citep[section II.2]{my:sharm} or spatial dependence \citep{ref:ochs21b, foot:ochs22} of the wave amplitude. This shows that homogeneous-plasma models that ignore ponderomotive effects cannot adequately describe the energy--momentum transfer between waves and plasma even when resonant absorption \perse occurs in a homogeneous-plasma region. The OC formalism presented here provides means to describe such processes rigorously, generally, and without cumbersome calculations.

\begin{table}
\begin{center}
\begin{tabular}{r@{\qquad}c@{\qquad}l}
Quantity & Notation & Interpretation\\[2pt]
\hline
\m{\int \dd\vec{p}\,\vec{p} F_s} & \m{\vec{P}_s} & OC momentum density\\[2pt]
\m{\int \dd\vec{p}\,H_{0s} F_s} & \m{\mc{E}_s} & OC energy density\\[2pt]
\m{\int \dd\vec{p}\, (\vec{p} \vec{v}_s + \Delta_s \matr{1}) F_s} & \m{\vec{\Pi}_s} & OC momentum flux density\\[2pt]
\m{\int \dd\vec{p}\, (H_{0s} + \Delta_s) \vec{v}_s F_s} & \m{\vec{Q}_s} & OC energy flux density\\[2pt]
\m{\int \dd\vec{k}\,\vec{k} J} & \m{\vec{P}_{\text{w}}} & wave momentum density\\[2pt]
\m{\int \dd\vec{k}\,w J} & \m{\mc{E}_{\text{w}}} & wave energy density\\[2pt]
\m{\int \dd\vec{k}\,\vec{k} \vvg J} & \m{\vec{\Pi}_{\text{w}}} & wave momentum flux density\\[2pt]
\m{\int \dd\vec{k}\,w \vvg J} & \m{\vec{Q}_{\text{w}}} & wave energy flux density\\[2pt]
\end{tabular}
\end{center}
\caption{Interpretation of the individual terms in \eq{eq:mcons} and \eq{eq:econs}. The wave energy--momentum is understood as the canonical (`Minkowski') energy--momentum, which must not be confused with the kinetic (`Abraham') energy--momentum \citep{my:amc, ref:dewar77}. Whether the terms with \m{\Delta_s F_s} should be attributed to OCs or to the wave is a matter of convention, because \m{\Delta_s F_s} scales linearly both with \m{F_s} and with \m{J}. In contrast, the wave energy density is defined unambiguously as \m{\mc{E}_s \doteq \int \dd\vec{p}\,H_{0s} F_s} and does not contain \m{\Delta_s}. This is because \m{\int \dd\vec{p}\,\Delta_s F_s} is a part of the wave energy density \m{\mc{E}_{\text{w}}} \citep{my:kchi}. Similarly, \m{\int \dd\vec{p}\,(\pd_{\vec{v}_s}\Delta_s) F_s} is a part of the wave momentum density \citep{my:amc}.}
\label{tab:inter}
\end{table}

\subsection{Summary of \Sec{sec:onshell}}
\label{sec:sumshell}

In summary, we have considered plasma interaction with general broadband single-mode on-shell waves (for examples, see \Sec{sec:examples}). Assuming a general response matrix \m{\matr{\Xi}}, these waves have a dispersion function \m{\Lambda(t, \vec{x}, \omega, \vec{k})} and polarization \(\vec{\eta}(t, \vec{x}, \omega, \vec{k})\) determined by
\begin{eqnarray}
\matr{\Xi}_\herm\vec{\eta} = \Lambda \vec{\eta},
\qquad
\Lambda = \vec{\eta}^\dag\matr{\Xi}_\herm\vec{\eta},
\end{eqnarray}
where the normalization \m{\vec{\eta}^\dag\vec{\eta} = 1} is assumed. Specifically for \m{\matr{\Xi}} given by \eq{eq:Xifinn}, one has
\begin{align}
\Lambda(t, \vec{x}, \omega, \vec{k}) 
= \vec{\eta}^\dag \matr{\Xi}_0 \vec{\eta}
& - \sum_s \int \dd\vec{p}\,
\vec{\eta}^\dag \matr{\wp}_s(\vec{p})\vec{\eta}\,F_s(\vec{p})
\notag\\
& 
+ \sum_s \fint \dd\vec{p}\, 
\frac{|\vec{\alpha}_s^{\dag}\vec{\eta}|^2 (\vec{p})}{\omega - \vec{k} \cdot \vec{v}_s(\vec{p})}\,
\vec{k} \cdot \frac{\pd F_s(\vec{p})}{\pd \vec{p}},
\end{align}
where the arguments \m{(t, \vec{x}, \omega, \vec{k})} are omitted for brevity. (Some notation is summarized in \Sec{sec:sumth}.) The wave frequency \m{\omega = w(t, \vec{x}, \vec{k})} satisfies
\begin{eqnarray}
\Lambda(t, \vec{x}, w(t, \vec{x}, \vec{k}), \vec{k}) = 0
\end{eqnarray}
and \m{w(t, \vec{x}, -\vec{k}) = -w(t, \vec{x}, \vec{k})}, where \(w\) is a real function at real arguments. The wave local linear growth rate \m{\gamma}, which is assumed to be small in this section, is
\begin{eqnarray}
\gamma (t, \vec{x}, \vec{k}) = -\left(
\frac{\vec{\eta}^\dag \matr{\Xi}_\aherm \vec{\eta}}{\pd_{\omega}\Lambda}
\right)_{(t, \vec{x}, w(t, \vec{x}, \vec{k}), \vec{k})},
\end{eqnarray}
or explicitly,
\begin{eqnarray}\notag
\gamma (t, \vec{x}, \vec{k}) = \upi \sum_s \int\dd\vec{p}\,
\frac{|\vec{\alpha}_s^\dag \vec{\eta}|^2}{\pd_{\omega}\Lambda(t, \vec{x}, w, \vec{k})}\,
\delta(w - \vec{k} \cdot \vec{v}_s(t, \vec{x}, \vec{p}))\,\vec{k} \cdot \frac{\pd F_s(t, \vec{x}, \vec{p})}{\pd \vec{p}},
\end{eqnarray}
where \(w \equiv w(t, \vec{x}, \vec{k})\) and \m{|\vec{\alpha}_s^\dag\vec{\eta}|^2 \equiv 
|\vec{\alpha}_s^\dag\vec{\eta}|^2(t, \vec{x}, w, \vec{k}; \vec{p})}. The nonlinear potentials \eq{eq:trfW} are expressed through the scalar function
\begin{eqnarray}
\avr{\st{W}}_s(t, \vec{x}, \omega, \vec{k}; \vec{p}) 
= |\vec{\alpha}_s^\dag\vec{\eta}|^2 (\varsigma_{\vec{k}} J(t, \vec{x}, \vec{k}) + \varsigma_{-\vec{k}}J(t, \vec{x}, -\vec{k}))
\,\delta(\Lambda(t, \vec{x}, \omega, \vec{k})),
\end{eqnarray}
where \m{\varsigma_{\vec{k}} \doteq \sgn(\pd_{\omega}\Lambda(t, \vec{x}, \omega, \vec{k}))} is evaluated at \(\omega = w(t, \vec{x}, \vec{k})\); see also \eq{eq:varsig}. The function \(J\) is the phase-space action density governed by the WKE:
\begin{eqnarray}\label{eq:wkesum}
\frac{\pd J}{\pd t} - \frac{\pd w}{\pd \vec{x}} \cdot \frac{\pd J}{\pd \vec{k}}
+ \frac{\pd w}{\pd \vec{k}} \cdot \frac{\pd J}{\pd \vec{x}} = 2 \gamma J,
\end{eqnarray}
where \m{\pd_\vec{k} w = \vvg} is the group velocity. Collisional dissipation is assumed small compared to collisionless dissipation, so it is neglected in \eq{eq:wkesum} but can be reintroduced by an \adhoc modification of \m{\gamma} (\Sec{sec:feq}). Unlike the field equation used in the standard QLT, \eq{eq:wkesum} exactly conserves the action of nonresonant waves, \ie those with \m{\gamma = 0}. The WKE must be solved together with the QL equation for the OC distribution \(F_s\),
\begin{eqnarray}\label{eq:Fsum}
\frac{\pd F_s}{\pd t}
- \frac{\pd \mc{H}_s}{\pd \vec{x}} \cdot \frac{\pd F_s}{\pd \vec{p}}
+ \frac{\pd \mc{H}_s}{\pd \vec{p}} \cdot \frac{\pd F_s}{\pd \vec{x}}
= \frac{\pd}{\pd \vec{p}} \cdot \left(\vecst{D}_s\,\frac{\pd F_s}{\pd \vec{p}}\right)
+ \collision_s,
\end{eqnarray}
because \(F_s\) determines the coefficients in \eq{eq:wkesum} and \m{J} determines the coefficients in \eq{eq:Fsum}. When \m{\matr{\Xi}_0} and \m{|\vec{\alpha}_s^\dag \vec{\eta}|^2} are independent of \m{t} and \m{\vec{x}}, \eq{eq:wkesum} and \eq{eq:Fsum} conserve the total momentum and energy of the system; specifically,
\begin{align}
\textstyle
\pd_t (\sum_s P_{s,i} + P_{\text{w},i}) 
+ \pd_j (\sum_s {\Pi_{s,i}}^j + {\Pi_{\text{w},i}}^j)
& = - \sum_s \int \dd\vec{p}\,F_s \pd_i H_{0s},\\
\textstyle
\pd_t (\sum_s \mc{E}_{s} + \mc{E}_{\text{w}}) 
+ \pd_j (\sum_s Q^j_s + Q^j_{\text{w}})
& = \sum_s \int \dd\vec{p}\,F_s \pd_t H_{0s}.
\end{align}
Here, the notation is as in \Tab{tab:inter}, or see \eq{eq:mcons} and \eq{eq:econs} instead.

\section{Thermal equilibrium}
\label{sec:Boltz}

In this section, we discuss, for completeness, the properties of plasmas in thermal equilibrium.

\subsection{Boltzmann--Gibbs distribution}
\label{sec:centr}

As discussed in \Sec{sec:cop}, collisions conserve the density of each species, the total momentum density, and the total energy density, while the plasma total entropy density \m{\entropy} either grows or remains constant. Let us search for equilibrium states in particular. At least one of the states in which \m{\sigma} remains constant is the one that maximizes the entropy density at fixed \m{\int \dd\vec{p}\,F_s}, \m{\sum_s \int \dd\vec{p}\,\vec{p}F_s}, and \m{\sum_s\int \dd\vec{p}\,\mc{H}_s F_s}. This `state of thermal equilibrium' can be found as an extremizer of
\begin{gather}
\sigma' \doteq \sigma 
- \sum_s \lambda_s^{(\mc{N})}\int \dd\vec{p}\,F_s 
- \vec{\lambda}^{(\vec{P})} \cdot \sum_s \int \dd\vec{p}\,\vec{p}F_s
- \lambda^{(\mc{H})} \sum_s\int \dd\vec{p}\,\mc{H}_s F_s
\end{gather}
considered as a functional of all \m{F_s}, where \m{\lambda_s^{(\mc{N})}}, \m{\vec{\lambda}^{(\vec{P})}}, and \m{\lambda^{(\mc{H})}} are Lagrange multipliers. Using \eq{eq:Hth2}, one finds that extremizers of \m{\sigma'} satisfy
\begin{gather}
0 = \frac{\delta \sigma'}{\delta F_s} = - \ln F_s - 1 - \lambda_s^{(\mc{N})} 
- \vec{\lambda}^{(\vec{P})} \cdot \vec{p}
- \lambda^{(\mc{H})} \mc{H}_s,
\end{gather}
whence
\begin{gather}\label{eq:Fext}
F_s = \text{const}_s \times \exp(- \vec{\lambda}^{(\vec{P})} \cdot \vec{p} - \lambda^{(\mc{H})}\mc{H}_s).
\end{gather}
The pre-exponential constant is determined by the given density of species \m{s}, while \m{\vec{\lambda}^{(\vec{P})}} and \m{\lambda^{(\mc{H})}} can be expressed through the densities of the plasma momentum and energy stored in the whole distribution. Because
\begin{gather}
\frac{\delta^2 \sigma'}{\delta F_s \delta F_s} = - \frac{1}{F_s} < 0,
\qquad
\frac{\delta^2 \sigma'}{\delta F_s \delta F_{s' \ne s}} = 0,
\end{gather}
the matrix \m{\delta^2 \sigma'/\delta F_s \delta F_{s'}} is positive-definite, so the entropy is maximal (as opposed to minimal) at the extremizer \eq{eq:Fext}.

The distribution \eq{eq:Fext} is known as the Boltzmann--Gibbs distribution, with \m{T \doteq 1/\lambda^{(\mc{H})}} being the temperature (common for all species). Also, let us introduce a new, rescaled Lagrange multiplier \m{\vec{\mcc{u}}} via \m{\vec{\lambda}^{(\vec{P})} = -\vec{\mcc{u}}/T}. Then,
\begin{eqnarray}
F_s(\vec{p}) = F_s^{(0)}\exp\left(-\frac{\mc{H}_s(\vec{p}) - \vec{\mcc{u}} \cdot \vec{p}}{T}\right),
\end{eqnarray}
where \m{F_s^{(0)}} is independent of \m{\vec{p}}. Correspondingly,
\begin{eqnarray}\label{eq:dfb0}
\frac{\pd F_s(\vec{p})}{\pd \vec{p}} 
= - F_s(\vec{p})\,\frac{\vec{v}_s - \vec{\mcc{u}}}{T},
\end{eqnarray}
where we used \eq{eq:vOC2}. From \eq{eq:dfb0}, one obtains
\begin{align}
\delta(&\vec{k} \cdot \vec{v}_s - \vec{k} \cdot \vec{v}'_{s'})\, \vec{k}\cdot
\left(\frac{\pd F_s(\vec{p})}{\pd \vec{p}}\,F_{s'}(\vec{p}')
- F_s(\vec{p})\,\frac{\pd F_{s'}(\vec{p}')}{\pd \vec{p}'}\right)
\notag\\
& = - \frac{1}{T}\,\delta(\vec{k} \cdot \vec{v}_s - \vec{k} \cdot \vec{v}'_{s'})\,
(\vec{k} \cdot \vec{v}_s - \vec{k} \cdot \vec{v}'_{s'})\,
F_s(\vec{p})F_{s'}(\vec{p}')
\notag\\
& = 0,
\end{align}
where \m{\mc{H}'_{s'} \doteq \mc{H}_{s'}(\vec{p}')}. Then, \eq{eq:cop2} yields that the collision operator vanishes on the Boltzmann--Gibbs distribution, and thus, expectedly, \m{(\dd\entropy/\dd t)_{\text{coll}} = 0}. One can also show that the Boltzmann--Gibbs distribution is \textit{the only} distribution (strictly speaking, a class of distributions parameterized by \m{T} and \m{\vec{\mcc{u}}}) for which the entropy density is conserved (\App{app:gibbs}).

The property \eq{eq:dfb0} of the thermal-equilibrium state also leads to other notable results that we derive below. In doing so, we will assume the reference frame where \m{\vec{\mcc{u}} = 0}, so the Boltzmann--Gibbs distribution has a better known form
\begin{eqnarray}\label{eq:dfb}
F_s(\vec{p}) = F_s^{(0)}\exp\left(-\frac{\mc{H}_s(\vec{p})}{T}\right),
\qquad
\frac{\pd F_s(\vec{p})}{\pd \vec{p}} 
= - F_s(\vec{p})\,\frac{\vec{v}_s}{T}.
\end{eqnarray}
(For \m{\mc{H}_s} isotropic in \m{\vec{p}}, this is the frame where the plasma total momentum density \m{\sum_s \int \dd\vec{p}\,\vec{p}F_s} is zero.) The generalizations to arbitrary \m{\vec{\mcc{u}}} are straightforward.

\subsection{Fluctuation--dissipation theorem}
\label{sec:fdt}

Let us describe microscopic fluctuations in equilibrium plasmas in terms of \m{\matrst{S}(\omega, \vec{k}) \doteq (2\upi)^{\stn}\,\matru(\omega, \vec{k})}, \ie
\begin{equation}\label{eq:Sft0}
\matrst{S}(\omega, \vec{k}) \doteq
\int \dd\tau \dd\vec{s}\, 
\avr{\Psiq(t + \tau/2, \vec{x} + \vec{s}/2) \Psiq^\dag(t - \tau/2, \vec{x} - \vec{s}/2)}\,\ee^{\ii \omega \tau - \ii \vec{k} \cdot \vec{s}},
\end{equation}
which can also be represented in terms of the Fourier image \m{\fourier{\vec{\Psi}}(\omega, \vec{k})} of the microscopic field \m{\Psiq(t, \vec{x})}:
\begin{equation}\label{eq:Sft}
\matrst{S}(\omega, \vec{k}) = \frac{\avr{
\fourier{\vec{\Psi}}(\omega, \vec{k})
\smash{\fourier{\vec{\Psi}}}^\dag(\omega, \vec{k})
}}{\mcu{T}\mcu{V}_n}.
\end{equation}
For statistically homogeneous fields that persist on time \m{\mcu{T} \to \infty} within volume \m{\mcu{V}_n \to \infty}, the Fourier transform is formally divergent; hence the appearance of the factors \m{\mcu{T}} and \m{\mcu{V}_n} in \eq{eq:Sft}.\footnote{To make \eq{eq:Sft} look more physical (local), one can absorb the global factors \m{\mcu{T}} and \m{\mcu{V}_n} in the definition of the Fourier transform; cf.\ \Sec{eq:homDL}.} Also, as seen from \eq{eq:LRu}, any quadratic function of the microscopic field can be expressed through \m{\matrst{S}} via
\begin{eqnarray}
\avr{(\boperst{L}\vec{\psi})(\boperst{R}\vec{\psi})^\dag}
\approx \int \frac{\dd\omega}{2\upi}\,\frac{\dd\vec{k}}{(2\upi)^n}\,
(\matrst{L}\matrst{S}\matrst{R}^\dag)(\omega, \vec{k}),
\end{eqnarray}
where \m{\boperst{L}} and \m{\boperst{R}} are any linear operators and \m{\matrst{L}} and \m{\matrst{R}} are their symbols.

From \eq{eq:ufina}, one finds that, in general,
\begin{equation}
\matrst{S}(\omega, \vec{k})
= 2\upi\, \sum_{s'}\int \dd\vec{p}'\,\delta(\omega - \vec{k} \cdot \vec{v}'_{s'})F_{s'}(\vec{p}')
\matr{\Xi}^{-1}(\omega, \vec{k})
(\vec{\alpha}_{s'}\vec{\alpha}_{s'}^\dag)(\omega, \vec{k};\vec{p}')
\matr{\Xi}^{-\dag}(\omega, \vec{k}).
\label{eq:mcWsp}
\end{equation}
For a thermal distribution in particular, which satisfies \eq{eq:dfb}, one can rewrite \eq{eq:Xiah} as follows:
\begin{align}
\matr{\Xi}_\aherm(\omega, \vec{k}) 
& \approx
\frac{\upi}{T} \sum_s \int \dd\vec{p}\,
\vec{\alpha}_s(\omega, \vec{k}; \vec{p})\vec{\alpha}_s^\dag(\omega, \vec{k}; \vec{p})
\delta(\omega - \vec{k} \cdot \vec{v}_s)\,(\vec{k} \cdot \vec{v}_s) F_s(\vec{p})
\notag\\
& =
\frac{\upi \omega}{T} \sum_s \int \dd\vec{p}\,
\vec{\alpha}_s(\omega, \vec{k}; \vec{p})\vec{\alpha}_s^\dag(\omega, \vec{k}; \vec{p})
\delta(\omega - \vec{k} \cdot \vec{v}_s) F_s(\vec{p}).
\label{eq:Xia2}
\end{align}
By comparing this with \eq{eq:mcWsp}, one also finds that
\begin{eqnarray}\label{eq:fdt0}
\matrst{S}(\omega, \vec{k}) = 
\frac{2T}{\omega}\,(\matr{\Xi}^{-1}\matr{\Xi}_\aherm\matr{\Xi}^{-\dag})(\omega, \vec{k}).
\end{eqnarray}
Due to \eq{eq:invXi}, this leads to the fluctuation--dissipation theorem in the following form:
\begin{eqnarray}
\matrst{S}(\omega, \vec{k}) = -\frac{2T}{\omega}\,(\matr{\Xi}^{-1})_\aherm(\omega, \vec{k}).
\label{eq:ufdt1}
\end{eqnarray}
For examples of \m{\matr{\Xi}} for specific systems, see \Sec{sec:examples}.

\subsection{Kirchhoff's law}
\label{sec:kirch}

Consider the power deposition via polarization drag:
\begin{eqnarray}
\mcc{P} =  \sum_s \int \dd\vec{p}\,(\vec{v}_s \cdot \vec{\mcc{F}}_s) F_s(\vec{p}).
\end{eqnarray}
Using \eq{eq:auxFd} for \m{\vec{\mcc{F}}_s}, \eq{eq:Xia2} for \m{\matr{\Xi}_\aherm}, and \eq{eq:ufdt1} for \m{\matrst{S}}, this can also be expressed as follows:
\begin{align}
\mcc{P} & 
\approx \sum_s
\int \frac{\dd\vec{k}}{(2\upi)^n}\,\dd\vec{p}\,(\vec{k} \cdot \vec{v}_s)\,
(\vec{\alpha}_s^\dag (\matr{\Xi}^{-1})_\aherm \vec{\alpha}_s)(\vec{k} \cdot \vec{v}_s, \vec{k}; \vec{p})
F_s(\vec{p})
\notag\\
& = \sum_s\int \frac{\dd\omega}{2\upi} \frac{\dd\vec{k}}{(2\upi)^n}\,\dd\vec{p}\,
2\upi\omega\,\delta(\omega - \vec{k} \cdot \vec{v}_s)
(\vec{\alpha}_s^\dag (\matr{\Xi}^{-1})_\aherm \vec{\alpha}_s)(\omega, \vec{k}; \vec{p})F_s(\vec{p})
\notag\\
& = 2T \int \frac{\dd\omega}{2\upi} \frac{\dd\vec{k}}{(2\upi)^n} 
\tr\bigg((\matr{\Xi}^{-1})_\aherm
\frac{\upi \omega}{T}\, \sum_s \int \dd\vec{p}\,\delta(\omega - \vec{k} \cdot \vec{v}_s)
(\vec{\alpha}_s\vec{\alpha}_s^\dag)(\omega, \vec{k}; \vec{p})F_s(\vec{p})\bigg)
\notag\\
& = -\int \frac{\dd\omega}{2\upi}\,\frac{\dd\vec{k}}{(2\upi)^n}\,\omega \tr(\matrst{S}\matr{\Xi}_\aherm)(\omega, \vec{k}).
\end{align}
Thus, the spectral density of the power deposition via polarization drag is given by
\begin{eqnarray}
\mcc{P}_{\omega, \vec{k}} = -\omega \tr(\matrst{S}\matr{\Xi}_\aherm),
\end{eqnarray}
which is a restatement of Kirchhoff's law \citep[section~11.4]{book:krall}. For examples of \m{\matr{\Xi}} for specific systems, see \Sec{sec:examples}.

\subsection{Equipartition theorem}
\label{sec:equi}

As flows from \Sec{sec:cons}, the energy of on-shell waves of a field \m{\vec{\osc{\Psi}}} in a homogeneous \m{n}-dimensional plasma of a given volume \m{\mcu{V}_n} can be written as
\begin{align}
\mcu{V}_n\mc{E}_{\text{w}}
& = \int \dd\vec{k}\,\mcu{V}_n w(\vec{k})J(\vec{k})
\notag\\
& = (2\upi)^n\int \frac{\mcu{V}_n \dd\vec{k}}{(2\upi)^n}\int_0^\infty \dd\omega\,
\omega\,\pd_\omega\Lambda(\omega, \vec{k})\,h(\vec{k})\,\delta(\omega - w(\vec{k}))
\notag\\
& = (2\upi)^n \sum_{\vec{k}} \int^{\infty}_0 \dd\omega\,\omega\, \pd_\omega\Lambda(\omega, \vec{k})\, (\vec{\eta}^\dag\matrU\vec{\eta})(\omega, \vec{k}).
\end{align}
To apply this to microscopic fluctuations, one can replace \m{\matrU} with \m{\matru} and substitute \m{\matru = (2\upi)^{-(n+1)}\matrst{S}}. Then, the total energy of a mode with given wavevector \m{\vec{k}} and polarization \m{\vec{\eta}} can be expressed as
\begin{eqnarray}
\mc{E}_{\vec{k}, \vec{\eta}} = \frac{1}{2\upi}  \int^{\infty}_0 \dd\omega\,
\omega\,(\pd_\omega\Lambda)\,\vec{\eta}^\dag \matrst{S} \vec{\eta},
\end{eqnarray}
where the arguments \m{(\omega, \vec{k})} are omitted for brevity. For thermal equilibrium, one can substitute \eq{eq:ufdt1} for \m{\matrst{S}}; then,
\begin{align}\label{eq:eqp}
\mc{E}_{\vec{k}, \vec{\eta}} 
= -\frac{T}{\upi}\, \im \int^{\infty}_0 \dd\omega\,(\pd_\omega\Lambda)\,
\vec{\eta}^\dag \matr{\Xi}^{-1} \vec{\eta}.
\end{align}
The integrand peaks at \m{\omega = w(\vec{k})}, where the mode eigenvalue \m{\Lambda} is small. Due to damping, the actual zero of \m{\Lambda} is slightly below the real axis in the complex-frequency space. Then, at infinitesimally small damping, \m{\vec{\eta}^\dag \matr{\Xi}^{-1} \vec{\eta}} can be approximated near \m{\omega = w(\vec{k})} as
\begin{align}
\vec{\eta}^\dag \matr{\Xi}^{-1} \vec{\eta}
\approx \frac{1}{\Lambda}
\approx \frac{1}{\pd_\omega\Lambda(\omega, \vec{k})}
\left(\pv\frac{1}{\omega - w(\vec{k})}
-\ii \upi \delta(\omega - w(\vec{k}))
\right).
\end{align}
This leads to the well-known equipartition theorem:
\begin{align}\label{eq:eqpt}
\mc{E}_{\vec{k}} = T.
\end{align}
Note that according to \eq{eq:eqpt}, the sum \m{\mcu{V}_n\mc{E}_{\text{w}} = \sum_{\vec{k}, \vec{\eta}}\mc{E}_{\vec{k}, \vec{\eta}}} is divergent. This indicates that not all modes can be classical and on-shell (weakly damped) simultaneously.

\subsection{Summary of \Sec{sec:Boltz}}
\label{sec:sumbol}

In thermal equilibrium, when all species have Boltzmann--Gibbs distributions with common temperature \m{T}, the collision operator vanishes, the entropy is conserved, and the spectrum of microscopic fluctuations \eq{eq:Sft0} satisfies the fluctuation--dissipation theorem:
\begin{eqnarray}\label{eq:dft2}
\matrst{S}(\omega, \vec{k}) = -\frac{2T}{\omega}\,(\matr{\Xi}^{-1})_\aherm(\omega, \vec{k}),
\end{eqnarray}
where \m{\matr{\Xi}} is the dispersion matrix \eq{eq:Xifinn} and \m{_\aherm} denotes the anti-Hermitian part (or the imaginary part in case of scalar fields). From here, it is shown that the spectral density of the power deposition via polarization drag is given by \m{\mcc{P}_{\omega, \vec{k}} = - \omega \tr(\matrst{S}\matr{\Xi}_\aherm)}, which is a restatement of Kirchhoff's law. For on-shell waves, \eq{eq:dft2} reduces to the equipartition theorem, which says that the energy per mode equals \m{T}. Applications to specific systems are discussed in \Sec{sec:examples}.

\section{Examples}
\label{sec:examples}

In this section, we show how to apply our general formulation to nonrelativistic electrostatic interactions (\Sec{sec:lang}), relativistic electromagnetic interactions (\Sec{sec:emrel}), Newtonian gravity (\Sec{sec:grav}), and relativistic gravity, including gravitational waves (\Sec{sec:gw}).

\subsection{Nonrelativistic electrostatic interactions}
\label{sec:lang}

\subsubsection{Main equations}

Let us show how our general formulation reproduces (and generalizes) the well-known results for electrostatic turbulence in nonmagnetized nonrelativistic plasma. In this case,
\begin{eqnarray}\label{eq:hamES}
H_s = \frac{p^2}{2m_s} + e_s \avr{\varphi} +  e_s\osc{\varphi},
\end{eqnarray}
where \(e_s\) is the electric charge, \(\varphi\) is the electrostatic potential, and \m{\avr{\varphi}} and \m{\osc{\varphi}} are its average and oscillating parts, respectively. Then, \m{H_{0s} = \avr{H}_s = p^2/(2m_s) + e_s \avr{\varphi}}, \m{\osc{H}_s = e_s \osc{\varphi}}, \m{\boper{\alpha}_s = e_s}, and \m{\boper{L}_s = \boper{R}_s = \boper{0}}, so \m{\matr{\wp}_s = \matr{0}}. The matrix \eq{eq:Usum} is a scalar (Wigner function) given by
\begin{equation}
\st{U}(t, \vec{x}, \omega, \vec{k}) 
= \int \frac{\dd\tau}{2\upi}\,\frac{\dd\vec{s}}{(2\upi)^{n}}\,\, 
\avr{
{\macro{\osc{\varphi}}}(t + \tau/2, \vec{x} + \vec{s}/2)\,
{\macro{\osc{\varphi}}}(t - \tau/2, \vec{x} - \vec{s}/2)
}
\,\ee^{\ii\omega \tau - \ii \vec{k} \cdot \vec{s}}.
\label{eq:dewU}
\end{equation}
(Underlining denotes the macroscopic part, \m{n \doteq \dim\vec{x}}, and the arguments \m{(t, \vec{x})} will be omitted from now on.) Correspondingly,
\begin{align}
\matrst{D}_s
& \approx e_s^2 \int \dd\vec{k}\, \upi\,\vec{k}\vec{k} \st{U}(\vec{k} \cdot \vec{v}_s, \vec{k}),
\label{eq:Des1}\\
\matr{\Uptheta}_s
& = e_s^2\,\frac{\pd}{\pd \vartheta} \fint \dd\omega\,\dd\vec{k}\,
 \left.
\frac{\vec{k}\vec{k} \st{U}(\omega, \vec{k})}{\omega - \vec{k} \cdot \vec{v}_s + \vartheta}
\right|_{\vartheta=0},
\label{eq:Ues}
\\
\Delta_s = \Phi_s 
& = \frac{e_s^2}{2}\frac{\pd}{\pd \vec{p}} \cdot
\fint \dd\omega\,\dd\vec{k}\,\frac{\vec{k} \st{U}(\omega, \vec{k})}{\omega - \vec{k} \cdot \vec{v}_s},
\label{eq:Phies}
\end{align}
and also
\begin{eqnarray}
\mc{H}_s = \frac{p^2}{2m_s} + e_s\avr{\varphi} + \Delta_s,
\qquad
\vec{v}_s = \frac{\vec{p}}{m_s} + \frac{\pd \Delta_s}{\pd \vec{p}}.
\end{eqnarray}

The Lagrangian density of a free electrostatic field is
\begin{eqnarray}
\mcc{L}_0 = \frac{1}{8\upi}\,\delta^{ij}(\pd_i\osc{\varphi})(\pd_j\osc{\varphi})
= \frac{\pd}{\pd x^i}\left(\frac{1}{8\upi}\,\delta^{ij} \osc{\varphi}(\pd_j\osc{\varphi})\right)
+ \frac{1}{2}\,\osc{\varphi}\left(-\frac{\delta^{ij}\pd_i \pd_j}{4\upi}\right)\,\osc{\varphi}.
\end{eqnarray}
The first term on the right-hand side does not contribute to the field action \m{S_0} and thus can be ignored. The second term is of the form \eq{eq:Lagr0} with \m{M = 1}, \m{\matr{\metric} = 1} (\Sec{sec:matr}), and \m{\boper{\Xi}_0 = \oper{k}^2/(4\upi)}, so \m{\matr{\Xi}_0 (\omega, \vec{k}) = k^2/(4\upi)}, where \m{k^2 \equiv \vec{k}^2 \equiv \delta^{ij}k_i k_j}. Then, \eq{eq:Xis} gives
\begin{eqnarray}
\matr{\Xi}(\omega, \vec{k}) = \Xi(\omega, \vec{k}) = \frac{k^2\epsilon_\parallel (\omega, \vec{k})}{4\upi},
\end{eqnarray}
where the arguments \m{t} and \m{\vec{x}} are omitted for brevity and \m{\epsilon_\parallel} is the parallel permittivity:
\begin{eqnarray}
\epsilon_\parallel(\omega, \vec{k}) =
1 + \sum_s \frac{4\upi e_s^2}{k^2} \int \dd\vec{p}\,\frac{\vec{k}}{\omega - \vec{k} \cdot \vec{v}_s + \ii 0} 
\cdot \frac{\pd F_s}{\pd \vec{p}}.
\label{eq:epspar}
\end{eqnarray}

\subsubsection{Collisions and fluctuations}
\label{sec:langmicro}

By \eq{eq:mcWsp}, the spectrum of microscopic oscillations of \m{\osc{\varphi}} is a scalar given by
\begin{eqnarray}
\st{S}(\omega, \vec{k}) = 2\upi\sum_{s} \left(\frac{4\upi e_s}{k^2 |\epsilon_\parallel (\omega, \vec{k})|}\right)^2 \int \dd\vec{p}\,
\delta(\omega - \vec{k} \cdot \vec{v}_{s})F_{s}(\vec{p}),
\label{eq:Ses}
\end{eqnarray}
where we substituted \(n = 3\) for three-dimensional plasma. For thermal equilibrium, \eq{eq:ufdt1} leads to the well-known formula \citep[section~51]{book:landau10}
\begin{eqnarray}
\st{S}(\omega, \vec{k}) = - \frac{2 T}{\omega}\im\left(\frac{1}{\Xi(\omega, \vec{k})}\right)
= -\frac{8\upi T}{\omega k^2}\im\left(\frac{1}{\epsilon_\parallel(\omega, \vec{k})}\right)
= \frac{8\upi T}{\omega k^2} \frac{\im \epsilon_\parallel(\omega, \vec{k})}{|\epsilon_\parallel(\omega, \vec{k})|^2}.
\end{eqnarray}
The spectrum \m{\st{S}_\rho} of charge-density fluctuations is found using Poisson's equation \m{\micro{\osc{\rho}} = \oper{k}^2\micro{\osc{\varphi}}/4\upi}, whence \m{\st{S}_\rho \approx (k^2/4\upi)^2\st{S}}. Fluctuations of other fields are found similarly. Also, \eq{eq:mcYsymb} leads to
\begin{eqnarray}
|\mc{X}_{ss'}(\omega, \vec{k}; \vec{p}, \vec{p}')|^2 = 
\left(\frac{4\upi e_s e_{s'}}{k^2 |\epsilon_\parallel(\omega, \vec{k})|}\right)^2.
\end{eqnarray}
Then, \eq{eq:cop2} yields the standard Balescu--Lenard collision operator:
\begin{align}
\collision_s
= \frac{\pd}{\pd \vec{p}} \cdot
\sum_{s'}\int \frac{\dd\vec{k}}{(2\upi)^3}\,\dd\vec{p}'\,& \frac{\upi \vec{k}\vec{k}}{|\epsilon_\parallel(\vec{k} \cdot \vec{v}_s, \vec{k})|^2}\,
\left(\frac{4\upi e_s e_{s'}}{k^2}\right)^2
\delta(\vec{k} \cdot \vec{v}_s - \vec{k} \cdot \vec{v}'_{s'})
\notag\\
& \cdot
\left(
\frac{\pd F_s(\vec{p})}{\pd \vec{p}}\,F_{s'}(\vec{p}')
- F_s(\vec{p})\,\frac{\pd F_{s'}(\vec{p}')}{\pd \vec{p}'}
\right).
\label{eq:BLen}
\end{align}
(As a reminder, the distribution functions are normalized such that \(\int \dd\vec{p}\,F_s\) is the local average density of species \m{s} \eq{eq:density}.)

\subsubsection{On-shell waves}
\label{sec:onshelles}

For on-shell waves, \eq{eq:UW} gives \m{\st{U}(\omega, \vec{k}) = (h(\vec{k}) + h(-\vec{k}))\delta(\omega - w(\vec{k}))}, where \m{w(\vec{k})} is determined by the dispersion relation
\begin{eqnarray}\label{eq:esdr}
\epsilon_{\parallel\herm}(w(\vec{k}), \vec{k})  = 0,
\end{eqnarray}
and \m{\epsilon_{\parallel\herm} \equiv \re \epsilon_\parallel} is given by
\begin{eqnarray}
\epsilon_{\parallel\herm}(\omega, \vec{k}) 
= 1 + \sum_s \frac{4\upi e_s^2}{k^2} \fint \dd\vec{p}\,\frac{\vec{k}}{\omega - \vec{k} \cdot \vec{v}_s} 
\cdot \frac{\pd F_s}{\pd \vec{p}}.
\end{eqnarray}
The phase-space density of the wave action, defined in \eq{eq:Jinc}, is
\begin{eqnarray}
J(\vec{k}) = h(\vec{k})\,\frac{\pd\Xi_\herm(\vec{k})}{\pd \omega} 
= h(\vec{k}) \,\frac{k^2}{4\upi}\frac{\pd\epsilon_{\parallel\herm}(w(\vec{k}), \vec{k})}{\pd \omega},
\label{eq:Jes}
\end{eqnarray}
and the dressing function \eq{eq:Ues} is given by
\begin{align}
\matr{\Uptheta}_{s}
 & = e_s^2\,\frac{\pd}{\pd \vartheta} \fint \dd\vec{k}\,(h(\vec{k}) + h(-\vec{k}))
 \left.
\frac{\vec{k} \vec{k}}{w(\vec{k}) - \vec{k} \cdot \vec{v}_s + \vartheta}
\right|_{\vartheta=0}
\notag\\
& = 2e_s^2\,\frac{\pd}{\pd \vartheta} \fint \dd\vec{k}\,h(\vec{k})
 \left.
\frac{\vec{k} \vec{k}}{w(\vec{k}) - \vec{k} \cdot \vec{v}_s + \vartheta}
\right|_{\vartheta=0}.
\label{eq:Thetes2}
\end{align}
Using these, one obtains (\App{sec:momes})
\begin{eqnarray}\label{eq:totmes}
\sum_s \int \dd\vec{p}\,\vec{p} F_s + \int \dd\vec{k}\,\vec{k} J = \sum_s \int \dd\vec{p}\,\vec{p} \avr{f}_s,
\end{eqnarray}
so the conserved quantity \eq{eq:mcons1} is the average momentum of the plasma (while the electrostatic field carries no momentum, naturally). Also (\App{sec:enes}),
\begin{eqnarray}
\sum_s \int \dd\vec{p}\,H_{0s} F_s + \int \dd\vec{k}\,w J 
= \sum_s \int \dd\vec{p}\,H_{0s}\avr{f}_s 
+ \frac{1}{8\upi}\,\avr{\smash{\osc{\vec{E}}}^\dag \osc{\vec{E}}},
\end{eqnarray}
so, expectedly, the conserved quantity \eq{eq:encons1} is the average particle energy plus the energy of the electrostatic field. In combination with our equations for \m{F_s} and \m{J} (\Sec{sec:sumshell}), these results can be considered as a generalization and concise restatement of the OC QLT by \citet{ref:dewar73}, which is rigorously reproduced from our general formulation as a particular case.

\subsubsection{Eikonal waves}
\label{sec:eikw}

As a particular case, let us consider an eikonal wave
\begin{eqnarray}
\macro{\osc{\varphi}} \approx \re(\ee^{\ii \theta}\env{\varphi}),
\qquad
\avr{\omega}\doteq -\pd_t\theta,
\qquad
\avr{\vec{k}}\doteq \pd_{\vec{x}} \theta,
\end{eqnarray}
which may or may not be on-shell. As seen from \Sec{sec:wigvec},
\begin{eqnarray}
\st{U} \approx \frac{|\env{\varphi}|^2}{4}\,\sum_{\pm} \delta(\omega \pm \avr{\omega})\,\delta(\vec{k} \pm \avr{\vec{k}}).
\end{eqnarray}
For nonresonant particles, the dressing function is well defined is found as follows:
\begin{align}
\matr{\Uptheta}_s
& \approx - \int \dd\omega\,\dd\vec{k}\,
\frac{e_s^2 \vec{k}\vec{k}|\env{\varphi}|^2}{4(\omega - \vec{k} \cdot \vec{v}_s)^2}\,\sum_{\pm} \delta(\omega \pm \avr{\omega})\,\delta(\vec{k} \pm \avr{\vec{k}})
\notag\\
& = - \frac{e_s^2 \avr{\vec{k}}\,\avr{\vec{k}} |\env{\varphi}|^2}{2(\avr{\omega} - \avr{\vec{k}} \cdot \vec{v}_s)^2}.
\end{align}
Similarly, the ponderomotive energy for nonresonant particles is
\begin{align}
\Delta_s
& \approx \frac{e_s^2 |\env{\varphi}|^2}{8m_s}
\frac{\pd}{\pd \vec{v}_s} \cdot \int \dd{\omega}\,\dd\vec{k}\,
\frac{\vec{k}}{\omega - \vec{k} \cdot \vec{v}_s}
\sum_{\pm} \delta(\omega \pm \avr{\omega})\,\delta(\vec{k} \pm \avr{\vec{k}})
\notag\\
& = \frac{e_s^2 |\env{\varphi}|^2}{8m_s} \int \dd{\omega}\,\dd\vec{k}\,
\frac{k^2}{(\omega - \vec{k} \cdot \vec{v}_s)^2}
\sum_{\pm} \delta(\omega \pm \avr{\omega})\,\delta(\vec{k} \pm \avr{\vec{k}})
\notag\\
& = \frac{e_s^2 \avr{k}^2 |\env{\varphi}|^2}{4m_s(\avr{\omega} - \avr{\vec{k}} \cdot \vec{v}_s)^2},
\end{align}
in agreement with \citep{ref:dewar72c, ref:cary77}. One can also express these functions in terms of the electric-field envelope \m{\env{\vec{E}} \approx -\ii \avr{\vec{k}} \env{\varphi}}:
\begin{eqnarray}
\matr{\Uptheta}_{s} \approx -\frac{e_s^2\env{\vec{E}}\smash{\env{\vec{E}}}^\dag}{2(\avr{\omega} - \avr{\vec{k}} \cdot \vec{v}_s)^2},
\qquad
\Delta_s \approx \frac{e_s^2 |\smash{\env{\vec{E}}}|^2}{4m_s (\avr{\omega} - \avr{\vec{k}} \cdot \vec{v}_s)^2}.
\end{eqnarray}

For on-shell in particular, one can use \eq{eq:Jes} together with \m{h(\vec{k}) = \frac{1}{4} |\env{\varphi}|^2 \delta(\vec{k} - \avr{\vec{k}})} (cf.\ \eq{eq:hdef}) to obtain the well-known expression for the wave action density \m{\mc{I} \doteq \int \dd\vec{k}\,J}:
\begin{eqnarray}\label{eq:Iesex}
\mc{I} = \frac{|\smash{\env{\vec{E}}}|^2}{16\upi}\frac{\pd\epsilon_{\parallel\herm}(\omega, \avr{\vec{k}})}{\pd \omega}\,\Big|_{\omega = w(\avr{\vec{k}})}.
\end{eqnarray}

For non-too-hot plasma, one has \m{\epsilon_{\parallel\herm}(\omega, \vec{k}) \approx 1 - \omega_p^2/\omega^2}, where \m{\omega_p \doteq \sum_s 4\upi \mc{N}_s e_s^2/m_s} is the plasma frequency. The corresponding waves are Langmuir waves. Their dispersion relation is \m{w(\avr{\vec{k}}) \approx \pm\omega_p}, so \m{\mc{I} \approx \pm |\env{\vec{E}}|^2/(8\upi\omega_p)} (and accordingly, the wave energy density is \m{\mc{E}_{\text{w}} = w\mc{I} \ge 0} for either sign). Remember, though, that this expression is only approximate. Using it instead of \eq{eq:Iesex} can result in violation of the exact conservation laws of QLT. Conservation of the Langmuir-wave action in non-stationary plasmas beyond the cold-plasma approximation is also discussed in \citep{my:dense, my:mquanta, my:langact}. 

\subsubsection{Homogeneous plasma}
\label{eq:homDL}

In homogeneous \(n\)-dimensional plasma of a given volume \(\mcu{V}_n\), the Wigner function \eq{eq:dewU} has the form \m{\st{U} = \mc{U}(t, \vec{k}) \delta(\omega - w(t, \vec{k}))}. The function \(\mc{U}\) is readily found using \eq{eq:Wfourr}:
\begin{eqnarray}
\mc{U}(t, \vec{k}) = \frac{1}{\mcu{V}_n} \int \dd\vec{x}\,\dd\omega\, U
= \frac{1}{\mcu{V}_n}\,|\avr{\fourier{\macro{\osc{\varphi}}}(t, \vec{k})|^2}
= \frac{1}{\mcu{V}_n}\,|\fourier{\macro{\osc{\varphi}}}(t, \vec{k})|^2.
\end{eqnarray}
Then,
\begin{eqnarray}
\matrst{D}_{s} 
\approx \frac{\upi e_s^2}{\mcu{V}_n} \int \dd\vec{k}\,
\vec{k}\vec{k}\,|\fourier{\macro{\osc{\varphi}}}(t, \vec{k})|^2\,
\delta(w(t, \vec{k}) - \vec{k} \cdot \vec{v}_s).
\end{eqnarray}
This coincides with the well-known formula for the QL-diffusion coefficient in homogeneous electrostatic plasma.\footnote{See, for example, equation (16.17) in \citep{book:stix}. The extra mass factor appears there because QL diffusion is considered in the velocity space instead of the momentum space.} The functions \m{\matr{\Uptheta}_{s}} and \m{\Delta_s} are also important in homogeneous turbulence in that they ensure the proper energy--momentum conservation; for example, see \citep[section~16.3]{book:stix} and \citep[section~II.2]{my:sharm}. These functions can be expressed through \m{\fourier{\macro{\osc{\varphi}}}} too. However, they have a simpler representation in terms of the Wigner function~\m{\st{U}}, as in \eq{eq:Ues} and \eq{eq:Phies}, respectively. This is because \m{\st{U}} is a \textit{local} property of the field, which makes it more fundamental than the amplitudes of global Fourier harmonics commonly used in the literature.

\subsection{Relativistic electromagnetic interactions}
\label{sec:emrel}

\subsubsection{Main equations}
\label{sec:mainem}

Let us extend the above results to relativistic electromagnetic interactions. In this case,
\begin{eqnarray}\label{eq:hamEM}
H_s = \sqrt{m_s^2 c^4 + (\vec{p}c - e_s \vec{A})^2} + e_s \varphi,
\end{eqnarray}
where \m{c} is the speed of light and \m{\vec{A}} is the vector potential. Let us adopt the Weyl gauge for the oscillating part of the electromagnetic field (\m{\osc{\varphi} = 0}) and Taylor-expand \m{H_s} to the second order in \m{\osc{\vec{A}}}. This leads to
\begin{eqnarray}
& \displaystyle
H_s \approx H_{0s} - e_s\vec{\beta}_s^\dag \osc{\vec{A}} + \frac{e_s^2}{2 c^2}\,\smash{\osc{\vec{A}}}^\dag \matr{\mu}_s^{-1} \osc{\vec{A}},
\\
& \displaystyle
H_{0s} = \sqrt{m_s^2 c^4 + (\vec{p}c - e_s \avr{\vec{A}})^2} + e_s \avr{\varphi}
\end{eqnarray}
(although plasma is assumed nonmagnetized, a \textit{weak} average magnetic field \m{\avr{\vec{B}} = \del \times \avr{\vec{A}}} is allowed, so \m{\avr{\vec{A}}} can be order-one and thus generally must be retained), where
\begin{eqnarray}
\vec{\beta}_s = \frac{1}{m_s c \gamma_s}\left(\vec{p} - \frac{e_s}{c}\,\avr{\vec{A}}\right),
\qquad
\matr{\mu}_s^{-1} = \frac{\matr{1} - \vec{\beta}_s\vec{\beta}_s^\dag}{m_s \gamma_s},
\end{eqnarray}
and \m{\gamma_s \doteq (1-\beta_s^2)^{-1/2}}. In the equations presented below, \m{\vec{\beta}_s = \vec{v}_s/c} (where \m{\vec{v}_s} is the OC velocity) is a sufficiently accurate approximation. Also, \m{\matr{\mu}_s \equiv (\pd^2_{\vec{p}\vec{p}} H_{0s})^{-1}} can be interpreted as the relativistic-mass tensor. 

Let us choose the field \m{\osc{\vec{\Psi}}} of our general theory to be the oscillating electric field \m{\osc{\vec{E}} = \ii \oper{\omega}\osc{\vec{A}}/c}; then (cf.\ \eq{eq:Hs}),
\begin{eqnarray}
\boper{\alpha}_s = \ii e_s\vec{v}_s \oper{\omega}^{-1},
\qquad
\boper{L}_s = e_s^2\oper{\omega}^{-1},
\qquad
\boper{R}_s = \matr{\mu}_s^{-1}\oper{\omega}^{-1}.
\end{eqnarray}
(Other ways to identify \m{\boper{L}_s} and \m{\boper{R}_s} are also possible and lead to the same results.) Then,
\begin{eqnarray}
\vec{\alpha}_s = \frac{\ii e_s\vec{v}_s}{\omega},
\qquad
\matr{\wp}_s = 
\frac{e_s^2}{\omega^2}\,\matr{\mu}_s^{-1}.
\end{eqnarray}
The average Wigner matrix of \m{\osc{\vec{E}}} is
\begin{equation}
\matrU(t, \vec{x}, \omega, \vec{k}) 
= \int \frac{\dd\tau}{2\upi}\,\frac{\dd\vec{s}}{(2\upi)^{3}}\,\, 
\avr{
\macro{\osc{\vec{E}}}(t + \tau/2, \vec{x} + \vec{s}/2)\,
\smash{\macro{\osc{\vec{E}}}}^\dag(t - \tau/2, \vec{x} - \vec{s}/2)
}
\,\ee^{\ii\omega \tau - \ii \vec{k} \cdot \vec{s}}
\label{eq:EMdewU}
\end{equation}
(the arguments \m{t} and \m{\vec{x}} are henceforth omitted), and the nonlinear potentials are
\begin{align}
\matrst{D}_s
& = \upi e_s^2 \int \dd\vec{k}\,\vec{k}\vec{k}\,\frac{\vec{v}_s^\dag \matrU(\vec{k} \cdot \vec{v}_s, \vec{k})\vec{v}_s}{(\vec{k} \cdot \vec{v}_s)^2},
\label{eq:emD1}\\
\matr{\Uptheta}_s
& = e_s^2\,\frac{\pd}{\pd \vartheta} \fint \dd\omega\,\dd\vec{k}\,
 \left.
\frac{\vec{k}\vec{k}}{\omega^2}
\frac{(\vec{v}_s^\dag \matrU \vec{v}_s)}{\omega - \vec{k} \cdot \vec{v}_s + \vartheta}
\right|_{\vartheta=0},
\label{eq:EMUes}
\\
\Delta_s 
& = \frac{e_s^2}{2}\frac{\pd}{\pd\vec{p}} \cdot 
\fint \dd\omega\,\dd\vec{k}\,\frac{\vec{k}}{\omega^2}\frac{(\vec{v}_s^\dag \matrU \vec{v}_s)}{\omega - \vec{k} \cdot \vec{v}_s}
+ \frac{e_s^2}{2} \int \dd\omega\,\dd\vec{k}\,\frac{\tr(\matrU \matr{\mu}_s^{-1})}{\omega^2}.
\label{eq:Deltaes1}
\end{align}
When plasma is nonrelativistic and the field is electrostatic (so \m{\matrU = \vec{k}\vec{k}^\dag \st{U}_\varphi}, where \m{\st{U}_\varphi} is scalar), \eq{eq:emD1} gives the same \m{\matrst{D}_s} as \eq{eq:Des1} and \eq{eq:Deltaes1} gives the same \m{\Delta_s} as \eq{eq:Phies}. For \m{\matr{\Uptheta}_s}, the equivalence between \eq{eq:EMUes} and \eq{eq:Ues} should not be expected because \m{\matr{\Uptheta}_s} is a part of a distribution function, which is not gauge-invariant. (Canonical momenta in the Weyl gauge are different from those in the electrostatic gauge.) But it is precisely the dressing function \eq{eq:EMUes} that leads to the correct expressions for the momentum and energy stored in the OC distribution (\Sec{sec:onshellem}).

The Lagrangian density of a free electromagnetic field is
\begin{eqnarray}
\mcc{L}_0 = \frac{\smash{\osc{\vec{E}}}^\dag\osc{\vec{E}} - \smash{\osc{\vec{B}}}^\dag\osc{\vec{B}}}{8\upi}.
\end{eqnarray}
From Faraday's law, one has \m{\osc{\vec{B}} = \oper{\omega}^{-1}c(\boper{k} \times \vec{E})}.\footnote{Here, the oscillating field \m{\osc{\vec{\Psi}} = \osc{\vec{E}}} has the same dimension as \m{\vec{x}}, so the standard vector notation (including the dot product and the cross product) is naturally extended to \m{\osc{\vec{\Psi}}}.} Then, \m{- \smash{\osc{\vec{B}}}^\dag\osc{\vec{B}}/c^2} can be represented as follows (up to a divergence, which is insignificant):
\begin{align}
(\osc{\vec{E}} \times \oper{\omega}^{-1}\boper{k}) \cdot (\oper{\omega}^{-1}\boper{k} \times \osc{\vec{E}})
& = \osc{\vec{E}} \cdot \oper{\omega}^{-2}(\boper{k} \times (\boper{k} \times \osc{\vec{E}}))
\notag\\
& = \smash{\osc{\vec{E}}}^\dag \oper{\omega}^{-2}(\boper{k}(\boper{k} \cdot \osc{\vec{E}}) - \osc{\vec{E}} \oper{k}^2)
\notag\\
& = \smash{\osc{\vec{E}}}^\dag \oper{\omega}^{-2}(\boper{k}\smash{\boper{k}}^\dag - \matr{1}\oper{k}^2)\osc{\vec{E}}.
\label{eq:BB}
\end{align}
Then, the vacuum dispersion operator can be written as (cf.\ \eq{eq:Lagr0})
\begin{eqnarray}
\boper{\Xi}_0(\omega, \vec{k}) = \frac{1}{4\upi}\,\left(\matr{1} + c^2 \oper{\omega}^{-2}\big(\boper{k}\smash{\boper{k}}^\dag - \matr{1}\oper{k}^2\big)\right).
\end{eqnarray}
The total dispersion matrix is readily found to be
\begin{eqnarray}\label{eq:Xiem}
\matr{\Xi}(\omega, \vec{k}) = \frac{1}{4\upi}\left(\matr{\epsilon}(\omega, \vec{k}) + \frac{c^2}{\omega^2}\,(\vec{k}\smash{\vec{k}}^\dag - \matr{1}k^2)\right),
\end{eqnarray}
where \m{\matr{\epsilon}} (not to be confused with the small parameter \m{\epsilon} that we introduced earlier) is the dielectric tensor:
\begin{eqnarray}
\matr{\epsilon}(\omega, \vec{k}) = \matr{1} - \frac{\matr{\mcc{w}}_p}{\omega^2} + \sum_s \frac{4\upi e_s^2}{\omega^2}\int \dd\vec{p}\,
\frac{\vec{v}_s \vec{v}_s^\dag}{\omega - \vec{k} \cdot \vec{v}_s + \ii 0}\,\vec{k} \cdot \frac{\pd F_s}{\pd\vec{p}}.
\label{eq:veceps}
\end{eqnarray}
Here, \m{\matr{\mcc{w}}_p} is the squared relativistic plasma frequency, which is a matrix, because the `masses' \m{\vec{\mu}_s} are matrices:
\begin{eqnarray}
\matr{\mcc{w}}_p \doteq \sum_s 4 \upi e_s^2 \int \dd\vec{p}\,F_s \vec{\mu}_s^{-1}.
\end{eqnarray}

\subsubsection{Collisions and fluctuations}
\label{sec:emmicro}

By \eq{eq:mcWsp}, the spectrum of microscopic oscillations of \m{\osc{\vec{E}}} is a matrix given by
\begin{equation}
\matrst{S}(\omega, \vec{k})
= 2\upi\, \sum_s \left(\frac{4\pi e_s}{\omega}\right)^2 \int \dd\vec{p}\,\delta(\omega - \vec{k} \cdot \vec{v}_{s})F_{s}(\vec{p})
\matr{\epsilon}^{-1}(\omega, \vec{k})\,
\vec{v}_{s}\vec{v}_{s}^\dag\,
\matr{\epsilon}^{-\dag}(\omega, \vec{k}).
\label{eq:Sem}
\end{equation}
In the electrostatic limit, one can replace \m{\matr{\epsilon}^{-1}} with \m{\epsilon_\parallel^{-1}\vec{k}\vec{k}^\dag/k^2}, where \m{\epsilon_\parallel} is the relativistic generalization of \eq{eq:epspar}; then \eq{eq:Sem} leads to \eq{eq:Ses} as a particular case. For thermal equilibrium, one can also use \eq{eq:ufdt1} and the following form of \m{\matr{\epsilon}^{-1}} for isotropic plasma:
\begin{equation}
\matr{\epsilon}^{-1} = \frac{1}{\epsilon_\perp}\bigg(\matr{1} - \frac{\vec{k}\vec{k}^\dag}{k^2}\bigg) 
+ \frac{1}{\epsilon_\parallel}\frac{\vec{k}\vec{k}^\dag}{k^2},
\end{equation}
where \m{\epsilon_\perp} is the (scalar) transverse permittivity. Also, \eq{eq:mcYsymb} leads to
\begin{eqnarray}
|\mc{X}_{ss'}(\omega, \vec{k}; \vec{p}, \vec{p}')|^2 \approx \left(\frac{4\upi e_s e_{s'}}{\omega^2}\right)^2
|\vec{v}_s^\dag \matr{\epsilon}^{-1}(\omega, \vec{k}) \vec{v}'_{s'}|^2.
\end{eqnarray}
Then the collision operator \eq{eq:cop2} is obtained in the form
\begin{align}
\collision_s
= \frac{\pd}{\pd \vec{p}}\cdot
\sum_{s'} 2 e_s^2 e_{s'}^2
\int & \frac{\dd\vec{k}}{(2\upi)^3}\,\dd\vec{p}'\,
\frac{|\vec{v}_s^\dag \matr{\epsilon}^{-1}(\vec{k} \cdot \vec{v}_s, \vec{k}) \vec{v}'_{s'}|^2}{(\vec{k} \cdot \vec{v}_s)^4}\,
\delta(\vec{k} \cdot \vec{v}_s - \vec{k} \cdot \vec{v}'_{s'})
\notag\\
& \times \vec{k}\vec{k} \cdot
\left(
\frac{\pd F_s(\vec{p})}{\pd \vec{p}}\,F_{s'}(\vec{p}')
- F_s(\vec{p})\,\frac{\pd F_{s'}(\vec{p}')}{\pd \vec{p}'}
\right),
\end{align}
which is in agreement with \citep{ref:hizanidis83, ref:silin61}. Replacing \m{\matr{\epsilon}^{-1}} with \m{\epsilon_\parallel^{-1}\vec{k}\vec{k}^\dag/k^2} leads to the standard Balescu--Lenard operator \eq{eq:BLen} as a particular case.

\subsubsection{On-shell waves}
\label{sec:onshellem}

Electromagnetic on-shell waves satisfy
\begin{eqnarray}\label{eq:emdr}
\left(\matr{\epsilon}_\herm(w(\vec{k}), \vec{k}) + \frac{c^2}{w(\vec{k})^2}\,(\vec{k}\smash{\vec{k}}^\dag - \matr{1}k^2)\right)
\env{\vec{E}} = 0,
\end{eqnarray}
where \m{\env{\vec{E}}} is the complex envelope vector parallel to the polarization vector \m{\vec{\eta}}; also,
\begin{eqnarray}
\matr{\epsilon}_\herm(\omega, \vec{k}) = \matr{1} - \frac{\matr{\mcc{w}}_p}{\omega^2} + \sum_s \frac{4\upi e_s^2}{\omega^2}\fint \dd\vec{p}\,
\frac{\vec{v}_s \vec{v}_s^\dag}{\omega - \vec{k} \cdot \vec{v}_s}\,\vec{k} \cdot \frac{\pd F_s}{\pd\vec{p}}.
\end{eqnarray}
This yields (see \eq{eq:BB})
\begin{eqnarray}\label{eq:emdr2}
\smash{\env{\vec{E}}}^\dag\matr{\epsilon}_\herm(w(\vec{k}), \vec{k})\env{\vec{E}}
=- \frac{c^2}{w(\vec{k})^2}\,\smash{\env{\vec{E}}}^\dag(\vec{k}\smash{\vec{k}}^\dag - \matr{1}k^2)\env{\vec{E}}
= \smash{\env{\vec{B}}}^\dag \env{\vec{B}}.
\end{eqnarray}
Then, the phase-space density of the wave action \eq{eq:Jinc} can be cast in the form
\begin{eqnarray}
J(\vec{k}) 
= h(\vec{k})\vec{\eta}^\dag\,\frac{\pd \matr{\Xi}_\herm(\omega, \vec{k})}{\pd \omega}\,\vec{\eta}\,\Big|_{\omega = w(\avr{\vec{k}})}
= \frac{h(\vec{k})}{4\upi\omega^2}\,\vec{\eta}^\dag\,\frac{\pd(\omega^2\matr{\epsilon}_{\herm}(\omega, \vec{k}))}{\pd \omega}\,\vec{\eta}\,\Big|_{\omega = w(\avr{\vec{k}})}
\label{eq:Jem}
\end{eqnarray}
(cf.\ \citep{my:quasiop1}), and the dressing function \eq{eq:EMUes} is given by
\begin{align}
\matr{\Uptheta}_s
& = e_s^2\,\frac{\pd}{\pd \vartheta} \fint \dd\vec{k}\,(h(\vec{k}) + h(-\vec{k}))
 \left.
\frac{\vec{k}\vec{k}}{w^2(\vec{k})}
\frac{(\vec{v}_s^\dag \vec{\eta} \vec{\eta}^\dag \vec{v}_s)}{w(\vec{k}) - \vec{k} \cdot \vec{v}_s + \vartheta}
\right|_{\vartheta=0}
\notag\\
& = 2e_s^2\,\frac{\pd}{\pd \vartheta} \fint \dd\vec{k}\,h(\vec{k})
 \left.
\frac{\vec{k}\vec{k}}{w^2(\vec{k})}
\frac{(\vec{\eta}^\dag \vec{v}_s \vec{v}_s^\dag \vec{\eta})}{w(\vec{k}) - \vec{k} \cdot \vec{v}_s + \vartheta}
\right|_{\vartheta=0}.
\label{eq:Thetem2}
\end{align}
Using these, one obtains (\App{sec:momem})
\begin{eqnarray}\label{eq:totmem}
\sum_s \int \dd\vec{p}\,\vec{p} F_s + \int \dd\vec{k}\,\vec{k} J 
= \vec{\mc{P}}^{(\text{kin})}
+ \frac{\avr{\osc{\vec{E}} \times \osc{\vec{B}}}}{4\upi c},
\label{eq:poy}
\end{eqnarray}
where \m{\vec{\mc{P}}^{(\text{kin})}} is the average density of the plasma \textit{kinetic} (up to \m{\avr{\vec{A}}}) momentum,
\begin{eqnarray}\label{eq:Pkin}
\vec{\mc{P}}^{(\text{kin})} \doteq \sum_s \int \dd\vec{p}\,\avr{(\vec{p} - e_s \osc{\vec{A}}/c)f_s}
= \sum_s \int \dd\vec{p}\,\vec{p}\,\avr{f_s^{\text{(kin)}}},
\end{eqnarray}
the functions \m{f_s^{\text{(kin)}}(\vec{p}) \doteq f_s(\vec{p} + e_s \osc{\vec{A}}/c)} are the distributions of kinetic (up to \m{\avr{\vec{A}}}) momenta, and the second term in \eq{eq:poy} is the well-known average momentum of electromagnetic field. Similarly (\App{sec:enem}),
\begin{eqnarray}\label{eq:toteem}
\sum_s \int \dd\vec{p}\,H_{0s} F_s + \int \dd\vec{k}\,w J 
= 
\mc{K}^{(\text{kin})}
+ \frac{1}{8\upi}\,\avr{\big(
\smash{\osc{\vec{E}}}^\dag \osc{\vec{E}}
+ \smash{\osc{\vec{B}}}^\dag \osc{\vec{B}}
\big)},
\end{eqnarray}
where \m{\mc{K}^{(\text{kin})}} is given by
\begin{eqnarray}\label{eq:Kkin}
\mc{K}^{(\text{kin})} \doteq \sum_s \int \dd\vec{p}\,H_{0s}\,\avr{f_s^{(\text{kin})}}.
\end{eqnarray}
In other words, the total momentum and energy of the system in the OC--wave representation are the same as those in the original particle--field variables.

\subsubsection{Eikonal waves}
\label{sec:eikwem}

As a particular case, let us consider an eikonal wave
\begin{eqnarray}
\label{eq:eikem}
\macro{\osc{\vec{E}}} \approx \re(\ee^{\ii \theta}\env{\vec{E}}),
\qquad
\avr{\omega}\doteq -\pd_t\theta,
\qquad
\avr{\vec{k}}\doteq \pd_{\vec{x}} \theta,
\end{eqnarray}
which may or may not be on-shell. Then, \eq{eq:EMUes} and \eq{eq:Deltaes1} lead to (cf.\ \Sec{sec:eikw})
\begin{eqnarray}
& \displaystyle
\matr{\Uptheta}_s
= -
\frac{\avr{\vec{k}}\,\avr{\vec{k}}}{\avr{\omega}^2}
\frac{e_s^2 |\vec{v}_s^\dag \env{\vec{E}}|^2}{2(\avr{\omega} - \avr{\vec{k}} \cdot \vec{v}_s)^2},
\\
& \displaystyle
\Delta_s 
= \frac{e_s^2\smash{\env{\vec{E}}}^\dag\matr{\mu}_s^{-1}\env{\vec{E}}}{4\avr{\omega}^2}
+ 
\frac{e_s^2\avr{\vec{k}}}{4\avr{\omega}^2} \cdot\frac{\pd}{\pd\vec{p}}
\bigg(\frac{|\vec{v}_s^\dag \env{\vec{E}}|^2}{\avr{\omega} - \avr{\vec{k}} \cdot \vec{v}_s}\bigg).
\end{eqnarray}
For on-shell waves in particular, one also obtains the action density in the form
\begin{align}
\mc{I} & = \frac{1}{16\upi\omega^2}\,
\smash{\env{\vec{E}}}^\dag\,
\frac{\pd(\omega^2\matr{\epsilon}_{\herm}(\omega, \vec{k}))}{\pd \omega}\,
\env{\vec{E}}\,\Big|_{\omega = w(\avr{\vec{k}})}
\notag\\
& = \frac{1}{16\upi\omega}\left(
\smash{\env{\vec{E}}}^\dag\,
\frac{\pd(\omega \matr{\epsilon}_{\herm}(\omega, \vec{k}))}{\pd \omega}\,
\env{\vec{E}}
+ \smash{\env{\vec{B}}}^\dag
\env{\vec{B}}
\right)\Big|_{\omega = w(\avr{\vec{k}})},
\end{align}
where we used \eq{eq:emdr2}.

\subsection{Newtonian gravity}
\label{sec:grav}

For Newtonian interactions governed by a gravitostatic potential \m{\varphi_g}, one has
\begin{eqnarray}\label{eq:hamGS}
H_s = \frac{p^2}{2m_s} + m_s \avr{\varphi}_g +  m_s\osc{\varphi}_g,
\qquad
\mcc{L}_0 = - \frac{(\del \tilde{\varphi}_g)^2}{8\upi \msf{G}},
\end{eqnarray}
where \m{\msf{G}} is the gravitational constant. This system is identical to that considered in \Sec{sec:lang} for nonrelativistic electrostatic interactions up to coefficients. Specifically, \m{e_s} are replaced with \m{m_s}, a factor \m{-\msf{G}^{-1}} appears in \m{\matr{\Xi}_0}, and the dispersion matrix becomes
\begin{eqnarray}
\matr{\Xi}(\omega, \vec{k}) = \Xi(\omega, \vec{k}) = -\frac{k^2\epsilon_g(\omega, \vec{k})}{4\upi \msf{G}}.
\end{eqnarray}
Thus, \m{\epsilon_\parallel} is replaced with \m{-\epsilon_g/\msf{G}}, where \m{\epsilon_g} is the gravitostatic permittivity given by
\begin{eqnarray}
\epsilon_g(\omega, \vec{k}) =
1 - \sum_s \frac{4\upi \msf{G} m_s^2}{k^2} \int \dd\vec{p}\,\frac{\vec{k}}{\omega - \vec{k} \cdot \vec{v}_s + \ii 0} 
\cdot \frac{\pd F_s}{\pd \vec{p}}.
\label{eq:epsg}
\end{eqnarray}
This readily yields, for example, kinetic theory of the Jeans instability \citep{ref:trigger04}, whose dispersion relation is given by \m{\epsilon_g(\omega, \vec{k}) = 0} (modulo the usual analytic continuation of the permittivity to modes with \m{\im\omega < 0}).

\subsection{Relativistic gravity}
\label{sec:gw}

\subsubsection{Main equations}
\label{sec:gwmain}

The dynamics of a relativistic neutral particle with mass \m{m} in a spacetime metric \m{g_{\alpha\beta}} with signature \m{(-+++)} is governed by a covariant Hamiltonian (see, for example, \citep{my:gwponder})
\begin{eqnarray}
\msf{H}(\vecst{x}, \vecst{p}) 
 = \frac{1}{2m}\left(m^2 + g^{\alpha\beta}(\vecst{x}) p_\alpha p_\beta\right)
\equiv \msf{H}(\matr{g}, \vecst{p}).
\end{eqnarray}
Here, \m{\vecst{x} \equiv (x^0, \vec{x})}, and \m{x^0 = t}, as usual. Also, \m{\vecst{p} \equiv (p_0, \vec{p})} is the index-free notation for the four-momentum~\m{p_\alpha}, \m{g^{\alpha\beta}} is the inverse metric, \m{\matr{g}} is the index-free notation for \m{g^{\alpha\beta}}, the units are such that \m{c = 1}, and the species index is omitted.\footnote{This section uses notation different from that used in the rest of the paper. In particular, the Greek indices span from 0 to 3, and the standard rules of index manipulations apply.} The corresponding Hamilton's equations, with \m{\tau} the proper time, are
\begin{eqnarray}\label{eq:gwheq}
\frac{\dd x^\alpha}{\dd \tau} = \frac{\pd \msf{H}}{\pd p_\alpha},
\qquad
\frac{\dd p_\alpha}{\dd \tau} = -\frac{\pd \msf{H}}{\pd x^\alpha}.
\end{eqnarray}
This dynamics is constrained to the shell \m{p_0 = P_0(t, \vec{x}, \vec{p})}, where \m{P_0} is the (negative) solution of
\begin{eqnarray}\label{eq:Hnce}
\msf{H}(\matr{g}, P_0(t, \vec{x}, \vec{p}), \vec{p}) = 0.
\end{eqnarray}
This means that the particle distribution in the \m{(\vecst{x}, \vecst{p})} space is delta-shaped and thus does not satisfy \eq{eq:asmff}. Hence, we will consider particles in the six-dimensional space \m{(\vec{x}, \vec{p})} instead. The corresponding dynamics is governed by the Hamiltonian
\begin{eqnarray}\label{eq:Hnc}
H = - P_0(t, \vec{x}, \vec{p}).
\end{eqnarray}
This is seen from the fact that
\begin{eqnarray}
\frac{\pd H}{\pd \placeholder} = - \frac{\pd P_0}{\pd \placeholder} = \frac{\pd \msf{H}/\pd \placeholder}{\pd \msf{H}/\pd p_0},
\end{eqnarray}
where \m{\placeholder \in \cbb{t, \vec{x}, \vec{p}}}, so Hamilton's equations obtained from \eq{eq:Hnc} are equivalent to \eq{eq:gwheq}:
\begin{eqnarray}\label{eq:heqnc}
\frac{\dd x^\alpha}{\dd t} = \frac{\pd H}{\pd p_\alpha} = \frac{\pd \msf{H}/\pd p_\alpha}{\pd \msf{H}/\pd p_0},
\qquad
\frac{\dd p_\alpha}{\dd t} = - \frac{\pd H}{\pd x^\alpha} = \frac{\pd \msf{H}/\pd x^\alpha}{\pd \msf{H}/\pd p_0}.
\end{eqnarray}

Let us decompose the metric into the average part and oscillations, \m{g_{\alpha\beta} = \avr{g}_{\alpha\beta} + \osc{g}_{\alpha\beta}}, and approximate the inverse metric to the second order in \m{\osc{\matr{g}}}:
\begin{eqnarray}
g^{\alpha\beta} \approx \avr{g}^{\alpha\beta} - \osc{g}^{\alpha\beta} + \osc{g}^{\alpha\gamma}\avr{g}_{\gamma\delta}\osc{g}^{\delta\beta}, 
\end{eqnarray}
where the indices of \m{\osc{\matr{g}}} are manipulated using the background metric \m{\avr{g}_{\alpha\beta}}. This gives
\begin{eqnarray}\label{eq:Hnc3}
\msf{H}
 = 
\frac{1}{2m}\left(m^2 +
\avr{g}^{\alpha\beta}p_\alpha p_\beta 
- \osc{g}^{\alpha\beta}p_\alpha p_\beta 
+ \osc{g}^{\alpha\beta}\avr{g}_{\beta\gamma}\osc{g}^{\gamma\delta}p_\alpha p_\delta\right).
\end{eqnarray}
The Hamiltonian \eq{eq:Hnc} is expanded in \m{\osc{\matr{g}}} as follows:
\begin{eqnarray}\label{eq:Hnc2}
H(\matr{g}, \vecst{p}) \approx
- P_0
- \frac{\pd P_0}{\pd \osc{g}^{\alpha\beta}}\,\osc{g}^{\alpha\beta}
- \frac{1}{2}\,\frac{\pd^2P_0}{\pd \osc{g}^{\alpha\beta} \pd \osc{g}^{\gamma\delta}}\,\osc{g}^{\alpha\beta}\osc{g}^{\gamma\delta},
\end{eqnarray}
where \m{P_0} and the derivatives on the right-hand side are evaluated on \m{(\avr{\matr{g}}, \vecst{p})}. To calculate these derivatives, let us differentiate \eq{eq:Hnce} and use \eq{eq:Hnc3} for \m{\msf{H}}. This gives
\begin{eqnarray}
0 = \frac{\pd\msf{H}}{\pd \osc{g}^{\alpha\beta}}
+ \frac{\pd\msf{H}}{\pd p_0} \frac{\pd P_0}{\pd \osc{g}^{\alpha\beta}}
= \frac{1}{2m}\left(-p_\alpha p_\beta + 2P^0 \frac{\pd P_0}{\pd \osc{g}^{\alpha\beta}}\right),
\end{eqnarray}
where the derivatives with respect to the oscillating metric are taken at fixed \m{p_\alpha} and at \m{\osc{\matr{g}} \to 0}, and \m{P^0 \equiv P^0(\matr{g}, \vecst{p}) = \avr{g}^{0\alpha}p_\alpha}; thus,
\begin{eqnarray}
\frac{\pd P_0}{\pd \osc{g}^{\alpha\beta}} = \frac{p_\alpha p_\beta}{2 P^0}.
\end{eqnarray}
Similarly, differentiating \eq{eq:Hnce} twice gives
\begin{align}
0 & = \frac{\pd^2\msf{H}}{\pd \osc{g}^{\alpha\beta}\pd \osc{g}^{\gamma\delta}}
+ \frac{\pd\msf{H}}{\pd p_0}\frac{\pd^2P_0}{\pd \osc{g}^{\alpha\beta}\pd \osc{g}^{\gamma\delta}}
+ \frac{\pd P_0}{\pd \osc{g}^{\alpha\beta}}\frac{\pd}{\pd p_0} \frac{\pd \msf{H}}{\pd \osc{g}^{\gamma\delta}}
+ \frac{\pd}{\pd p_0}\left(
\frac{\pd\msf{H}}{\pd \osc{g}^{\alpha\beta}}
+ \frac{\pd\msf{H}}{\pd p_0}\frac{\pd P_0}{\pd \osc{g}^{\alpha\beta}}
\right)\frac{\pd P_0}{\pd g^{\gamma\delta}}
\notag\\
& = \frac{\pd^2\msf{H}}{\pd \osc{g}^{\alpha\beta}\pd \osc{g}^{\gamma\delta}}
+ \frac{\pd\msf{H}}{\pd p_0} \frac{\pd^2P_0}{\pd \osc{g}^{\alpha\beta}\pd \osc{g}^{\gamma\delta}}
+ \frac{\pd P_0}{\pd \osc{g}^{\alpha\beta}}\frac{\pd}{\pd p_0}\frac{\pd \msf{H}}{\pd \osc{g}^{\gamma\delta}}
+ \frac{\pd P_0}{\pd \osc{g}^{\gamma\delta}}\frac{\pd}{\pd p_0} \frac{\pd\msf{H}}{\pd \osc{g}^{\alpha\beta}}
+ \frac{\pd^2\msf{H}}{\pd p_0\pd p_0}\frac{\pd P_0}{\pd \osc{g}^{\alpha\beta}}\frac{\pd P_0}{\pd \osc{g}^{\gamma\delta}}
\notag\\
& = \frac{1}{2m}\left(\avr{g}_{\beta\gamma}p_\alpha p_\delta + \avr{g}_{\delta\alpha} p_\gamma p_\beta
+ 2P^0 \frac{\pd^2P_0}{\pd \osc{g}^{\alpha\beta}\pd \osc{g}^{\gamma\delta}}
- \frac{1}{2P^0}\frac{\pd(p_\alpha p_\beta p_\gamma p_\delta)}{\pd p_0} 
+ \avr{g}^{00}\,\frac{p_\alpha p_\beta p_\gamma p_\delta}{2(P^0)^2}\right),
\notag
\end{align}
whence
\begin{eqnarray}
\frac{\pd^2P_0}{\pd \osc{g}^{\alpha\beta}\pd \osc{g}^{\gamma\delta}}
=
- \frac{1}{2P^0}\,(\avr{g}_{\beta\gamma}p_\alpha p_\delta + \avr{g}_{\delta\alpha} p_\gamma p_\beta)
+ \frac{1}{4(P^0)^2}\frac{\pd(p_\alpha p_\beta p_\gamma p_\delta)}{\pd p_0} 
- \avr{g}^{00}\,\frac{p_\alpha p_\beta p_\gamma p_\delta}{4(P^0)^3}.
\notag
\end{eqnarray}
Then, \eq{eq:Hnc2} yields
\begin{eqnarray}
H \approx
H_0 + \alpha_{\alpha\beta}\,\osc{g}^{\alpha\beta}
+ \frac{1}{2}\,\osc{g}_{\alpha\beta} \wp^{\alpha\beta}{}_{\gamma\delta} \osc{g}^{\gamma\delta},
\label{eq:hamGW}
\end{eqnarray}
where we introduced \m{H_0 = - P_0}, \m{\alpha^{\alpha\beta} = p^\alpha p^\beta/(2P^0)}, and
\begin{eqnarray}
\wp^{\alpha\beta}{}_{\gamma\delta} = \frac{\delta^{\beta}_{\gamma} p^\alpha p_\delta + \delta_{\delta}^{\alpha} p^\beta p_\gamma}{2P^0}
- \frac{1}{4(P^0)^2}\frac{\pd(p^\alpha p^\beta p_\gamma p_\delta)}{\pd p_0} 
+ \avr{g}^{00}\,\frac{p^\alpha p^\beta p_\gamma p_\delta}{4(P^0)^3}.
\end{eqnarray}

\subsubsection{Nonlinear potentials}

Let us treat \m{\osc{\matr{g}}} as a 16-dimensional vector \citep{foot:mygwinv}, so \m{\alpha_{\alpha\beta}} serves as \m{\matr{\alpha}^\dag} and \m{\wp^{\alpha\beta}{}_{\gamma\delta}} serves as \m{\matr{\wp}}. (Because these operators happen to be local in the \m{\vecst{x}}~representation, here we do not distinguish them from their symbols.) Let us also introduce
\begin{eqnarray}
\mcc{E} \doteq p_\alpha p_\beta p_\gamma p_\delta \st{U}^{\alpha\beta\gamma\delta}
\end{eqnarray}
and notice that \m{v^i \approx \dot{x}^i = p^i/p^0} (see \eq{eq:heqnc}), so \m{\omega - \vec{k} \cdot \vec{v} = - k_\rho p^\rho/P^0} and \(\smash{\delta(\omega - \vec{k} \cdot \vec{v})} = \smash{P^0\delta(k^\rho p_\rho)}\). Then, one finds from \eq{eq:trfW} that (\App{app:gw})
\begin{align}
\matrst{D}
& = \frac{\upi}{4P^0} \int \dd\vecst{k}\,\vec{k}\vec{k} \mcc{E}\,\delta(k^\rho p_\rho),
\label{eq:Dgw}
\\
\matr{\Uptheta} 
& = \frac{1}{4P^0}\frac{\pd}{\pd \vartheta} \fint \dd\vecst{k}
\left.
\frac{\vec{k}\vec{k}\mcc{E}}{\vartheta P^0 - k^\rho p_\rho}
\right|_{\vartheta=0},
\\
\Delta & = 
\frac{p_\alpha p_\beta}{2P^0}\int \dd\vecst{k}\,\st{U}^\alpha{}_\gamma{}^{\gamma\beta}
- \frac{1}{8P^0}
\frac{\pd}{\pd p_\lambda}
\fint \dd\vecst{k}\,
\frac{k_\lambda\mcc{E}}{k^\rho p_\rho}.
\label{eq:deltadw}
\end{align}
Equation \eq{eq:deltadw} (where one takes \m{p_0 = P_0} \textit{after} the differentiation) is in agreement with the result that was obtained for quasimonochromatic waves in \citep{my:gwponder}. The derivation of the dispersion matrix \m{\matr{\Xi}} for relativistic gravitational interactions in matter is cumbersome, so it is not presented here, but see \citep{foot:mygweq}. The collision integral and fluctuations for relativistic gravitational interactions are straightforward to obtain from the general formulas presented in \Secs{sec:sumth} and~\ref{sec:Boltz}. This can be used to describe QL interactions of gravitational waves, including not only the usual vacuum modes\footnote{Vacuum gravitational waves satisfy \m{\omega^2 = |\vec{k}|^2}. Hence, satisfying the resonance condition \m{k^\rho p_\rho = 0} requires \m{|\vec{k} \cdot \vec{v}| = |\vec{k}|}, which requires particle speeds not smaller than the speed of light (remember that \m{c = 1} in our units). For massive particles, this cannot be satisfied, so \m{\matrst{D}} vanishes for vacuum gravitational waves. However, such waves can still produce adiabatic ponderomotive effects determined by \m{\Delta} \citep{my:gwponder}.} but also waves coupled with matter, for example, the relativistic Jeans mode.

Also notice that the OC Hamiltonian \m{\mc{H} = H_0 + \Delta} can be put in a covariant form as follows. Like in the original system (\Sec{sec:gwmain}), \m{\mc{H}} determines the ponderomotively modified shell \m{p_0 = \mc{P}_0(t, \vec{x}, \vec{p})} via \m{\mc{H} = -\mc{P}_0}. On one hand, the covariant OC Hamiltonian \m{\mc{H}'} vanishes on this shell,\footnote{The covariant Hamiltonian is the dispersion function of particles as quantum waves in the semiclassical limit \citep{my:gwponder}.} so it can be Taylor-expanded as follows:
\begin{eqnarray}
\mc{H}' \approx (p_0 - \mc{P}_0)\,\frac{\pd \mc{H}}{\pd p_0}\bigg|_{p = \mc{P}_0} 
\approx 
(p_0 - P_0 + \Delta)\lambda,
\qquad
\lambda \doteq \frac{\pd \msf{H}}{\pd p_0}\bigg|_{p = P_0}.
\label{eq:sh1}
\end{eqnarray}
On the other hand, it can also be represented as \m{\mc{H}' = \msf{H}(\avr{\matr{g}}, \vecst{p}) + \Delta'} (here \m{\Delta'} is the ponderomotive term yet to be found) and Taylor-expanded around the unperturbed shell \m{p_0 = P_0(t, \vec{x}, \vec{p})} as
\begin{eqnarray}
\mc{H}' \approx \Delta' + (p_0 - P_0)\lambda = (p_0 - P_0 + \Delta'/\lambda)\lambda.
\label{eq:sh2}
\end{eqnarray}
By comparing \eq{eq:sh1} with \eq{eq:sh2}, one finds that \m{\Delta' = \lambda\Delta}. Because \m{\lambda = P^0/m}, this leads to the following covariant Hamiltonian for OCs:
\begin{eqnarray}
& \displaystyle
\mc{H}' = \frac{1}{2m}\left(m^2 + 
g_{\rm eff}^{\alpha\beta} p_\alpha p_\beta
- \frac{1}{4}
\frac{\pd}{\pd p_\lambda}
\fint \dd\vecst{k}\,
\frac{k_\lambda\mcc{E}}{k^\rho p_\rho}
\right),
\label{eq:hprime}
\\
& \displaystyle
g_{\rm eff}^{\alpha\beta} \doteq \avr{g}^{\alpha\beta} + \int \dd\vecst{k}\,\st{U}^\alpha{}_\gamma{}^{\gamma\beta}.
\end{eqnarray}

\subsubsection{Gauge invariance}

As shown in \citep{foot:mygwquasi, foot:mygwinv} adiabatic QL interactions via gravitational waves (\ie those determined by \m{\matr{\Uptheta}} and \m{\Delta}) can be formulated in a form invariant with respect to gauge transformations
\begin{eqnarray}
\label{eq:gauge}
\osc{g}^{\alpha\beta} \to \osc{g}'^{\alpha\beta} = \osc{g}^{\alpha\beta} + \del^{(\alpha} \osc{\xi}^{\beta)},
\end{eqnarray}
where \m{\del} is the covariant derivative associated with the background metric \m{\avr{\matr{g}}}, \m{\osc{\xi}^\mu} is an arbitrary vector field, and \(\smash{\psi^{(\alpha} \eta^{\beta)}} \equiv \smash{(\psi^{\alpha} \eta^{\beta} + \psi^{\beta} \eta^{\alpha})/2}\). Let us show that this also extends to resonant interactions. Recall that within the assumed accuracy the nonlinear potentials are supposed to be calculated only to the zeroth order in the geometrical-optics parameter. Then, the modification of the average Wigner matrix of the metric oscillations under the transformation \eq{eq:gauge} can be written as
\begin{align}
\st{U}'^{\alpha\beta\gamma\delta} - \st{U}^{\alpha\beta\gamma\delta}
& = 
\wsymbx{\Big(
\ii
\oper{k}^{(\alpha}\,
\avr{\ket{\osc{\xi}^{\beta)}}\bra{\osc{g}^{\gamma\delta}}}
-\ii 
\avr{\ket{\osc{g}^{\alpha\beta}}\bra{\osc{\xi}^{(\gamma}}}\,
\oper{k}^{\delta)}
+ \oper{k}^{(\alpha}
\avr{\ket{\osc{\xi}^{\beta)}}\bra{\osc{\xi}^{(\gamma}}}
\,\oper{k}^{\delta)}
\Big)}
\notag\\
& =
\ii k^{(\alpha} \mc{W}^{\beta)\gamma\delta}
- \ii k^{(\delta} \mc{W}^{\gamma)\alpha\beta *}
+ k^{(\alpha} \msf{W}_{\osc{\xi}}^{\beta)(\gamma} k^{\delta)},
\end{align}
where \m{\mc{W}^{\beta\gamma\delta} \doteq \avr{\wsymbx{\ket{\osc{\xi}^{\beta}}\bra{\osc{g}^{\gamma\delta}}}}} and \m{\msf{W}_{\osc{\xi}}^{\beta\gamma}} is the average Wigner matrix of \m{\osc{\xi}^\alpha}. The corresponding change of \m{\mcc{E}} is
\begin{eqnarray}\notag
\mcc{E}' - \mcc{E} = (k^\rho p_\rho) \Big(
\ii p_\beta p_\gamma p_\delta\mc{W}^{\beta\gamma\delta}
- \ii p_\alpha p_\beta p_\gamma \mc{W}^{\gamma\alpha\beta *}
+ k^\lambda p_\lambda p_\beta p_\gamma \msf{W}_{\osc{\xi}}^{\beta\gamma}
\Big) \equiv (k^\rho p_\rho)A.
\end{eqnarray}
Then, the difference in the diffusion coefficients \eq{eq:Dgw} is
\begin{eqnarray}
\matrst{D}' - \matrst{D}
= \frac{\upi}{4P^0} \int \dd\vecst{k}\,\vec{k}\vec{k}\,\delta(k^\rho p_\rho)\,(k^\rho p_\rho)\,A = 0,
\end{eqnarray}
because \m{\delta(k^\rho p_\rho)\,(k^\rho p_\rho) = 0}. In particular, this rules out QL diffusion via `coordinate waves'.

\subsubsection{Lorenz gauge and effective metric}

Let us consider gravitational waves in the Lorenz gauge, \m{\del_\alpha \osc{g}^{\alpha\beta} = 0}. In this case,
\begin{eqnarray}
k_\alpha \st{U}^{\alpha\beta\gamma\delta}
= k_\beta \st{U}^{\alpha\beta\gamma\delta}
= k_\gamma \st{U}^{\alpha\beta\gamma\delta}
= k_\delta \st{U}^{\alpha\beta\gamma\delta} = 0,
\end{eqnarray}
and thus \m{k_\lambda \pd \mcc{E}/\pd p^\lambda = 0}. Then,
\begin{eqnarray}
\frac{\pd}{\pd p_\lambda}
\fint \dd\vecst{k}\,
\frac{k_\lambda\mcc{E}}{k^\rho p_\rho}
= \frac{\pd}{\pd \vartheta}\fint \dd\vecst{k}\left.\frac{(k_\lambda k^\lambda)\mcc{E}}{k^\rho p_\rho + \vartheta}
\right|_{\vartheta = 0}.
\label{eq:termd}
\end{eqnarray}
This simplifies the expression \eq{eq:deltadw} for \m{\Delta} and \eq{eq:hprime} for \m{\mc{H}'}. Furthermore, if the waves are not significantly affected by matter, so the vacuum dispersion \m{k_\lambda k^\lambda = 0} can be assumed, the term \eq{eq:termd} vanishes completely. Then, \eq{eq:hprime} becomes
\begin{eqnarray}
\mc{H}' = \frac{1}{2m}\left(m^2 + g_{\rm eff}^{\alpha\beta} p_\alpha p_\beta\right)
\end{eqnarray}
and QL diffusion disappears, because particles cannot resonate with waves. This shows that the only average QL effect of vacuum gravitational waves on particles is the modification of the spacetime metric by \m{\int \dd\vecst{k}\,\st{U}_{\alpha\gamma}{}^{\gamma}{}_{\beta} = \mc{O}(\varepsilon^2)}. For quasimonochromatic waves, this effect is discussed in further detail in \citep{my:gwponder}.

\section{Summary}
\label{sec:conclusions}

In summary, we have presented quasilinear theory for classical plasma interacting with inhomogeneous turbulence in the presence of background fields. Because we use the Weyl symbol calculus, global-mode decomposition is not invoked, so our formulation is local and avoids the usual issues with complex-frequency modes. Also, the particle Hamiltonian is kept general, so the results are equally applicable to relativistic, electromagnetic, and even non-electromagnetic (for example, gravitational) interactions. Because our approach is not bounded by the limitations of variational analysis either, effects caused by collisional and collisionless dissipation are also included naturally.

Our main results are summarized in \Secs{sec:Deq}, \ref{sec:sumth}, \ref{sec:sumshell}, \ref{sec:sumbol} and are as follows. Starting from the Klimontovich equation, we derive a Fokker--Planck model for the dressed oscillation-center distribution. This model captures quasilinear diffusion, interaction with the background fields, and ponderomotive effects simultaneously. The local diffusion coefficient is manifestly positive-semidefinite. Waves are allowed to be off-shell (not constrained by a dispersion relation), and a collision integral of the Balescu--Lenard type emerges in a form that is not restricted to any particular Hamiltonian. This operator conserves particles, momentum, and energy, and it also satisfies the \m{H}-theorem, as usual. As a spin-off, a general expression for the spectrum (average Wigner matrix) of microscopic fluctuations of the interaction field is derived. For on-shell waves, which satisfy a quasilinear wave-kinetic equation, our theory conserves the momentum and energy of the wave--plasma system. Dewar's oscillation-center quasilinear theory of electrostatic turbulence \citep{ref:dewar73} is proven formally as a particular case and given a concise formulation. Also discussed as examples are relativistic electromagnetic and gravitational interactions, and quasilinear theory for gravitational waves is proposed.

Aside from having the aesthetic appeal of a rigorous local theory, our formulation can help, for example, better understand and model quasilinear plasma heating and current drive. First of all, it systematically accounts for the wave-driven evolution of the nonresonant-particle distribution and for the ponderomotive effects caused by plasma inhomogeneity in both time and space. As discussed above (\Sec{sec:cons}), this is generally important for adequately calculating the energy--momentum transfer between waves and plasma even when resonant absorption \perse occurs in a homogeneous-plasma region. Second, our formulation provides general formulas that equally hold in any canonical variables and for any Hamiltonians that satisfy our basic assumptions (\Sec{sec:basm}). Therefore, our results can be applied to various plasma models immediately. This eliminates the need for \adhoc calculations, which can be especially cumbersome beyond the homogeneous-plasma approximation. Discussing specific models of applied interest, however exciting, is beyond the scope of this paper and is left to future work.

\section*{Funding}

This work was supported by the U.S.\ DOE through Contract DE-AC02-09CH11466. It is also based upon the work supported by National Science Foundation under the grant No.\ PHY~1903130. 

\section*{Declaration of interests}
The author reports no conflict of interest.

\appendix

\section{Average Wigner matrices}
\label{sec:avrwig}

\subsection{Positive semidefinitness}
\label{app:possemi}

As known from \citep{ref:cartwright76}, the average Wigner function of any scalar field on the real axis is non-negative if the averaging is done over a sufficiently large phase-space volume. Here, we extend this theorem to average Wigner matrices of vector fields in a multi-dimensional space, \m{\vec{\psi}(\vecst{x})}, and show that such matrices are positive-semidefinite.

For any given function \m{h(\vec{z}) \equiv h(\vecst{x}, \vecst{k})}, we define its local phase-space average as the following convolution integral:\footnote{This ensures that \m{\pd_\vec{z}\avr{h} = \avr{\pd_\vec{z}h}}, as readily seen from \eq{eq:Gh} using integration by parts.}
\begin{eqnarray}
\avr{h}(\vec{z}) \doteq 
\int \dd\vec{z}'\,
\mc{G}(\vec{z} - \vec{z}')\,h(\vec{z}')
\equiv
\int \dd\vecst{x}'\,\dd\vecst{k}'\,
\mc{G}(\vecst{x} - \vecst{x}', \vecst{k} - \vecst{k}')\,h(\vecst{x}', \vecst{k}')
\label{eq:Gh}
\end{eqnarray}
with a Gaussian window function
\begin{eqnarray}
\mc{G}(\vecst{x}, \vecst{k}) \doteq \frac{1}{(2\upi \sigmax \sigmak)^\stn}\,
\exp\left(-\frac{|\vecst{x}|^2}{2\sigmax^2} - \frac{|\vecst{k}|^2}{2\sigmak^2}\right)
\end{eqnarray}
and positive constants \m{\sigmax} and \m{\sigmak} yet to be specified. Unlike in \Sec{sec:bc}, the following notation will be assumed for the `scalar product' for variables with upper, lower, and mixed indices:
\begin{eqnarray}
& \vecst{x}' \cdot \vecst{x}'' \doteq \delta_{ij} \st{x}'^i \st{x}''^j,
\qquad
\vecst{k}' \cdot \vecst{k}'' \doteq \delta^{ij} \st{k}'_i \st{k}''_j,
\qquad
\vecst{k} \cdot \vecst{x} \doteq \st{k}_i \st{x}^i.
\end{eqnarray}
(The Latin indices in this appendix range from 0 to \m{n}, \m{\delta_{ij}} and \m{\delta^{ij}} are unit matrices, and summation over repeating indices is assumed.) In particular, note that \m{|\vecst{x}|^2 \doteq \vecst{x} \cdot \vecst{x} \ge 0} and must not be confused with the squared spacetime interval, which can have either sign. Likewise, \m{|\vecst{k}|^2 \doteq \vecst{k} \cdot \vecst{k} \ge 0} must not be confused with \m{\st{k}_i\st{k}^i = -\omega^2 + \vec{k}^2}.

The average Wigner matrix of any given vector field \m{\vec{\psi}} is given by
\begin{align}
\avr{\matrst{W}}_{\vec{\psi}}(\vecst{x}, \vecst{k}) 
= \frac{1}{(2\upi)^\stn}\frac{1}{(2\upi \sigmax \sigmak)^\stn}
 \int & \dd\vecst{s}\,\dd\vecst{x}'\,\dd\vecst{k}'\,
\vec{\psi}(\vecst{x}' + \vecst{s}/2) \vec{\psi}^\dag(\vecst{x}' - \vecst{s}/2)
\notag\\
& \times \exp\left(-\frac{|\vecst{x} - \vecst{x}'|^2}{2\sigmax^2} - \frac{|\vecst{k} - \vecst{k}'|^2}{2\sigmak^2}-\ii \vecst{k}' \cdot \vecst{s}\right).
\label{eq:aW}
\end{align}
The integral over \m{\vecst{k}'} can be readily taken:
\begin{eqnarray}
\int \dd\vecst{k}'\,\exp\left( - \frac{|\vecst{k} - \vecst{k}'|^2}{2\sigmak^2}-\ii \vecst{k}' \cdot \vecst{s}\right)
= (2\upi)^{\stn/2} \sigmak^\stn \exp\left(- \frac{\sigmak^2|\vecst{s}|^2}{2}-\ii \vecst{k}\cdot\vecst{s}\right).
\end{eqnarray}
Then, using the variables \m{\vecst{x}_{1,2} \doteq \vecst{x}' \pm \vecst{s}/2}, one can rewrite \eq{eq:aW} as follows:
\begin{eqnarray}
\avr{\matrst{W}}_{\vec{\psi}}(\vecst{x}, \vecst{k}) 
= \frac{1}{(2\upi)^{3\stn/2}\sigmax^\stn}
 \int \dd\vecst{x}_1\,\dd\vecst{x}_2\,\vec{\psi}(\vecst{x}_1) \vec{\psi}^\dag(\vecst{x}_2)\,\ee^{-\phi},
\label{eq:auxWa}\\
\phi 
= \frac{|\vecst{x} - (\vecst{x}_1 + \vecst{x}_2)/2|^2}{2\sigmax^2}
+ \frac{\sigmak^2|\vecst{x}_1 - \vecst{x}_2|^2}{2}
+ \ii \vecst{k}\cdot (\vecst{x}_1 - \vecst{x}_2).
\end{eqnarray}
The function \m{\phi} can also be expressed as \m{\phi = |\vecst{x}|^2/(2\sigmax^2) + \phi(\vecst{x}_1) + \phi^*(\vecst{x}_2) - \lambda \vecst{x}_1 \cdot \vecst{x}_2}, where
\begin{eqnarray}
\phi(\vecst{y}) \doteq \frac{|\vecst{y}|^2}{2}\left(\frac{1}{4\sigmax^2} + \sigmak^2\right)
- \frac{\vecst{x} \cdot \vecst{y}}{2\sigmax^2}
+ \ii \vecst{k} \cdot \vecst{y}
\end{eqnarray}
and \m{\lambda \doteq \sigmak^2 - (4\sigmax^2)^{-1}}. Then, using \m{\vec{\xi}(\vecst{y}) \doteq \vec{\psi}(\vecst{y})\ee^{-\phi(\vecst{y})}}, one obtains from \eq{eq:auxWa} that
\begin{eqnarray}
\avr{\matrst{W}}_{\vec{\psi}}(\vecst{x}, \vecst{k}) 
= \frac{\ee^{-|\vecst{x}|^2/(2\sigmax^2)}}{(2\upi)^{3\stn/2}\sigmax^\stn}
 \int \dd\vecst{x}_1\,\dd\vecst{x}_2\,\vec{\xi}(\vecst{x}_1) \vec{\xi}^\dag(\vecst{x}_2)\,\ee^{\lambda \vecst{x}_1 \cdot \vecst{x}_2}.
\end{eqnarray}
By Taylor-expanding \m{\ee^{\lambda \vecst{x}_1 \cdot \vecst{x}_2}}, one obtains
\begin{eqnarray}
\avr{\matrst{W}}_{\vec{\psi}}(\vecst{x}, \vecst{k}) = \frac{\ee^{-|\vecst{x}|^2/(2\sigmax^2)}}{(2\upi)^{3\stn/2}\sigmax^\stn} \sum_{m=0}^\infty\frac{\lambda^m}{m!}\,\matrst{J}_m,
\end{eqnarray}
where \m{\matrst{J}_m \doteq \int \dd\vecst{x}_1\,\dd\vecst{x}_2\,(\vecst{x}_1 \cdot \vecst{x}_2)^m \vec{\xi}(\vecst{x}_1) \vec{\xi}^\dag(\vecst{x}_2)}. Note that
\begin{eqnarray}
(\vecst{x}_1 \cdot \vecst{x}_2)^m = \sum_{\vec{\mu}(m)} \prod_{i=1}^{\stn} (\st{x}_1^i\st{x}^i_2)^{m_i},
\end{eqnarray}
where the summation is performed over all combinations \(\vec{\mu}(m) \equiv \{m_1, m_2, \ldots, m_\stn\}\) of integers \m{m_i \ge 0} such that \m{\sum_i m_i = m}. Thus,
\begin{eqnarray}
\matrst{J}_m = \sum_{\vec{\mu}(m)} \matr{\mc{J}}_{\vec{\mu}}\matr{\mc{J}}^\dag_{\vec{\mu}},
\qquad
\matr{\mc{J}}_{\vec{\mu}} = \int \dd\vecst{y}\,\vec{\xi}(\vecst{y})\prod_{i=1}^{\stn}(\st{y}^i)^{m_i}.
\end{eqnarray}
Because each \m{\matrst{J}_m} is positive-semidefinite, the Wigner matrix \m{\avr{\matrst{W}}_{\vec{\psi}}} is positive-semidefinite when \m{\lambda \ge 0}, or equivalently, when \m{\sigmax\sigmak > 1/2}. This condition is assumed to be satisfied for the phase-space averaging of \m{\matrst{W}_{\vec{\psi}}} used in the main text. Loosely, this means that the averaging is done over the phase-space volume \m{\Delta\vecst{x}\,\Delta\vecst{k} \sim (\sigmax\sigmak)^\stn \gtrsim 1}.

\subsection{Invariant limit for eikonal fields}
\label{sec:wfune}

For eikonal fields \eq{eq:Psic}, one has
\begin{eqnarray}
\vec{\psi}(\vecst{x} + \vecst{s}/2) \vec{\psi}^\dag(\vecst{x} - \vecst{s}/2)
\approx  
\big(\matr{A}(\vecst{x})\,\ee^{\ii\avr{\vecst{k}}(\vecst{x}) \cdot \vecst{s}} + \text{c.c.}\big)
+ \big(\matr{B}(\vecst{x})\,\ee^{2\ii\theta(\vecst{x})} + \text{c.c.}\big).
\end{eqnarray}
Here, \m{\matr{A} \doteq \vec{\eta}\vec{\eta}^\dag|\env{a}|^2/4}, \m{\matr{B} \doteq \vec{\eta}\vec{\eta}^\intercal\env{a}^2/4}, `c.c.' stands for complex conjugate, we used the linear approximation \m{\theta(\vecst{x} \pm \vecst{s}/2) \approx \theta(\vecst{x}) \pm \avr{\vecst{k}}(\vecst{x})\cdot\vecst{s}/2}, with \m{\avr{\vecst{k}} \equiv (-\avr{\omega}, \avr{\vec{k}})}. Then,
\begin{eqnarray}
\matrst{W}_{\vec{\psi}}(\vecst{x}, \vecst{k})
\approx  
\matr{A}(\vecst{x})\,\delta(\vecst{k} - \avr{\vecst{k}}(\vecst{x}))
+ \matr{A}^*(\vecst{x})\,\delta(\vecst{k} + \avr{\vecst{k}}(\vecst{x}))
+ 2\re\big(\matr{B}(\vecst{x})\ee^{2\ii\theta(\vecst{x})}\delta(\vecst{k})\big).
\end{eqnarray}
Let us adopt \m{\sigmax \ll l_c}, where \m{l_c} is the least characteristic scale of \m{\env{a}}, \m{\vec{\eta}}, and \m{\avr{\vecst{k}}}. Then,
\begin{eqnarray}
\avr{\matrst{W}}_{\vec{\psi}}(\vecst{x}, \vecst{k})
\approx  
\matr{A}(\vecst{x})\,\mc{G}_{\st{k}}(\vecst{k} - \avr{\vecst{k}})
+ \matr{A}^*(\vecst{x})\,\mc{G}_{\st{k}}(\vecst{k} + \avr{\vecst{k}})
+ 2\re\big(\matr{B}(\vecst{x})\mc{G}_{\st{k}}(\vecst{k})\zeta \ee^{2\ii\theta(\vecst{x})}\big).
\end{eqnarray}
Here, \m{\mc{G}_{\st{k}}(\vecst{k})} are normalized Gaussians that can be replaced with delta functions if \m{\sigmak} is small compared to any scale of interest in the \m{\vecst{k}} space:
\begin{eqnarray}
\mc{G}_{\st{k}}(\vecst{k}) \doteq \frac{1}{(\sqrt{2\upi}\sigmak)^\stn}\,
\exp\left( - \frac{|\vecst{k}|^2}{2\sigmak^2}\right) \to \delta(\vecst{k}).
\end{eqnarray}
Also, the function
\begin{eqnarray}
\zeta \approx \frac{1}{(\sqrt{2\upi}\sigmax)^\stn}
\int \dd\vecst{x}'\,
\exp\left(-\frac{|\vecst{x}' - \vecst{x}|^2}{2\sigmax^2} + 2\ii\avr{\vecst{k}}(\vecst{x}) \cdot (\vecst{x}' - \vecst{x})\right)
= \ee^{- 2|\avr{\vecst{k}}(\vecst{x})|^2 \sigmax^2}
\end{eqnarray}
can be made exponentially small by adopting \m{\sigmax \gg |\avr{\vecst{k}}|^{-1}}.\footnote{Even though \m{\sigmax} has been assumed small compared to \m{l_c}, the smallness of the geometrical-optics parameter \m{\epsilon \doteq (|\avr{\vecst{k}}|l_c)^{-1} \ll 1} allows choosing \m{\sigmax} in the interval \m{|\avr{\vecst{k}}|^{-1} \ll \sigmax \ll l_c}.} In this limit, the average Wigner matrix of an eikonal field is independent of \m{\sigmax} and \m{\sigmak}:
\begin{eqnarray}
\avr{\matrst{W}}_{\vec{\psi}}(\vecst{x}, \vecst{k})
\approx  
\matr{A}(\vecst{x})\,\delta(\vecst{k} - \avr{\vecst{k}}(\vecst{x}))
+ \matr{A}^*(\vecst{x})\,\delta(\vecst{k} + \avr{\vecst{k}}(\vecst{x})).
\end{eqnarray}
This \m{\avr{\matrst{W}}_{\vec{\psi}}} is also Hermitian and positive-semidefinite (in agreement with the general theory from \Sec{app:possemi}),  because so are \m{\matr{A}} and \m{\matr{A}^*}. The same properties pertain to the Wigner matrix of an ensemble of randomly phased eikonal fields, because it equals the sum of the Wigner matrices of the individual components (see also \Sec{sec:broad}).

\section{Auxiliary proofs}
\label{app:aux}

\subsection{Proof of \eq{eq:LRs}}
\label{app:LRs}

Like in the case of \eq{eq:LR1}, one finds that
\begin{align}
(\boperst{L}\vec{\psi}(\vecst{x}))^i (\boperst{R}\vec{\psi}(\vecst{x}))^{*j}
& = 
\braket{\vecst{x}|\operst{L}^i{}_{i'}|\psi^{i'}}
\braket{\psi^{j'}|(\operst{R}^j{}_{j'})^\dag|\vecst{x}} 
\notag\\
& = (2\upi)^\stn 
\braket{\vecst{x}|\operst{L}^i{}_{i'}\operst{W}_{\vec{\psi}}^{i'j'}(\operst{R}^\dag)_{j'}{}^j|\vecst{x}}
\notag\\
& = (2\upi)^\stn 
\braket{\vecst{x}|(\boperst{L}\boperst{W}_\psi\smash{\boperst{R}}^\dag)^{ij}|\vecst{x}}
\notag\\
& = \textstyle \int \dd\vecst{k}\,
\big(\matrst{L}(\vecst{x}, \vecst{k}) \star \matrst{W}_{\vec{\psi}}(\vecst{x}, \vecst{k}) \star \matrst{R}^\dag(\vecst{x}, \vecst{k})\big)^{ij}.
\label{eq:LRproof}
\end{align}
This proves \eq{eq:LR2}. At \m{\epsilon \to 0}, when \m{\star} becomes the usual product, \eq{eq:LRproof} gives
\begin{eqnarray}
(\boperst{L}\vec{\psi})(\boperst{R}\vec{\psi})^\dag
= \textstyle \int \dd\vecst{k}\,\matrst{L} \matrst{W}_{\vec{\psi}} \matrst{R}^\dag,
\label{eq:auxtrtrQ}
\end{eqnarray}
and in particular, taking the trace of \eq{eq:auxtrtrQ} yields
\begin{eqnarray}
\textstyle 
(\boperst{R}\vec{\psi})^\dag  (\boperst{L}\vec{\psi}) = \tr \big((\boperst{L}\vec{\psi})(\boperst{R}\vec{\psi})^\dag\big)
= \int \dd\vecst{k}\,\tr(\matrst{L} \matrst{W}_{\vec{\psi}} \matrst{R}^\dag)
= \int \dd\vecst{k}\,\tr(\matrst{W}_{\vec{\psi}} \matrst{Q}).
\label{eq:auxQ}
\end{eqnarray}
Here, \m{\matrst{Q} \doteq \matrst{R}^\dag \matrst{L}}, and we used that \m{\tr(\matrst{A}\matrst{B}) = \tr(\matrst{B}\matrst{A})} for any matrices \m{\matrst{A}} and \m{\matrst{B}}. 

For real fields, one can also replace the integrand with
\begin{eqnarray}
\tr\big(\matrst{W}_{\vec{\psi}}^* \matrst{Q}^*\big) 
= \tr\big(\matrst{Q}^\dag\matrst{W}_{\vec{\psi}}^\dag \big) 
= \tr\big(\matrst{Q}^\dag\matrst{W}_{\vec{\psi}}\big)
= \tr\big(\matrst{W}_{\vec{\psi}}\matrst{Q}^\dag\big),
\end{eqnarray}
where we used \m{\tr\matrst{A}^\intercal = \tr\matrst{A}}, \m{(\matrst{A}\matrst{B})^\intercal = \matrst{B}^\intercal\matrst{A}^\intercal}, \m{\matrst{W}_{\vec{\psi}}^\dag = \matrst{W}_{\vec{\psi}}}, and, again, \m{\tr(\matrst{A}\matrst{B}) = \tr(\matrst{B}\matrst{A})}, respectively. In summary then,
\begin{eqnarray}
\textstyle 
(\boperst{R}\vec{\psi})^\dag (\boperst{L}\vec{\psi})
= \int \dd\vecst{k}\,\tr(\matrst{W}_{\vec{\psi}} \matrst{Q})
= \int \dd\vecst{k}\,\tr\big(\matrst{W}_{\vec{\psi}}\matrst{Q}^\dag\big),
\end{eqnarray}
so the anti-Hermitian part of \m{\matrst{Q}} does not contribute to the integrals. Thus,
\begin{eqnarray}
\textstyle 
(\boperst{R}\vec{\psi})^\dag (\boperst{L}\vec{\psi}) 
= \int \dd\vecst{k}\,\tr(\matrst{W}_{\vec{\psi}}(\matrst{R}^\dag \matrst{L})_\herm).
\end{eqnarray}
Because \m{\boperst{L}} and \m{\boperst{R}} are arbitrary, they can as well be swapped; then one obtains \eq{eq:LRtr}.

\subsection{Proof of \eq{eq:dJ}}
\label{app:dJ}

Suppose the dominant term in \(\vec{\mu}\) in \eq{eq:TsymbM0} has the form \(\vec{\mu}_h \tau^h\), where \(h\) is a natural number and \m{\mu_h = \mc{O}(\epsilon^2)}. (Here \(h\) is a power index, so no summation over \(h\) is assumed.) Let us Taylor-expand \(\mc{J}[A, G]\) in \m{\vec{\mu}_h}:
\begin{align}
\mc{J}[A, G] & - \mc{J}[A, G_0] - \mc{O}(\epsilon^2)
\notag\\
& \approx \vec{\mu}_h \cdot \frac{\pd}{\pd \vec{\mu}_h}\left(
\int \dd\vec{K}\,A(\vec{X}, \vec{K}) 
\lim_{\nu\to 0+}
\int_0^{\infty}\dd\tau\,\ee^{-\nu \tau + \ii \Omega \tau + \ii \vec{K} \cdot \vec{\mu}_h \tau^h}
\right)_{\vec{\mu}_h = \vec{0}}
\notag\\
& \approx \vec{\mu}_h \cdot \int \dd\vec{K}\,A(\vec{X}, \vec{K})
\lim_{\nu \to 0+} \frac{\pd}{\pd \vec{\mu}_h} \left(
\int_0^{\infty} \dd\tau\,\ee^{-\nu \tau + \ii\Omega \tau + \ii \vec{K} \cdot \vec{\mu}_h \tau^h}
\right)_{\vec{\mu}_h = \vec{0}}
\notag\\
& \approx \ii \vec{\mu}_h \cdot \int \dd\vec{K}\,\vec{K} A(\vec{X}, \vec{K})
\lim_{\nu \to 0+}\int_0^{\infty} \dd\tau\,\tau^h \ee^{-\nu \tau + \ii \Omega \tau}
\notag\\
& \approx \ii^{1-h} \vec{\mu}_h \cdot \int \dd\vec{K}\,\vec{K} A(\vec{X}, \vec{K}) \,\frac{\dd^h G_0(\Omega (\vec{X}, \vec{K}))}{\dd\Omega^h}
\notag\\
& \approx \ii^{1-h} \vec{\mu}_h \cdot \frac{\eth^h}{\pd \Omega^h}
\int \dd\vec{K}\,\vec{K} A(\vec{X}, \vec{K}) G_0(\Omega(\vec{X}, \vec{K})).
\end{align}
Provided that \(A\) is sufficiently smooth and well behaved, the overall coefficient here is \(\mc{O}(1)\), so \m{\mc{J}[A, G] - \mc{J}[A, G_0] = \mc{O}(\mu_h) + \mc{O}(\epsilon^2)}. Because \(\mu_h = \mc{O}(\epsilon^2)\), this proves \eq{eq:dJ}.

\subsection{Proof of \eq{eq:Dconv}}
\label{app:Dconv}

Here, we show that
\begin{align}
\wsymbx(&\oper{u}^{\alpha}\oper{G}\oper{u}^{\beta})
\notag\\
& = \frac{1}{(2\upi)^N} \int \dd\vec{S}\,
\ee^{-\ii \vec{K} \cdot \vec{S}}
\braket{\vec{X} + \vec{S}/2 | \oper{u}^{\alpha}\oper{G}\oper{u}^{\beta} | \vec{X} - \vec{S}/2}
\notag\\
& = \frac{1}{(2\upi)^N} \int \dd\vec{X}'\,\dd\vec{K}''\,\dd\vec{K}'\,\dd\vec{S}\,\dd\vec{S}'\,
{W}_{\vec{u}}^{\alpha\beta}(\vec{X}', \vec{K}')
G(\vec{X}', \vec{K}'')\,
\ee^{-\ii \vec{K} \cdot \vec{S} + \ii (\vec{K}' + \vec{K}'') \cdot \vec{S}'}
\notag\\
& \hspace{2cm}\times
\delta(\vec{X} + \vec{S}/2 - \vec{X}' - \vec{S}'/2)
\delta(\vec{X} - \vec{S}/2 - \vec{X}' + \vec{S}'/2)
\notag\\
& = \frac{1}{(2\upi)^N} \int \dd\vec{X}'\,\dd\vec{K}'\,\dd\vec{K}''\,\dd\vec{S}\,\dd\vec{S}'\,
{W}_{\vec{u}}^{\alpha\beta}(\vec{X}', \vec{K}')
G(\vec{X}', \vec{K}'')\,
\ee^{-\ii \vec{K} \cdot \vec{S} + \ii (\vec{K}' + \vec{K}'') \cdot \vec{S}'}
\notag\\
& \hspace{2cm}\times
\delta (\vec{S} - \vec{S}')
\delta (\vec{X} - \vec{S}/2 - \vec{X}' + \vec{S}'/2)
\notag\\
& = \frac{1}{(2\upi)^N} \int \dd\vec{X}'\,\dd\vec{K}'\,\dd\vec{K}''\,\dd\vec{S}\,
{W}_{\vec{u}}^{\alpha\beta}(\vec{X}', \vec{K}')
G(\vec{X}', \vec{K}'')\,
\ee^{\ii (\vec{K}' + \vec{K}'' - \vec{K}) \cdot \vec{S}}
\delta(\vec{X} - \vec{X}')
\notag\\
& = \frac{1}{(2\upi)^N} \int \dd\vec{K}'\,\dd\vec{K}''\,\dd\vec{S}\,
{W}_{\vec{u}}^{\alpha\beta}(\vec{X}, \vec{K}')
G(\vec{X}, \vec{K}'')\,
\ee^{\ii (\vec{K}' + \vec{K}'' - \vec{K}) \cdot \vec{S}}
\notag\\
& = \int \dd\vec{K}'\,
{W}_{\vec{u}}^{\alpha\beta}(\vec{X}, \vec{K}')
G(\vec{X}, \vec{K} - \vec{K}').
\end{align}

\subsection{Proof of \eq{eq:DTheta}}
\label{app:DTheta}

Using \eq{eq:Gstar}, \eq{eq:Gstar2}, and \eq{eq:Wtstar} in application to \(W_{\vec{u}}^{\alpha\beta}\), one finds that
\begin{align}
D_0^{\alpha\beta}(\vec{X})
& \doteq \int \dd\vec{K}\,\avr{W}_{\vec{u}}^{\alpha\beta}(\vec{X}, \vec{K}) G^*(\vec{X}, \vec{K})
\notag\\
& = \int \dd\vec{K}\,\avr{W}_{\vec{u}}^{\alpha\beta}(\vec{X}, - \vec{K}) G^*(\vec{X},-\vec{K})
\notag\\
& = \int \dd\vec{K}\,\avr{W}_{\vec{u}}^{\alpha\beta}(\vec{X}, - \vec{K}) G(\vec{X}, \vec{K})
\notag\\
& = \int \dd\vec{K}\,\avr{W}_{\vec{u}}^{\alpha\beta *}(\vec{X}, \vec{K}) G(\vec{X}, \vec{K})
\notag\\
& = (D_0^{\alpha\beta}(\vec{X}))^*
\end{align}
and also
\begin{align}
\Uptheta^{\alpha \beta c}(\vec{X})
& \doteq -\int \dd\vec{K}\,\avr{W}_{\vec{u}}^{\alpha\beta}(\vec{X}, \vec{K})(G^{|c}(\vec{X}, \vec{K}))^*
\notag\\
& =\int \dd\vec{K}\,\avr{W}_{\vec{u}}^{\alpha\beta}(\vec{X}, \vec{K}) G^{|c}(\vec{X}, - \vec{K})
\notag\\
& =\int \dd\vec{K}\,\avr{W}_{\vec{u}}^{\alpha\beta}(\vec{X}, - \vec{K}) G^{|c}(\vec{X}, \vec{K})
\notag\\
& =\int \dd\vec{K}\,(\avr{W}_{\vec{u}}^{\alpha\beta}(\vec{X}, \vec{K}))^* G^{|c}(\vec{X}, \vec{K})
\notag\\
& = - (\Uptheta^{\alpha\beta c}(\vec{X}))^*.
\end{align}

\subsection{Proof of \eq{eq:rhoest}}
\label{app:contr}

Let us estimate
\begin{eqnarray}
\mc{L}^{(1)} \avr{f}
\doteq \frac{\pd}{\pd z^{\alpha}}
\left(
J^{\alpha \mu} J^{\beta \nu} \mc{P}_{\mu \nu}^{(1)}\,
\frac{\pd\avr{f}}{\pd z^{\beta}}
\right), 
\end{eqnarray}
where \(\mc{P}_{\mu \nu}^{(1)}\) has the form
\begin{eqnarray}
\mc{P}_{\mu \nu}^{(1)} \doteq \int \dd\vec{K}\,q_{\nu}\,
\frac{\pd \avr{W}(\vec{X}, \vec{K})}{\pd z^{\mu}}\,
G(\Omega (\vec{X}, \vec{K})).
\end{eqnarray}
First, notice that
\begin{align}
\mc{P}_{\mu \nu}^{(1)}
& = \frac{\pd}{\pd z^{\mu}} \int \dd\vec{K}\,q_{\nu}\avr{W}G
- \int \dd\vec{K}\,q_{\nu}\avr{W}\,\frac{\pd G}{\pd z^{\mu}}
\notag\\
& = \frac{\pd}{\pd z^{\mu}} \int \dd\vec{K}\,q_{\nu}\avr{W}G
+ \frac{\pd v^{\lambda}}{\pd z^{\mu}} \int \dd\vec{K}\,q_{\nu}q_{\lambda}\avr{W}G'
\notag\\
& = \frac{\pd}{\pd z^{\mu}} \int \dd\vec{K}\,q_{\nu}\avr{W}G
+ \frac{\pd v^{\lambda}}{\pd z^{\mu}} \frac{\eth}{\pd\Omega} \int \dd\vec{K}\, q_{\nu}q_{\lambda}\avr{W}G
\notag\\
& \equiv 
\frac{\pd \mc{Q}_{\nu}^{(1)}}{\pd z^{\mu}}
+ \frac{\pd v^{\lambda}}{\pd z^{\mu}}\,\mc{R}_{\lambda \nu}^{(1)}.
\end{align}
Because \(\mc{Q}_{\nu}^{(1)}\) and \(\mc{R}_{\lambda \nu}^{(1)}\) are \(\mc{O}(\varepsilon^2)\), one has \m{\mc{P}_{\mu \nu}^{(1)} \sim \kappa_{\mu} \varepsilon^2}, where \(\kappa_{\mu}\) is the characteristic inverse scale along the \(\mu\)th phase-space axis. Thus,
\begin{eqnarray}\label{eq:aL1}
\mc{L}^{(1)} \avr{f} \sim 
(J^{\alpha \mu} \kappa_{\alpha} \kappa_{\mu})
J^{\beta \nu} \kappa_{\beta}\varepsilon^2\avr{f}
= \mc{O}(\epsilon\varepsilon^2), 
\end{eqnarray}
where we used (see \eq{eq:Jqq} and \eq{eq:order})
\begin{eqnarray}\label{eq:Jkappas}
J^{\alpha \mu}\kappa_{\alpha}\kappa_{\mu} \sim \kappa_x\kappa_p = \mc{O}(\epsilon).
\end{eqnarray}
The first part of \eq{eq:rhoest} is obtained by considering \(\im\mc{L}^{(1)} \avr{f}\) and using \eq{eq:aL1}.

Let us also estimate
\begin{eqnarray}
\mc{L}^{(2)} \avr{f} \doteq 
\frac{\pd}{\pd z^{\alpha}} \left(
J^{\alpha \mu} J^{\beta \nu} \mc{P}_{\mu \nu}^{(2)}\,\frac{\pd\avr{f}}{\pd z^{\beta}}
\right),
\end{eqnarray}
where \(\mc{P}_{\mu \nu}^{(2)}\) has the form
\begin{eqnarray}
\mc{P}_{\mu \nu}^{(2)} \doteq \int \dd\vec{K}\,
\frac{\pd^2\avr{W}}{\pd z^{\mu}\pd z^{\nu}}\,G(\Omega (\vec{X}, \vec{K})).
\end{eqnarray}
First, note that
\begin{align}
\mc{P}_{\mu \nu}^{(2)}
= & \int \dd\vec{K}\,\frac{\pd^2\avr{W}}{\pd z^{\mu}\pd z^{\nu}}\,G
\notag\\
= & \frac{\pd}{\pd z^{\mu}} \int \dd\vec{K}\,\frac{\pd \avr{W}}{\pd z^{\nu}}\,G
- \int \dd\vec{K}\,\frac{\pd \avr{W}}{\pd z^{\nu}}\frac{\pd G}{\pd z^{\mu}}
\notag\\
= & \frac{\pd^2}{\pd z^{\mu}\pd z^{\nu}} \int \dd\vec{K}\, \avr{W} G
- \frac{\pd}{\pd z^{\mu}} \int \dd\vec{K}\,\avr{W}\, \frac{\pd G}{\pd z^{\nu}}
- \frac{\pd}{\pd z^{\nu}} \int \dd\vec{K}\,\avr{W}\,\frac{\pd G}{\pd z^{\mu}}
+ \int \dd\vec{K}\,\avr{W}\,\frac{\pd^2G}{\pd z^{\mu}\pd z^{\nu}}
\notag\\
= & \frac{\pd^2}{\pd z^{\mu}\pd z^{\nu}} \int \dd\vec{K}\,\avr{W}\,G
- \frac{\pd}{\pd z^{\mu}} \int \dd\vec{K}\,\avr{W}\,G'
  \left(- q_{\lambda}\,\frac{\pd v^{\lambda}}{\pd z^{\nu}}\right)
\notag\\ &
- \frac{\pd}{\pd z^{\nu}} \int \dd\vec{K}\,\avr{W}\,G'
  \left(- q_{\lambda}\,\frac{\pd v^{\lambda}}{\pd z^{\mu}}\right)
+ \int \dd\vec{K}\,\avr{W}\,\frac{\pd^2G}{\pd z^{\mu}\pd z^{\nu}}
\notag\\
= & \frac{\pd^2}{\pd z^{\mu}\pd z^{\nu}} \int \dd\vec{K}\,\avr{W}G
+ \frac{\pd}{\pd z^{\mu}} 
  \left(\frac{\pd v^{\lambda}}{\pd z^{\nu}} \frac{\eth}{\pd\Omega} \int \dd\vec{K}\,q_{\lambda}\avr{W}G\right)
\notag\\ &
+ \frac{\pd}{\pd z^{\nu}} 
  \left(\frac{\pd v^{\lambda}}{\pd z^{\mu}} \frac{\eth}{\pd\Omega} \int \dd\vec{K}\,q_{\lambda}\avr{W}G\right)
+ \int \dd\vec{K}\,\avr{W}\frac{\pd^2G}{\pd z^{\mu}\pd z^{\nu}}.
\end{align}
Next, note that
\begin{align}
\int \dd\vec{K}\,\avr{W}\,\frac{\pd^2G}{\pd z^{\mu}\pd z^{\nu}}
& = \int \dd\vec{K}\,\avr{W}\,\frac{\pd}{\pd z^{\mu}}
\left(- q_{\lambda}G'\frac{\pd v^{\lambda}}{\pd z^{\nu}}\right)
\notag\\
& = - \frac{\pd v^{\lambda}}{\pd z^{\nu}} \int \dd\vec{K}\,
q_{\lambda}\avr{W}\,\frac{\pd G'}{\pd z^{\mu}}
- \frac{\pd^2 v^{\lambda}}{\pd z^{\mu}\pd z^{\nu}}
\int \dd\vec{K}\,q_{\lambda}\avr{W}G'
\notag\\
& = \frac{\pd v^{\lambda}}{\pd z^{\nu}} \frac{\pd v^{\delta}}{\pd z^{\mu}}
\int \dd\vec{K}\,q_{\lambda}q_{\delta} \avr{W} G''
- \frac{\pd^2 v^{\lambda}}{\pd z^{\mu}\pd z^{\nu}} 
\int \dd\vec{K}\,q_{\lambda}\avr{W} G'
\notag\\
& = \frac{\pd v^{\lambda}}{\pd z^{\nu}} \frac{\pd v^{\delta}}{\pd z^{\mu}}
\frac{\eth^2}{\pd\Omega^2} \int \dd\vec{K}\,q_{\lambda}q_{\delta}\avr{W}G
- \frac{\pd^2v^{\lambda}}{\pd z^{\mu}\pd z^{\nu}} \frac{\eth}{\pd\Omega}\int \dd\vec{K}\,q_{\lambda}\avr{W} G.
\end{align}
Assuming the notation
\begin{align}
\mc{S}^{(2)} & \doteq \int \dd\vec{K}\, \avr{W} G = \mc{O}\left(\varepsilon^2\right),
\notag\\
\mc{Q}_{\lambda}^{(2)} & \doteq \frac{\eth}{\pd\Omega}\int \dd\vec{K}\, q_{\lambda}\avr{W} G = \mc{O}\left(\varepsilon^2\right),
\notag\\
\mc{R}_{\lambda \delta}^{(2)} & \doteq \frac{\eth^2}{\pd\Omega^2}
\int \dd\vec{K}\, q_{\lambda}q_{\delta}\avr{W}G
= \mc{O}\left(\varepsilon^2\right), 
\end{align}
one can then rewrite \(\mc{P}_{\mu \nu}^{(2)}\) as follows:
\begin{align}
\mc{P}_{\mu \nu}^{(2)}
& = \frac{\pd^2\mc{S}^{(2)}}{\pd z^{\mu}\pd z^{\nu}}
+ \frac{\pd}{\pd z^{\mu}}\left(\frac{\pd v^{\lambda}}{\pd z^{\nu}}\,\mc{Q}_{\lambda}^{(2)}\right)
+ \frac{\pd}{\pd z^{\nu}}\left(\frac{\pd v^{\lambda}}{\pd z^{\mu}}\,\mc{Q}_{\lambda}^{(2)}\right)
- \frac{\pd^2v^{\lambda}}{\pd z^{\mu}\pd z^{\nu}}\,\mc{Q}_{\lambda}^{(2)}
+ \frac{\pd v^{\lambda}}{\pd z^{\nu}}\frac{\pd v^{\delta}}{\pd z^{\mu}}\,\mc{R}_{\lambda \delta}^{(2)}
\notag\\
& = \frac{\pd^2\mc{S}^{(2)}}{\pd z^{\mu}\pd z^{\nu}}
+ \frac{\pd^2 v^{\lambda}}{\pd z^{\mu}\pd z^{\nu}}\,\mc{Q}_{\lambda}^{(2)}
+ \frac{\pd v^{\lambda}}{\pd z^{\nu}}\frac{\pd \mc{Q}_{\lambda}^{(2)}}{\pd z^{\mu}}
+ \frac{\pd v^{\lambda}}{\pd z^{\mu}}\frac{\pd \mc{Q}_{\lambda}^{(2)}}{\pd z^{\nu}}
+ \frac{\pd^2v^{\lambda}}{\pd z^{\mu}\pd z^{\nu}}\,\mc{Q}_{\lambda}^{(2)}
+ \frac{\pd v^{\lambda}}{\pd z^{\nu}}\frac{\pd v^{\delta}}{\pd z^{\mu}}\,\mc{R}_{\lambda\delta}^{(2)}.
\notag
\end{align}
Each term on the right-hand side of this equation scales as \(\varepsilon^2\kappa_{\mu}\kappa_{\nu}\), so
\begin{eqnarray}\label{eq:aL2}
\mc{L}^{(2)} \avr{f} \sim 
(J^{\alpha \mu}\kappa_{\alpha}\kappa_{\mu})
(J^{\beta \nu}\kappa_{\beta}\kappa_{\nu})
\varepsilon^2 \avr{f}
\sim \epsilon^2\varepsilon^2 \avr{f},
\end{eqnarray}
where we again used \eq{eq:Jkappas}. The second part of \eq{eq:rhoest} is obtained by considering \(\re\mc{L}^{(2)} \avr{f}\) and using \eq{eq:aL2}.

\subsection{Proof of \eq{eq:DDD}}
\label{app:DDD}

Using \eq{eq:aux11} and assuming the notation \(\dd_t \doteq \pd_t + v^{\gamma}\pd_{\gamma}\), one finds that
\begin{align}
\pd_{\alpha}& (\oper{D}^{\alpha\beta}\pd_{\beta}\avr{f}) 
- \pd_{\alpha}((\st{D}^{\alpha\beta}+\varrho^{\alpha\beta})\pd_{\beta}\avr{f})
\notag \\
= & - \pd_{\alpha}\left(\Uptheta^{\alpha\beta}\dd_t\pd_{\beta}\avr{f}
+ \frac{1}{2}\,(\dd_t\Uptheta^{\alpha\beta}) \pd_{\beta}\avr{f}\right)
\notag \\
= & - \pd_{\alpha}
\left(
\frac{1}{2}\,\Uptheta^{\alpha\beta}\dd_t \pd_{\beta}\avr{f}
+ \frac{1}{2}\,\dd_t(\Uptheta^{\alpha\beta}\pd_{\beta}\avr{f})
\right)
\notag \\
= & -\pd_{\alpha}
\left(
\frac{1}{2}\,\Uptheta^{\alpha\beta}\pd_{\beta}\dd_t\avr{f}
- \frac{1}{2}\,\Uptheta^{\alpha\beta}(\pd_{\beta}v^{\gamma})\pd_{\gamma}\avr{f}
\right)
- \pd_{\alpha}\left(
\frac{1}{2}\,\dd_t(\Uptheta^{\alpha\beta}\pd_{\beta}\avr{f})
\right)
\notag \\
= & -\pd_{\alpha}\left(
\frac{1}{2}\,\Uptheta^{\alpha\beta}\pd_{\beta}\dd_t\avr{f}
\right)
+ \pd_{\alpha}\left(\frac{1}{2}\,\Uptheta^{\alpha\beta}(\pd_{\beta}v^{\gamma})\pd_{\gamma}\avr{f}\right)
\notag \\
& - \dd_t \left(
\frac{1}{2}\,\pd_{\alpha}(\Uptheta^{\alpha\beta}\pd_{\beta}\avr{f})
\right)
- (\pd_{\alpha}v^{\gamma}) \pd_{\gamma}\left(
\frac{1}{2}\,\Uptheta^{\alpha\beta}\pd_{\beta}\avr{f}
\right).
\label{eq:aux12}
\end{align}
Because \(\Uptheta^{\alpha\beta} = \mc{O}(\varepsilon^2)\) and \(\dd_t \avr{f} = \mc{O}(\varepsilon^2)\), the first term on the right-hand side of \eq{eq:aux12} is negligible. Also note that due to \eq{eq:VV}, the factor \m{\pd_{\alpha}v^{\gamma}} in the last term on the right-hand side of \eq{eq:aux12} commutes with \m{\pd_\gamma}. Hence, one obtains
\begin{align}
\pd_{\alpha} (\oper{D}^{\alpha\beta}\pd_{\beta}\avr{f})
& - \pd_{\alpha}(\st{D}^{\alpha\beta}\pd_{\beta}\avr{f})
+ \dd_t \left(\frac{1}{2}\,\pd_{\alpha}(\Uptheta^{\alpha \beta}\pd_{\beta}\avr{f})\right)
\notag\\
& = \pd_{\alpha}
\left(
\varrho^{\alpha\beta} \pd_{\beta}\avr{f}
+ \frac{1}{2}\,\Uptheta^{\alpha\beta} \left(\pd_{\beta}v^{\gamma}\right) \pd_{\gamma}\avr{f}
\right)
- \pd_{\gamma}
\left(
\frac{1}{2}\,\Uptheta^{\alpha\beta} (\pd_{\alpha}v^{\gamma}) \pd_{\beta}\avr{f}
\right)
\notag\\
& = \pd_{\alpha}\left(
\varrho^{\alpha\beta} \pd_{\beta} \avr{f}
+ \frac{1}{2}
\left(
\Uptheta^{\alpha \gamma}(\pd_{\gamma}v^{\beta})
-\Uptheta^{\gamma \beta}(\pd_{\gamma}v^{\alpha})
\right)\pd_{\beta}\avr{f}
\right)
\notag\\
& \equiv \pd_{\alpha} (U^{\alpha\beta} \pd_{\beta}\avr{f}).
\label{eq:aux14}
\end{align}
Next, notice that
\begin{align}
\varrho^{\alpha\beta}
& = - \frac{1}{2}\, J^{\alpha \mu}J^{\beta \nu}\fint \dd\vec{K}\left(
q_{\nu}\,\frac{\pd\avr{W}}{\pd z^\mu}
- q_{\mu}\,\frac{\pd\avr{W}}{\pd z^\nu}
\right)
\frac{1}{\Omega}
\notag\\
& =-\frac{1}{2}\,J^{\alpha \mu}J^{\beta \nu} \fint \dd\vec{K}\,
\frac{q_{\nu}}{\Omega}\frac{\pd\avr{W}}{\pd z^\mu}
+ \frac{1}{2}\,J^{\alpha \mu}J^{\nu \beta}\fint \dd\vec{K}\,
\frac{q_{\mu}}{\Omega}\frac{\pd\avr{W}}{\pd z^\nu}
\notag\\
& = -\frac{1}{2}\,J^{\alpha \mu}J^{\beta\nu}\left(
\frac{\pd}{\pd z^\mu}\fint \dd\vec{K}\, \frac{q_{\nu}\avr{W}}{\Omega}
-\fint \dd\vec{K}\,q_{\nu}\avr{W}\,\frac{\pd}{\pd z^\mu}\frac{1}{\Omega}
\right.
\notag\\
& \hspace{2.5cm}
\left.
-\frac{\pd}{\pd z^\nu}\fint \dd\vec{K}\,\frac{q_{\mu}\avr{W}}{\Omega}
+\fint \dd\vec{K}\,q_{\mu}\avr{W}\frac{\pd}{\pd z^\nu}\frac{1}{\Omega}\right)
\notag\\
& = -\frac{1}{2}\,J^{\alpha\mu}J^{\beta \nu}
\left(
\frac{\pd}{\pd z^\mu} \fint \dd\vec{K}\,\frac{q_{\nu}\avr{W}}{\Omega}
+ \frac{\pd v^{\lambda}}{\pd z^\mu} \frac{\eth}{\pd \Omega}
\fint \dd\vec{K}\,\frac{q_{\lambda}q_{\nu}\avr{W}}{\Omega}
\right.
\notag\\
& \hspace{2.5cm}
\left.
- \frac{\pd}{\pd z^\nu} \fint \dd\vec{K}\,\frac{q_{\mu}\avr{W}}{\Omega}
- \frac{\pd v^{\lambda}}{\pd z^\nu} \frac{\eth}{\pd \Omega}
\fint \dd\vec{K}\, \frac{q_{\lambda}q_{\mu}\avr{W}}{\Omega}
\right).
\label{eq:aux13}
\end{align}
Assuming the notation
\begin{eqnarray}
Q_{\mu} \doteq \frac{1}{2} \fint \dd\vec{K}\,\frac{q_{\mu}\avr{W}}{\Omega},
\qquad
R_{\mu \nu}\doteq \frac{1}{2}\frac{\eth}{\pd \Omega}\fint \dd\vec{K}\,
\frac{q_{\mu}q_{\nu}\avr{W}}{\Omega},
\end{eqnarray}
one can rewrite \eq{eq:aux13} compactly as follows:
\begin{eqnarray}
\varrho^{\alpha\beta}
= J^{\alpha \mu}J^{\beta \nu} (\pd_{\nu}Q_{\mu} - \pd_{\mu}Q_{\nu})
+ J^{\alpha \mu}J^{\beta \nu} 
((\pd_{\nu}v^{\lambda})R_{\lambda \mu} - (\pd_{\mu}v^{\lambda})R_{\lambda \nu}).
\end{eqnarray}
Notice also that \(\Uptheta^{\alpha\beta} = 2J^{\alpha \mu}J^{\beta \nu}R_{\mu \nu}\). Hence, for \m{U^{\alpha\beta}} introduced in \eq{eq:aux14}, one has
\begin{align}
U &^{\alpha\beta} - J^{\alpha \mu} J^{\beta \nu} 
(\pd_{\nu} Q_{\mu} - \pd_{\mu} Q_{\nu})
\notag\\
& = \varrho^{\alpha\beta} - J^{\alpha \mu} J^{\beta \nu}
(\pd_{\nu}Q_{\mu} - \pd_{\mu}Q_{\nu})
- (
\Uptheta^{\gamma \beta} (\pd_{\gamma}v^{\alpha}) 
- \Uptheta^{\alpha \gamma}(\pd_{\gamma}v^{\beta})
)/2
\notag\\
& = J^{\alpha \mu} J^{\beta \nu} 
((\pd_{\nu} v^{\lambda}) R_{\lambda \mu} - (\pd_{\mu}v^{\lambda}) R_{\lambda \nu})
+ J^{\alpha \mu} J^{\gamma \nu} (\pd_{\gamma} v^{\beta}) R_{\mu \nu} 
- J^{\gamma \mu}J^{\beta \nu} (\pd_{\gamma}v^{\alpha}) R_{\mu \nu}
\notag\\
& = 
  J^{\alpha \mu} J^{\beta \nu} (\pd_{\nu}v^{\lambda}) R_{\lambda \mu}
- J^{\alpha \mu} J^{\beta \nu} (\pd_{\mu}v^{\lambda}) R_{\lambda \nu}
+ J^{\alpha \mu} J^{\gamma \nu} (\pd_{\gamma}v^{\beta}) R_{\mu \nu}
- J^{\gamma \mu} J^{\beta \nu} (\pd_{\gamma}v^{\alpha})R_{\mu \nu}
\notag\\
& = 
  J^{\alpha \mu} J^{\beta \lambda} (\pd_{\lambda}v^{\nu}) R_{\mu \nu}
- J^{\alpha \lambda} J^{\beta \nu}(\pd_{\lambda} v^{\mu}) R_{\mu \nu}
+ J^{\alpha \mu} J^{\gamma \nu} (\pd_{\gamma}v^{\beta}) R_{\mu \nu}
- J^{\gamma \mu}J^{\beta \nu}(\pd_{\gamma}v^{\alpha}) R_{\mu \nu}
\notag\\
& = (
  J^{\alpha \mu} J^{\beta \lambda} J^{\nu \gamma}
- J^{\alpha \lambda} J^{\beta \nu} J^{\mu \gamma}
+ J^{\alpha \mu}J^{\gamma \nu}J^{\beta \lambda}
- J^{\gamma \mu}J^{\beta \nu}J^{\alpha \lambda}
) (\pd_{\gamma \lambda}^2\avr{H}) R_{\mu \nu}
\notag\\
& = (
  J^{\alpha \mu} J^{\beta \lambda} J^{\nu \gamma}
- J^{\alpha \lambda} J^{\beta \nu} J^{\mu \gamma}
- J^{\alpha \mu} J^{\gamma \nu} J^{\lambda \beta}
+ J^{\mu \gamma} J^{\beta \nu} J^{\alpha \lambda}
) (\pd_{\gamma \lambda}^2\avr{H}) R_{\mu \nu}
\notag\\
& = 0,
\end{align}
where we used \eq{eq:uv} for \(v^\alpha\) and the anti-symmetry of \(J^{\alpha\beta}\). Therefore,
\begin{eqnarray}
U^{\alpha\beta}
= J^{\alpha \mu} J^{\beta \nu} (\pd_{\nu}Q_{\mu} - \pd_{\mu}Q_{\nu})
= (J^{\alpha \mu} J^{\beta \nu} - J^{\alpha \nu} J^{\beta \mu}) \pd_{\nu}Q_{\mu}
= - U^{\beta \alpha}, 
\end{eqnarray}
and accordingly,
\begin{align}
\pd_{\alpha}U^{\alpha\beta}
& = (J^{\alpha \mu}J^{\beta \nu} - J^{\alpha \nu}J^{\beta \mu})\pd_{\nu \alpha}^2 Q_{\mu}
= J^{\alpha \mu} J^{\beta \nu} \pd_{\nu \alpha}^2 Q_{\mu}
\notag\\
& = J^{\nu \mu} J^{\beta \alpha} \pd_{\nu \alpha}^2 Q_{\mu}
= J^{\beta \alpha} \pd_{\alpha} (J^{\nu \mu} \pd_{\nu} Q_{\mu})
= - J^{\alpha\beta} \pd_{\alpha} (J^{\mu \nu} \pd_{\mu}Q_{\nu})
\equiv J^{\alpha\beta} \pd_{\alpha}\Phi.
\label{eq:divU}
\end{align}
Here, \m{\Phi \doteq -J^{\mu \nu}\pd_{\mu}Q_{\nu}}, which is equivalent to \eq{eq:phi}. From \eq{eq:divU} and the fact that \m{U^{\alpha\beta}\pd_{\alpha\beta} = 0} due to the anti-symmetry of \m{U^{\alpha\beta}}, one has
\begin{eqnarray}
\pd_{\alpha} (U^{\alpha\beta}\pd_{\beta}\avr{f})
= J^{\alpha\beta} (\pd_{\alpha}\Phi) (\pd_{\beta}\avr{f})
= \poissonz{\Phi, \avr{f}}.
\end{eqnarray}
Hence, \eq{eq:aux14} leads to \eq{eq:DDD}.

\subsection{Proof of \eq{eq:mccG}}
\label{app:Wmcc}

The correlation function
\begin{eqnarray}
\mcc{C}_{ss'}(t, \vec{x}, \tau, \vec{s}; \vec{p}, \vec{p}') \doteq \avr{
g_{s}(t + \tau/2, \vec{x} + \vec{s}/2, \vec{p})
g_{s'}(t - \tau/2, \vec{x} - \vec{s}/2, \vec{p}')
}
\end{eqnarray}
can be readily expressed as 
\begin{align}\notag
\mcc{C}_{ss'} =
\bigg(\delta_{ss'}\!\!\!\sum_{\sigma_s = \sigma'_{s'}} + \sum_{\sigma_s \ne \sigma'_{s'}}\bigg)\,
\langle
& 
\delta(\vec{x} + \vec{s}/2 - \avr{\vec{x}}_{\sigma_s}(t + \tau/2))\delta(\vec{p} - \avr{\vec{p}}_{\sigma_s}(t + \tau/2))
\notag\\
& \times \delta(\vec{x} - \vec{s}/2 - \avr{\vec{x}}_{\sigma'_{s'}}(t - \tau/2))\delta(\vec{p}' - \avr{\vec{p}}_{\sigma'_{s'}}(t - \tau/2))
\rangle
- \mcc{C}_{\avr{\mcc{f}}}.
\notag
\end{align}
Here, \m{\macrob{\ldots}} is another (in addition to overbar) notation for averaging used in this appendix, the dependence of \m{\avr{\mcc{f}}_s} on \m{(t, \vec{x})} is neglected, and `\m{\sigma_s \ne \sigma'_{s'}}' denotes that excluded are the terms that have \m{s' = s} and \m{\sigma_s = \sigma'_{s'}} simultaneously. Aside from this, the summations over \m{\sigma_s} are taken over all \m{N_s \gg 1} particles of type \m{s}, and the summations over \m{\sigma_{s'}} are taken over all \m{N_{s'} \gg 1} particles of type \m{s'}. Also,
\begin{eqnarray}
\mcc{C}_{\avr{\mcc{f}}} \doteq \macrob{
\avr{\mcc{f}}_{s}(t + \tau/2, \vec{x} + \vec{s}/2, \vec{p})
\avr{\mcc{f}}_{s'}(t - \tau/2, \vec{x} - \vec{s}/2, \vec{p}')
}.
\end{eqnarray}
To the leading order, pair correlations can be neglected. Then,
\begin{align}\notag
\sum_{\sigma_s \ne \sigma'_{s'}} \macrob{\ldots}
& =  
\sum_{\sigma_s \ne \sigma'_{s'}} 
\underbrace{\macrob{\delta(\vec{x} + \vec{s}/2 - \avr{\vec{x}}_{\sigma_s}(t + \tau/2))\delta(\vec{p} - \avr{\vec{p}}_{\sigma_s}(t + \tau/2))}}_{\avr{\mcc{f}}_s(t + \tau/2, \vec{x} + \vec{s}/2, \vec{p})/N_s}
\notag\\
& \hphantom{\sum_{\sigma_s \ne \sigma'_{s'}} } \times 
\underbrace{\macrob{\delta(\vec{x} - \vec{s}/2 - \avr{\vec{x}}_{\sigma'_{s'}}(t - \tau/2))\delta(\vec{p}' - \avr{\vec{p}}_{\sigma'_{s'}}(t - \tau/2))}}_{\avr{\mcc{f}}_{s'}(t - \tau/2, \vec{x} - \vec{s}/2, \vec{p}')/N_{s'}}
\notag\\
& 
= \frac{\mcc{C}_{\avr{\mcc{f}}}}{N_s N_{s'}}\sum_{\sigma_s, \sigma'_{s'}} (1 - \delta_{ss'}\delta_{\sigma_s\sigma'_{s'}})
= (1 - N_s^{-1} \delta_{ss'}) \mcc{C}_{\avr{\mcc{f}}} \approx \mcc{C}_{\avr{\mcc{f}}}.
\end{align}
Let us also use \m{\avr{\vec{p}}_{\sigma_s}(t + \tau/2) \approx \avr{\vec{p}}_{\sigma_s}(t)}. Then,
\begin{align}\notag
\mcc{C}_{ss'} \approx
\delta_{ss'}\delta(\vec{p} - \vec{p}') \sum_{\sigma=1}^{N_s}
\langle 
& \delta(\vec{x} + \vec{s}/2 - \avr{\vec{x}}_{\sigma}(t + \tau/2))
\notag\\
& \times \delta(\vec{x} - \vec{s}/2 - \avr{\vec{x}}_{\sigma}(t - \tau/2)) \delta(\vec{p} - \avr{\vec{p}}_{\sigma}(t))
\rangle.
\end{align}
Next, notice that 
\begin{align}
& 
\macrob{
\delta(\vec{x} + \vec{s}/2 - \avr{\vec{x}}_{\sigma}(t + \tau/2))
\delta(\vec{x} - \vec{s}/2 - \avr{\vec{x}}_{\sigma}(t - \tau/2))
\delta(\vec{p} - \avr{\vec{p}}_{\sigma}(t))
}
\notag\\
& = 
\macrob{
\delta(\vec{s} + \avr{\vec{x}}_{\sigma}(t - \tau/2) - \avr{\vec{x}}_{\sigma}(t + \tau/2))
\delta(\vec{x} - \vec{s}/2 - \avr{\vec{x}}_{\sigma}(t - \tau/2))
\delta(\vec{p} - \avr{\vec{p}}_{\sigma}(t))
}
\notag\\
& = 
\macrob{
\delta(\vec{s} + \avr{\vec{x}}_{\sigma}(t - \tau/2) - \avr{\vec{x}}_{\sigma}(t + \tau/2))
\delta(\vec{x} - (\avr{\vec{x}}_{\sigma}(t + \tau/2) + \avr{\vec{x}}_{\sigma}(t - \tau/2))/2)
\delta(\vec{p} - \avr{\vec{p}}_{\sigma}(t))
}
\notag\\
& \approx 
\macrob{
\delta(\vec{s} - \vec{v}_s(t, \avr{\vec{x}}_\sigma, \avr{\vec{p}}_\sigma) \tau)
\delta(\vec{x} - \avr{\vec{x}}_{\sigma}(t))
\delta(\vec{p} - \avr{\vec{p}}_{\sigma}(t))
}
\notag\\
& \approx 
\delta(\vec{s} - \vec{v}_s(t, \vec{x}, \vec{p})\tau)
\macrob{
\delta(\vec{x} - \avr{\vec{x}}_{\sigma}(t))
\delta(\vec{p} - \avr{\vec{p}}_{\sigma}(t))
}
\notag\\
& = \delta(\vec{s} - \vec{v}_s(t, \vec{x}, \vec{p})\tau)
\avr{\mcc{f}}_{s}(t, \vec{x}, \vec{p})/N_s.
\end{align}
Hence,
\begin{eqnarray}
\mcc{C}_{ss'} = 
\delta_{ss'}\delta(\vec{p} - \vec{p}') 
\delta(\vec{s} - \vec{v}_s(t, \vec{x}, \vec{p})\tau)F_{s}(t, \vec{x}, \vec{p}),
\end{eqnarray}
where we used \m{\avr{\mcc{f}}_{s} \approx F_s}. Therefore,
\begin{align}\notag
\mcc{G}_{ss'}(t, \vec{x}, \omega, \vec{k}; \vec{p}, \vec{p}') 
& =
\int \frac{\dd\tau}{2\upi}\frac{\dd\vec{s}}{(2\upi)^n}\,\ee^{\ii\omega\tau - \ii\vec{k}\cdot\vec{s}}\,
\mcc{C}_{ss'}(t, \vec{x}, \tau, \vec{s}; \vec{p}, \vec{p}')
\notag\\
& \approx \frac{1}{(2\upi)^n}\,\delta_{ss'}\delta(\vec{p} - \vec{p}') F_{s}(t, \vec{x}, \vec{p}).
\end{align}

\subsection{Proof of \eq{eq:deltadw}}
\label{app:gw}

Using the symmetry \m{\st{U}^{\alpha\beta\gamma\delta} = \st{U}^{\beta\alpha\gamma\delta} = \st{U}^{\beta\alpha\delta\gamma}}, one readily obtains from \eq{eq:Phifin} that
\begin{eqnarray}
& \displaystyle
\Delta = \frac{1}{2P^0}\int \dd\vecst{k}\,\avr{g}_{\beta\gamma} p_\alpha p_\delta \st{U}^{\alpha\beta\gamma\delta} 
+ \frac{1}{8} \fint \dd\vecst{k}\,\mc{J},
\label{eq:vpia}
\\
& \displaystyle
\mc{J} 
= - \frac{\pd'}{\pd \vec{p}} \cdot
\left(\frac{\vec{k}\mcc{E}}{P^0 \varpi}\right)
- \frac{1}{(P^0)^2}\frac{\pd \mcc{E}}{\pd p_0}
+ \frac{\avr{g}^{00} \mcc{E}}{(P^0)^3},
\end{eqnarray}
where \m{\varpi \doteq k_\rho p^\rho = P^0(\vec{k} \cdot \vec{v} - \omega)} and the prime in \m{\pd'} denotes that \m{p_0} is considered as a function of \m{\vec{p}} at differentiation. One can also write this as follows:
\begin{eqnarray}
\mc{J} = 
\frac{\vec{k}}{\varpi} \cdot 
\left(\frac{\pd' P_0}{\pd \vec{p}}\right)
\frac{\mcc{E}}{(P^0)^2}
- \frac{1}{P^0}\frac{\pd'}{\pd \vec{p}} \cdot
\left(\frac{\vec{k}\mcc{E}}{\varpi}\right)
- \frac{1}{(P^0)^2}\frac{\pd \mcc{E}}{\pd p_0}
+ \frac{\avr{g}^{00} \mcc{E}}{(P^0)^3}.
\end{eqnarray}
As shown in \citep[appendix~B]{my:gwponder}, the following equality is satisfied:
\begin{eqnarray}
\frac{\vec{k}}{\varpi} \cdot \left(\frac{\pd' P_0}{\pd \vec{p}}\right) = \frac{1}{\varpi}\frac{\pd\varpi}{\pd p_0} - \frac{\avr{g}^{00}}{P^0}.
\end{eqnarray}
Also notice that 
\begin{align}
\frac{\pd'}{\pd \vec{p}} \cdot
\left(\frac{\vec{k}\mcc{E}}{\varpi}\right)
& = \frac{\pd}{\pd \vec{p}} \cdot
\left(\frac{\vec{k}\mcc{E}}{\varpi}\right)
+ \frac{\pd P_0}{\pd \vec{p}} \cdot\frac{\pd}{\pd p_0}
\left(\frac{\vec{k}\mcc{E}}{\varpi}\right)
\notag\\
& = \frac{\pd}{\pd \vec{p}} \cdot
\left(\frac{\vec{k}\mcc{E}}{\varpi}\right)
- \vec{k} \cdot \vec{v}\, \frac{\pd}{\pd p_0}
\left(\frac{\mcc{E}}{\varpi}\right),
\end{align}
where we used Hamilton's equation \m{\pd_{\vec{p}} P_0 = -\pd_{\vec{p}} H = -\vec{v}}. Therefore,
\begin{eqnarray}
\mc{J}
= \frac{1}{\varpi}\frac{\pd\varpi}{\pd p_0}
\frac{\mcc{E}}{(P^0)^2}
- \frac{1}{P^0}
\frac{\pd}{\pd \vec{p}} \cdot
\left(\frac{\vec{k}\mcc{E}}{\varpi}\right)
+ 
\frac{\vec{k} \cdot \vec{v}}{P^0}\frac{\pd}{\pd p_0}
\left(\frac{\mcc{E}}{\varpi}\right)
- \frac{1}{(P^0)^2}\frac{\pd \mcc{E}}{\pd p_0}.
\end{eqnarray}
The first and the last terms can be merged; then, one obtains
\begin{align}
\mc{J} &
= - \frac{1}{P^0}
\frac{\pd}{\pd \vec{p}} \cdot
\left(\frac{\vec{k}\mcc{E}}{\varpi}\right)
+ 
\frac{\vec{k} \cdot \vec{v}}{P^0}\frac{\pd}{\pd p_0}
\left(\frac{\mcc{E}}{\varpi}\right)
- \frac{\varpi}{(P^0)^2}\frac{\pd}{\pd p_0}\left(\frac{\mcc{E}}{\varpi}\right)
\notag\\
& = - \frac{1}{P^0}
\frac{\pd}{\pd \vec{p}} \cdot
\left(\frac{\vec{k}\mcc{E}}{\varpi}\right)
+ \frac{\omega}{P^0}\frac{\pd}{\pd p_0}\left(\frac{\mcc{E}}{\varpi}\right)
\notag\\
& = - \frac{1}{P^0}
\frac{\pd}{\pd p_\lambda}
\left(\frac{k_\lambda\mcc{E}}{\varpi}\right).
\end{align}
In combination with \eq{eq:vpia}, this leads to \eq{eq:deltadw}.

\section{Properties of the collision operator}
\label{app:ccons}

Here, we prove the properties of the collision operator discussed in \Sec{sec:cop}. To shorten the calculations, we introduce two auxiliary functions,
\begin{eqnarray}\notag
& \displaystyle
\mc{Z}_{ss'}(\vec{k}; \vec{p}, \vec{p}') \doteq \upi\,\delta(\vec{k} \cdot \vec{v}_s - \vec{k} \cdot \vec{v}'_{s'})
\mc{Q}_{ss'}(\vec{k} \cdot \vec{v}_s, \vec{k}; \vec{p}, \vec{p}'),\\
& \displaystyle
\mc{F}_{ss'}(\vec{p}, \vec{p}') \doteq \frac{\pd F_s(\vec{p})}{\pd p_j}\,F_{s'}(\vec{p}')
- F_s(\vec{p})\,\frac{\pd F_{s'}(\vec{p}')}{\pd p_j'},
\end{eqnarray}
which have the following properties:
\begin{eqnarray}\label{eq:ZF}
\mc{Z}_{ss'}(\vec{k}; \vec{p}, \vec{p}') = \mc{Z}_{s's}(\vec{k}; \vec{p}', \vec{p}),
\qquad
\mc{F}_{ss'}(\vec{p}, \vec{p}') = - \mc{F}_{s's}(\vec{p}', \vec{p}).
\end{eqnarray}

\subsection{Momentum conservation}

Momentum conservation is proven as follows. Using integration by parts, one obtains
\begin{align}
& \sum_s \int \dd\vec{p}\,p_l\collision_s
\notag\\
& =
\sum_{s,s'} \int \dd\vec{p}\,p_l\,
\frac{\pd}{\pd p_i}
\int \frac{\dd\vec{k}}{(2\upi)^n}\,\dd\vec{p}'\,k_i k_j\,
\mc{Z}_{ss'}(\vec{k}; \vec{p}, \vec{p}') \mc{F}_{ss'}(\vec{p}, \vec{p}')
\notag\\
& = -\sum_{s,s'} \int \frac{\dd\vec{k}}{(2\upi)^n}\,\dd\vec{p}\,\dd\vec{p}'\,k_l k_j\,
\mc{Z}_{ss'}(\vec{k}; \vec{p}, \vec{p}') \mc{F}_{ss'}(\vec{p}, \vec{p}').
\label{eq:aux61}
\end{align}
Now we swap the dummy variables \m{s \leftrightarrow s'} and \m{\vec{p} \leftrightarrow \vec{p}'} and then apply \eq{eq:ZF}:
\begin{align}
& \sum_s \int \dd\vec{p}\,p_l\collision_s
\notag\\
& = -\sum_{s',s} \int \frac{\dd\vec{k}}{(2\upi)^n}\,\dd\vec{p}'\,\dd\vec{p}\,k_l k_j\,
\mc{Z}_{s's}(\vec{k}; \vec{p}', \vec{p}) \mc{F}_{s's}(\vec{p}', \vec{p})
\notag\\
& = \sum_{s',s} \int \frac{\dd\vec{k}}{(2\upi)^n}\,\dd\vec{p}'\,\dd\vec{p}\,k_l k_j\,
\mc{Z}_{ss'}(\vec{k}; \vec{p}, \vec{p}') \mc{F}_{ss'}(\vec{p}, \vec{p}').
\label{eq:aux62}
\end{align}
The expression on the right-hand side of \eq{eq:aux62} is minus that in \eq{eq:aux61}. Hence, both are zero, which proves that \m{\sum_s \int \dd\vec{p}\,p_l\collision_s = 0}.

\subsection{Energy conservation}

Energy conservation is proven similarly, using that \m{v_s^i = \pd\mc{H}_s/\pd p_i} and the fact that \m{\vec{k} \cdot \vec{v}_s} and \m{\vec{k} \cdot \vec{v}'_{s'}} are interchangeable due to the presence of \m{\delta(\vec{k} \cdot \vec{v}_s - \vec{k} \cdot \vec{v}'_{s'})} in \m{\mc{Z}_{ss'}}:
\begin{align}
& \sum_s \int \dd\vec{p}\,\mc{H}_s\collision_s
\notag\\
& =
\sum_{s,s'} \int \dd\vec{p}\,\mc{H}_s\,
\frac{\pd}{\pd p_i}
\int \frac{\dd\vec{k}}{(2\upi)^n}\,\dd\vec{p}'\,k_i k_j\,
\mc{Z}_{ss'}(\vec{k}; \vec{p}, \vec{p}') \mc{F}_{ss'}(\vec{p}, \vec{p}')
\notag\\
& = -\sum_{s,s'} \int \frac{\dd\vec{k}}{(2\upi)^n}\,\dd\vec{p}\,\dd\vec{p}'\,(\vec{k} \cdot \vec{v}_s) k_j\,
\mc{Z}_{ss'}(\vec{k}; \vec{p}, \vec{p}') \mc{F}_{ss'}(\vec{p}, \vec{p}')
\notag\\
& = -\sum_{s',s} \int \frac{\dd\vec{k}}{(2\upi)^n}\,\dd\vec{p}'\,\dd\vec{p}\,(\vec{k} \cdot \vec{v}'_{s'}) k_j\,
\mc{Z}_{s's}(\vec{k}; \vec{p}', \vec{p}) \mc{F}_{s's}(\vec{p}', \vec{p})
\notag\\
& = \sum_{s',s} \int \frac{\dd\vec{k}}{(2\upi)^n}\,\dd\vec{p}'\,\dd\vec{p}\,(\vec{k} \cdot \vec{v}_s) k_j\,
\mc{Z}_{ss'}(\vec{k}; \vec{p}, \vec{p}') \mc{F}_{ss'}(\vec{p}, \vec{p}').
\end{align}
Like in the previous case, the third and the fifth lines are minus each other, whence \m{\sum_s \int \dd\vec{p}\,\mc{H}_s\collision_s = 0}.

\subsection{\mt{H}-theorem}
\label{app:Hth}

From \eq{eq:Hth1} and \eq{eq:Hth2}, one has
\begin{eqnarray}
\left(\frac{\dd\entropy}{\dd t}\right)_{\text{coll}} 
= - \sum_s \int \dd\vec{p}\,(1 + \ln F_s(\vec{p}))\collision_s
= - \sum_s \int \dd\vec{p}\,\ln F_s(\vec{p})\collision_s,
\end{eqnarray}
where we used particle conservation, \m{\int \dd\vec{p}\,\collision_s = 0}. Then,
\begin{align}
\left(\frac{\dd\entropy}{\dd t}\right)_{\text{coll}} 
& = - \sum_{ss'} \int \dd\vec{p}\,\ln F_s(\vec{p})\,\frac{\pd}{\pd p_i}
\int \frac{\dd\vec{k}}{(2\upi)^n}\,\dd\vec{p}'\,k_i k_j 
\mc{Z}_{ss'}(\vec{k}; \vec{p}, \vec{p}') \mc{F}_{ss'}(\vec{p}, \vec{p}')
\notag\\
& = \sum_{ss'} \int \frac{\dd\vec{k}}{(2\upi)^n}\, \dd\vec{p}\,\dd\vec{p}'\,
k_i k_j\,
\frac{\pd \ln F_s(\vec{p})}{\pd p_i}\,\mc{Z}_{ss'}(\vec{k}; \vec{p}, \vec{p}') \mc{F}_{ss'}(\vec{p}, \vec{p}').
\label{eq:auxS1}
\end{align}
Let us swap the dummy variables \m{s \leftrightarrow s'} and \m{\vec{p} \leftrightarrow \vec{p}'} and then apply \eq{eq:ZF} to obtain
\begin{eqnarray}
\left(\frac{\dd\entropy}{\dd t}\right)_{\text{coll}} = -\sum_{ss'} \int \frac{\dd\vec{k}}{(2\upi)^n}\,\int\dd\vec{p}\,\dd\vec{p}'\,
k_i k_j\,
\frac{\pd \ln F_{s'}(\vec{p}')}{\pd p_i'}\,\mc{Z}_{ss'}(\vec{k}; \vec{p}, \vec{p}') \mc{F}_{ss'}(\vec{p}, \vec{p}').
\label{eq:auxS2}
\end{eqnarray}
Upon comparing \eq{eq:auxS2} with \eq{eq:auxS1}, one can put the result in a symmetrized form:
\begin{align}
\left(\frac{\dd\entropy}{\dd t}\right)_{\text{coll}} = 
\frac{1}{2}\sum_{ss'} 
\int \frac{\dd\vec{k}}{(2\upi)^n}\,\dd\vec{p}\,\dd\vec{p}'\, k_i k_j 
& \mc{Z}_{ss'}(\vec{k}; \vec{p}, \vec{p}') \mc{F}_{ss'}(\vec{p}, \vec{p}')
\notag\\
& \times \left(\frac{\pd \ln F_s(\vec{p})}{\pd p_i} - \frac{\pd \ln F_{s'}(\vec{p}')}{\pd p_i'}\right).
\end{align}
But notice that
\begin{eqnarray}
\mc{F}_{ss'}(\vec{p}, \vec{p}') = 
\left(
\frac{\pd \ln F_s(\vec{p})}{\pd p_j} - \frac{\pd \ln F_{s'}(\vec{p}')}{\pd p_j'}
\right) F_s(\vec{p})F_{s'}(\vec{p}').
\end{eqnarray}
Thus,
\begin{align}\label{eq:dsdt}
\left(\frac{\dd\entropy}{\dd t}\right)_{\text{coll}} = 
\frac{1}{2}\sum_{ss'} 
\int \frac{\dd\vec{k}}{(2\upi)^n}\, \dd\vec{p}\,\dd\vec{p}'\,&
\mc{Z}_{ss'}(\vec{k}; \vec{p}, \vec{p}') F_s(\vec{p})F_{s'}(\vec{p}')
\notag\\
& \times \left(\vec{k}\cdot
\frac{\pd \ln F_s(\vec{p})}{\pd \vec{p}}
-
\vec{k}\cdot\frac{\pd \ln F_{s'}(\vec{p}')}{\pd \vec{p}'}
\right)^2 \ge 0.
\end{align}

\section{Conservation laws for on-shell waves}
\label{app:emcons}

Here, we prove the momentum-conservation theorem \eq{eq:mcons} and the energy-conservation theorem \eq{eq:econs} for QL interactions of plasmas with on-shell waves.

\subsection{Momentum conservation}
\label{app:emcons1}

Let us multiply \eq{eq:divJ} by \(k_l\) and integrate over \(\vec{k}\). Then, one obtains
\begin{align}
0 
= & \int \dd\vec{k}\,k_l\,\frac{\pd J}{\pd t}
+ \int \dd\vec{k}\,k_l\,\frac{\pd(\vg^i J)}{\pd x^i}
- \int \dd\vec{k}\,k_l\frac{\pd}{\pd k_i}\left(\frac{\pd w}{\pd x^i}\,J\right)
- 2\int \dd\vec{k}\,k_l\gamma J
\notag\\
= &\, \frac{\pd}{\pd t} \int \dd\vec{k}\,k_l J
+ \frac{\pd}{\pd x^i} \int\dd\vec{k}\,k_l \vg^i J
+ \int \dd\vec{k}\,\frac{\pd w}{\pd x^l}\,J
- 2\int \dd\vec{k}\,w \gamma J.
\label{eq:aux81}
\end{align}
Similarly, multiplying \eq{eq:divF} by \m{\mc{H}_s} and integrating over \m{\vec{p}} yields
\begin{align}
0 
= & \int\dd\vec{p}\,p_l\,\frac{\pd F_s}{\pd t}
+ \int\dd\vec{p}\,p_l\,\frac{\pd (v_s^i F_s)}{\pd x^i}
- \int\dd\vec{p}\,p_l\,\frac{\pd}{\pd p_i} \left(\frac{\pd \mc{H}_s}{\pd x^i} \,F_s\right)
\notag\\
& - \int\dd\vec{p}\,p_l\,\frac{\pd}{\pd p_i} \left(\st{D}_{s,ij}\,\frac{\pd F_s}{\pd p_j}\right)
- \int\dd\vec{p}\,p_l\,\collision_s
\notag\\
= & \,\frac{\pd}{\pd t} \int\dd\vec{p}\,p_l F_s 
+ \frac{\pd}{\pd x^i}\int\dd\vec{p}\, p_l v_s^i F_s
+ \frac{\pd}{\pd x^l}\int\dd\vec{p}\,\Delta_s F_s
+ \int\dd\vec{p}\,\frac{\pd H_{0s}}{\pd x^l} \,F_s
\notag\\
& - \int\dd\vec{p}\,\Delta_s\,\frac{\pd F_s}{\pd x^l}
+ \int\dd\vec{p}\,\st{D}_{s,lj}\,\frac{\pd F_s}{\pd p_j}
- \int\dd\vec{p}\,p_l\,\collision_s.
\label{eq:aux82}
\end{align}
Let us sum up \eq{eq:aux82} over species and also add it with \eq{eq:aux81}. The contribution of the collision integral disappears due to \eq{eq:copcl}, so one obtains
\begin{align}
0 = &\, \frac{\pd}{\pd t} \left(
\sum_s \int\dd\vec{p}\,p_l F_s +
\int \dd\vec{k}\,k_l J
\right)
+ \frac{\pd}{\pd x^i}
\left(
\sum_s \int\dd\vec{p}\, p_l v_s^i F_s
+ \int \dd\vec{k}\,k_l \vg^i J
\right)
\notag\\
& 
+ \sum_s\frac{\pd}{\pd x^l}\int\dd\vec{p}\,\Delta_s F_s
+ \sum_s\int\dd\vec{p}\,\frac{\pd H_{0s}}{\pd x^l}\,F_s
\notag\\
& 
+ \sum_s\int\dd\vec{p}\,\st{D}_{s,lj}\,\frac{\pd F_s}{\pd p_j}
- 2\int \dd\vec{k}\,k_l \gamma J
\notag\\
& 
- \sum_s \int\dd\vec{p}\,\Delta_s\,\frac{\pd F_s}{\pd x^l}
+ \int \dd\vec{k}\,J\,\frac{\pd w}{\pd x^l}.
\label{eq:aux83}
\end{align}
Next, notice that
\begin{align}
2 \int\dd\vec{k}\,k_l \gamma J
& = 2 \upi \sum_s \int\dd\vec{p}\,\dd\vec{k}\,
k_l k_j\,\frac{|\vec{\alpha}_s^\dag \vec{\eta}|^2}{\pd_{\omega}\Lambda}\,J 
\delta(w - \vec{k} \cdot \vec{v}_s)\,\frac{\pd F_s}{\pd p_j}
\notag \\
& 
= \sum_s \int\dd\vec{p}\,\st{D}_{s,l j}\,\frac{\pd F_s}{\pd p_j}.
\label{eq:aux84}
\end{align}
Also, assuming that \m{\matr{\Xi}_0}, \m{|\vec{\alpha}_s^\dag \vec{\eta}|^2}, and \m{\vec{\eta}^\dag\matr{\wp}_s\vec{\eta}} are independent of \m{\vec{x}} and using \eq{eq:aeta}, one gets
\begin{align}
& \sum_s \int \dd\vec{p}\,\Delta_s\frac{\pd F_s}{\pd x^l}
\notag \\
& = \sum_s \int \dd\vec{p}\,\frac{\pd F_s}{\pd x^l} \bigg(
\frac{\pd}{\pd p_i} \fint \dd\omega\,\dd\vec{k}\,\frac{k_i}{2(\omega - \vec{k} \cdot \vec{v}_s)}\,
|\vec{\alpha}_s^\dag \vec{\eta}|^2 (h(\vec{k}) + h(-\vec{k}))\,
\delta(\omega - w(\vec{k}))
\notag \\
& 
\hphantom{= \sum_s \int \dd\vec{p}\,\frac{\pd F_s}{\pd x^l} \bigg(}
+ \frac{1}{2} \int \dd\omega\,\dd\vec{k}\,(\vec{\eta}^\dag\matr{\wp}_s\vec{\eta})
(h(\vec{k}) + h(-\vec{k}))\,\delta(\omega - w(\vec{k}))
\bigg)
\notag \\
& = - \frac{1}{2}\sum_s \fint \dd\omega\,\dd\vec{k}\,\dd\vec{p}\,
(h(\vec{k}) + h(-\vec{k}))\,
\delta(\omega - w(\vec{k}))\,
\frac{k_i |\vec{\alpha}_s^\dag \vec{\eta}|^2}{\omega - \vec{k} \cdot \vec{v}_s}\,
\frac{\pd^2 F_s}{\pd x^l\pd p_i}
\notag \\
& 
\hphantom{\,=\,}
+ \frac{1}{2}\sum_s \fint \dd\omega\,\dd\vec{k}\,\dd\vec{p}\,
(h(\vec{k}) + h(-\vec{k}))\,\delta(\omega - w(\vec{k}))\,
\frac{\pd}{\pd x^l}\,(\vec{\eta}^\dag\matr{\wp}_s\vec{\eta} F_s)
\notag \\
& = - \sum_s \int \dd\omega\,\dd\vec{k}\,
h(\vec{k})\,
\delta(\omega - w(\vec{k}))\,\frac{\pd}{\pd x^l}
\fint\dd\vec{p}\,\left(
\frac{k_i |\vec{\alpha}_s^\dag \vec{\eta}|^2}{\omega - \vec{k} \cdot \vec{v}_s}
\frac{\pd F_s}{\pd p_i}
- \vec{\eta}^\dag\matr{\wp}_s\vec{\eta}F_s
\right)
\notag \\
& = - \sum_s \int \dd\omega\,\dd\vec{k}\,
h(\vec{k})\,
\delta(\omega - w(\vec{k}))\,\frac{\pd (\vec{\eta}^\dag\matr{\Xi}\vec{\eta})}{\pd x^l}
\notag \\
& = - \sum_s \int \dd\vec{k}\,
h(\vec{k})\,\frac{\pd \Lambda(w(\vec{k}), \vec{k})}{\pd x^l}
\notag \\
& = \sum_s \int \dd\vec{k}\,
J\,\frac{\pd w}{\pd x^l},
\label{eq:aux85}
\end{align}
where we also used \eq{eq:Jhwx}. Substituting \eq{eq:aux84} and \eq{eq:aux85} into \eq{eq:aux83} leads to \eq{eq:mcons}.

\subsection{Energy conservation}
\label{app:emcons2}

Let us multiply \eq{eq:divJ} by \(w\) and integrate over \(\vec{k}\). Then, one obtains
\begin{align}
0 
= & \int \dd\vec{k}\,w\,\frac{\pd J}{\pd t}
+ \int \dd\vec{k}\,w\,\frac{\pd(\vg^i J)}{\pd x^i}
- \int \dd\vec{k}\,w\,\frac{\pd}{\pd k_i}\left(\frac{\pd w}{\pd x^i}J\right)
- 2\int \dd\vec{k}\,w \gamma J
\notag\\
= &\, \frac{\pd}{\pd t} \int \dd\vec{k}\,w J
- \int \dd\vec{k}\,\frac{\pd w}{\pd t}\,J
+ \frac{\pd}{\pd x^i} \int\dd\vec{k}\,w \vg^i J
- \int \dd\vec{k}\,\frac{\pd w}{\pd x^i}\, \vg^i J
\notag\\
&
+ \int \dd\vec{k}\,\vg^i\,\frac{\pd w}{\pd x^i}\,J
- 2\int \dd\vec{k}\,w \gamma J
\notag\\
= &\, \frac{\pd}{\pd t} \int \dd\vec{k}\,w J
+ \frac{\pd}{\pd x^i}\int \dd\vec{k}\,w \vg^i J
- \int \dd\vec{k}\,\frac{\pd w}{\pd t}\,J
- 2\int \dd\vec{k}\,w \gamma J.
\label{eq:aux711}
\end{align}
Similarly, multiplying \eq{eq:divF} by \m{\mc{H}_s} and integrating over \m{\vec{p}} yields
\begin{align}
0 
= & \int\dd\vec{p}\,\mc{H}_s\,\frac{\pd F_s}{\pd t}
+ \int\dd\vec{p}\,\mc{H}_s\,\frac{\pd (v_s^i F_s)}{\pd x^i}
- \int\dd\vec{p}\,\mc{H}_s\,\frac{\pd}{\pd p_i} \left(\frac{\pd \mc{H}_s}{\pd x^i} \,F_s\right)
\notag\\
& - \int\dd\vec{p}\,\mc{H}_s\,\frac{\pd}{\pd p_i} \left(\st{D}_{s,ij}\,\frac{\pd F_s}{\pd p_j}\right)
- \int\dd\vec{p}\,\mc{H}_s\,\collision_s
\notag\\
= & \,\frac{\pd}{\pd t} \int\dd\vec{p}\,\mc{H}_s F_s 
- \int\dd\vec{p}\,\frac{\pd \mc{H}_s}{\pd t}\,F_s
+ \frac{\pd}{\pd x^i}\int\dd\vec{p}\, \mc{H}_s v_s^i F_s
- \int\dd\vec{p}\,\frac{\pd \mc{H}_s }{\pd x^i}\,v_s^i F_s
\notag\\
& + \int\dd\vec{p}\,v_s^i\,\frac{\pd \mc{H}_s}{\pd x^i} \,F_s
+ \int\dd\vec{p}\,v_s^i \st{D}_{s,ij}\,\frac{\pd F_s}{\pd p_j}
- \int\dd\vec{p}\,\mc{H}_s\,\collision_s
\notag\\
= & \,\frac{\pd}{\pd t} \int\dd\vec{p}\,\mc{H}_s F_s 
+ \frac{\pd}{\pd x^i}\int\dd\vec{p}\, \mc{H}_s v_s^i F_s
- \int\dd\vec{p}\,\frac{\pd H_{0s}}{\pd t}\,F_s
- \int\dd\vec{p}\,\frac{\pd \Delta_s}{\pd t}\,F_s
\notag\\
& + \int\dd\vec{p}\,v_s^i \st{D}_{s,ij}\,\frac{\pd F_s}{\pd p_j}
- \int\dd\vec{p}\,\mc{H}_s\,\collision_s
\notag\\
= & \,\frac{\pd}{\pd t} \int\dd\vec{p}\,H_{0s} F_s 
+ \frac{\pd}{\pd x^i}\int\dd\vec{p}\, H_{0s} v_s^i F_s
+ \frac{\pd}{\pd x^i}\int\dd\vec{p}\, \Delta_s v_s^i F_s
- \int\dd\vec{p}\,\frac{\pd H_{0s}}{\pd t}\,F_s
\notag\\
& 
+ \int\dd\vec{p}\,\Delta_s\,\frac{\pd F_s}{\pd t}
+ \int\dd\vec{p}\,v_s^i \st{D}_{s,ij}\,\frac{\pd F_s}{\pd p_j}
- \int\dd\vec{p}\,\mc{H}_s\,\collision_s.
\label{eq:aux72}
\end{align}
Let us sum up \eq{eq:aux72} over species and also add it with \eq{eq:aux711}. The contribution of the collision integral disappears due to \eq{eq:copcl}, so one obtains
\begin{align}
0 = &\, \frac{\pd}{\pd t} \left(
\sum_s \int\dd\vec{p}\,H_{0s} F_s +
\int \dd\vec{k}\,w J
\right)
+ \frac{\pd}{\pd x^i}
\left(
\sum_s \int\dd\vec{p}\, H_{0s} v_s^i F_s
+ \int \dd\vec{k}\,w \vg^i J
\right)
\notag\\
& 
+ \frac{\pd}{\pd x^i}\sum_s \int\dd\vec{p}\, \Delta_s v_s^i F_s
- \sum_s\int\dd\vec{p}\,\frac{\pd H_{0s}}{\pd t}\,F_s
\notag\\
& 
+ \sum_s \int\dd\vec{p}\,v_s^i \st{D}_{s,ij}\,\frac{\pd F_s}{\pd p_j}
- 2\int \dd\vec{k}\,w \gamma J
\notag\\
& 
+ \sum_s \int\dd\vec{p}\,\Delta_s\,\frac{\pd F_s}{\pd t}
- \int \dd\vec{k}\,J\,\frac{\pd w}{\pd t}.
\label{eq:aux73}
\end{align}
Next, notice that
\begin{align}
2 \int\dd\vec{k}\,w \gamma J
& = 2 \upi \sum_s \int\dd\vec{p}\,\dd\vec{k}\,
w k_j\,\frac{|\vec{\alpha}_s^\dag \vec{\eta}|^2}{\pd_{\omega}\Lambda}\,J 
\delta(w - \vec{k} \cdot \vec{v}_s)\,\frac{\pd F_s}{\pd p_j}
\notag \\
& 
= \sum_s \int\dd\vec{p}\,v_s^i \st{D}_{s,i j}\,\frac{\pd F_s}{\pd p_j}.
\label{eq:aux74}
\end{align}
Also, assuming that \m{\matr{\Xi}_0}, \m{|\vec{\alpha}_s^\dag \vec{\eta}|^2}, and \m{\vec{\eta}^\dag\matr{\wp}_s\vec{\eta}} are independent of \m{t} and using \eq{eq:aeta}, one gets
\begin{align}
& \sum_s \int \dd\vec{p}\,\Delta_s\frac{\pd F_s}{\pd t}
\notag \\
& = \sum_s \int \dd\vec{p}\,\frac{\pd F_s}{\pd t} \,\bigg(
\frac{\pd}{\pd p_i} \fint \dd\omega\,\dd\vec{k}\,\frac{k_i}{2(\omega - \vec{k} \cdot \vec{v}_s)}\,
|\vec{\alpha}_s^\dag \vec{\eta}|^2 (h(\vec{k}) + h(-\vec{k}))\,
\delta(\omega - w(\vec{k}))
\notag \\
& 
\hphantom{= \sum_s \int \dd\vec{p}\,\frac{\pd F_s}{\pd t}  \bigg(}
+ \frac{1}{2}\int \dd\omega\,\dd\vec{k}\,(\vec{\eta}^\dag\matr{\wp}_s\vec{\eta})
(h(\vec{k}) + h(-\vec{k}))\,\delta(\omega - w(\vec{k}))
\bigg)
\notag \\
& = - \frac{1}{2}\sum_s \fint \dd\omega\,\dd\vec{k}\,\dd\vec{p}\,
(h(\vec{k}) + h(-\vec{k}))\,\delta(\omega - w(\vec{k}))\,
\frac{k_i |\vec{\alpha}_s^\dag \vec{\eta}|^2}{\omega - \vec{k} \cdot \vec{v}_s}\frac{\pd^2 F_s}{\pd t\pd p_i}
\notag \\
& \hphantom{\,=\,}
+ \frac{1}{2}\sum_s \fint \dd\omega\,\dd\vec{k}\,\dd\vec{p}\,
(h(\vec{k}) + h(-\vec{k}))\,\delta(\omega - w(\vec{k}))\,
\frac{\pd}{\pd t}\,(\vec{\eta}^\dag\matr{\wp}_s\vec{\eta} F_s)
\notag \\
& = - \sum_s \int \dd\omega\,\dd\vec{k}\,
h(\vec{k})\,
\delta(\omega - w(\vec{k}))\,\frac{\pd}{\pd t}
\fint\dd\vec{p}\left( 
\frac{k_i |\vec{\alpha}_s^\dag \vec{\eta}|^2}{\omega - \vec{k} \cdot \vec{v}_s}
\frac{\pd F_s}{\pd p_i}
- \vec{\eta}^\dag\matr{\wp}_s\vec{\eta} F_s
\right)
\notag \\
& = - \sum_s \int \dd\omega\,\dd\vec{k}\,
h(\vec{k})\,
\delta(\omega - w(\vec{k}))\,\frac{\pd (\vec{\eta}^\dag\matr{\Xi}\vec{\eta})}{\pd t}
\notag \\
& = - \sum_s \int \dd\vec{k}\,
h(\vec{k})\,\frac{\pd \Lambda(w(\vec{k}), \vec{k})}{\pd t}
\notag \\
& = \sum_s \int \dd\vec{k}\,
J\,\frac{\pd w}{\pd t},
\label{eq:aux75}
\end{align}
where we also used \eq{eq:Jhwt}. Substituting \eq{eq:aux74} and \eq{eq:aux75} into \eq{eq:aux73} leads to \eq{eq:econs}.

\section{Uniqueness of the entropy-preserving distribution}
\label{app:gibbs}

Here, we prove that the Boltzmann--Gibbs distribution is the only distribution for which the entropy density \m{\sigma} is conserved. According to \eq{eq:dsdt}, \m{\sigma} is conserved when
\begin{eqnarray}\label{eq:kG}
\delta(\vec{k} \cdot (\vec{v}_s - \vec{v}'_{s'}))\,
(\vec{k}\cdot\vec{G}_{ss'}(\vec{p}, \vec{p}'))^2 = 0
\end{eqnarray}
(for all \m{\vec{p}}, \m{\vec{p}'}, and \m{\vec{k}}, as well as all \m{s} and \m{s'}), where
\begin{eqnarray}\label{eq:vGdef}
\vec{G}_{ss'}(\vec{p}, \vec{p}') \doteq \frac{\pd \ln F_s(\vec{p})}{\pd \vec{p}} - \frac{\pd \ln F_{s'}(\vec{p}')}{\pd \vec{p}'}.
\end{eqnarray}
Let us decompose the vector \m{\vec{G}_{ss'}(\vec{p}, \vec{p}')} into components parallel and perpendicular to the vector \m{\vec{v}_s - \vec{v}'_{s'}}:
\begin{eqnarray}\label{eq:Gess}
\vec{G}_{ss'}(\vec{p}, \vec{p}') = \lambda_{ss'}(\vec{v}_s, \vec{v}'_{s'})\,(\vec{v}_s - \vec{v}'_{s'}) + \vec{G}_{ss'}^\perp(\vec{p}, \vec{p}'),
\end{eqnarray}
where \m{\lambda_{ss'}(\vec{v}_s, \vec{v}'_{s'})} is a scalar function. (Because the velocities are functions of the momenta, one can as well consider \m{\lambda_{ss'}} as a function of \m{\vec{p}} and \m{\vec{p}'}.) Due to the presence of the delta function in \eq{eq:kG}, the contribution of the first term to \eq{eq:kG} is zero, so \eq{eq:kG} can be written as 
\begin{eqnarray}\label{eq:kG2}
\delta(\vec{k} \cdot (\vec{v}_s - \vec{v}'_{s'}))\,
(\vec{k}\cdot\vec{G}_{ss'}^\perp(\vec{p}, \vec{p}'))^2 = 0.
\end{eqnarray}
By considering this formula for \m{\vec{k}} parallel to \m{\vec{G}_{ss'}^\perp(\vec{p}, \vec{p}')} (and thus perpendicular to \m{\vec{v}_s - \vec{v}'_{s'}}), one finds that \m{\vec{G}_{ss'}^\perp(\vec{p}, \vec{p}') = 0}. Combined with \eq{eq:vGdef} and \eq{eq:Gess}, this yields
\begin{eqnarray}\label{eq:FFss}
\frac{\pd \ln F_s(\vec{p})}{\pd \vec{p}}
-
\frac{\pd \ln F_{s'}(\vec{p}')}{\pd \vec{p}'}
= \lambda_{ss'}(\vec{v}_s, \vec{v}'_{s'})\,(\vec{v}_s - \vec{v}'_{s'}).
\end{eqnarray}
Also, by swapping \m{\vec{p} \leftrightarrow \vec{p}'} and \m{s \leftrightarrow s'}, one finds that 
\begin{eqnarray}\label{eq:lamss}
\lambda_{ss'}(\vec{v}_s, \vec{v}'_{s'}) = \lambda_{s's}(\vec{v}'_{s'}, \vec{v}_s).
\end{eqnarray}

Equation \eq{eq:FFss} yields, in particular, that\footnote{The idea of this argument was brought to author's attention by G.~W.\ Hammett and is taken from \citep{foot:landreman17}, where it is applied to single-species plasmas with a specific~\m{\mc{H}_s}.}
\begin{subequations}
\begin{eqnarray}
\frac{\pd \ln F_s(\vec{p})}{\pd p_2}
-
\frac{\pd \ln F_{s'}(\vec{p}')}{\pd p_2'}
= \lambda_{ss'}(\vec{v}_s, \vec{v}'_{s'})\,(v_{s,2} - v'_{s',2}),
\\
\frac{\pd \ln F_s(\vec{p})}{\pd p_3}
-
\frac{\pd \ln F_{s'}(\vec{p}')}{\pd p_3'}
= \lambda_{ss'}(\vec{v}_s, \vec{v}'_{s'})\,(v_{s,3} - v'_{s',3}),
\end{eqnarray}
\end{subequations}
where we have assumed some coordinate axes in the momentum and velocity space labeled  \m{\cb{1, 2, 3, \ldots}}. Then,
\begin{subequations}\label{eq:LamLam}
\begin{eqnarray}\label{eq:LamLam1}
\frac{\pd^2\ln F_s(\vec{p})}{\pd p_2 \pd v_{s,1}}
= \frac{\pd \lambda_{ss'}(\vec{v}_s, \vec{v}'_{s'})}{\pd v_{s,1}}\,(v_{s,2} - v'_{s',2}),
\\
\frac{\pd^2 \ln F_s(\vec{p})}{\pd p_3 \pd v_{s,1}}
= \frac{\pd \lambda_{ss'}(\vec{v}_s, \vec{v}'_{s'})}{\pd v_{s,1}}\,(v_{s,3} - v'_{s',3}),
\end{eqnarray}
\end{subequations}
where the derivative with respect to \m{v_{s,1}} is taken at fixed \m{v_{s, i \ne 1}} and at fixed \m{\vec{v}'_{s'}}. Due to \eq{eq:FFss}, \m{\lambda_{ss'}(\vec{v}_s, \vec{v}'_{s'})} is continuous for all \m{F_s} and \m{F_{s'}}. (Here we consider only physical distributions, which are always differentiable.) Then, \eq{eq:LamLam} leads to
\begin{eqnarray}
\frac{1}{v_{s,2} - v'_{s',2}}\frac{\pd^2\ln F_s(\vec{p})}{\pd p_2 \pd v_{s,1}}
= \frac{1}{v_{s,3} - v'_{s',3}}\frac{\pd^2 \ln F_s(\vec{p})}{\pd p_3 \pd v_{s,1}}.
\end{eqnarray}
By differentiating this with respect to \m{v'_{s',2}}, one obtains
\begin{eqnarray}
\frac{\pd^2\ln F_s(\vec{p})}{\pd p_2 \pd v_{s,1}} = 0,
\end{eqnarray}
whence \eq{eq:LamLam1} yields
\begin{eqnarray}\label{eq:lvsz}
\frac{\pd \lambda_{ss'}(\vec{v}_s, \vec{v}'_{s'})}{\pd v_{s,1}} = 0.
\end{eqnarray}
By repeating this argument for other axes and for \m{\vec{v}'} instead of \m{\vec{v}}, one can also extend \eq{eq:lvsz} to
\begin{eqnarray}
\frac{\pd \lambda_{ss'}(\vec{v}_s, \vec{v}'_{s'})}{\pd \vec{v}} = 0,
\qquad
\frac{\pd \lambda_{ss'}(\vec{v}_s, \vec{v}'_{s'})}{\pd \vec{v}'} = 0.
\end{eqnarray}
Hence, \m{\lambda_{ss'}(\vec{v}_s, \vec{v}'_{s'})} is actually independent of the velocities; \ie \m{\lambda_{ss'}(\vec{v}_s, \vec{v}'_{s'}) = \lambda_{ss'}}. Using this along with \eq{eq:lamss}, one also finds that
\begin{eqnarray}\label{eq:lssp}
\lambda_{ss'} = \lambda_{s's}.
\end{eqnarray}

Let us rewrite \eq{eq:FFss} as follows:
\begin{eqnarray}
\frac{\pd \ln F_s(\vec{p})}{\pd \vec{p}} - \lambda_{ss'} \vec{v}_s
= \frac{\pd \ln F_{s'}(\vec{p}')}{\pd \vec{p}'} - \lambda_{s's} \vec{v}'_{s'}.
\end{eqnarray}
Here, the left-hand side is independent of \m{\vec{p}'} and the right-hand side is independent of \m{\vec{p}}, so both must be equal to some vector
\begin{equation}\label{eq:mussp}
\vec{\mu}_{ss'} = \vec{\mu}_{s's}
\end{equation}
that is independent of both \m{\vec{p}} and \m{\vec{p}'}. Because \m{\vec{v}_s = \pd_{\vec{p}}\mc{H}_s}, this is equivalent to
\begin{subequations}\label{eq:FFsub}
\begin{eqnarray}
\ln F_s(\vec{p}) - \lambda_{ss'} \mc{H}_s(\vec{p}) = \vec{\mu}_{ss'} \cdot \vec{p} + \eta_{ss'}
\end{eqnarray}
(and similarly for \m{\vec{p}'}), where the integration constant \m{\eta_{ss'}} is independent of both~\m{\vec{p}} and~\m{\vec{p}'}. This is supposed to hold for any \m{s'}, so one can also write
\begin{eqnarray}
\ln F_s(\vec{p}) - \lambda_{ss''} \mc{H}_s(\vec{p}) = \vec{\mu}_{ss''} \cdot \vec{p} + \eta_{ss''},
\end{eqnarray}
\end{subequations}
where \m{s''} is any other species index. Subtracting equations \eq{eq:FFsub} from each other gives
\begin{eqnarray}\label{eq:l1}
(\lambda_{ss'} - \lambda_{ss''}) \mc{H}_s(\vec{p}) = (\vec{\mu}_{ss'} - \vec{\mu}_{ss''}) \cdot \vec{p} + \eta_{ss'} - \eta_{ss''}.
\end{eqnarray}
By differentiating this with respect to \m{\vec{p}}, one finds
\begin{eqnarray}\label{eq:l2}
(\lambda_{ss'} - \lambda_{ss''}) \vec{v}_s = \vec{\mu}_{ss'} - \vec{\mu}_{ss''}.
\end{eqnarray}
By differentiating this further with respect to \m{\vec{v}_s}, one obtains \m{\lambda_{ss'} = \lambda_{ss''}}. Then, \eq{eq:l2} yields \m{\vec{\mu}_{ss'} = \vec{\mu}_{ss''}}, and \eq{eq:l1} yields \m{\eta_{ss'} = \eta_{ss''}}. In other words, the functions \m{\lambda_{ss'}}, \m{\vec{\mu}_{ss'}}, and \m{\eta_{ss'}} are independent of their second index and thus can as well be written as 
\begin{eqnarray}
\lambda_{ss'} = \lambda_{s},
\qquad 
\vec{\mu}_{ss'} = \vec{\mu}_s,
\qquad
\eta_{ss'} = \eta_s.
\end{eqnarray}
But then, \eq{eq:lssp} and \eq{eq:mussp} also yield \m{\lambda_s = \lambda_{s'} \equiv \lambda} and \m{\vec{\mu}_s = \vec{\mu}_{s'} \equiv \vec{\mu}}. Therefore, \eq{eq:FFsub} can be written as
\begin{eqnarray}
F_s(\vec{p}) = \const_s \times \exp(\lambda \mc{H}_s(\vec{p}) + \vec{\mu} \cdot \vec{p}),
\end{eqnarray}
which is the Boltzmann--Gibbs distribution (\Sec{sec:centr}). This proves that a plasma that conserves its entropy density necessarily has the Boltzmann--Gibbs distribution.

\section{Total momentum and energy}
\label{app:enmom}

Here, we show that the total momentum and energy in the OC--wave representation equals the total momentum and energy in the particle--field representation.

\subsection{Nonrelativistic electrostatic interactions}
\label{sec:enmomes}

\subsubsection{Momentum}
\label{sec:momes}

Assuming the notation \m{\mc{P}_l \doteq \sum_s \int \dd\vec{p}\,p_l \avr{f}_s} and using \eq{eq:Thetes2} for \m{\vec{\Uptheta}_s}, one can represent the OC momentum density as follows:
\begin{align}
\sum_s \int \dd\vec{p}\,p_l F_s
& = \mc{P}_l 
+ \frac{1}{2}\sum_s \int \dd\vec{p}\,p_l\,\frac{\pd}{\pd p_i}\left(\Uptheta_{s,ij}\,\frac{\pd \avr{f}_s}{\pd p_j}\right)
\notag\\
& \approx \mc{P}_l 
- \frac{1}{2}\sum_s \int \dd\vec{p}\,\Uptheta_{s,lj}\,\frac{\pd F_s}{\pd p_j}
\notag\\
& = \mc{P}_l 
-\int \dd\vec{k}\,h(\vec{k})\left.\frac{\pd}{\pd \vartheta}\sum_s e_s^2 \fint \dd\vec{p}\,
\frac{k_l}{w(\vec{k}) - \vec{k} \cdot \vec{v}_s + \vartheta}\,
\vec{k}\cdot\frac{\pd F_s}{\pd \vec{p}}
\right|_{\vartheta=0}
\notag\\
& = \mc{P}_l 
-\int \dd\vec{k}\,
k_l h(\vec{k})\left.\frac{\pd}{\pd \vartheta}\left(\frac{k^2(\epsilon_{\parallel\herm}(w(\vec{k}) + \vartheta, \vec{k})-1)}{4\upi}\right)
\right|_{\vartheta=0}
\notag\\
& = \mc{P}_l - \int \dd\vec{k}\,k_l J,
\end{align}
where we substituted \eq{eq:Jes}. This leads to \eq{eq:totmes}.

\subsubsection{Energy}
\label{sec:enes}

Assuming the notation \m{\mc{K} \doteq \sum_s \int \dd\vec{p}\,H_{0s} \avr{f}_s} and using \eq{eq:Thetes2} for \m{\matr{\Uptheta}_s}, one can represent the OC energy density as follows:
\begin{align}
\sum_s \int \dd\vec{p}\,H_{0s} F_s
& = \mc{K} 
+ \frac{1}{2}\sum_s \int \dd\vec{p}\,\frac{p^2}{2m_s}\,\frac{\pd}{\pd p_i}\left(\Uptheta_{s,ij}\,\frac{\pd \avr{f}_s}{\pd p_j}\right)
\notag\\
& \approx \mc{K}
- \frac{1}{2}\sum_s \int \dd\vec{p}\,v_s^i\Uptheta_{s,ij}\,\frac{\pd F_s}{\pd p_j}
\notag\\
& = \mc{K} 
-\int \dd\vec{k}\,h(\vec{k})\,\frac{\pd}{\pd \vartheta}\sum_s e_s^2 \fint \dd\vec{p}
\left.
\frac{\vec{k}\cdot{\vec{v}}}{w(\vec{k}) - \vec{k} \cdot \vec{v}_s + \vartheta}\,
\vec{k}\cdot\frac{\pd F_s}{\pd \vec{p}}
\right|_{\vartheta=0}.
\label{eq:auxK}
\end{align}
Notice that 
\begin{align}
\frac{\pd}{\pd \vartheta}\frac{\vec{k}\cdot{\vec{v}}}{w(\vec{k}) - \vec{k} \cdot \vec{v}_s + \vartheta}
& = \frac{\pd}{\pd \vartheta}\left(-1 + \frac{w(\vec{k}) + \vartheta}{w(\vec{k}) - \vec{k} \cdot \vec{v}_s + \vartheta}\right)
\notag\\
& = \frac{1}{w(\vec{k}) - \vec{k} \cdot \vec{v}_s + \vartheta}
+ w(\vec{k})\,\frac{\pd}{\pd \vartheta} \frac{1}{w(\vec{k}) - \vec{k} \cdot \vec{v}_s + \vartheta}.
\label{eq:kwkw}
\end{align}
Then,
\begin{align}
\sum_s \int \dd\vec{p}\,H_{0s} F_s
= & \,\mc{K} -\int \dd\vec{k}\,w(\vec{k}) h(\vec{k})\,\frac{\pd}{\pd \vartheta}
\sum_s e_s^2 \fint \dd\vec{p}
\left.
\frac{\vec{k}}{w(\vec{k}) - \vec{k} \cdot \vec{v}_s + \vartheta}\,
\cdot\frac{\pd F_s}{\pd \vec{p}}
\right|_{\vartheta=0}
\notag\\
& -\int \dd\vec{k}\,h(\vec{k}) \sum_s e_s^2 \fint \dd\vec{p}\,
\frac{\vec{k}}{w(\vec{k}) - \vec{k} \cdot \vec{v}_s}\,
\cdot\frac{\pd F_s}{\pd \vec{p}}
\notag\\
= & \,\mc{K} -\int \dd\vec{k}\,w(\vec{k}) h(\vec{k})
\left.\frac{\pd}{\pd \vartheta}\left(\frac{k^2(\epsilon_{\parallel\herm}(w(\vec{k}) + \vartheta, \vec{k})-1)}{4\upi}\right)
\right|_{\vartheta=0}
\notag\\
& -\int \dd\vec{k}\,h(\vec{k})\,
\frac{k^2(\epsilon_{\parallel\herm}(w(\vec{k}), \vec{k})-1)}{4\upi}.
\end{align}
Using \eq{eq:esdr} and \eq{eq:Jes}, one obtains that the sum of the OC and wave energy is given by
\begin{align}
\sum_s \int \dd\vec{p}\,H_{0s} F_s + \int \dd\vec{k}\,w J 
& = \mc{K} + \int \dd\vec{k}\,h(\vec{k})\,\frac{k^2}{4\upi}
\notag\\
& = \mc{K} + \sum_\sigma\frac{\avr{k}_\sigma^2 |\env{\varphi}_\sigma|^2}{16\upi}
\notag\\
& = \sum_s \int \dd\vec{p}\left(\frac{p^2}{2m_s} + e_s\avr{\varphi}_s\right)\avr{f}_s 
+ \frac{1}{8\upi}\,\avr{\smash{\osc{\vec{E}}}^\dag \osc{\vec{E}}},
\end{align}
where we also substituted \eq{eq:hdef}.

\subsection{Relativistic electromagnetic interactions}
\label{sec:enmomem}

\subsubsection{Momentum}
\label{sec:momem}

Let us assume the notation \m{\vec{\mc{P}} \doteq \sum_s \int \dd\vec{p}\,\vec{p} \avr{f}_s} and
\begin{eqnarray}\label{eq:chia}
\matr{\chi}(\omega, \vec{k}) \doteq \sum_s \frac{4\upi e_s^2}{\omega^2}\fint \dd\vec{p}\,
\frac{\vec{v}_s \vec{v}_s^\dag}{\omega - \vec{k} \cdot \vec{v}_s}\,\vec{k} \cdot \frac{\pd F_s}{\pd\vec{p}}
= \matr{\epsilon}(\omega, \vec{k}) - \matr{1} + \frac{\matr{\mcc{w}}_p}{\omega^2}.
\end{eqnarray}
Then, using \eq{eq:Thetem2} for \m{\vec{\Uptheta}_s}, one can represent the OC momentum density as follows:
\begin{align}
\sum_s \int \dd\vec{p}\,p_l F_s
& \approx \mc{P}_l 
- \frac{1}{2}\sum_s \int \dd\vec{p}\,\Uptheta_{s,lj}\,\frac{\pd F_s}{\pd p_j}
\notag\\
& = \mc{P}_l 
-\int \dd\vec{k}\,k_l h(\vec{k})\left.\frac{\pd}{\pd \vartheta}\sum_s e_s^2 \fint \dd\vec{p}\,
\frac{1}{w^2(\vec{k})}
\frac{(\vec{\eta}^\dag \vec{v}_s \vec{v}_s^\dag \vec{\eta})}{w(\vec{k}) - \vec{k} \cdot \vec{v}_s + \vartheta}\,\vec{k} \cdot \frac{\pd F_s}{\pd p_j}
\right|_{\vartheta=0}
\notag\\
& = \mc{P}_l 
-\int \dd\vec{k}\,
\frac{k_l h(\vec{k})}{4\upi w^2(\vec{k})}
\left.\vec{\eta}^\dag\frac{\pd}{\pd \omega}\left(
\omega^2\matr{\chi}(\omega, \vec{k})
\right)\vec{\eta}
\right|_{\omega = w(\vec{k})}
\notag\\
& = \mc{P}_l 
+ \int \dd\vec{k}\,\frac{k_l h(\vec{k})}{2\upi w(\vec{k})}
-\int \dd\vec{k}\,
\frac{k_l h(\vec{k})}{4\upi w^2(\vec{k})}
\left.\vec{\eta}^\dag\frac{\pd}{\pd \omega}\left(
\omega^2\matr{\epsilon}(\omega, \vec{k})
\right)\vec{\eta}
\right|_{\omega = w(\vec{k})}
\notag\\
& = \mc{P}_l + \int \dd\vec{k}\,\frac{k_l}{4\upi w(\vec{k})}\,(h(\vec{k}) + h(-\vec{k})) - \int \dd\vec{k}\,k_l J,
\label{eq:auxFP}
\end{align}
where we substituted \eq{eq:Jem} and used \eq{eq:wsym}. Next, let us rewrite \eq{eq:Pkin} as
\begin{eqnarray}
\vec{\mc{P}}
= \vec{\mc{P}}^{(\text{kin})} + \avr{(\osc{\vec{A}}/c)\sum_s e_s \textstyle \int \dd\vec{p}\,f_s}
= \vec{\mc{P}}^{(\text{kin})} + \frac{1}{4 \upi c}\, \avr{\osc{\vec{A}} (\del \cdot \osc{\vec{E}})},
\end{eqnarray}
where the last equality is due to Gauss's law. This gives
\begin{eqnarray}
\mc{P}_l - \mc{P}^{(\text{kin})}_l
= \frac{1}{4 \upi}\, \avr{(-\ii\oper{\omega}^{-1}\osc{E}_l)(\pd_j \osc{E}^j)}
= -\frac{\ii}{4 \upi}\, \avr{(\oper{\omega}^{-1}\osc{E}_l)(\pd_j \osc{E}^j)^*}.
\end{eqnarray}
Then, using \eq{eq:LRs} and also \eq{eq:UW} for \m{\matrU}, one obtains
\begin{align}
\mc{P}_l - \mc{P}^{(\text{kin})}_l
& \approx -\frac{\ii}{4 \upi}\, 
\int \dd\omega\,\dd\vec{k}\,\omega^{-1} \st{U}_l{}^j(\omega, \vec{k})(\ii k_j)^*
\notag\\
& \approx - 
\int \dd\vec{k}\,\frac{\vec{k} \cdot \vec{\eta}^*}{4\upi w(\vec{k})}\,\eta_l
(h(\vec{k}) + h(-\vec{k})),
\end{align}
and thus \eq{eq:auxFP} can be written as follows:
\begin{align}
\sum_s \int \dd\vec{p}\,\vec{p} F_s & + \int \dd\vec{k}\,\vec{k} J
\approx \vec{\mc{P}} + \int \dd\vec{k}\,\frac{\vec{k}}{4\upi w(\vec{k})}\,(h(\vec{k}) + h(-\vec{k}))
\notag\\
& \approx \vec{\mc{P}}^{(\text{kin})} 
+ \int \dd\vec{k}\,\frac{1}{4\upi w(\vec{k})}\,
(\vec{k} - \vec{\eta} (\vec{k} \cdot \vec{\eta}^*))\,
(h(\vec{k}) + h(-\vec{k}))
\notag\\
& = \vec{\mc{P}}^{(\text{kin})}
+ \re \int \dd\vec{k}\,\frac{h(\vec{k})}{2\upi w(\vec{k})}\,
\vec{\eta}^* \times (\vec{k} \times \vec{\eta}),
\end{align}
where we used \eq{eq:etastar} and \eq{eq:wsym} again. For an eikonal wave \eq{eq:eikem}, which has \m{h(\vec{k}) = \delta(\vec{k} - \avr{\vec{k}})|\env{\vec{E}}|^2/4} (\Sec{sec:wigvec}), this gives
\begin{align}
\re \int \dd\vec{k}\,\frac{h(\vec{k})}{2\upi w(\vec{k})}\,
\vec{\eta}^* \times (\vec{k} \times \vec{\eta})
= \frac{1}{8\upi c} \,\re \left(\smash{\env{\vec{E}}}^* \times \left(\frac{c\avr{\vec{k}}}{\avr{\omega}} \times \env{\vec{E}}\right)\right)
= \frac{\avr{\osc{\vec{E}} \times \osc{\vec{B}}}}{4\upi c}.
\end{align}
In case of a broadband spectrum, the same equality applies as well, because contributions of the individual eikonal waves to both left-hand side and the right-hand side are additive. (Alternatively, one can invoke \eq{eq:LRs} again.) This leads to \eq{eq:totmem}.

\subsubsection{Energy}
\label{sec:enem}

Assuming the notation \m{\mc{K} \doteq \sum_s \int \dd\vec{p}\,H_{0s} \avr{f}_s} and using \eq{eq:Thetem2} for \m{\matr{\Uptheta}_s}, one can represent the OC energy density as follows:
\begin{align}
\sum_s \int & \dd\vec{p}\,H_{0s} F_s
\approx \mc{K}
- \frac{1}{2}\sum_s \int \dd\vec{p}\,v_s^i\Uptheta_{s,ij}\,\frac{\pd F_s}{\pd p_j}
\notag\\
& \approx \mc{K}
- \int \dd\vec{k}\,h(\vec{k})\,\frac{\pd}{\pd \vartheta}\sum_s e_s^2 \fint \dd\vec{p}
\left.
\frac{(\vec{k} \cdot \vec{v}_s)}{w^2(\vec{k})}
\frac{(\vec{\eta}^\dag \vec{v}_s \vec{v}_s^\dag \vec{\eta})}{w(\vec{k}) - \vec{k} \cdot \vec{v}_s + \vartheta}\,
\vec{k} \cdot \frac{\pd F_s}{\pd \vec{p}}
\right|_{\vartheta=0}.
\notag
\end{align}
Using \eq{eq:kwkw} and \eq{eq:chia} for \m{\matr{\chi}}, one further obtains
\begin{align}
\sum_s \int & \dd\vec{p}\,H_{0s} F_s
\notag\\
= & \,\mc{K} 
-\int \dd\vec{k}\,\frac{h(\vec{k})}{4\upi w(\vec{k})}
\left.
\vec{\eta}^\dag\,\frac{\pd}{\pd \vartheta}\bigg(
\sum_s 4\upi e_s^2 \fint \dd\vec{p}\,
\frac{\vec{v}_s \vec{v}_s^\dag}{w(\vec{k}) - \vec{k} \cdot \vec{v}_s + \vartheta}\,
\vec{k} \cdot \frac{\pd F_s}{\pd \vec{p}}
\bigg)\vec{\eta}
\right|_{\vartheta=0}
\notag\\
& -\int \dd\vec{k}\,\frac{h(\vec{k})}{4\upi}\,
\vec{\eta}^\dag\,\bigg(
\sum_s \frac{4\upi e_s^2}{w^2(\vec{k})} \fint \dd\vec{p}\,
\frac{\vec{v}_s \vec{v}_s^\dag}{w(\vec{k}) - \vec{k} \cdot \vec{v}_s}\,
\vec{k} \cdot \frac{\pd F_s}{\pd \vec{p}}
\bigg)\vec{\eta}
\notag\\
= & \,\mc{K} 
-\int \dd\vec{k}\,\frac{h(\vec{k})}{4\upi w(\vec{k})}
\left.
\vec{\eta}^\dag \frac{\pd(\omega^2\matr{\chi}(\omega, \vec{k}))}{\pd\omega} \vec{\eta}
\right|_{\omega = w(\vec{k})}
-\int \dd\vec{k}\,\frac{h(\vec{k})}{4\upi}\,\vec{\eta}^\dag\matr{\chi}\vec{\eta}
\notag\\
= & \,\mc{K} 
-\int \dd\vec{k}\,\frac{h(\vec{k})}{4\upi w(\vec{k})}
\left.
\vec{\eta}^\dag \frac{\pd(\omega^2\matr{\epsilon}(\omega, \vec{k}))}{\pd\omega} \vec{\eta}
\right|_{\omega = w(\vec{k})}
+ \int \dd\vec{k}\,\frac{h(\vec{k})}{2\upi}
\notag\\
& - \int \dd\vec{k}\,\frac{h(\vec{k})}{4\upi}\,\vec{\eta}^\dag\left(
\matr{\epsilon}(w(\vec{k}), \vec{k}) - \matr{1} + \frac{\matr{\mcc{w}}_p}{w^2(\vec{k})}
\right)\vec{\eta}
\notag\\
= & \,\mc{K} 
-\int \dd\vec{k}\,w J
+ \int \dd\vec{k}\,\frac{3h(\vec{k})}{4\upi}
- \int \dd\vec{k}\,\frac{h(\vec{k})}{4\upi}\,\vec{\eta}^\dag\matr{\epsilon}(w(\vec{k}), \vec{k})\vec{\eta}
- \int \dd\vec{k}\,\frac{h(\vec{k})}{4\upi w^2(\vec{k})}\,\vec{\eta}^\dag\matr{\mcc{w}}_p\vec{\eta}.
\notag
\end{align}
Using \eq{eq:emdr2} and proceeding as in \Sec{sec:momem}, one can also cast this as follows:
\begin{eqnarray}
\sum_s \int \dd\vec{p} H_{0s} F_s + \int \dd\vec{k}\,w J
= \mc{K} 
+ \frac{1}{8\upi}\,\big(3\avr{\smash{\osc{\vec{E}}}^\dag \osc{\vec{E}}} -\avr{\smash{\osc{\vec{B}}}^\dag \osc{\vec{B}}}\big)
- \frac{1}{8\upi c^2}\, \avr{\smash{\osc{\vec{A}}}^\dag\matr{\mcc{w}}_p\osc{\vec{A}}}.
\label{eq:Kaux}
\end{eqnarray}
Now, notice that
\begin{align}
\mc{K} 
& = \sum_s \int\dd\vec{p}\,\avr{H_{0s}(\vec{p})f_s^{(\text{kin})}(\vec{p} - e_s\osc{\vec{A}}/c)}
\notag\\
& = \sum_s \int\dd\vec{p}\,\avr{H_{0s}(\vec{p} + e_s\osc{\vec{A}}/c)f_s^{(\text{kin})}(\vec{p})}
\notag\\
& \approx \mc{K}^{(\text{kin})}
 + \avr{(\osc{\vec{A}}/c) \cdot \sum_s e_s \textstyle \int\dd\vec{p}\,\vec{v}_s f_s^{(\text{kin})}}
 + \sum_s \frac{e_s^2}{2c^2} \int\dd\vec{p}\,\avr{(\osc{\vec{A}} \matr{\mu}_s^{-1} \osc{\vec{A}})\,f_s^{(\text{kin})}}
\notag\\
& \approx \mc{K}^{(\text{kin})}
 + \frac{1}{c}\,\avr{\osc{\vec{A}} \cdot \osc{\vec{j}}}
 + \frac{1}{8\upi c^2}\, \avr{\smash{\osc{\vec{A}}}^\dag\matr{\mcc{w}}_p\osc{\vec{A}}},
\label{eq:Kaux2}
\end{align}
where \m{\osc{\vec{j}}} is the oscillating-current density. From Ampere's law,
\begin{align}
\frac{1}{c}\,\avr{\osc{\vec{A}} \cdot \osc{\vec{j}}}
& = \avr{\frac{(-\ii \oper{\omega}^{-1}\osc{\vec{E}})^\dag}{4\upi}
\big(\ii c \boper{k} \times \osc{\vec{B}} + \ii \oper{\omega}\osc{\vec{E}}\big)}
\notag\\
& \approx 
-\avr{\frac{\smash{\osc{\vec{E}}}^\dag \osc{\vec{E}}}{4\upi}}
-\avr{\frac{\osc{\vec{E}}}{4\upi} \cdot\bigg(\frac{c \boper{k}}{\oper{\omega}} \times \osc{\vec{B}}\bigg)}
\notag\\
& \approx 
-\avr{\frac{\smash{\osc{\vec{E}}}^\dag \osc{\vec{E}}}{4\upi}}
-\avr{\bigg(\osc{\vec{E}} \times \frac{c \boper{k}}{\oper{\omega}}\bigg) \cdot \frac{\osc{\vec{B}}}{4\upi}}
\notag\\
& \approx 
\frac{1}{4\upi}\,\big(\avr{\smash{\osc{\vec{B}}}^\dag \osc{\vec{B}}} - \avr{\smash{\osc{\vec{E}}}^\dag \osc{\vec{E}}}\big).
\label{eq:Kaux3}
\end{align}
Substituting \eq{eq:Kaux2} and \eq{eq:Kaux3} into \eq{eq:Kaux} leads to \eq{eq:toteem}.

\section{Selected notation}
\label{app:notation}

This paper uses the following notation (also see \Sec{sec:primer} for the index convention):
\begin{longtable}{lll}
\hline Symbol & Definition & Explanation\\
\hline\\
\m{\placeholder} & & placeholder\\
\m{\placeholder^*} & & complex conjugate\\
\m{\placeholder^{-1}} & & inverse\\
\m{\placeholder^\dag} & & Hermitian adjoint\\
\m{\placeholder^{-\dag}} & \m{(\placeholder^\dag)^{-1}} & inverse Hermitian adjoint\\
\m{\placeholder^\intercal} & & transpose\\
\m{\placeholder^{|a}} & \Sec{sec:G} & auxiliary notation\\
\m{\placeholder^{(\mu)}} & \Sec{sec:trc} & contribution from the microscopic part\\
\m{\placeholder^{\text{(m)}}} & \Sec{sec:trc} & contribution from the macroscopic part\\
\m{\avr{\placeholder}} & & average part or, for eikonal waves, a quantity evaluated\\
&& on the local wavevector\\
\m{\osc{\placeholder}} & & oscillatory part\\
\m{\macro{\placeholder}} & & macroscopic part\\
\m{\micro{\placeholder}} & & microscopic part\\
\m{\oper{\placeholder}} & & operator\\
\m{\dot{\placeholder}} &  & time derivative\\
\m{\fourier{\placeholder}} & \eq{eq:psik} & Fourier image\\
\m{\env{\placeholder}} & & envelope of an eikonal (or monochromatic) wave\\
\m{\placeholder_\aherm} & \Sec{sec:matr} & anti-Hermitian part\\
\m{\placeholder_\herm} & \Sec{sec:matr} & Hermitian part\\
\m{\pd_\placeholder} & \m{\pd/\pd\placeholder} & partial derivative (but \m{\pd_i \doteq \pd/\pd x^i}, \m{\pd_\alpha \doteq \pd/\pd z^\alpha}, \m{\pd_a \doteq \pd/\pd X^a})\\
\m{\pd^a, \pd^i} & \m{\pd/\pd K_a, \pd/\pd p_i} & partial derivative with respect to a lower-index quantity\\
\m{\eth_\placeholder} & \eq{eq:eth} & auxiliary notation\\
\m{\dd_t} & \eq{eq:ddt}, \eq{eq:conv2} & convective time derivative\\
\m{\poissonx{\placeholder, \placeholder}} & \eq{eq:poissonx} & Poisson bracket on \m{\cb{\vecst{x}, \vecst{k}}}\\
\m{\poissonz{\placeholder, \placeholder}} & \eq{eq:poissonz} & Poisson bracket on \m{\cb{\vec{x}, \vec{k}}}\\
\m{[\placeholder, \placeholder]} & & commutator\\
\m{\braket{\placeholder | \placeholder}} & \eq{eq:inner1}, \eq{eq:inner2} & inner product on \m{\hilspacex} or on \m{\hilspaceX}\\
\m{\doteq} & & definition\\
\m{\cdot} & \Sec{sec:braket} & scalar product\\
\m{\star} & \eq{eq:moyalx} & Moyal product on \m{\cb{\vecst{x}, \vecst{k}}}\\
\m{\bigstar} & \eq{eq:bigstar} & Moyal product on \m{\cb{\vec{X}, \vec{K}}}\\
\m{\ii 0} & \Sec{sec:G} & \m{\ii}~times an infinitesimally small positive number\\
\m{\fint} & & principal-value integral\\

\m{\text{eigv}} && eigenvalue\\
\m{\im} & & imaginary part\\
\m{\pv} & \eq{eq:pv} & auxiliary notation\\
\m{\re} & & real part\\
\m{\woperX{}} && operator corresponding to a Weyl symbol on \m{\cb{\vec{X}, \vec{K}}}\\
\m{\woperx{}} && operator corresponding to a Weyl symbol on \m{\cb{\vecst{x}, \vecst{k}}}\\
\m{\wsymb{}} && same as \m{\wsymbX{}} or \m{\wsymbx{}} when the two are equal\\
\m{\wsymbX{}} && Weyl symbol of an operator on \m{\hilspaceX}\\
\m{\wsymbx{}} && Weyl symbol of an operator on \m{\hilspacex}\\
\m{\sgn} & & sign\\
\m{\tr} & & trace\\
\\
\m{\Gamma, \Gamma_s} & \Sec{sec:pd} & part of a collision operator\\
\m{\Delta_s} & \Sec{sec:Phi} & particle's total ponderomotive energy in on-shell waves\\
\m{\Theta^{\alpha\beta c}} & \eq{eq:T0} & auxiliary notation\\
\m{\Uptheta^{\alpha\beta}} & \eq{eq:Upth} & dressing function (since \Sec{sec:phi})\\
\m{\Uptheta_{ij}, \matr{\Uptheta}} & \eq{eq:uptheta2} & dressing function (a part of \m{\Uptheta^{\alpha\beta}})\\
\m{\Lambda} & \Sec{sec:mono} & dispersion function (one of \m{\Lambda_b})\\
\m{\Lambda_b} & \Sec{sec:mono} & \m{b}th eigenvalue of \m{\matr{\Xi}}\\
\m{\vec{\Pi}_s} & \Tab{tab:inter} & OC momentum flux density of species \m{s}\\
\m{\vec{\Pi}_{\text{w}}} & \Tab{tab:inter} & wave momentum flux density\\
\m{\matr{\Xi}} & \Sec{sec:Xi} & dispersion matrix\\
\m{\boper{\Xi}} & \eq{eq:Xidef}, \eq{eq:Xiop2} & dispersion operator\\
\m{\matr{\Xi}_0} & \Sec{sec:Xi} & vacuum dispersion matrix\\
\m{\boper{\Xi}_0} & \eq{eq:Lagr0} & vacuum dispersion operator\\
\m{\matr{\Xi}_p} & \eq{eq:Xiwps} & Weyl symbol of \m{\boper{\Xi}_p}\\
\m{\boper{\Xi}_p} & \Sec{sec:plasmaL} & auxiliary operator\\
\m{\Phi, \Phi_s} & \eq{eq:phi}, \eq{eq:Phisc} & ponderomotive energy\\
\m{\Psi^i, \vec{\Psi}} & \Sec{sec:plasmaL} & generic interaction field\\
\m{\osc{\Psi}^i_{\text{c}}, \Psiosc_{\text{c}}} & \eq{eq:Psic} & complexified interaction field\\
\m{\Omega} & \eq{eq:Oms} & auxiliary notation\\
\\
\m{\alpha_{s,i}, \vec{\alpha}_s} & \Sec{sec:plasmaL} & Weyl symbols of \m{\oper{\alpha}_{s,i}} and \m{\boper{\alpha}_s}\\
\m{\oper{\alpha}_{s,i}, \boper{\alpha}_s} & \Sec{sec:plasmaL} & coupling operators\\
\m{\gamma} & \eq{eq:gamma1}, \eq{eq:gamma2} & linear dissipation rate as a function of \m{\cb{t, \vec{x}, \vec{k}}}\\
\m{\avr{\gamma}} & \Sec{sec:eikonal} & local linear dissipation rate of an eikonal wave\\
\m{\delta} && Kronecker symbol or delta function\\
\m{\epsilon} & \Sec{sec:ordering} & geometrical-optics parameter\\
\m{\matr{\epsilon}} & \eq{eq:veceps} & dielectric tensor\\
\m{\epsilon_\parallel, \epsilon_\perp} & \eq{eq:epspar} & parallel and transverse permittivity\\
\m{\varepsilon} & \Sec{sec:ordering} & small parameter proportional to the oscillation amplitude\\
\m{\vec{\eta}} & \Sec{sec:mono} & polarization vector (one of \m{\vec{\eta}_b})\\
\m{\vec{\eta}_b} & \Sec{sec:mono} & \m{b}th eigenvector of \m{\matr{\Xi}}\\
\m{\theta} && eikonal phase\\
\m{\kappa} & \Sec{sec:wenm} & auxiliary notation\\
\m{\kappa_x, \kappa_p} & \eq{eq:order} & characteristic inverse scales in \m{\vec{x}} and \m{\vec{p}}, respectively\\
\m{\rho_s} && charge density of species \m{s}\\
\m{\varrho^{\alpha\beta}} & \eq{eq:varrho} & auxiliary notation\\
\m{\matr{\wp}_s} & \eq{eq:defwp} & Weyl symbol of \m{\boper{\wp}_s}\\
\m{\boper{\wp}_s} & \eq{eq:defwp} & coupling operator\\
\m{\entropy} & \eq{eq:Hth2} & entropy density\\
\m{\varsigma_{\vec{k}}} & \eq{eq:varsig} & sign of the action density\\
\m{\varphi} & & electrostatic potential\\
\m{\psi, \vec{\psi}} & & any field\\
\m{\omega} & & coordinate in the frequency space dual to \m{t}\\
\m{\avr{\omega}} & \m{-\pd_t\theta} & local frequency of an eikonal wave\\
\m{\oper{\omega}} & \m{\ii\pd_t} & frequency operator\\
\\
\m{C} & \eq{eq:avCH} & Fourier image of \m{W}\\
\m{C_\placeholder, \matr{C}_\placeholder} & \eq{eq:Ccor}, \eq{eq:Ccor2} & Fourier images of \m{W_\placeholder} and \m{\matr{W}_\placeholder}\\
\m{\collision_s} & \Sec{sec:cop} & collision operator of species \m{s}\\
\m{\st{C}_\placeholder}, \m{\matrst{C}_\placeholder} & \eq{eq:C1}, \eq{eq:C2} & Fourier images of \m{\st{W}_\placeholder} and \m{\matrst{W}_\placeholder}\\
\m{D^{\alpha\beta}} & \Sec{sec:D} & Weyl symbol of \m{\oper{D}^{\alpha\beta}}\\
\m{\st{D}^{\alpha\beta}} & \eq{eq:stD} & phase-space-diffusion coefficient\\
\m{\st{D}_{ij}, \matrst{D}_s} & \eq{eq:Dsc}, \eq{eq:Dpdiff} & momentum-diffusion coefficient (part of \m{\st{D}^{\alpha\beta}})\\
\m{D_0^{\alpha\beta}} & \eq{eq:D0} & auxiliary notation\\
\m{\oper{D}^{\alpha\beta}} & \eq{eq:GD} & diffusion operator on \m{\hilspaceX}\\
\m{\mc{E}_s} & \Tab{tab:inter} & OC energy density of species \m{s}\\
\m{\mc{E}_{\text{w}}} & \Tab{tab:inter} & wave energy density (also see \eq{eq:eikenmom} for eikonal waves)\\
\m{F, F_s} & \eq{eq:Fdef}, \eq{eq:Fdef2} & OC distribution functions\\
\m{\mcc{F}_s} & \eq{eq:drag} & polarization drag for species \m{s}\\
\m{G,G_s} & \Sec{sec:G} & Weyl symbols of \m{\oper{G}} and \m{\oper{G}_s}\\
\m{G_0} & \eq{eq:G00} & approximation of \m{G} to the zeroth order in \m{\epsilon}\\
\m{\oper{G}, \oper{G}_s} & \eq{eq:greeneff} & effective Green's operators on \m{\hilspaceX}\\
\m{\mcc{G}_{ss'}} & \eq{eq:mccG} & spectrum of the correlations between \m{g_s} and \m{g_{s'}}\\
\m{\oper{\mcu{G}}} & \Sec{sec:ft} & Green's operator on \m{\hilspaceX}\\
\m{\mc{I}} & \eq{eq:mcIdef} & action density of an eikonal wave\\
\m{J} & \Sec{sec:wke} & phase-space action density\\
\m{J^{\alpha\beta}, \matr{J}} & \eq{eq:Jdef} & canonical Poisson structure\\
\m{\mc{J}^i, \vec{\mc{J}}} & \eq{eq:mcJdef} & action flux density of an eikonal wave\\
\m{H, H_s} &  & particle Hamiltonian\\
\m{\mc{H}_s} & \eq{eq:mcH} & OC Hamiltonian of species \m{s}\\
\m{\hilspacex} & & Hilbert space formed by functions on \m{\vecst{x}}\\
\m{\hilspaceX} & & Hilbert space formed by functions on \m{\vec{X}}\\
\m{\vec{K}} & \m{\cb{-\omega, \vec{q}}} & coordinate in the wavevector space dual to \m{\vec{X}}\\
\m{\boper{K}} & \m{\cb{-\oper{\omega}, \boper{q}}} & wavevector operator on \m{\hilspaceX}\\
\m{\oper{L}} & \eq{eq:Loper} & extended Liouvillian (up to a factor \m{\ii})\\
\m{\boper{L}_s} & \eq{eq:Hs} & coupling operator\\
\m{\oper{\mc{L}}_X} & \eq{eq:poissonX} & same as the Poisson bracket on \m{\cb{\vec{X}, \vec{K}}}\\
\m{\oper{\mc{L}}_\indexst} & \eq{eq:poissonx} & same as the Poisson bracket on \m{\cb{\vecst{x}, \vecst{k}}}\\
\m{\mcc{L}} & \eq{eq:abr} & \m{\vec{\Psi}}-dependent part of the plasma Lagrangian density\\
\m{\mcc{L}_0} & \eq{eq:Lagr0} & Lagrangian density of \m{\vec{\Psi}} in vacuum\\
\m{M} & & number of components of \m{\vec{\Psi}} or of another vector field\\
\m{N} & \m{2n + 1} & dimension of the extended phase space \m{\vec{X}}\\
\m{\mc{N}, \mc{N}_s} & \eq{eq:density} & OC density\\
\m{\mc{O}} && big~O (`at most of the order of')\\
\m{P_s^i, \vec{P}_s} & \Tab{tab:inter} & OC momentum density of species \m{s}\\
\m{P^i_{\text{w}}, \vec{P}_{\text{w}}} & \Tab{tab:inter} & wave momentum density (also see \eq{eq:eikenmom} for eikonal waves)\\
\m{Q_s^i, \vec{Q}_s} & \Tab{tab:inter} & OC energy flux density of species \m{s}\\
\m{Q^i_{\text{w}}, \vec{Q}_{\text{w}}} & \Tab{tab:inter} & wave energy flux density\\
\m{\mc{Q}_{ss'}} & \Sec{sec:cop} & symmetrized coefficient in the collision operator\\
\m{\boper{R}_s} & \eq{eq:Hs} & coupling operator\\
\m{\mathbb{R}} && real axis\\
\m{S} & \Sec{sec:mono} & action integral\\
\m{S_{\text{ad}}} & \eq{eq:sad} & adiabatic action integral\\
\m{\matrst{S}} & \eq{eq:Sft0} & spectrum of the macroscopic oscillations\\
\m{\oper{T}_\tau} & \eq{eq:Tr2} & shift operator (see also \Sec{sec:shift})\\
\m{\st{U}^{ij}, \matrU} & \eq{eq:Usum} & average Wigner function of the macroscopic field \m{\Psim(\vecst{x})}\\
\m{\matrU_{\text{c}\pm}, \matrU_{\text{c}}} & \Sec{sec:wifinc} & average Wigner matrix of \m{\Psiosc_{\text{c}}} and \m{\Psiosc_{\text{c}}^*} (\m{\matrU_{\text{c}} \equiv \matrU_{\text{c}+}})\\
\m{V^a, \vec{V}} & \Sec{sec:ft} & unperturbed velocity in the \m{\vec{X}} space\\
\m{\mcu{V}_n} &  & volume of \m{n}-dimensional homogeneous plasma\\
\m{W} & \eq{eq:WH} & Wigner function of \m{\osc{H}(\vec{X})}\\
\m{W_\placeholder}, \m{\matr{W}_\placeholder} & \eq{eq:Wigf}, \eq{eq:wigtX} & Weyl symbol of \m{\oper{W}_\placeholder} (Wigner function or matrix)\\
\m{\oper{W}} & \eq{eq:WHH} & density operator on \m{\hilspaceX} of \m{\osc{H}}\\
\m{\oper{W}_\placeholder}, \m{\boper{W}_\placeholder} & \eq{eq:densopX}, \eq{eq:boperW} & density operator on \m{\hilspaceX} of a given field\\
\m{\st{W}, \st{W}_s} & \Sec{sec:Deq} & Wigner functions of \m{\osc{H}} and \m{\osc{H}_s} with \m{\vec{p}} as a parameter\\
\m{\st{W}_\placeholder, \matrst{W}_\placeholder} & \eq{eq:aux10}, \eq{eq:Wtenx} & Weyl symbol of \m{\operst{W}_\placeholder} (Wigner function or matrix)\\
\m{\operst{W}_\placeholder, \boperst{W}_\placeholder} & \eq{eq:stWdop}, \eq{eq:boperstW} & density operator on \m{\hilspaceX} of a given field\\
\m{\matru} & \Sec{sec:WPsi} & average Wigner matrix of the microscopic field \m{\Psiq(\vecst{x})}\\
\m{X^a, \vec{X}} & \m{\cb{t, \vec{z}}} & coordinate in the extended phase space\\
\m{\oper{X}^a, \boper{X}} & \m{\cb{\oper{t}, \boper{z}}} & operator of the position in the extended phase space\\
\m{\mc{X}_{ss'}} & \eq{eq:mcYsymb} & Weyl symbol of \m{\oper{\mc{X}}_{ss'}}\\
\m{\oper{\mc{X}}_{ss'}} & \eq{eq:opUps} & coupling operators on \m{\hilspacex} that enter \m{\micro{\osc{H}}}\\
\\
\m{\dd} & & differential\\
\m{e_s} & & charge of species \m{s}\\
\m{f, f_s} &  & distribution function\\
\m{g, g_s} &  \Sec{sec:g} & initial conditions for \m{\osc{f}} and \m{\osc{f}_s}\\
\m{h} & \eq{eq:hdef}, \eq{eq:Jinc}  & rescaled phase-space action density\\
\m{h_{\text{c}\pm}, h_{\text{c}}} &  \Sec{sec:wifinc} & auxiliary notation (\m{h_{\text{c}+} \equiv h_{\text{c}}})\\
\m{k_i, \vec{k}} && coordinate in the wavevector space dual to \m{\vec{x}}\\
\m{\avr{k}_i, \avr{\vec{k}}} & \m{\pd_i\theta, \pd_\vec{x}\theta} & local wavevector of an eikonal wave\\
\m{\oper{k}_i, \boper{k}} & \m{-\ii\pd_i, -\ii\pd_\vec{x}} & wavevector operator\\
\m{\vecst{k}} & \m{\cb{-\omega, \vec{k}}} & coordinate in the wavevector space dual to \m{\vecst{x}}\\
\m{\avr{\vecst{k}}} & \m{\cb{-\avr{\omega}, \avr{\vec{k}}}} & local spacetime-wavevector of an eikonal wave\\
\m{\boperst{k}} & \m{-\ii\pd_\vecst{x}} & spacetime-wavevector operator\\
\m{\ket{\vecst{k}}} & \eq{eq:ketxk}, \eq{eq:ketxknorm} & eigenvector of \m{\boperst{k}} corresponding to the eigenvalue \m{\vecst{k}}\\
\m{\vec{\ell}_\tau} & \Sec{eq:shift1} & displacement in \m{\vec{X}} along unperturbed characteristics\\
\m{m_s} & & mass of species \m{s}\\
\m{n} & \m{\dim\vec{x}} & number of spatial dimensions\\
\m{\st{n}} & \m{n + 1} & number of spacetime dimensions\\
\m{p_i, \vec{p}} && coordinate in the momentum space\\
\m{\oper{p}_i, \boper{p}} & \m{\vec{p}} & position operator corresponding to the coordinate \m{\vec{p}}\\
\m{q^i, \vec{q}} & \m{\cb{\vec{k}, \vec{r}}}& coordinate in the wavevector space dual to \m{\vec{z}}\\
\m{\oper{q}^i, \boper{q}} & \m{\cb{\boper{k}, \boper{r}}} & wavevector operator corresponding to the coordinate \m{\vec{z}}\\
\m{r_i, \vec{r}} & & coordinate in the wavevector space dual to \m{\vec{p}}\\
\m{\oper{r}^i, \boper{r}} & \m{-\ii\pd^i, -\ii\pd_\vec{p}} & wavevector operator corresponding to the coordinate \m{\vec{p}}\\
\m{s} && species index\\
\m{t} && time\\
\m{\oper{t}} & \m{t} & time operator\\
\m{u^\alpha} & \eq{eq:uv} & oscillating part of the phase-space velocity\\
\m{\oper{u}^\alpha} & \eq{eq:u} & \m{u^\alpha} as an operator on \m{\hilspaceX}\\
\m{v^\alpha, v^i, \vec{v}} & \eq{eq:uv} & average velocity in phase space or in physical space,\\
 & \eq{eq:vOC} & or, since \Sec{sec:OC}, OC velocity\\
\m{\vg^i, \vvg} & \eq{eq:vvg2} & group velocity as a function of \m{\cb{t, \vec{x}, \vec{k}}}\\
\m{\avg^i, \avvg} & \eq{eq:avvg} & local group velocity of an eikonal wave\\
\m{w} & \eq{eq:wdef} & eikonal-wave frequency as a function of \m{\cb{t, \vec{x}, \vec{k}}}\\
\m{x^i, \vec{x}} & & coordinate in space\\
\m{\avr{x}^i, \avr{\vec{x}}} & & ray coordinate in space\\
\m{\oper{x}^i, \boper{x}} & \m{\vec{x}} & space-position operator\\
\m{\st{x}^i, \vecst{x}} & \m{\cb{t, \vec{x}}} & coordinate in spacetime\\
\m{\operst{x}^i, \boperst{x}} & \m{\vecst{x}} & spacetime-position operator\\
\m{\ket{\vecst{x}}} & \eq{eq:ketxk}, \eq{eq:ketxknorm} & eigenvector of \m{\boperst{x}} corresponding to the eigenvalue \m{\vecst{x}}\\
\m{z^\alpha, \vec{z}} & \m{\cb{\vec{x}, \vec{p}}} & coordinate in phase space\\
\m{\oper{z}^\alpha, \boper{z}} & \m{\cb{\boper{x}, \boper{p}}} & phase-space position operator\\
&&\\
\hline
\end{longtable}


\begin{thebibliography}{97}
\expandafter\ifx\csname natexlab\endcsname\relax\def\natexlab#1{#1}\fi
\def\au#1{#1} \def\ed#1{#1} \def\yr#1{#1}\def\at#1{#1}\def\jt#1{\textit{#1}}
  \def\bt#1{#1}\def\bvol#1{\textbf{#1}} \def\vol#1{#1} \def\pg#1{#1}
  \def\publ#1{#1}\def\arxiv#1{#1}\def\org#1{#1}\def\st#1{\textit{#1}}

\bibitem[Andersson {\em et~al.\/}(2021)Andersson, Joudioux, Oancea \&
  Raj]{ref:andersson21}
{\sc \au{Andersson, L.}, \au{Joudioux, J.}, \au{Oancea, M.~A.} \& \au{Raj, A.}}
  \yr{2021}  \at{Propagation of polarized gravitational waves}.  \jt{Phys. Rev.
  D}  \bvol{103},  \pg{044053}.

\bibitem[Balakin {\em et~al.\/}(2016)Balakin, Dodin, Fraiman \&
  Fisch]{my:shpulse}
{\sc \au{Balakin, A.~A.}, \au{Dodin, I.~Y.}, \au{Fraiman, G.~M.} \& \au{Fisch,
  N.~J.}} \yr{2016}  \at{Backward {Raman} amplification of broad-band pulses}.
  \jt{Phys. Plasmas}  \bvol{23},  \pg{083115}.

\bibitem[Besse {\em et~al.\/}(2011)Besse, Elskens, Escande \&
  Bertrand]{ref:besse11}
{\sc \au{Besse, N.}, \au{Elskens, Y.}, \au{Escande, D.~F.} \& \au{Bertrand,
  P.}} \yr{2011}  \at{Validity of quasilinear theory: refutations and new
  numerical confirmation}.  \jt{Plasma Phys. Control. Fusion}  \bvol{53},
  \pg{025012}.

\bibitem[Binney \& Tremaine(2008)]{book:binney}
{\sc \au{Binney, J.} \& \au{Tremaine, S.}} \yr{2008} {\em Galactic Dynamics:
  Second Edition\/}.  \publ{Princeton: Princeton Univ. Press}.

\bibitem[Bliokh {\em et~al.\/}(2015)Bliokh, Rodr{\'{\i}}guez-Fortu{\~{n}}o,
  Nori \& Zayats]{ref:bliokh15}
{\sc \au{Bliokh, K.~Y.}, \au{Rodr{\'{\i}}guez-Fortu{\~{n}}o, F.~J.}, \au{Nori,
  F.} \& \au{Zayats, A.~V.}} \yr{2015}  \at{Spin-orbit interactions of light}.
  \jt{Nat. Photonics}  \bvol{9},  \pg{796}.

\bibitem[Brizard {\em et~al.\/}(1993)Brizard, Cook \& Kaufman]{ref:brizard93}
{\sc \au{Brizard, A.~J.}, \au{Cook, D.~R.} \& \au{Kaufman, A.~N.}} \yr{1993}
  \at{Wave-action conservation for pseudo-{Hermitian} fields}.  \jt{Phys. Rev.
  Lett.}  \bvol{70},  \pg{521}.

\bibitem[Cartwright(1976)]{ref:cartwright76}
{\sc \au{Cartwright, N.~D.}} \yr{1976}  \at{A non-negative {Wigner}-type
  distribution}.  \jt{Physica A: Stat. Mech. Appl.}  \bvol{83},  \pg{210}.

\bibitem[Cary \& Brizard(2009)]{ref:cary09}
{\sc \au{Cary, J.~R.} \& \au{Brizard, A.~J.}} \yr{2009}  \at{Hamiltonian theory
  of guiding-center motion}.  \jt{Rev. Mod. Phys.}  \bvol{81},  \pg{693}.

\bibitem[Cary \& Kaufman(1977)]{ref:cary77}
{\sc \au{Cary, J.~R.} \& \au{Kaufman, A.~N.}} \yr{1977}  \at{Ponderomotive
  force and linear susceptibility in {Vlasov} plasma}.  \jt{Phys. Rev. Lett.}
  \bvol{39},  \pg{402}.

\bibitem[Cary \& Kaufman(1981)]{ref:cary81}
{\sc \au{Cary, J.~R.} \& \au{Kaufman, A.~N.}} \yr{1981}  \at{Ponderomotive
  effects in collisionless plasma: a {Lie} transform approach}.  \jt{Phys.
  Fluids}  \bvol{24},  \pg{1238}.

\bibitem[Catto {\em et~al.\/}(2017)Catto, Lee \& Ram]{ref:catto17}
{\sc \au{Catto, P.~J.}, \au{Lee, J.} \& \au{Ram, A.~K.}} \yr{2017}  \at{A
  quasilinear operator retaining magnetic drift effects in tokamak geometry}.
  \jt{J. Plasma Phys.}  \bvol{83},  \pg{905830611}.

\bibitem[Chavanis(2012)]{ref:chavanis12}
{\sc \au{Chavanis, P.-H.}} \yr{2012}  \at{Kinetic theory of long-range
  interacting systems with angle--action variables and collective effects}.
  \jt{Physica A}  \bvol{391},  \pg{3680}.

\bibitem[Crews \& Shumlak(2022)]{ref:crews22}
{\sc \au{Crews, D.~W.} \& \au{Shumlak, U.}} \yr{2022}  \at{On the validity of
  quasilinear theory applied to the electron bump-on-tail instability}.
  \jt{Phys. Plasmas}  \bvol{29},  \pg{043902}.

\bibitem[Dewar(1972)]{ref:dewar72c}
{\sc \au{Dewar, R.~L.}} \yr{1972}  \at{A {Lagrangian} theory for nonlinear wave
  packets in a collisionless plasma}.  \jt{J. Plasma Phys.}  \bvol{7},
  \pg{267}.

\bibitem[Dewar(1973)]{ref:dewar73}
{\sc \au{Dewar, R.~L.}} \yr{1973}  \at{Oscillation center quasilinear theory}.
  \jt{Phys. Fluids}  \bvol{16},  \pg{1102}.

\bibitem[Dewar(1977)]{ref:dewar77}
{\sc \au{Dewar, R.~L.}} \yr{1977}  \at{Energy-momentum tensors for dispersive
  electromagnetic waves}.  \jt{Austral. J. Phys.}  \bvol{30},  \pg{533}.

\bibitem[Dodin(2014)]{my:itervar}
{\sc \au{Dodin, I.~Y.}} \yr{2014}  \at{On variational methods in the physics of
  plasma waves}.  \jt{Fusion Sci. Tech.}  \bvol{65},  \pg{54}.

\bibitem[Dodin \& Fisch(2010{\natexlab{{\em a\/}}})]{my:kchi}
{\sc \au{Dodin, I.~Y.} \& \au{Fisch, N.~J.}} \yr{2010{\natexlab{{\em a\/}}}}
  \at{On generalizing the {$K$-$\chi$} theorem}.  \jt{Phys. Lett. A}
  \bvol{374},  \pg{3472}.

\bibitem[Dodin \& Fisch(2010{\natexlab{{\em b\/}}})]{my:mquanta}
{\sc \au{Dodin, I.~Y.} \& \au{Fisch, N.~J.}} \yr{2010{\natexlab{{\em b\/}}}}
  \at{On the evolution of linear waves in cosmological plasmas}.  \jt{Phys.
  Rev. D}  \bvol{82},  \pg{044044}.

\bibitem[Dodin \& Fisch(2012)]{my:amc}
{\sc \au{Dodin, I.~Y.} \& \au{Fisch, N.~J.}} \yr{2012}  \at{Axiomatic
  geometrical optics, {Abraham}--{Minkowski} controversy, and photon properties
  derived classically}.  \jt{Phys. Rev. A}  \bvol{86},  \pg{053834}.

\bibitem[Dodin \& Fisch(2014)]{my:lens}
{\sc \au{Dodin, I.~Y.} \& \au{Fisch, N.~J.}} \yr{2014}  \at{Ponderomotive
  forces \textit{on} waves in modulated media}.  \jt{Phys. Rev. Lett.}
  \bvol{112},  \pg{205002}.

\bibitem[Dodin {\em et~al.\/}(2009)Dodin, Geyko \& Fisch]{my:dense}
{\sc \au{Dodin, I.~Y.}, \au{Geyko, V.~I.} \& \au{Fisch, N.~J.}} \yr{2009}
  \at{Langmuir wave linear evolution in inhomogeneous nonstationary anisotropic
  plasma}.  \jt{Phys. Plasmas}  \bvol{16},  \pg{112101}.

\bibitem[Dodin {\em et~al.\/}(2019)Dodin, Ruiz, Yanagihara, Zhou \&
  Kubo]{my:quasiop1}
{\sc \au{Dodin, I.~Y.}, \au{Ruiz, D.~E.}, \au{Yanagihara, K.}, \au{Zhou, Y.} \&
  \au{Kubo, S.}} \yr{2019}  \at{Quasioptical modeling of wave beams with and
  without mode conversion. {I}. {Basic} theory}.  \jt{Phys. Plasmas}
  \bvol{26},  \pg{072110}.

\bibitem[Dodin {\em et~al.\/}(2017)Dodin, Zhmoginov \& Ruiz]{my:nonloc}
{\sc \au{Dodin, I.~Y.}, \au{Zhmoginov, A.~I.} \& \au{Ruiz, D.~E.}} \yr{2017}
  \at{Variational principles for dissipative (sub)systems, with applications to
  the theory of linear dispersion and geometrical optics}.  \jt{Phys. Lett. A}
  \bvol{381},  \pg{1411}.

\bibitem[Drummond \& Pines(1962)]{ref:drummond62}
{\sc \au{Drummond, W.~E.} \& \au{Pines, D.}} \yr{1962}  \at{Non-linear
  stability of plasma oscillations}.  \jt{Nucl. Fusion Suppl.}  \bvol{3},
  \pg{1049}.

\bibitem[Eriksson \& Helander(1994)]{ref:eriksson94}
{\sc \au{Eriksson, L.-G.} \& \au{Helander, P.}} \yr{1994}  \at{{Monte} {Carlo}
  operators for orbit-averaged {Fokker}--{Planck} equations}.  \jt{Phys.
  Plasmas}  \bvol{1},  \pg{308}.

\bibitem[Escande {\em et~al.\/}(2018)Escande, B{\'{e}}nisti, Elskens, Zarzoso
  \& Doveil]{ref:escande18}
{\sc \au{Escande, D.~F.}, \au{B{\'{e}}nisti, D.}, \au{Elskens, Y.},
  \au{Zarzoso, D.} \& \au{Doveil, F.}} \yr{2018}  \at{Basic microscopic plasma
  physics from ${N}$-body mechanics}.  \jt{Rev. Mod. Plasma Phys.}  \bvol{2},
  \pg{1},  \arxiv{arXiv: 1805.11408}.

\bibitem[Fetterman \& Fisch(2008)]{ref:fetterman08}
{\sc \au{Fetterman, A.~J.} \& \au{Fisch, N.~J.}} \yr{2008}  \at{$\alpha$
  channeling in a rotating plasma}.  \jt{Phys. Rev. Lett.}  \bvol{101},
  \pg{205003}.

\bibitem[Fisch(1987)]{ref:fisch87}
{\sc \au{Fisch, N.~J.}} \yr{1987}  \at{Theory of current drive in plasmas}.
  \jt{Rev. Mod. Phys.}  \bvol{59},  \pg{175}.

\bibitem[Fisch \& Rax(1992)]{ref:fisch92}
{\sc \au{Fisch, N.~J.} \& \au{Rax, J.~M.}} \yr{1992}  \at{Interaction of
  energetic alpha-particles with intense lower hybrid waves}.  \jt{Phys. Rev.
  Lett.}  \bvol{69},  \pg{612}.

\bibitem[Fraiman \& Kostyukov(1995)]{ref:fraiman95}
{\sc \au{Fraiman, G.~M.} \& \au{Kostyukov, I.~Yu.}} \yr{1995}  \at{Influence of
  external inhomogeneous static fields on interaction between beam of
  charged-particles and packet of electromagnetic waves}.  \jt{Phys. Plasmas}
  \bvol{2},  \pg{923}.

\bibitem[Galeev \& Sagdeev(1985)]{book:galeev85}
{\sc \au{Galeev, A.~A.} \& \au{Sagdeev, R.~Z.}} \yr{1985} {\em Theory of Weakly
  Turbulent Plasma, Part 4 in `Basic Plasma Physics I'\/}.  \publ{New York:
  North--Holland}, edited by A. A. Galeev and R. N. Sudan.

\bibitem[Gaponov \& Miller(1958)]{ref:gaponov58}
{\sc \au{Gaponov, A.~V.} \& \au{Miller, M.~A.}} \yr{1958}  \at{Potential wells
  for charged particles in a high-frequency electromagnetic field}.  \jt{Zh.
  Eksp. Teor. Fiz.}  \bvol{34},  \pg{242}.

\bibitem[Garg \& Dodin(2020)]{my:gwponder}
{\sc \au{Garg, D.} \& \au{Dodin, I.~Y.}} \yr{2020}  \at{Average nonlinear
  dynamics of particles in gravitational pulses: effective {Hamiltonian},
  secular acceleration, and gravitational susceptibility}.  \jt{Phys. Rev. D}
  \bvol{102},  \pg{064012}.

\bibitem[Garg \& Dodin(2021{\natexlab{{\em a\/}}})]{foot:mygwquasi}
{\sc \au{Garg, G.} \& \au{Dodin, I.~Y.}} \yr{2021{\natexlab{{\em a\/}}}}
  Gauge-invariant gravitational waves in matter beyond linearized gravity,
  \arxiv{arXiv: 2106.05062}.

\bibitem[Garg \& Dodin(2021{\natexlab{{\em b\/}}})]{foot:mygwinv}
{\sc \au{Garg, G.} \& \au{Dodin, I.~Y.}} \yr{2021{\natexlab{{\em b\/}}}} Gauge
  invariants of linearized gravity with a general background metric,
  \arxiv{arXiv: 2105.04680}.

\bibitem[Garg \& Dodin(2022)]{foot:mygweq}
{\sc \au{Garg, G.} \& \au{Dodin, I.~Y.}} \yr{2022} Gravitational wave modes in
  matter,  \arxiv{arXiv: 2204.09095}.

\bibitem[Hamilton(2020)]{ref:hamilton20}
{\sc \au{Hamilton, C.}} \yr{2020}  \at{A simple, heuristic derivation of the
  {Balescu}-{Lenard} kinetic equation for stellar systems}.  \jt{Mon. Notices
  Royal Astron. Soc.}  \bvol{501},  \pg{3371}.

\bibitem[Hayes(1973)]{ref:hayes73}
{\sc \au{Hayes, W.~D.}} \yr{1973}  \at{Group velocity and nonlinear dispersive
  wave propagation}.  \jt{Proc. R. Soc. Lond. A}  \bvol{332},  \pg{199}.

\bibitem[Hizanidis {\em et~al.\/}(1983)Hizanidis, Molvig \&
  Swartz]{ref:hizanidis83}
{\sc \au{Hizanidis, K.}, \au{Molvig, K.} \& \au{Swartz, K.}} \yr{1983}  \at{A
  retarded time superposition principle and the relativistic collision
  operator}.  \jt{J. Plasma Phys.}  \bvol{30},  \pg{223}.

\bibitem[Kaufman(1972)]{ref:kaufman72}
{\sc \au{Kaufman, A.~N.}} \yr{1972}  \at{Quasilinear diffusion of an
  axisymmetric toroidal plasma}.  \jt{Phys. Fluids}  \bvol{15},  \pg{1063}.

\bibitem[Kaufman(1987)]{ref:kaufman87}
{\sc \au{Kaufman, A.~N.}} \yr{1987}  \at{Phase-space-{Lagrangian} action
  principle and the generalized {$K$-$\chi$} theorem}.  \jt{Phys. Rev. A}
  \bvol{36},  \pg{982}.

\bibitem[Kaufman \& Holm(1984)]{ref:kaufman84}
{\sc \au{Kaufman, A.~N.} \& \au{Holm, D.~D.}} \yr{1984}  \at{The
  {Lie}-transformed {Vlasov} action principle: relativistically covariant wave
  propagation and self-consistent ponderomotive effects}.  \jt{Phys. Lett. A}
  \bvol{105},  \pg{277}.

\bibitem[Kennel \& Engelmann(1966)]{ref:kennel66}
{\sc \au{Kennel, C.~F.} \& \au{Engelmann, F.}} \yr{1966}  \at{Velocity space
  diffusion from weak plasma turbulence in a magnetic field}.  \jt{Phys.
  Fluids}  \bvol{9},  \pg{2377}.

\bibitem[Kentwell(1987)]{ref:kentwell87b}
{\sc \au{Kentwell, G.~W.}} \yr{1987}  \at{Oscillation-center theory at
  resonance}.  \jt{Phys. Rev. A}  \bvol{35},  \pg{4703}.

\bibitem[Kentwell \& Jones(1987)]{ref:kentwell87}
{\sc \au{Kentwell, G.~W.} \& \au{Jones, D.~A.}} \yr{1987}  \at{The
  time-dependent ponderomotive force}.  \jt{Phys. Rep.}  \bvol{145},  \pg{319}.

\bibitem[Krall \& Trivelpiece(1973)]{book:krall}
{\sc \au{Krall, N.~A.} \& \au{Trivelpiece, A.~W.}} \yr{1973} {\em Principles of
  Plasma Physics\/}.  \publ{New York: McGraw-Hill}.

\bibitem[Landau \& Lifshitz(1976)]{book:landau1}
{\sc \au{Landau, L.~D.} \& \au{Lifshitz, E.~M.}} \yr{1976} {\em Mechanics\/}.
  \publ{Oxford: Butterworth--Heinemann}.

\bibitem[Landreman(2017)]{foot:landreman17}
{\sc \au{Landreman, M.}} \yr{2017} The {H} theorem for the
  {Landau}--{Fokker}--{Planck} collision operator. Unpublished.

\bibitem[Lee {\em et~al.\/}(2018)Lee, Smithe, Wright \& Bonoli]{ref:lee18}
{\sc \au{Lee, J.}, \au{Smithe, D.}, \au{Wright, J.} \& \au{Bonoli, P.}}
  \yr{2018}  \at{A positive-definite form of bounce-averaged quasilinear
  velocity diffusion for the parallel inhomogeneity in a tokamak}.  \jt{Plasma
  Phys. Control. Fusion}  \bvol{60},  \pg{025007}.

\bibitem[Lichtenberg \& Lieberman(1992)]{book:lichtenberg}
{\sc \au{Lichtenberg, A.~J.} \& \au{Lieberman, M.~A.}} \yr{1992} {\em Regular
  and Chaotic Dynamics\/}.  \publ{New York: Springer--Verlag}, second edition.

\bibitem[Lifshitz \& Pitaevskii(1981)]{book:landau10}
{\sc \au{Lifshitz, E.~M.} \& \au{Pitaevskii, L.~P.}} \yr{1981} {\em Physical
  Kinetics\/}.  \publ{New York: Pergamon Press}.

\bibitem[Littlejohn(1979)]{ref:littlejohn79}
{\sc \au{Littlejohn, R.~G.}} \yr{1979}  \at{A guiding center {Hamiltonian}: a
  new approach}.  \jt{J. Math. Phys.}  \bvol{20},  \pg{2445}.

\bibitem[Littlejohn(1981)]{ref:littlejohn81}
{\sc \au{Littlejohn, R.~G.}} \yr{1981}  \at{Hamiltonian formulation of guiding
  center motion}.  \jt{Phys. Fluids}  \bvol{24},  \pg{1730}.

\bibitem[Littlejohn(1983)]{ref:littlejohn83}
{\sc \au{Littlejohn, R.~G.}} \yr{1983}  \at{Variational principles of guiding
  centre motion}.  \jt{J. Plasma Phys.}  \bvol{29},  \pg{111}.

\bibitem[Littlejohn(1986)]{ref:littlejohn86}
{\sc \au{Littlejohn, R.~G.}} \yr{1986}  \at{The semiclassical evolution of wave
  packets}.  \jt{Phys. Rep.}  \bvol{138},  \pg{193}.

\bibitem[Liu \& Dodin(2015)]{my:sharm}
{\sc \au{Liu, C.} \& \au{Dodin, I.~Y.}} \yr{2015}  \at{Nonlinear frequency
  shift of electrostatic waves in general collisionless plasma: unifying theory
  of fluid and kinetic nonlinearities}.  \jt{Phys. Plasmas}  \bvol{22},
  \pg{082117}.

\bibitem[Magorrian(2021)]{ref:magorrian21}
{\sc \au{Magorrian, J.}} \yr{2021}  \at{Stellar dynamics in the periodic cube}.
   \jt{Mon. Notices Royal Astron. Soc.}  \bvol{507},  \pg{4840}.

\bibitem[McDonald(1988)]{ref:mcdonald88}
{\sc \au{McDonald, S.~W.}} \yr{1988}  \at{Phase-space representations of wave
  equations with applications to the eikonal approximation for short-wavelength
  waves}.  \jt{Phys. Rep.}  \bvol{158},  \pg{337}.

\bibitem[McDonald(1991)]{ref:mcdonald91}
{\sc \au{McDonald, S.~W.}} \yr{1991}  \at{Wave kinetic equation in a
  fluctuating medium}.  \jt{Phys. Rev. A}  \bvol{43},  \pg{4484}.

\bibitem[McDonald {\em et~al.\/}(1985)McDonald, Grebogi \&
  Kaufman]{ref:mcdonald85b}
{\sc \au{McDonald, S.~W.}, \au{Grebogi, C.} \& \au{Kaufman, A.~N.}} \yr{1985}
  \at{Locally coupled evolution of wave and particle distribution in general
  magnetoplasma geometry}.  \jt{Phys. Lett. A}  \bvol{111},  \pg{19}.

\bibitem[McDonald \& Kaufman(1985)]{ref:mcdonald85}
{\sc \au{McDonald, S.~W.} \& \au{Kaufman, A.~N.}} \yr{1985}  \at{Weyl
  representation for electromagnetic waves: The wave kinetic equation}.
  \jt{Phys. Rev. A}  \bvol{32},  \pg{1708}.

\bibitem[Motz \& Watson(1967)]{ref:motz67}
{\sc \au{Motz, H.} \& \au{Watson, C. J.~H.}} \yr{1967}  \at{The radio-frequency
  confinement and acceleration of plasmas}.  \jt{Adv. Electron. Electron Phys.}
   \bvol{23},  \pg{153}.

\bibitem[Moyal(1949)]{ref:moyal49}
{\sc \au{Moyal, J.~E.}} \yr{1949}  \at{Quantum mechanics as a statistical
  theory}.  \jt{Proc. Cambridge Philosoph. Soc.}  \bvol{45},  \pg{99}.

\bibitem[Mynick(1988)]{ref:mynick88}
{\sc \au{Mynick, H.~E.}} \yr{1988}  \at{The generalized {Balescu}-{Lenard}
  collision operator}.  \jt{J. Plasma Phys.}  \bvol{39},  \pg{303}.

\bibitem[Oancea {\em et~al.\/}(2020)Oancea, Joudioux, Dodin, Ruiz, Paganini \&
  Andersson]{my:spinhall}
{\sc \au{Oancea, M.~A.}, \au{Joudioux, J.}, \au{Dodin, I.~Y.}, \au{Ruiz,
  D.~E.}, \au{Paganini, C.~F.} \& \au{Andersson, L.}} \yr{2020}
  \at{Gravitational spin {Hall} effect of light}.  \jt{Phys. Rev. D}
  \bvol{102},  \pg{024075}.

\bibitem[Ochs(2021)]{phd:ochs21}
{\sc \au{Ochs, I.~E.}} \yr{2021}  \at{Controlling and exploiting perpendicular
  rotation in magnetized plasmas}. PhD thesis, Princeton University.

\bibitem[Ochs \& Fisch(2021{\natexlab{{\em a\/}}})]{ref:ochs21}
{\sc \au{Ochs, I.~E.} \& \au{Fisch, N.~J.}} \yr{2021{\natexlab{{\em a\/}}}}
  \at{Nonresonant diffusion in alpha channeling}.  \jt{Phys. Rev. Lett.}
  \bvol{127},  \pg{025003}.

\bibitem[Ochs \& Fisch(2021{\natexlab{{\em b\/}}})]{ref:ochs21b}
{\sc \au{Ochs, I.~E.} \& \au{Fisch, N.~J.}} \yr{2021{\natexlab{{\em b\/}}}}
  \at{Wave-driven torques to drive current and rotation}.  \jt{Phys. Plasmas}
  \bvol{28},  \pg{102506}.

\bibitem[Ochs \& Fisch(2022)]{foot:ochs22}
{\sc \au{Ochs, I.~E.} \& \au{Fisch, N.~J.}} \yr{2022} Momentum conservation in
  current drive and alpha-channeling-mediated rotation drive,  \arxiv{arXiv:
  2201.07853}.

\bibitem[Pinsker(2001)]{ref:pinsker01}
{\sc \au{Pinsker, R.~I.}} \yr{2001}  \at{Introduction to wave heating and
  current drive in magnetized plasmas}.  \jt{Phys. Plasmas}  \bvol{8},
  \pg{1219}.

\bibitem[Rigas {\em et~al.\/}(2011)Rigas, S{\'{a}}nchez-Soto, Klimov,
  {\v{R}}eh{\'{a}}{\v{c}}ek \& Hradil]{ref:rigas11}
{\sc \au{Rigas, I.}, \au{S{\'{a}}nchez-Soto, L.~L.}, \au{Klimov, A.},
  \au{{\v{R}}eh{\'{a}}{\v{c}}ek, J.} \& \au{Hradil, Z.}} \yr{2011}  \at{Orbital
  angular momentum in phase space}.  \jt{Ann. Phys.}  \bvol{326},  \pg{426}.

\bibitem[Rogister \& Oberman(1968)]{ref:rogister68}
{\sc \au{Rogister, A.} \& \au{Oberman, C.}} \yr{1968}  \at{On the kinetic
  theory of stable and weakly unstable plasma. {Part}~1}.  \jt{J. Plasma Phys.}
   \bvol{2},  \pg{33}.

\bibitem[Rogister \& Oberman(1969)]{ref:rogister69}
{\sc \au{Rogister, A.} \& \au{Oberman, C.}} \yr{1969}  \at{On the kinetic
  theory of stable and weakly unstable plasma. {Part}~2}.  \jt{J. Plasma Phys.}
   \bvol{3},  \pg{119}.

\bibitem[Rostoker(1964)]{ref:rostoker64}
{\sc \au{Rostoker, N.}} \yr{1964}  \at{Superposition of dressed test
  particles}.  \jt{Phys. Fluids}  \bvol{7},  \pg{479}.

\bibitem[Ruiz(2017)]{phd:ruiz17}
{\sc \au{Ruiz, D.~E.}} \yr{2017}  \at{Geometric theory of waves and its
  applications to plasma physics}. PhD thesis, Princeton University,
  arXiv:1708.05423.

\bibitem[Ruiz \& Dodin(2015{\natexlab{{\em a\/}}})]{my:qdiel}
{\sc \au{Ruiz, D.~E.} \& \au{Dodin, I.~Y.}} \yr{2015{\natexlab{{\em a\/}}}}
  \at{First-principles variational formulation of polarization effects in
  geometrical optics}.  \jt{Phys. Rev. A}  \bvol{92},  \pg{043805}.

\bibitem[Ruiz \& Dodin(2015{\natexlab{{\em b\/}}})]{my:qlagr}
{\sc \au{Ruiz, D.~E.} \& \au{Dodin, I.~Y.}} \yr{2015{\natexlab{{\em b\/}}}}
  \at{On the correspondence between quantum and classical variational
  principles}.  \jt{Phys. Lett. A}  \bvol{379},  \pg{2623}.

\bibitem[Ruiz \& Dodin(2017{\natexlab{{\em a\/}}})]{my:covar}
{\sc \au{Ruiz, D.~E.} \& \au{Dodin, I.~Y.}} \yr{2017{\natexlab{{\em a\/}}}}
  \at{Extending geometrical optics: a {Lagrangian} theory for vector waves}.
  \jt{Phys. Plasmas}  \bvol{24},  \pg{055704}.

\bibitem[Ruiz \& Dodin(2017{\natexlab{{\em b\/}}})]{my:qponder}
{\sc \au{Ruiz, D.~E.} \& \au{Dodin, I.~Y.}} \yr{2017{\natexlab{{\em b\/}}}}
  \at{Ponderomotive dynamics of waves in quasiperiodically modulated media}.
  \jt{Phys. Rev. A}  \bvol{95},  \pg{032114}.

\bibitem[Ruiz {\em et~al.\/}(2019)Ruiz, Glinsky \& Dodin]{my:wcol}
{\sc \au{Ruiz, D.~E.}, \au{Glinsky, M.~E.} \& \au{Dodin, I.~Y.}} \yr{2019}
  \at{Wave kinetic equation for inhomogeneous drift-wave turbulence beyond the
  quasilinear approximation}.  \jt{J. Plasma Phys.}  \bvol{85},
  \pg{905850101}.

\bibitem[Schlickeiser \& Yoon(2014)]{ref:schlickeiser14}
{\sc \au{Schlickeiser, R.} \& \au{Yoon, P.~H.}} \yr{2014}  \at{Quasilinear
  theory of general electromagnetic fluctuations in unmagnetized plasmas}.
  \jt{Phys. Plasmas}  \bvol{21},  \pg{092102}.

\bibitem[Schmit {\em et~al.\/}(2010)Schmit, Dodin \& Fisch]{my:langact}
{\sc \au{Schmit, P.~F.}, \au{Dodin, I.~Y.} \& \au{Fisch, N.~J.}} \yr{2010}
  \at{Controlling hot electrons by wave amplification and decay in compressing
  plasma}.  \jt{Phys. Rev. Lett.}  \bvol{105},  \pg{175003}.

\bibitem[Silin(1961)]{ref:silin61}
{\sc \au{Silin, V.~P.}} \yr{1961}  \at{Collision integral for charged
  particles}.  \jt{Zh. Eksp. Teor. Fiz.}  \bvol{40},  \pg{1768}.

\bibitem[Stix(1992)]{book:stix}
{\sc \au{Stix, T.~H.}} \yr{1992} {\em Waves in Plasmas\/}.  \publ{New York:
  AIP}, second edition.

\bibitem[Tracy {\em et~al.\/}(2014)Tracy, Brizard, Richardson \&
  Kaufman]{book:tracy}
{\sc \au{Tracy, E.~R.}, \au{Brizard, A.~J.}, \au{Richardson, A.~S.} \&
  \au{Kaufman, A.~N.}} \yr{2014} {\em Ray Tracing and Beyond: Phase Space
  Methods in Plasma Wave Theory\/}.  \publ{New York: Cambridge University
  Press}.

\bibitem[Trigger {\em et~al.\/}(2004)Trigger, Ershkovich, van Heijst \&
  Schram]{ref:trigger04}
{\sc \au{Trigger, S.~A.}, \au{Ershkovich, A.~I.}, \au{van Heijst, G. J.~F.} \&
  \au{Schram, P. P. J.~M.}} \yr{2004}  \at{Kinetic theory of {Jeans}
  instability}.  \jt{Phys. Rev. E}  \bvol{69},  \pg{066403}.

\bibitem[Vedenov {\em et~al.\/}(1961)Vedenov, Velikhov \&
  Sagdeev]{ref:vedenov61b}
{\sc \au{Vedenov, A.~A.}, \au{Velikhov, E.~P.} \& \au{Sagdeev, R.~Z.}}
  \yr{1961}  \at{Nonlinear oscillations of rarified plasma}.  \jt{Nucl. Fusion}
   \bvol{1},  \pg{82}.

\bibitem[Weibel(1981)]{ref:weibel81}
{\sc \au{Weibel, E.~S.}} \yr{1981}  \at{Quasi-linear theory without the random
  phase approximation}.  \jt{Phys. Fluids}  \bvol{24},  \pg{413}.

\bibitem[Whitham(1974)]{book:whitham}
{\sc \au{Whitham, G.~B.}} \yr{1974} {\em Linear and Nonlinear Waves\/}.
  \publ{New York: Wiley}.

\bibitem[Wong(2000)]{ref:wong00}
{\sc \au{Wong, H.~V.}} \yr{2000}  \at{Particle canonical variables and guiding
  center {Hamiltonian} up to second order in the {Larmor} radius}.  \jt{Phys.
  Plasmas}  \bvol{7},  \pg{73}.

\bibitem[Yasseen(1983)]{ref:yasseen83}
{\sc \au{Yasseen, F.}} \yr{1983}  \at{Quasilinear theory of inhomogeneous
  magnetized plasmas}.  \jt{Phys. Fluids}  \bvol{26},  \pg{468}.

\bibitem[Yasseen \& Vaclavik(1986)]{ref:yasseen86}
{\sc \au{Yasseen, F.} \& \au{Vaclavik, J.}} \yr{1986}  \at{Quasilinear theory
  of uniformly magnetized inhomogeneous plasmas: electromagnetic fluctuations}.
   \jt{Phys. Fluids}  \bvol{29},  \pg{450}.

\bibitem[Ye \& Kaufman(1992)]{ref:ye92b}
{\sc \au{Ye, H.} \& \au{Kaufman, A.~N.}} \yr{1992}  \at{Self-consistent theory
  for ion gyroresonance}.  \jt{Phys. Fluids B}  \bvol{4},  \pg{1735}.

\bibitem[Yoon {\em et~al.\/}(2016)Yoon, Ziebell, Kontar \&
  Schlickeiser]{ref:yoon16}
{\sc \au{Yoon, P.~H.}, \au{Ziebell, L.~F.}, \au{Kontar, E.~P.} \&
  \au{Schlickeiser, R.}} \yr{2016}  \at{Weak turbulence theory for collisional
  plasmas}.  \jt{Phys. Rev. E}  \bvol{93},  \pg{033203}.

\bibitem[Zakharov {\em et~al.\/}(1992)Zakharov, L'vov \&
  Falkovich]{book:zakharov-b}
{\sc \au{Zakharov, V.~E.}, \au{L'vov, V.~S.} \& \au{Falkovich, G.}} \yr{1992}
  {\em Kolmogorov Spectra of Turbulence I: Wave Turbulence\/}.  \publ{New York:
  Springer-Verlag}.

\bibitem[Zhu \& Dodin(2021)]{my:wkeadv}
{\sc \au{Zhu, H.} \& \au{Dodin, I.~Y.}} \yr{2021}  \at{Wave-kinetic approach to
  zonal-flow dynamics: recent advances}.  \jt{Phys. Plasmas}  \bvol{28},
  \pg{032303}.

\end{thebibliography}

\end{document}